\documentclass{aa}
\usepackage{aalongtable}
\usepackage{graphicx}
\usepackage{amssymb}
\usepackage{natbib}
\bibpunct{(}{)}{;}{a}{}{,}
\usepackage{supertabular}
\usepackage{rotating}
\usepackage{setspace}
\usepackage{lscape}

\def\omc{$\Omega/\Omega_{\rm{c}}$}

\begin{document}
%
\title{A slitless spectroscopic survey for H$\alpha$ emission-line 
objects in SMC clusters}

\titlerunning{H$\alpha$ emission-line objects in SMC clusters}

\author{
Christophe Martayan \inst{1,2}
\and  Dietrich Baade \inst{3}
\and  Juan Fabregat  \inst{4}
}
\offprints {C. Martayan}
\mail{Christophe.Martayan@eso.org}
\institute{European Organisation for Astronomical Research in the Southern 
Hemisphere, Alonso de Cordova 3107, Vitacura, Casilla 19001, Santiago 19, Chile 
\and GEPI, Observatoire de Paris, CNRS, Universit\'e Paris Diderot, 
5 place Jules Janssen, 92195 Meudon Cedex, France
\and European Organisation for Astronomical Research in the Southern 
Hemisphere, Karl-Schwarzschild-Str.\ 2, 85748 Garching b.\ M\"unchen,
Germany
\and Observatorio Astron\'omico de Valencia, edifici 
Instituts d'investigaci\'o, 
Poligon la Coma, 46980 Paterna Valencia, Spain }
\date{Received / Accepted}
\abstract
{}
{This paper checks on the roles of metallicity and evolutionary age in the appearance of the so-called Be phenomenon.}  
{Slitless CCD spectra were obtained covering the bulk of the Small Magellanic Cloud.  
For H$\alpha$ line emission twice as
strong as the ambient continuum, the survey is complete to spectral type B2/B3 on the main sequence.  About 8,120 spectra
of 4,437 stars were searched for emission lines in 84 open clusters. 370 emission-line stars were found, 
among them at least 231 near the main sequence.  For 176 of them, photometry could be
found in the OGLE database.  For comparison with a higher-metallicity environment, the 
Galactic sample of the photometric H$\alpha$ survey by McSwain \& Gies (2005) was used.} 
{Among early spectral sub-types, Be stars are more frequent by a factor $\sim$3-5 in the SMC than in the Galaxy.  The
distribution with spectral type is similar in both galaxies, i.e. not strongly dependent on metallicity. The fraction of
Be stars does not seem to vary with local star density.  The Be phenomenon mainly sets in towards the end of the
main-sequence evolution (this trend \textit{may} be more pronounced in the SMC); but some Be stars already form with
Be-star characteristics.}  
{In all probability, the fractional critical angular rotation rate, \omc, is one of the main parameters 
governing the occurrence of the Be phenomenon.  If the Be character is only acquired during the course of evolution, 
the key circumstance is the evolution of \omc, which not only is dependent on metallicity but differently so for different 
mass ranges.}

\maketitle

\section{Introduction}
The so-called Be phenomenon manifests itself through the single,
intermittent, or permanent occurrence of emission lines in
main-sequence stars with spectral types O through A.  This line
emission arises from rotationally supported disks formed from matter
lost (often ejected) by the central star.  An excellent introduction
to the complex idiosyncrasies of Be stars is given by
\citet{porter03}.  While extremely rapid rotation is a necessary
condition for the Be phenomenon, it is not clear whether it is also
sufficient or which other conditions need to be fulfilled.  There are
reports suggesting that the Be phenomenon also depends on metallicity
\citep{maeder99} and evolutionary phase
\citep[e.g.,][]{fabregat2000,marta2007a}.  An obvious method to
investigate the latter possibility is the study of open star clusters 
with a suitable range of ages.
In the Galaxy, it has been applied various times but mostly using
photometric techniques.  A very good discussion of the subject and
previous work can be found in \citet{mcs2005}.

In the Small Magellanic Cloud (SMC), a major spectroscopic survey for
emission-line objects was performed by \citet{ma1993}.  Although it
identified 1898 emission-line objects, the photographic nature of the
data did not permit too many emission-line stars to be found near the
main sequence or in crowded areas such as open clusters.  This paper
reports on the first digital slitless-spectroscopy survey of the SMC
in search for Oe/Be/Ae stars (a second paper will concern the LMC).
During its execution, some 3 million spectra were obtained.

\section{Observations}

The observations\footnote{Observations at the European Southern Observatory, 
Chile under project number 069.D-0275(A)} were carried out (by JF) on September 25 and 26, 2002
with the Wide Field Imager (WFI) attached to the 2.2m MPG Telescope at
ESO's observatory on Cerro La Silla in Chile \citep[see][]{WFI}.  
Due to bad weather, the second night was only partly useful.

In its slitless spectroscopic mode the WFI has a crudely circular
field of view of diameter 0.31 degrees. The R50 grism yields a
dispersion of 54 nm/mm or 0.811 nm/pixel.  The nominal resolution at
optimal focus and seeing is 5.1 nm at H$\alpha$.  In order to reduce
crowding and overlapping spectra, the length of the spectra was
limited by means of a filter with a full width at half maximum of 7.4
nm centered on H$\alpha$.  This means that spectra are acceptably
separated if the stars are 2\arcsec~apart perpendicular to the
dispersion direction and 6\arcsec~apart along this direction.  These
numbers vary slightly with the quality of the focus.  But in general
they do not require an adaptation of the parameters for the automatic
extraction by software of the spectra.

\begin{figure*}[h!t]
\centering
\resizebox{\hsize}{!}{\includegraphics[height=12cm, angle=-90]{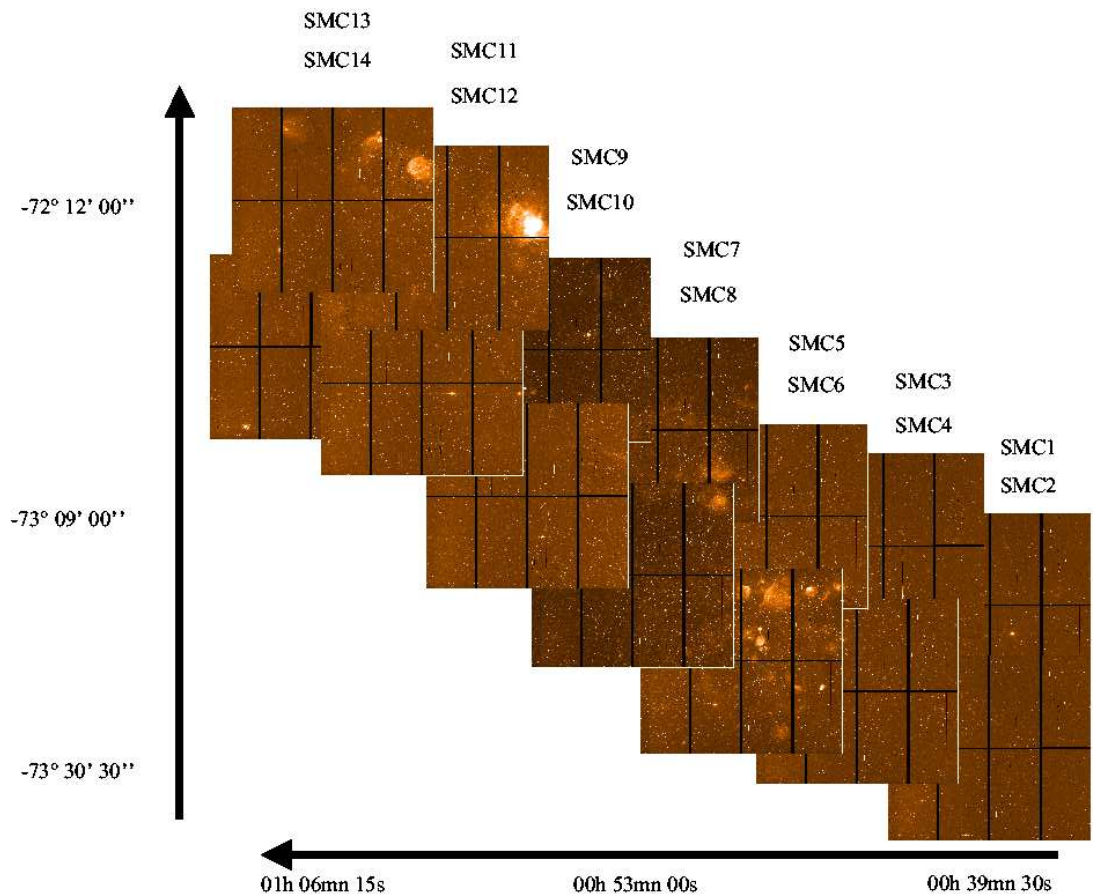}}
\caption{The tiling by WFI frames of the Small Magellanic Cloud.}
\label{fig1}       
\end{figure*}

The exposure time per field was set to 600 s.  The coverage of the SMC
achieved with 14 telescope pointings is shown in Fig.~ \ref{fig1}.

For a similar study with the WFI but in the Galaxy and with a broader
filter, see \citet{marta2008a}.

\section{Data reduction, extraction of spectra, and identification of 
emission-line stars}

\begin{figure*}[!ht]
\begin{tabular}{cc}
    \centering \includegraphics[angle=0, width=8cm]{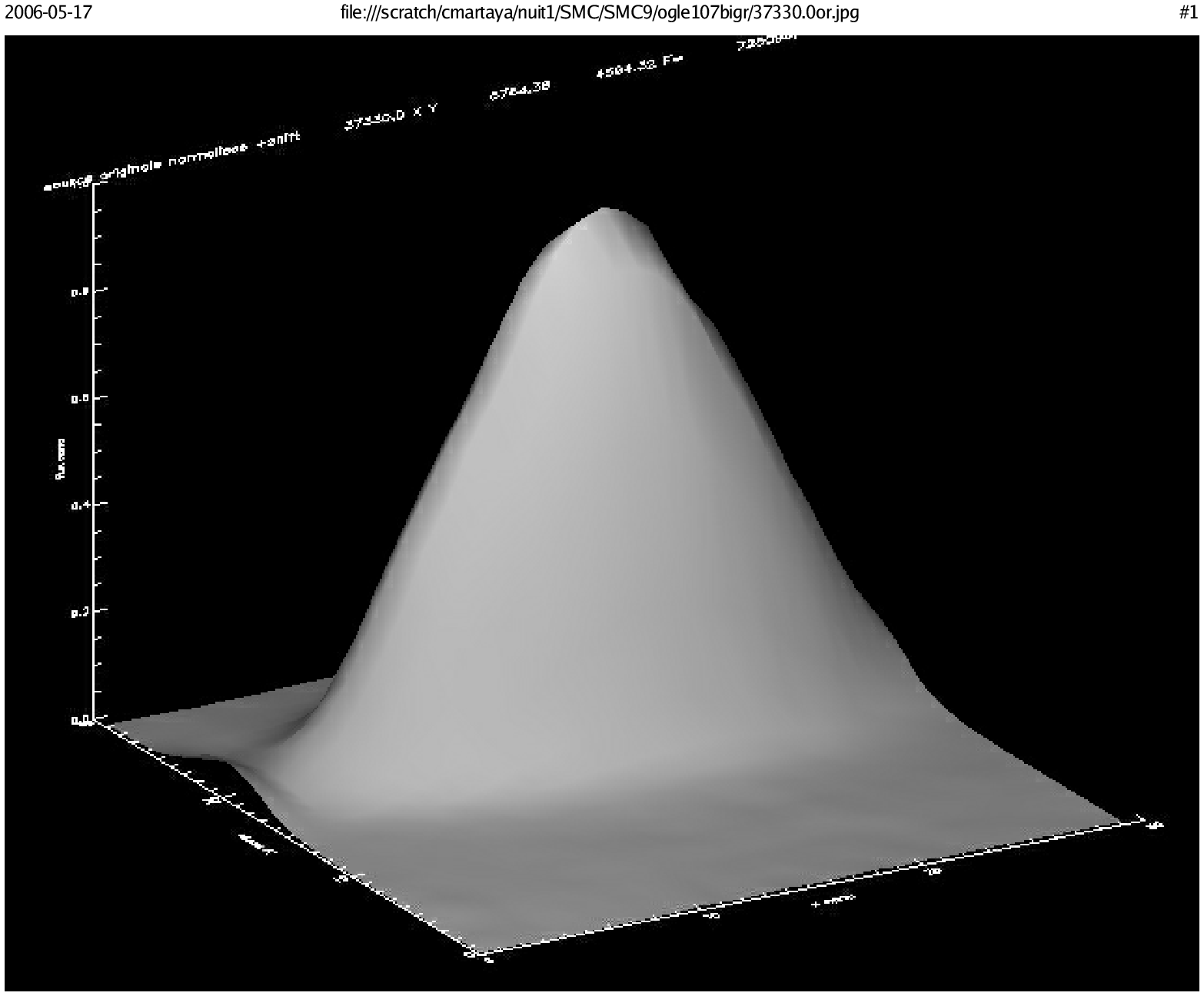} &
    \includegraphics[angle=0, width=8cm]{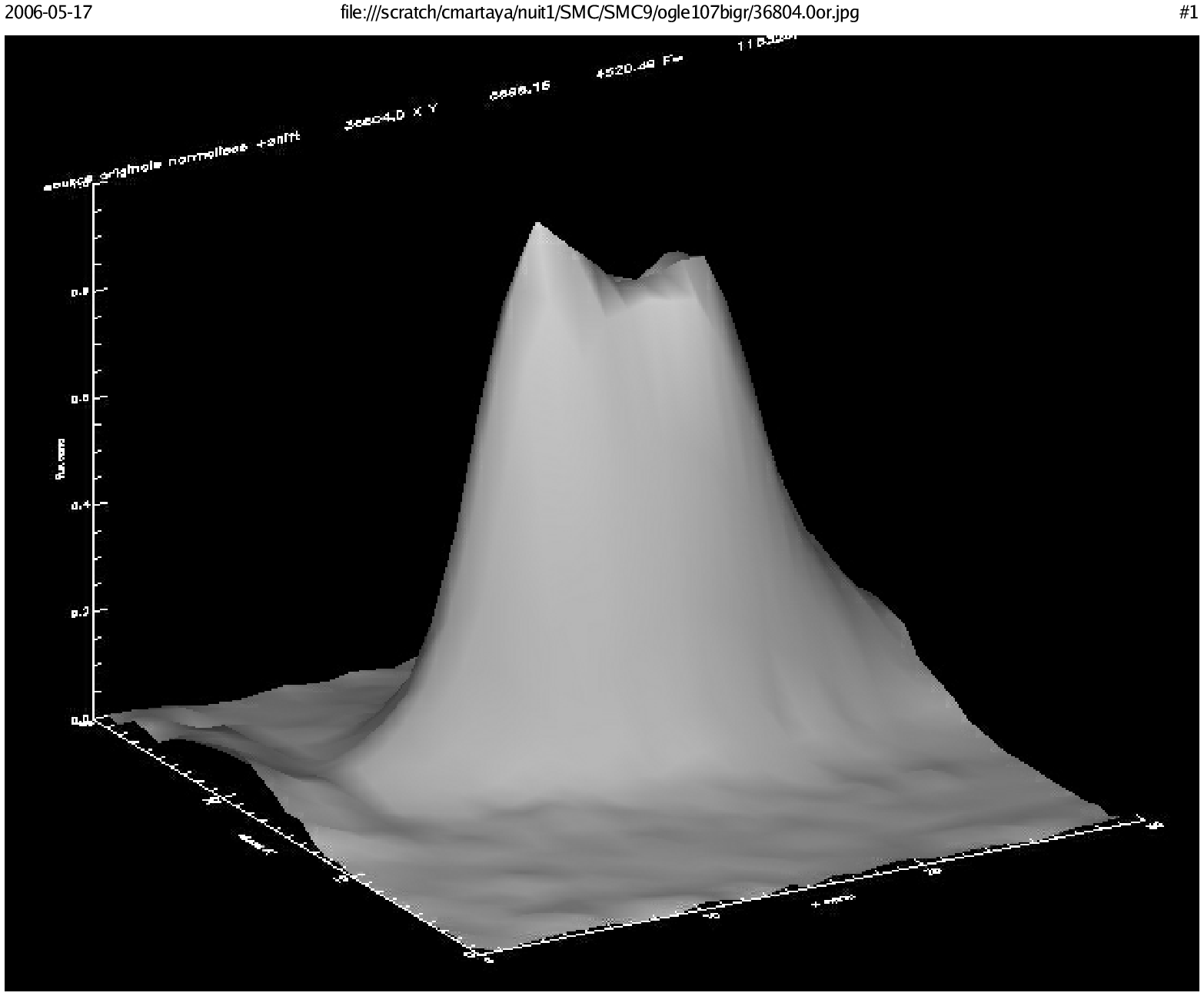} \\
    \includegraphics[angle=0, width=8cm]{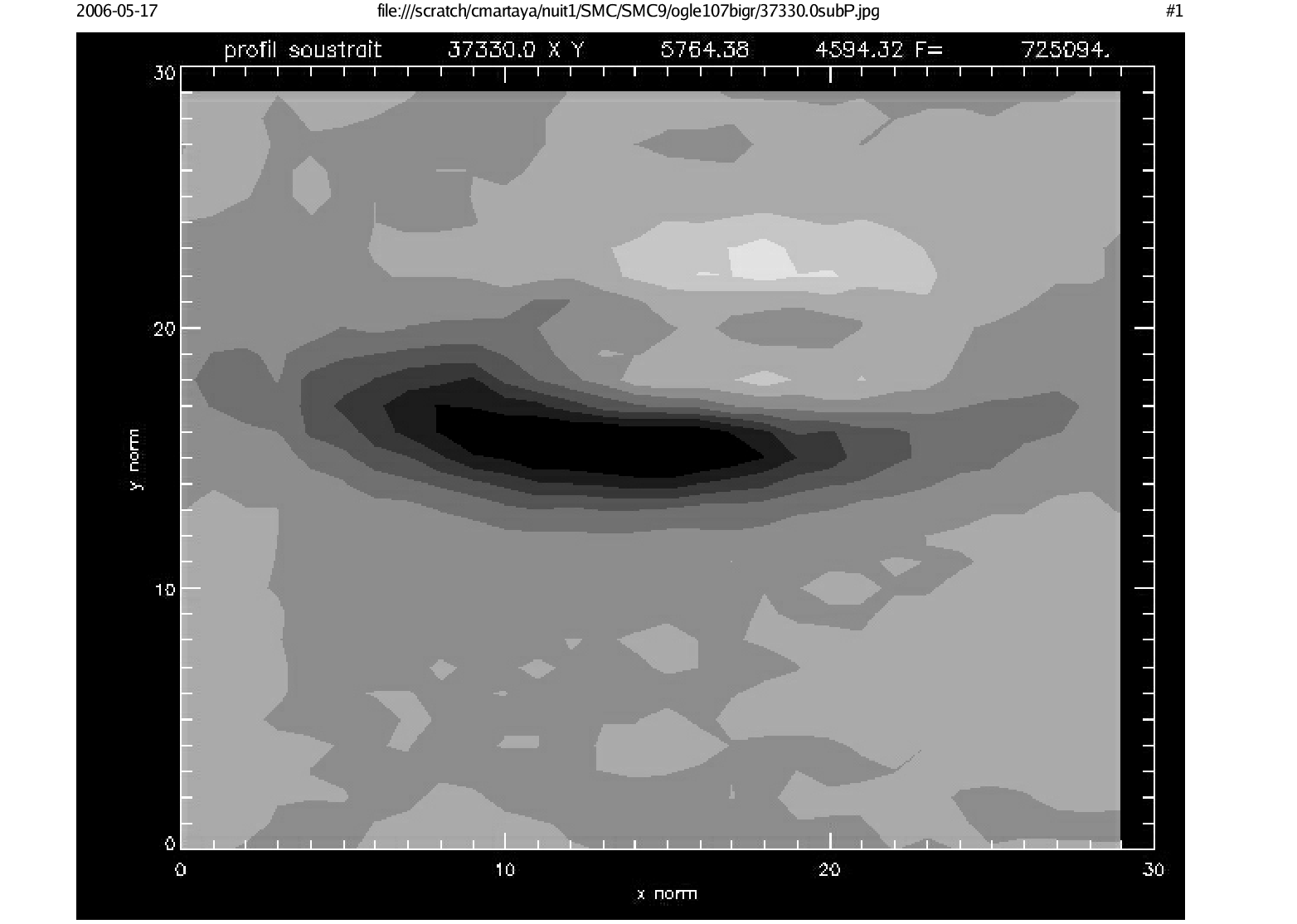} &
    \includegraphics[angle=0, width=8cm]{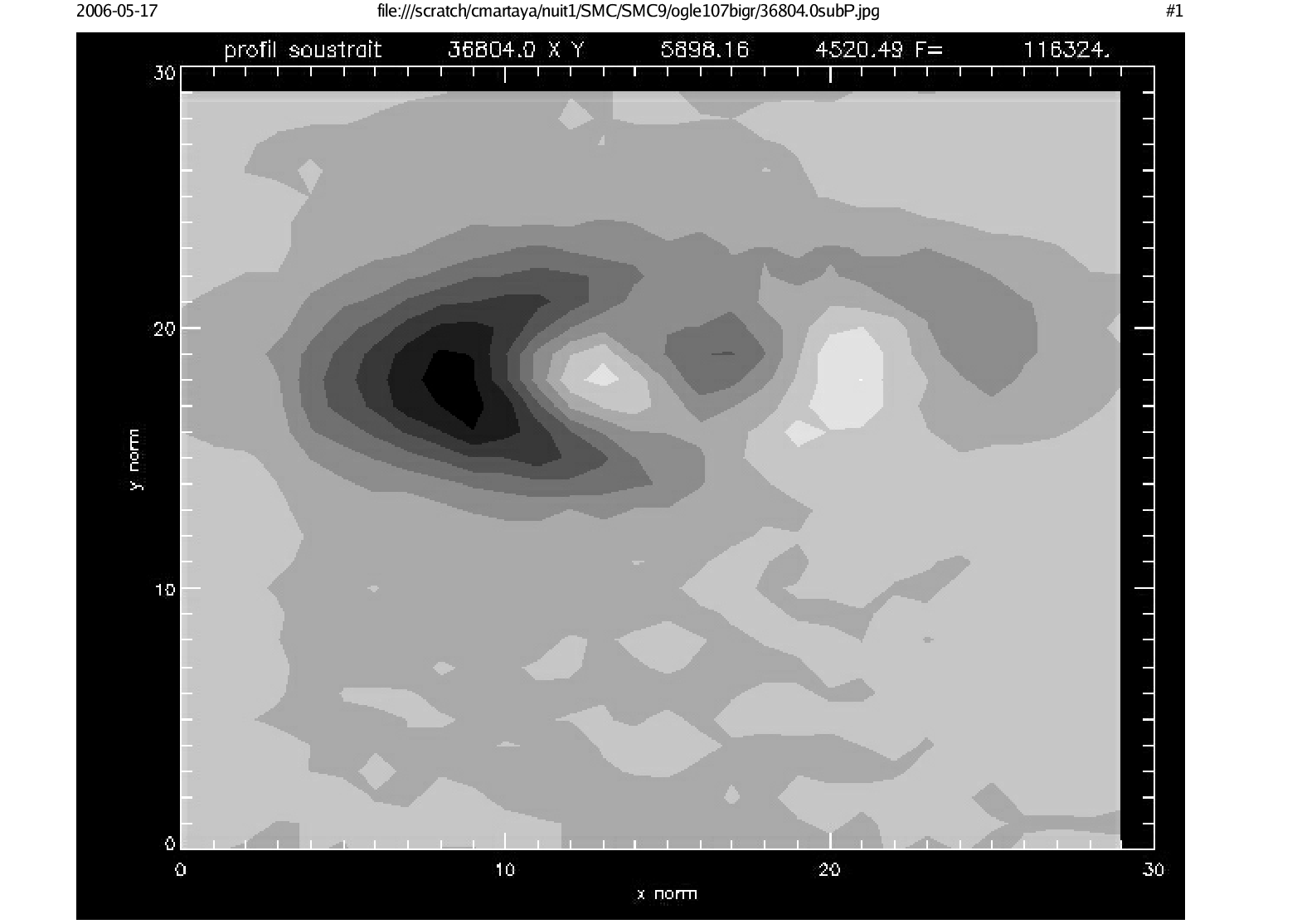} \\
\end{tabular}
\caption{The appearance of objects with and without H$\alpha$ line emission 
in defocused slitless WFI spectra.  Left: Non-emission-line star,
right: emission-line star.  Top: 3-D presentations of the original
flux distributions, bottom: 2-D projections of the residuals after
subtraction of the mean, scaled point-spread function (PSF) of pure continuum sources.  While a
pure continuum source is just blurred by the defocus, a (nearly
unresolved) emission line is effectively imaged like a point source,
yielding a roughly donut (or horseshoe) like image of the pupil.
The bright excess above the mean of the image corresponding to the emission peak is visible
in the middle-left of the horseshoe (bottom right figure).}
\label{albumfigs}
\end{figure*} 

\subsection{Data reduction}
\label{reduction}

The elementary CCD image processing was performed with
IRAF\footnote{IRAF is distributed by the US National Optical Astronomy
Observatory, which is operated by the Association of Universities
for Research in Astronomy (AURA), Inc., under cooperative agreement
with the US National Science Foundation} tasks and the MSCRED package.

The subsequent extraction of the bias- and flat field-corrected
spectra posed considerable technical challenges:
\begin{itemize}
\item 
All images suffer from substantial and non-homogeneous defocus, which
increases from the field center to the edges.  The reasons could not
be reconstructed.  (However, special techniques permit continuum and
emission-line objects to be distinguished at a fairly acceptable level
of confidence, turning the problem almost into an advantage - see
Fig.\ \ref{albumfigs}.)  
\item 
Each object appears in several spectral orders.  The 0th order is
undispersed and corresponds to the point-spread function (PSF).  The
order suitable for the extraction of the H$\alpha$ spectra is -1,
which in the following will be referred to as ``the spectrum''.  In
the case of bright stars, up to 7 or even 8 orders are visible,
significantly increasing the probability of spectra being
contaminated.  While this implies the possible non-detection of some
emission-line objects, it will not lead to false detections because
even the 0th order has a very different PSF.
\item 
There are parasitic spectra resulting from scattered light from stars
outside the direct field of view.
\item 
The position angle of the dispersion direction varies (by $\le$
$\pm$5$\degr$) from the center to the edges of the images.
\end{itemize} 

In principle, spurious detections of emission-line objects may arise
from particle events or hot pixels.  However, their very different
PSFs make such confusions rather unlikely. The extraction algorithm
used tries to automatically reject particle hits.  On the other hand, the
completeness of the survey is reduced by spectra falling onto the gaps
(amounting to 3.1\ \% in area) between the 4x2 CCDs of the WFI.

After careful, realistic tests with representative subsets of the data
it was decided that it is not just most time effective but also still
safe to let properly tuned software search for spectra automatically.
For this task and the 2-D extraction of the spectra SExtractor
\citep{bertin1996} was employed with specially adapted convolution
masks (E.\ Bertin, private communication).  About 3 million spectra
were extracted in the whole SMC (14 images, see Fig.~\ref{fig1}), of
which about 1 million are useful.

A comparison of such extracted spectra with spectra counted by eye
suggests that the extraction efficiency in clusters is on average
around 75 to 80\%, depending on the area density of the stars.  With
position of the cluster in the WFI field, i.e.\ mainly the level of
de-focus, this mean value may range from 60\% to 100\%.

Because of the need to work on samples with defined ages, only the
areas of 83 clusters listed in the OGLE database \citep{oglemapsphoto}
and a number of neighboring comparison fields (14 fields, 1 or 2 per
target field and with diameters of 2 to 4\arcmin, located close to the
open clusters treated) were retained for the classification as stars
with and without line emission.  The open clusters selected in this
way \citep[from][]{ageSMcocl} are compiled in Table~\ref{indcl}; other
parameters (total number of stars, number of emission-line stars,
etc.) are included (where applicable, separately for multiple
observations of a given cluster). This process led to a sample of
7,867 spectra to be examined for the presence of line emission.

\begin{figure}[!h]
\centering
\resizebox{\hsize}{!}{\includegraphics[angle=-90]{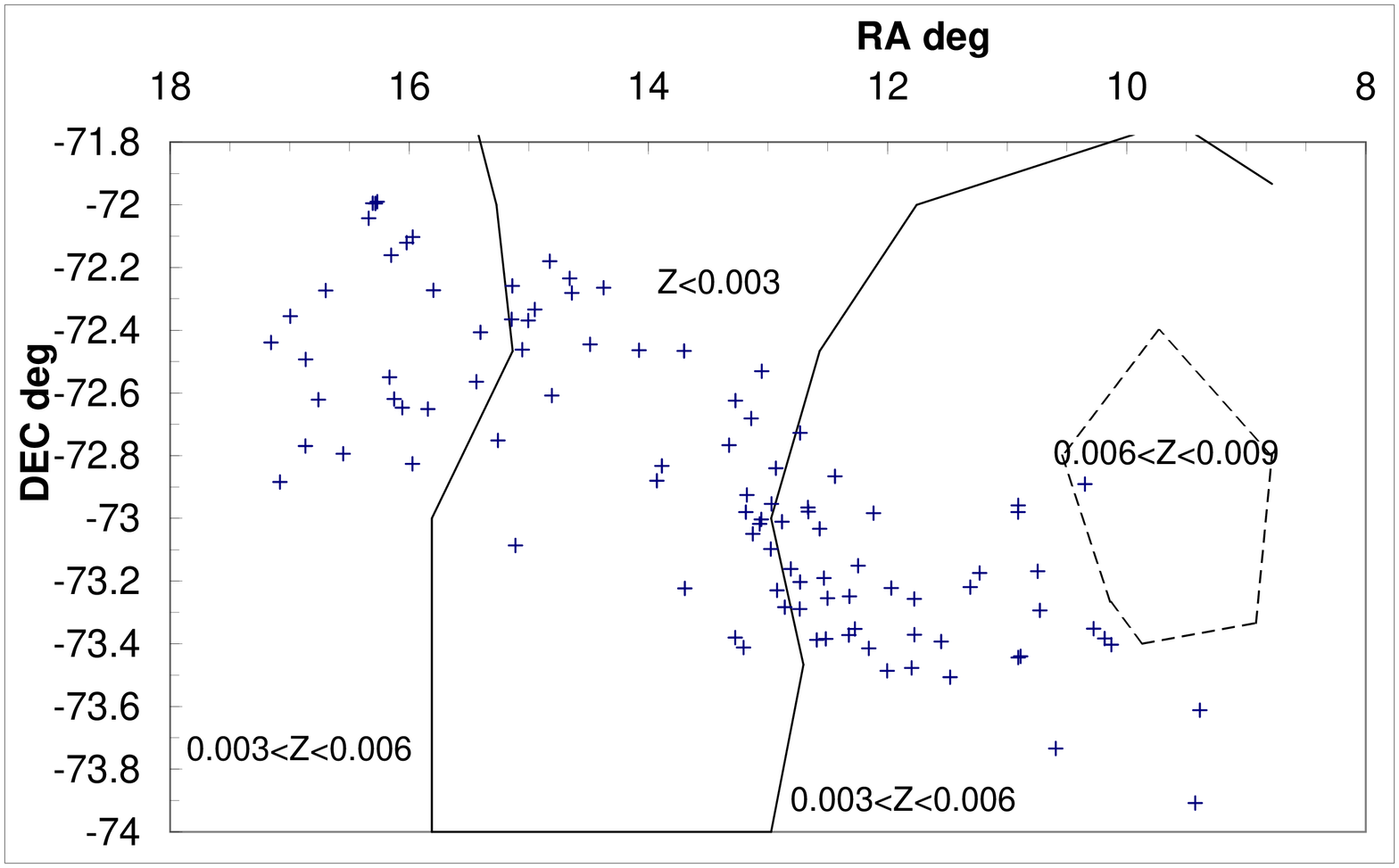}}
\caption{The positions of the WFI SMC clusters superimposed on spatial 
iso-metallicity curves \citep[from][]{cioni06}.}
\label{Zmap}
\end{figure}

The WFI database includes an 84th open cluster: NGC\,346, which is a
complex young structure, probably consisting of a number of
sub-aggregates. But OGLE photometry is not available for this dense
field, which also suffers from extended nebular emission.  Details are
in Sect.\ \ref{comocl}.
Fig.\ \ref{Zmap} shows the distribution of studied open clusters in the SMC 
as well as the metallicity areas by \citet{cioni06}. 
They confirm that the open clusters are in metallicity environment significantly 
lower than the Galactic one.


\subsection{Identification of emission-line stars: the {\rm Album} code}
\label{identifELS}

The exploratory tests alluded to above also showed that it would not
be safe to let software distinguish without human supervision between
stars with and without H$\alpha$ line emission.  However, the visual
classification of several thousand spectra could be greatly
facilitated by transforming them into an easy-to-classify format.  To
this effect, the {\it Album} package was written (in {\it IDL}).

{\it Album} starts out from the assumption that the 2-D PSF is only
slowly varying with position in the frame.  To compute
the latter, typically 50-250 spectra were registered (by cross
correlation), co-added, and normalized.  This step is
operator-supervised; ill-suited stars can be rejected.  
In the first step, all obvious emission-line stars, 
apparent binaries, too closely spaced sources, spectra with severe particle  
hits or otherwise reduced quality are rejected and a new regional 
mean spectrum is computed.  It was empiricially established that the 
inclusion, at the $\leq$5\% level, of emission-line objects  
only insignificantly modifies the regional mean spectrum profile.
The resulting
regional template spectrum was subtracted (after cross correlation and
shift) from each normalized 2-D spectrum (see
Fig.~\ref{albumfigs}-left) to be checked for H$\alpha$ line emission.
{\it Album} also rejects automatically artifacts such as ghosts or
particle events by applying a suite of tests to the shapes of the
spectra.

In the case of emission-line stars, the 2-D spectra show a secondary
peak (see Fig.~\ref{albumfigs}).  But after subtraction of the mean
PSF the resulting difference images display a more characteristic and
conspicuous ring-like structure.  This is due to the large defocus,
which affects the at most marginally resolved line emission like a
point source.  Therefore, while the continuum flux is just blurred by
the defocus, the line emission takes on roughly the shape of a donut
or horseshoe (i.e. the telescope pupil).

In a properly prepared and homogeneous album (hence the name {\it
Album}) of images, this peculiar structure is conveniently and more
readily and reliably recognized by the human eye than by software
developed with the same amount of effort.

Using this scheme, all stars were classified into three categories: 
definite emission-line stars, candidate emission-line stars, and stars without 
H$\alpha$ line emission.  An emission-line star is considered definite, if its 
flux distribution shows a significant secondary peak (cf.\ Fig.~\ref{albumfigs}) 
at a position consistent with H$\alpha$.  Obviously, this depends on the 
signal-to-noise ratio but also on the location within the frame and the associated 
defocus.  If the purity of this signature is potentially diluted by particle 
events or noise spikes, the object is called a candidate emission-line star. 
For example, for relatively bright objects from V=14 to 17, 100 to 80 \% of the 
emission-line stars found are classified as definite emission-line stars. 
However, towards lower brightness and lower signal-to-noise ratio, 
the fraction of definite emission-line stars drops from 55 to 13 \%.

\subsection{Efficiency of H$\alpha$ emission detection}
\label{deteclim}

\begin{figure}[h!]
\centering
\resizebox{\hsize}{!}{\includegraphics[angle=-90]{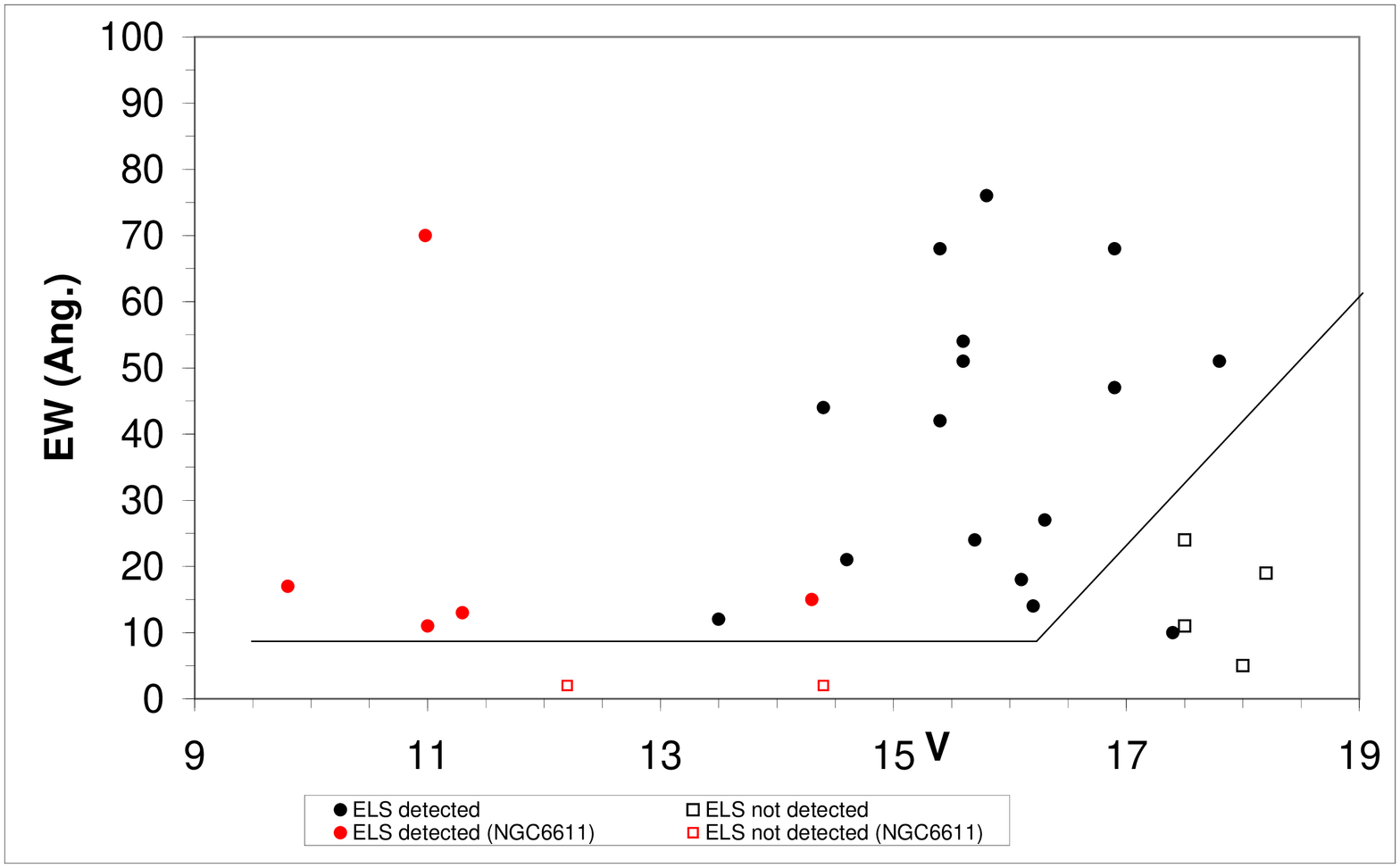}} 
\resizebox{\hsize}{!}{\includegraphics[angle=-90]{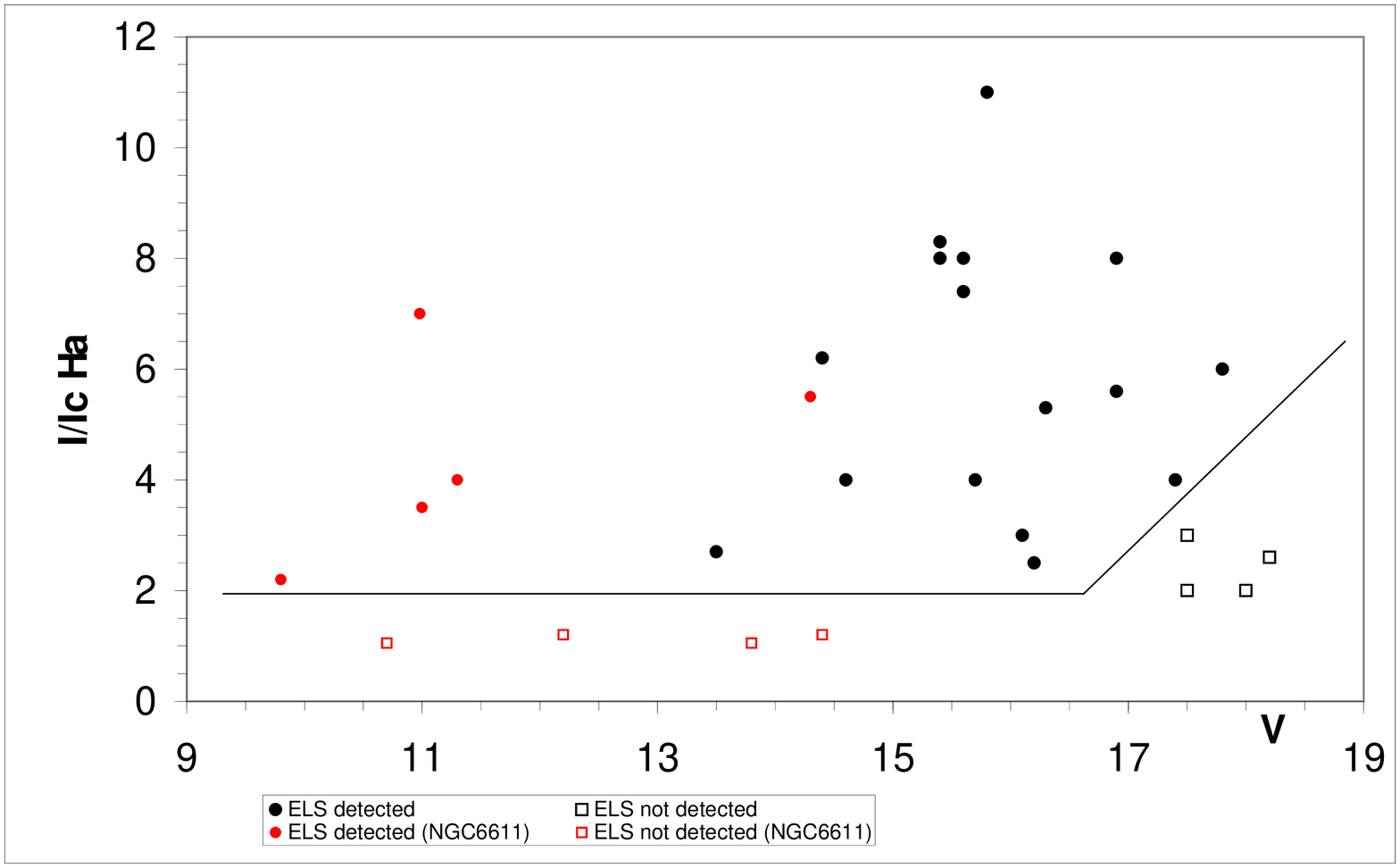}}
\caption{Detection efficiency in equivalent width (top) and peak line 
strength relative to the continuum (bottom) of H$\alpha$
emission with WFI slitless spectra.  The H$\alpha$ data are from
\citet{hummel99,hummel01} and \citet{marta2007b,marta2008a} for NGC\,330 (black)
and NGC\,6611 (red).  Emission-line stars detected in the WFI data are
represented by filled circles, open squares denote the misses. 
Note that, contrary to the conventional definition, positive values 
indicate a net line emission.}
\label{effdetect}
\end{figure}

In order to determine the thresholds for the detection of H$\alpha$
emission, {\it Album} was applied to WFI observations of the open
cluster NGC\,330 and its well-studied population of Be stars.  With
the help of {\it SIMBAD} the previously known Be stars and the ones
found by {\it Album} were compared, and the H$\alpha$ equivalent
widths and line strengths were taken from
\citet{hummel99,hummel01,marta2007b}, similar to the WFI
observations of NGC\,6611 by \citet{marta2008a}.  The results are
shown in Figure~\ref{effdetect} for the detection efficiency in terms
of H$\alpha$ line emission equivalent width (top) and strength
(bottom).  The {\it Album}-based procedure found slightly less than
80\% of the Be stars known in NGC\,330.

Fig.~\ref{effdetect} suggests that our SMC survey technique detects
H$\alpha$ line emission, when the equivalent width is higher than 10
\AA~or the peak intensity is more than twice the one of 
the underlying continuum, down to V$\sim$ 16.5-17 mag, which
corresponds to spectral types later than B2-B3
\citep[see][]{marta2007a}.  
Note that, contrary to the conventional definition, positive values 
indicate a net line emission.
For fainter stars, the signal-to-noise 
ratio is lower and the detection thresholds increase.

A further check is provided by the following: \citet{marta2007a}
observed 31 WFI stars preselected from the emission-line stars found with
the WFI in the field of NGC\,330.  At a spectral resolving power of
8,600 with the VLT, all of the 31 emission-line stars were confirmed
as true H$\alpha$ emission-line objects, 28 of which could be
classified as Be stars.  The remaining 3 turned out to be of a
different nature (compact planetary nebulae, supergiants, B[e]
or Herbig B[e] stars).

The case of NGC\,330 is also useful to guess the false-alarm
probablity, which evidently is low.  For a more precise constraint,
one would need to have a control sample with B/Be classifications that
are true at the time of the WFI observations.  This does not exist, and 
some instrument-independent uncertainty is introduced by the
comparison of observations made in different years because the
Oe/Be/Ae characteristics of a star are (sometimes: highly) 
time dependent and not seldom only intermittently present.

\section{Astrometry, photometry, and spectral classification}
For the detected emission-line stars to be put into an evolutionary
context, intrinsic colours and magnitudes are needed.  In the following
subsections, they are derived separately for the WFI emission-line
stars in the SMC and for a comparison data set in the Galaxy.

\subsection{WFI SMC data}
As the largest homogeneous source of photometry in SMC clusters we
chose OGLE \citep{oglemapsphoto}.  The first step for the
cross-identification is astrometry.

\subsubsection{Astrometry and correlation with OGLE}

The ASTROM package \citep{wallace2003} was applied to the extracted
spectra of the -1st order.  Per WFI image, 30 to 80 astrometric reference
stars from the GSC2.2 and/or UCAC2, USNO catalogues were utilized.  In
this way, the coordinates of the 7,867 SMC stars mentioned in
Sect.~\ref{reduction} (plus 55 in NGC\,346) were determined with an
accuracy of 0.5-1$\arcsec$.  Cross-correlating these WFI positions
with the OGLE catalogues \citet{oglemapsphoto} revealed a systematic
global offset of $-0.3\arcsec$ in right ascension and $+0.3\arcsec$ in
declination. In order to maximize the probability of identifying the
WFI emission-line stars in OGLE, these shifts were applied for the
extraction of the photometric data from OGLE (note that the WFI
coordinates provided in this paper do not include these offsets).

For each cluster, any multiple observations and identifications were
merged into one per star.  Table~\ref{mergcl} provides a summary of
the results for each cluster.  On average, 73.7\% of all WFI stars and
79.7\% of the WFI emission-line stars were found in OGLE. The
incompleteness is explained mostly on the part of OGLE, which is
undercomplete in crowded areas (i.e., clusters) and in the presence of
extended nebulosities.  But imperfect WFI coordinates also add their
share.  Note also that only about 80\% of the $\sim$ 3 square degrees 
of this WFI survey are covered by OGLE\,II.  This especially affects 
large complexes like the one of NGC\,346.

\subsubsection{Photometric spectral classification}
\label{SMCspecclass}
The resulting apparent colours and magnitudes need to be converted to
absolute ones, and absolute luminosities and spectral types need to be
derived so that inter-cluster comparisons within the SMC but also
between SMC and Galaxy become possible.  V$_{0}$, (B-V)$_{0}$,
(V-I)$_{0}$ were derived by means of the per-cluster E$_{B-V}$
reddenings from \citet{ageSMcocl}.  The absolute $M_{V}$ of each star
was calculated from the resulting V$_{0}$ and the SMC distance modulus
provided by \citet{distSMC}.  That is, for all clusters the same
effective distance was adopted.

Using the HR diagramme in Fig.\ \ref{globalHRSMC}, the following
regions were assumed to delineate the main sequence:
\newline
O stars: $M_{V}<-4.2$ and $-0.3\le$(B-V)$_{0}$$\le$+0.1 and
$-0.3\le$(V-I)$_{0}$$\le$+0.3
\newline
B stars: $-4.2\le$M$_{V}$$\le+0.43$ and $-0.4\le$(B-V)$_{0}$$\le$+0.1
and $-0.35\le$(V-I)$_{0}$$\le$+0.2
\newline
A stars: 0.43$<M_{V}\le2.55$ and $-0.4\le$(B-V)$_{0}$$\le$+0.25 and
$-0.38\le$(V-I)$_{0}$$\le$+0.25. 
\newline
Individual spectral types were assigned by applying the calibration of
\citet{lang1992} and \citet[][ and references therein]{wis2006} as 
shown in Table~\ref{caliblang} and Fig.\
\ref{globalHRSMC}.  The break-down of the full sample by spectral
types and emission-line characteristics is given in
Table~\ref{starssamples}.

Apart from the global uncertainties of spectral types derived from
photometry, differential reddening across a cluster and erroneous
membership assignments will introduce individual errors.  However,
they will only dilute but not probably falsify general trends derived
from the database at large.

Sect.\ \ref{infosngc346} compares for 12 stars in NGC\,346 spectral
types derived as described above and spectroscopic classifcations from
the literature.  The average difference only amounts to 1 spectral
sub-type.

\begin{table}[h]
\centering
\caption[]{Adopted ranges in absolute $V$ magnitude per spectral 
sub-type, following the calibration of \citet{lang1992} and \citet[][ and
references therein]{wis2006} for main-sequence stars.}
\centering
\begin{tabular}{llll}
\hline
\hline	
ST  & Mv Range & ST  & Mv Range \\
\hline	
Hot O &  $<$-6.0     &  B6  &  [-0.6; -0.25[ \\
O3  &  [-6.0; -5.9[  &  B7  &  [-0.25; 0.025[ \\
O4  &  [-5.9; -5.7[  &  B8  &  [0.025; 0.285[\\
O5  &  [-5.7; -5.5[  &  B9  &  [0.285; 0.43] \\
O6  &  [-5.5; -5.2[  &  A0  &  ]0.43; 1.0[\\
O7  &  [-5.2; -4.9[  &  A1  &  [1.0; 1.3[ \\
O8  &  [-4.9; -4.5[  &  A2  &  [1.3; 1.5[\\
O9  &  [-4.5; -4.2[  &  A3-A4  &  [1.5; 1.95[\\
B0  &  [-4.2; -3.25[  &  A5-A6  &  [1.95; 2.2[\\
B1  &  [-3.25; -2.55[ &  A7  &  [2.2; 2.4[\\
B2  &  [-2.55; -1.8[ &  A8-A9  &  [2.4; 2.55[\\
B3  &  [-1.8; -1.4[  &  F0-F1  &  [2.55; 3.6[\\
B4  &  [-1.4; -0.95[  &  cool F  &  $\ge$3.6 \\
B5  &  [-0.95; -0.6[  &  \\
\hline
\end{tabular}
\label{caliblang}
\end{table}
%

The results of the photometric and spectral classification 
and much additional information are compiled in a number of tables.
For basic data of Be stars and their absolute photometric data and
spectral types, see Tables \ref{tableBe1} and \ref{tableBe2},
respectively.  For mere candidate-Be stars, Tables \ref{tableBecand1}
and \ref{tableBecand2} are the equivalents.  Tables \ref{tableAeOe1}
and \ref{tableAeOe2} concern Oe and Ae stars.  Emission-line objects
well outside the main sequence are covered in Tables~\ref{table1otherels}
and~\ref{table2otherels}.  Table~\ref{primels} compiles all WFI-based
data (coordinates, etc.) for emission-line stars without counter part
in the OGLE catalogues \citep{oglemapsphoto}.  Where applicable, the
tables also contain information extracted from SIMBAD within a search
radius of 2\ \arcsec about each emission-line star.  Because of the
high density of objects especially in the cluster cores, this
information will inevitably suffer from misidentifications.

Similar tables for the 3,792 SMC non-emission line stars are available
on request.

\subsection{Comparison data for the Galaxy}
\label{intheGalaxy}
The most recent and comprehensive photometric survey for Be stars in
Galactic clusters is the one by \citet{mcs2005}.  It comprises 52
definite Be stars and 116 Be candidates in 48 of the 54 clusters
studied.  Using data from \citet{mcs2008}, we estimate that the
detection limit for H$\alpha$ line emission in that sample is ~7 \AA,
similar to the one of our spectroscopic study in the SMC.

\citet{mcs2005} published the Stroemgren parameters y, (b-y), E(b-y) 
but not m$_{1}$.  For a comparison with the SMC sample, the conversion
to the Johnson UBV system was performed as follows:
\begin{list}{--}{\itemsep=0cm\topsep=0cm}
\item First, (B-V) was derived from the relation \citet{warren77}:
$\mbox{(B-V)=1.668$\times$(b-y)-0.030}$ \\
\item Second, the $y$ magnitudes given by \citet{mcs2005}
in the system defined by \citet{cousins87} transform to standard V
magnitudes by means of the relation: $\mbox{V=y+0.038$\times$(B-V)}$
\citep{cousins85}.\\
\item Third, E[B-V] results from $\mbox{E[B-V]=E[b-y]/0.745}$, 
where E[b-y] is taken from \citet{mcs2005}.\\
\item Fourth, V$_{0}$ follows from $\mbox{V$_{0}$=V-3.1$\times$E[B-V]}$.\\
\item Fifth, absolute magnitudes, M$_{V}$, were calculated from 
$\mbox{M$_{V}$=V$_{0}$-$\mu$}$, using the distance moduli, $\mu$,
given by \citet{mcs2005} for each cluster.\\
\item Sixth, $\mbox{(B-V)$_{0}$=(B-V)-E[B-V]}$.\\
\end{list}
From this point on, spectral types of main-sequence stars were derived
in the same way as for the WFI SMC sample (Sect.\ \ref{SMCspecclass}).

\begin{table}[h]
\centering
\caption[]{Breakdown of the WFI SMC sample (col.\ 2) and the Galactic sample  
from \citet{mcs2005} (col.\ 3) by numbers of open clusters, spectral
types, and emission-line characteristics. ELS denotes emission-line stars.
The first line after the titles gives the number of open clusters used.}
\centering
\begin{tabular}{lll}
\hline
\hline	
Type   & Number & Number \\
       & (SMC) & (Galaxy) \\
\hline	
Open clusters & 84 & 54\\ 
\hline	
O & 25 & 13\\ 
Oe+Oe? & 6 & 3\\ 
B & 1384 & 1741\\ 
Be & 109 & 52\\ 
Be? & 54 & 116 \\ 
A & 250 & 495\\ 
Ae+Ae? & 7 & 57\\ 
Other non-ELS & 2408 & 17845\\ 
ELS outside main sequence & 90 & \\ 
Unclassified ELS & 49 & \\ 
NGC\,346 (Be, HBe, WR, etc) & 55 & \\ 
Total & 4437 & 20322 \\
\hline
\end{tabular}
\label{starssamples}
\end{table}
%

\section{Results and discussion}

\begin{figure*}[ht!]
\begin{tabular}{cc}
    \centering
    \includegraphics[angle=-90, width=8cm]{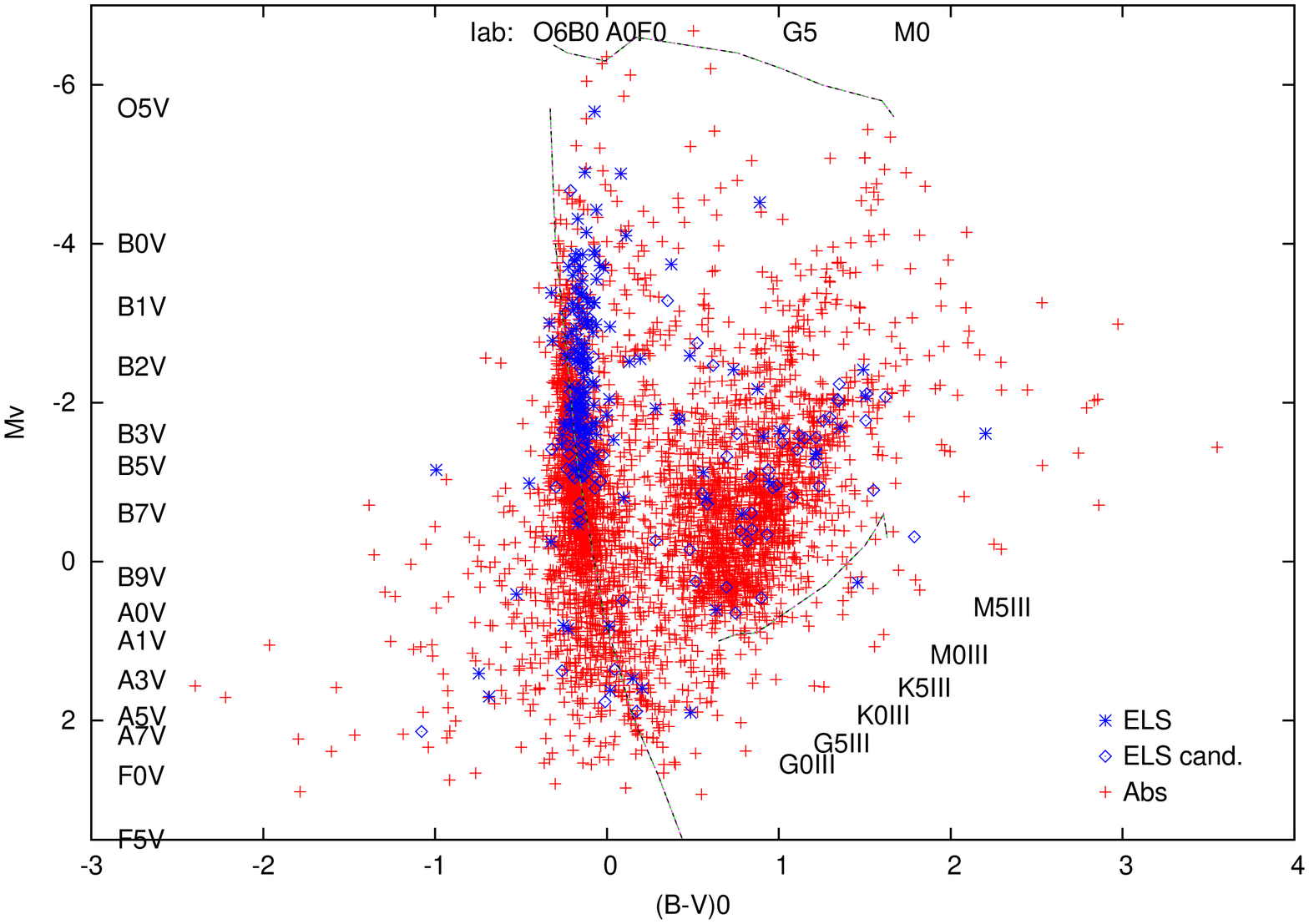} &  \includegraphics[angle=-90, width=8cm]{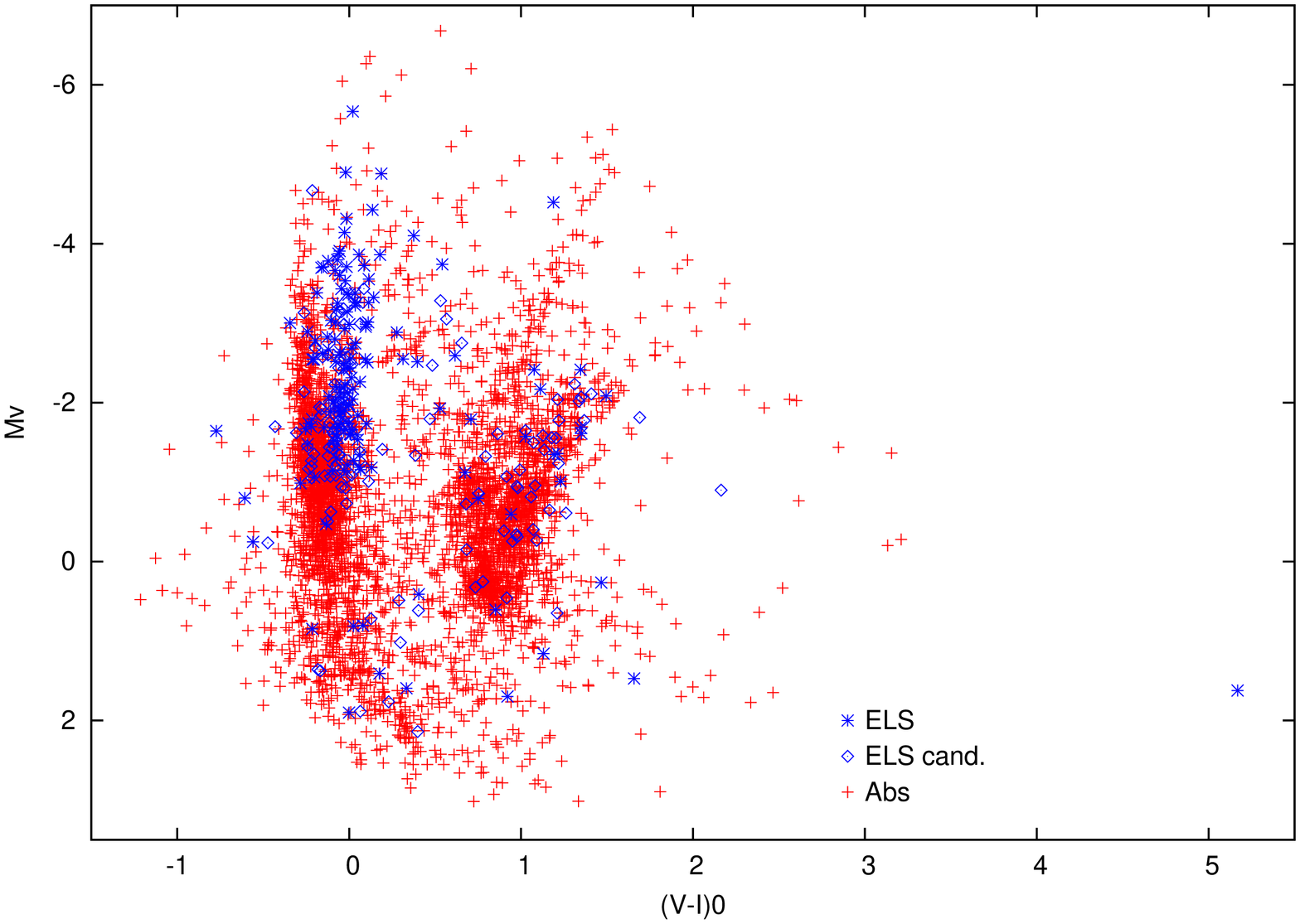} \\
\end{tabular}
\caption{Global colour-magnitude diagrams for stars in SMC open clusters.
Left: B--V; right: V--I, absolute $V$ magnitude vs. dereddened colour.  
Red crosses (+) indicate non-emission line stars, blue asterisks (*) 
mark emission-line stars (``ELS''), and candidate emission-line stars 
(``ELS cand.'') are plotted as diamonds.  The spectral calibration 
sequences corresponding to the lines displayed in the figure 
are from \citet{lang1992}.}
\label{globalHRSMC}
\end{figure*}

\subsection{Topology of global HR diagrams}

Reddening-free colour-magnitude diagrams incorporating all
WFI stars in SMC clusters are shown in Fig.~\ref{globalHRSMC}.
Emission-line stars are found on, or close to, the main sequence, the
red giant branch, and the asymptotic giant branch.  As expected from
Fig.\ \ref{effdetect}, the number of stars seems to become visibly
incomplete below $V=17$ mag (corresponding to $M_{V} \simeq -2.1$).

The emission-line stars near the main sequence are mainly Be stars.
While their spread in colour is not larger than the one of the apparent
zero-age main sequence they are significantly displaced towards redder
colours.  This topology is discussed in more detail in
Sect.~\ref{colourexc}.

\begin{figure}[ht!]
    \centering
    \resizebox{\hsize}{!}{\includegraphics[angle=-90]{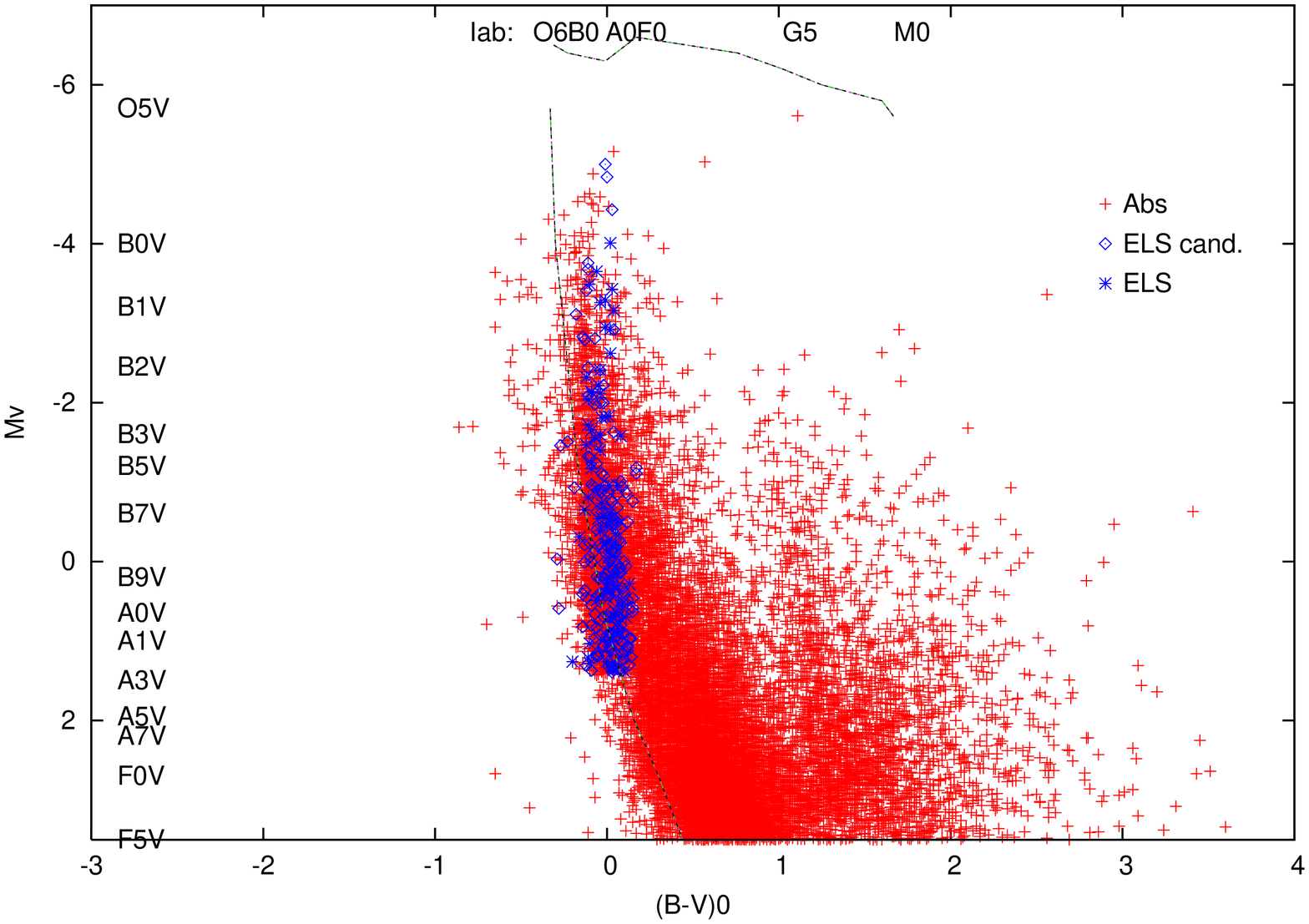}}    
    \caption{Global absolute $V$ magnitude vs.\ dereddened $(B-V)$ colour of the Galactic open cluster
    stars from \citet{mcs2005}.  The spectral calibration sequences corresponding to the lines displayed in the figure 
are from \citet{lang1992}.
    The symbols are the same as in Fig.\ \ref{globalHRSMC}. }
    \label{globalHRGalaxy}
\end{figure}

An analogous diagram ($M_{V}$ vs.\ (B-V)$_{0}$) of the Galactic data
from \citet{mcs2005} is depicted in Fig.~\ref{globalHRGalaxy}.

For all open SMC clusters with emission line stars and data for a
total of at least 10 stars, separate colour-magnitude diagrams are also
available (see Figures~\ref{HR1} to \ref{HR8} in Sect.~\ref{HRind}).

\subsection{Colour excesses of emission-line stars}
\label{colourexc}

Table~\ref{bmvCE} collects the mean colour offset per spectral sub-type  
between emission and non-emission line stars 
(separately for SMC and Galaxy).  Not only are emission-line stars
redder than non-emission lines stars but they possibly even delineate
a separate red sequence.  This is already on average more prominent in
$(V-I)_{0}$ than in $(B-V)_{0}$ but some individual stars deviate much
more strongly in $(V-I)$ than in $(B-V)$.  In $(B-V)$, the Galactic Be
stars seem to differ more strongly from the normal main sequence than
in the SMC.  Since for early-type stars the colour-magnitude diagrams
are degenerate in colour, nothing can be said about any systematic
differences in luminosity.

\begin{figure}[!ht]
\centering
\resizebox{\hsize}{!}{\includegraphics[angle=-90]{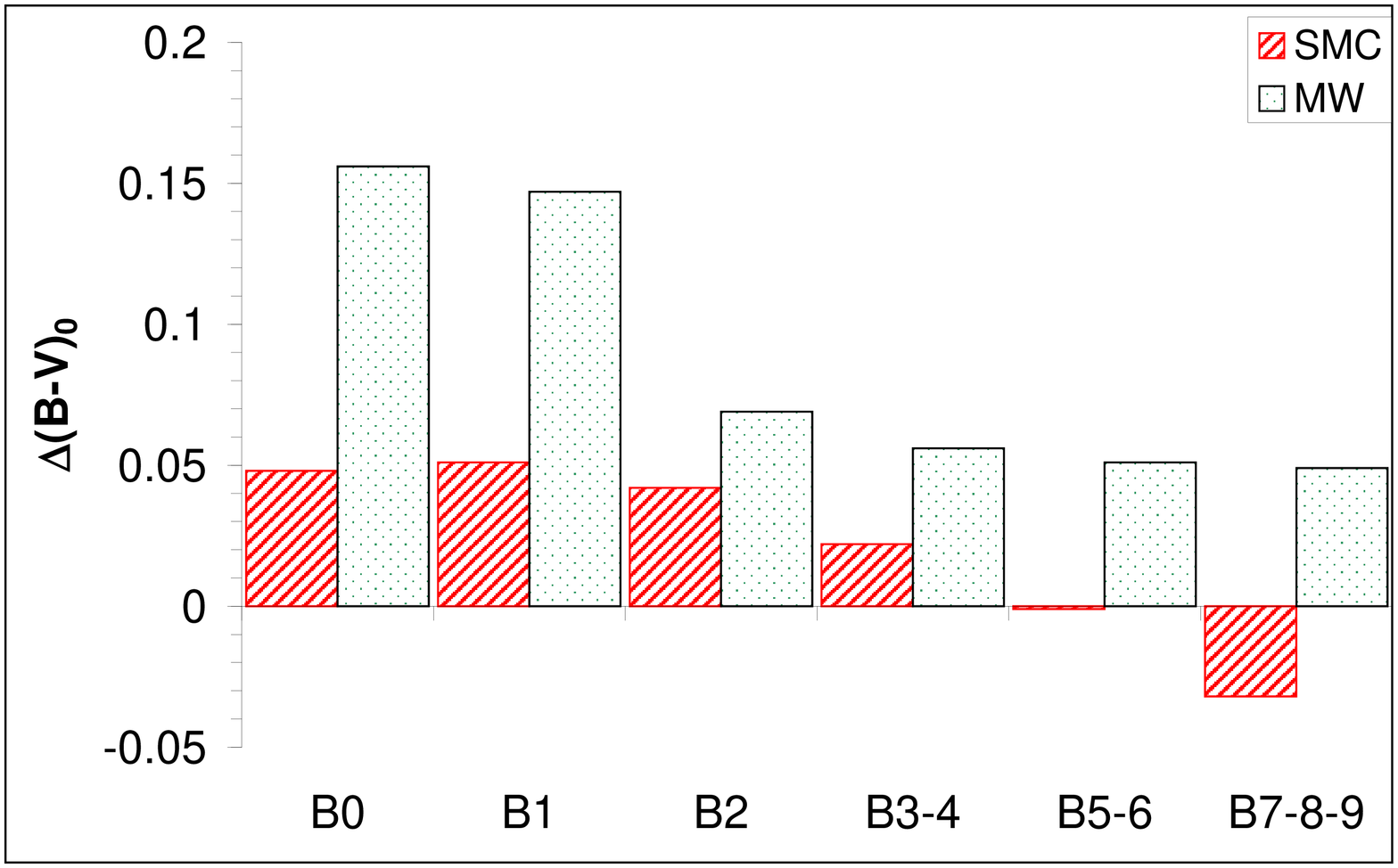}}
\caption{Comparison between normal stars and 
emission-line star of the reddening excess in $(B-V)_{0}$ in the SMC (left) 
and Galaxy \citep[right, from data of ][]{mcs2005}.
Note that the sample is not complete starting with spectral type B3.}
\label{statIC1}
\end{figure}

\begin{figure}[!ht]
\centering
\resizebox{\hsize}{!}{\includegraphics[angle=-90]{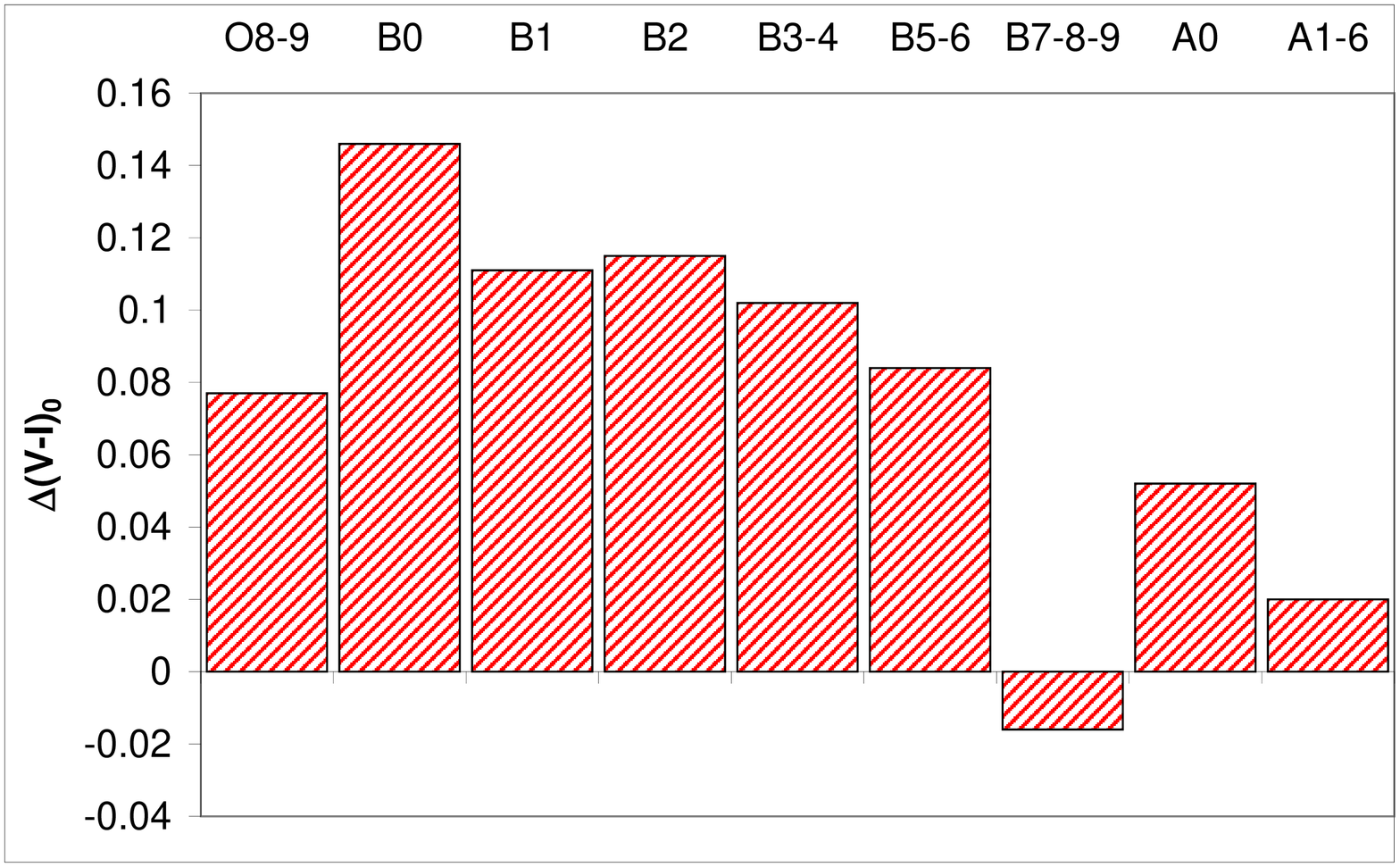}}
\caption{Comparison between normal stars and 
emission-line star of the reddening excess in $(V-I)_{0}$ in the SMC.}
\label{statIC2}
\end{figure}

The excess reddening
seems to reach a maximum around spectral types B0-B2 in both SMC and
Galaxy.  Figs.~\ref{statIC1} and \ref{statIC2} illustrate the reddening excess between
emission and non-emission line stars in $(B-V)_{0}$ and $(V-I)_{0}$ in
the SMC and Galaxy.  
Similar segregations of Be and normal B stars were also found by 
\citet{keller99b}.  This could be due to one or more of stellar evolution, light scattering in
the circumstellar disk \citep{dachs1988}, and
gravitational darkening linked to the fast rotation of Be stars 
\citep{fremat2005}.  Of
these, evolutionary differences are the least likely since the
comparisons are made for stars in the same open clusters and for
similar spectral type.  (The evolutionary status is discussed in more
detail in Sect.~\ref{Beevol}.)

This leaves fast rotation and circumstellar disks as candidate
explanations of the extra reddening in Be stars: At low metallicity
(SMC), Be stars seem to rotate faster than at high metallicity
\citep{marta2007a}, so it is expected that the fast-rotation effects
are larger in the SMC than in the Galaxy.  In theory \citep{maeder01}, 
this is partly compensated by the radii of low-metallicity stars 
being smaller by 15 to 20 \%
which can affect the luminosity of the stars. 
On the other hand, work by \citet{trundle2007} and \citet{evans2008msn}, 
indicates that, in the SMC, the class V stars of a given spectral 
sub-type are hotter than their counterparts in the Galaxy.  But 
their luminosity is about the same: for B0V, L$_{SMC}$/L$_{MW}$=0.9; 
for B1V L$_{SMC}$/L$_{MW}$=1.01; for B2V L$_{SMC}$/L$_{MW}$=1.04.

Circumstellar disks of Be stars
have been reported \citep{wis2006, wis2007, marta2007b} to be closer
to the central star at lower metallicity so that the circumstellar
extinction towards low-inclination Be stars would be increased.  But,
as explained above, the radii of the stars are smaller, thereby partly
offsetting the claimed difference in geometry.

\begin{table*}[h!t]
\footnotesize{
\centering
\caption[]{ Averaged values of $M_{V}$, $(B-V)_{0}$ (cols.\ 3, 8), 
(V-I)$_{0}$ (col.\ 5), and difference in colour index between emission
and non-emission line stars (cols.\ 4 and 9 for $(B-V)_{0}$ and col.\
6 for $(V-I)_{0}$.  Cols.\ 8 and 9 provide data from \citet{mcs2005}
stars in the Galaxy (``Galaxy'').  Crude error estimates (1 $\sigma$)
are 0.054 mag for the SMC $(B-V)_{0}$ values, and 0.030 mag for the
Galactic $(B-V)_{0}$ values.}
\centering
\begin{tabular}{llllllllll}
\hline
\hline	
ST & Mv & (B-V)$_{0}$ &  $\Delta$(B-V)$_{0}$ & (V-I)$_{0}$ & $\Delta$(V-I)$_{0}$  & N & \vline & Galaxy (B-V)$_{0}$ & Galaxy $\Delta$(B-V)$_{0}$\\  
\hline	
O8-O9  & -4.522 & -0.123 &       & -0.063 &       & 16 & \vline & -0.227 &       \\
O8-O9e & -4.638 & -0.098 & 0.025 & 0.014  & 0.077 & 5  & \vline &        &       \\
B0     & -3.569 & -0.190 &       & -0.154 &       & 49 & \vline & -0.286 &       \\
B0e    & -3.569 & -0.142 & 0.048 & -0.008 & 0.146 & 27 & \vline & -0.130 & 0.156 \\
B1     & -2.781 & -0.211 &       & -0.177 &       & 90 & \vline & -0.238 &       \\
B1e    & -2.799 & -0.160 & 0.051 & -0.066 & 0.111 & 37 & \vline & -0.091 & 0.147 \\
B2     & -1.947 & -0.190 &       & -0.184 &       & 258 & \vline & -0.238 &       \\
B2e    & -1.934 & -0.148 & 0.042 & -0.069 & 0.115 & 56 & \vline & -0.169 & 0.069 \\
B3-4   & -1.381 & -0.174 &       & -0.162 &       & 236 & \vline & -0.209 &       \\
B3-4e  & -1.414 & -0.152 & 0.022 & -0.060 & 0.102 & 22 & \vline & -0.153 & 0.056 \\
B5-6   & -0.916 & -0.156 &       & -0.144 &       & 386 & \vline & -0.194 &       \\
B5-6e  & -1.050 & -0.157 & -0.001& -0.060 & 0.084 & 19 & \vline & -0.143 & 0.051 \\
B7-8-9 & -0.193 & -0.132 &       & -0.114 &       & 365 & \vline & -0.153 &       \\
B7-8-9e& -0.490 & -0.164 & -0.032& -0.130 & -0.016 & 2 & \vline & -0.104 & -0.049\\
A0     & 0.718  & -0.062 &       & -0.088 &       & 116 & \vline & -0.124 &       \\
A0e    & 0.820  & -0.155 & 0.093 & -0.036 & 0.052 & 3 & \vline & -0.076 & 0.048 \\
A1--6  & 1.432  & -0.011 &       & -0.034 &       & 129 & \vline & -0.103 &       \\
A1--6e & 1.596  & -0.014 & 0.003 & -0.014 & 0.020 & 4 & \vline & -0.165: & -0.062\\
\hline
\end{tabular}
\label{bmvCE}
}
\end{table*}


\subsection{Frequency of Be stars as a function of local star density}
\label{locdense}
Various studies in the Galaxy \citep[for example][]{keller2004}
suggest that the rotational velocities of B stars are higher in open
clusters than in the field.  By contrast, investigations in the LMC
and SMC using statistical tests \citep{marta2006b,marta2007a} did not
find significantly different rotational velocities in clusters and the
field.
In the Galaxy, \citet{huang06} found more slow rotators in 
the field than in open clusters. They also concluded that the 
more massive B stars spin down during their main sequence phase and 
suggested that some of the rapid rotators found may have been spun 
up by mass-transfer in close binary systems.  These authors ascribe
the difference between open clusters and fields to a difference in the 
evolutionary phase of the stars (the older, the slower are the stars).  

The same difference is attributed by \citet{wolff07} to 
the difference in number density.
They argue that stars in low-density environments could retain their 
pre-main sequence disk for a longer time. Star-disk locking would, 
then, preserve the angular speed so that such stars cannot become young fast 
rotators.

However, the clusters studied are mainly young ones, among them NGC\,6611. 
\citet{marta2008a} have shown that in this cluster 
some early-type objects are still on the pre-main sequence.  They 
also found that the rotational velocities of these two kinds of objects
differ by about 20\%, with ZAMS stars rotating more slowly.  The 
theoretical models from \citet{meynet2000} explain this decrease by 
an internal redistribution of the angular momentum at the ZAMS.

\begin{figure}[]
\centering
\resizebox{\hsize}{!}{\includegraphics[angle=-90]{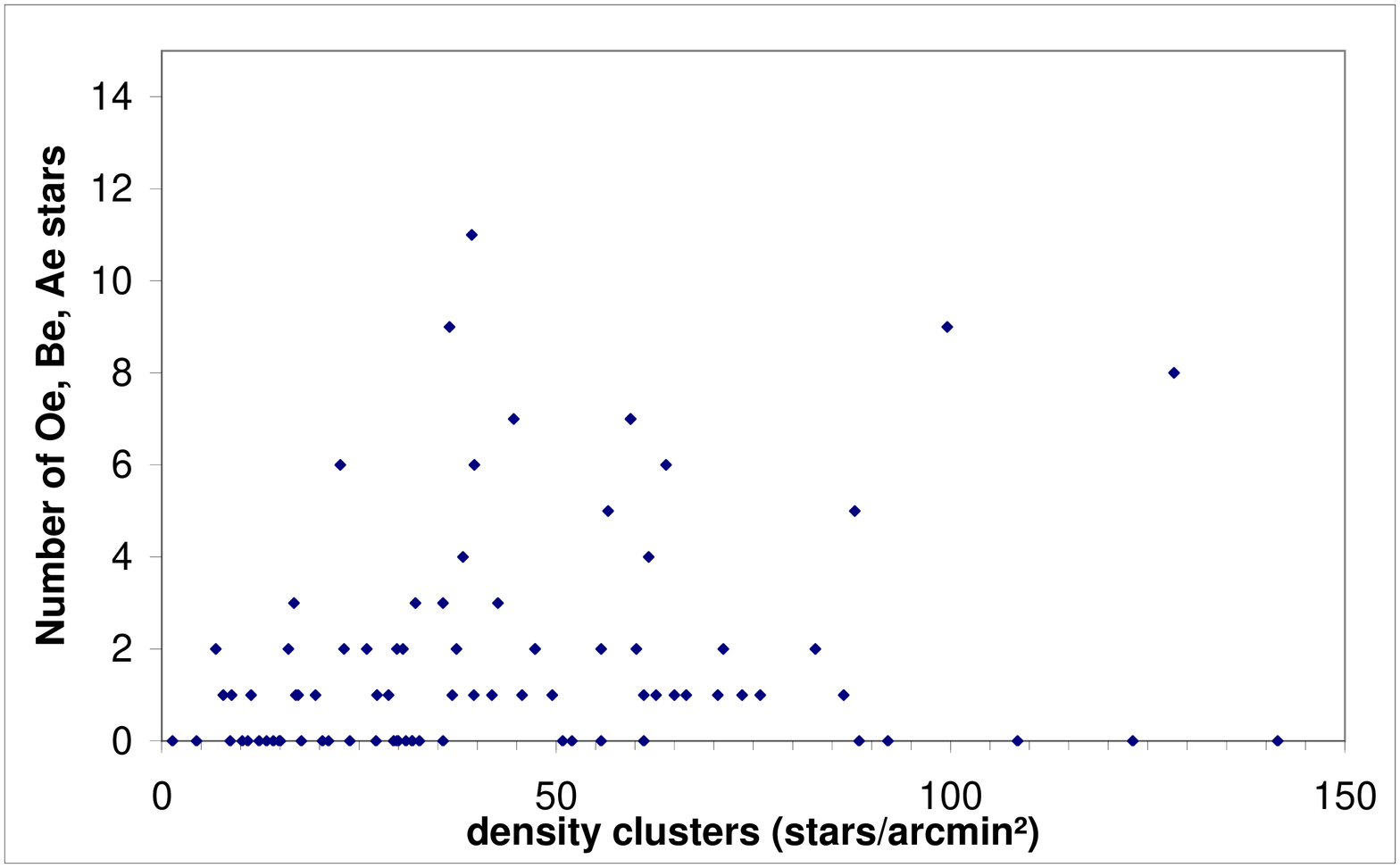}}
\resizebox{\hsize}{!}{\includegraphics[angle=-90]{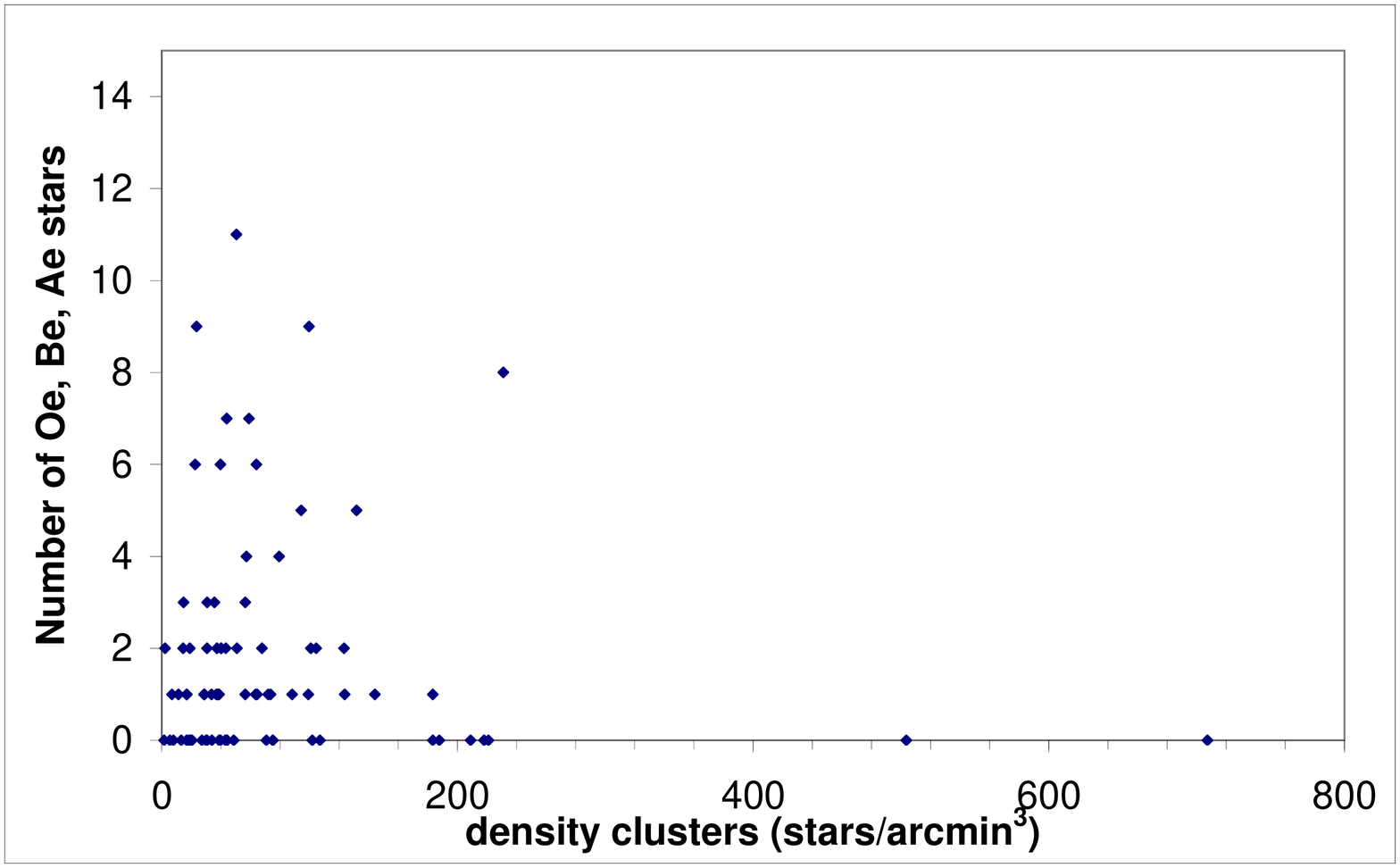}} 
\caption{Number of emission-line stars (Oe, Be, Ae) vs.\ area (top) or 
volume (bottom) density of SMC open clusters.}
\label{density1}
\end{figure}

If a difference between the rotational velocity of cluster and field stars 
is not due to biased sampling of evolutionary effects but related to 
local stellar density, the expected frequency
of Be stars should be higher in open clusters than in the field at
large and also higher in higher-density fields. 
The WFI SMC database permits such a comparison to be made.  
This hypothesis will be checked in a following paper dealing with SMC field stars. 
Accordingly, the star surface
and space density of each open cluster in the sample was calculated
and compared to its contents of main-sequence emission-line stars
(Oe, Be, Ae).  No correlation became evident (Fig.~\ref{density1}),
in agreement with the findings of \citet{mcs2005} for Galactic
clusters.

\subsection{The Be phenomenon:  SMC vs.\ Galaxy}

\begin{figure}[h!]
\resizebox{\hsize}{!}{\includegraphics[angle=-90, width=8cm]{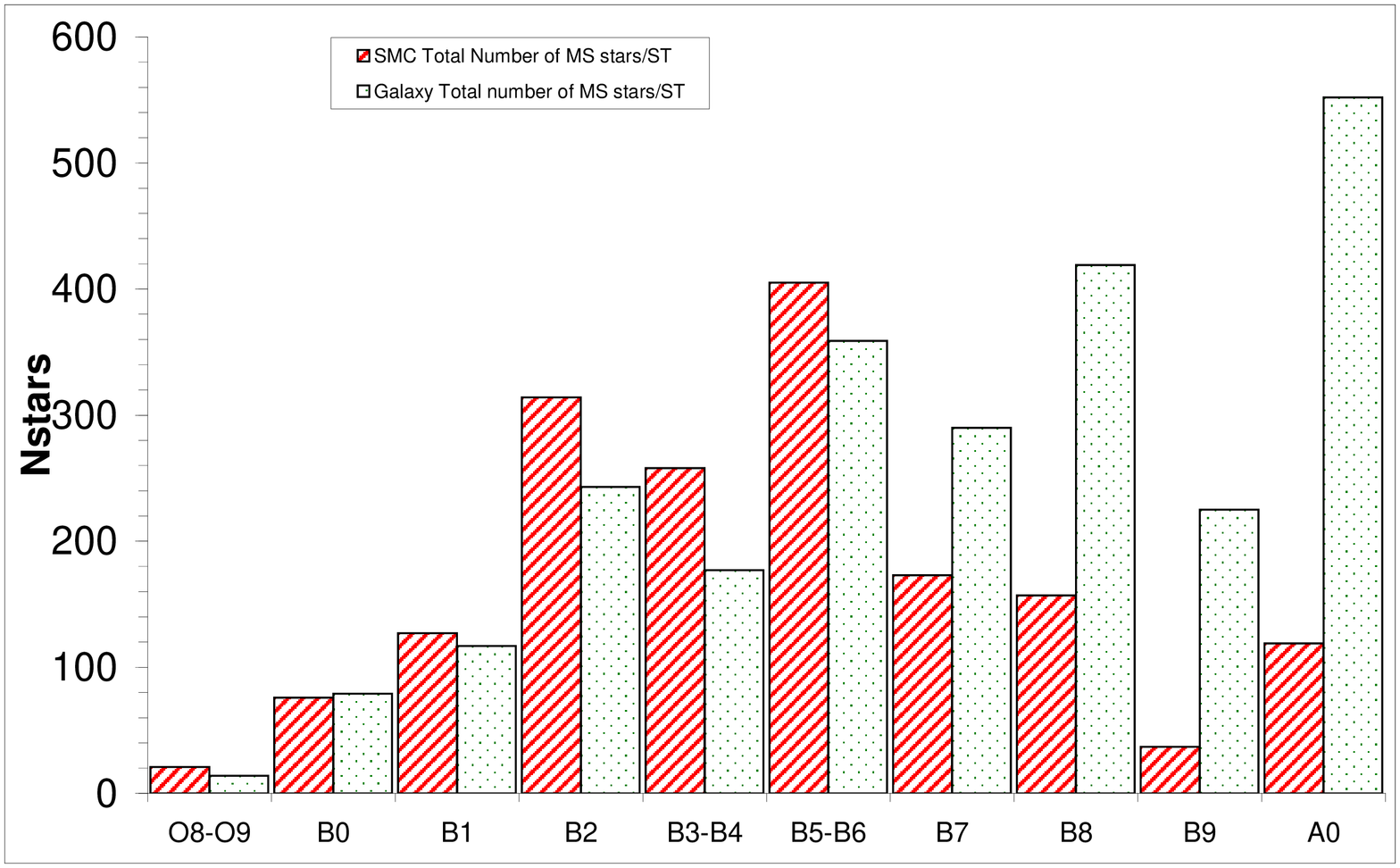}}
\caption{Global fractions of main-sequence stars with and without emission 
lines as a function of spectral type (from O to early A stars). 
Left bars: SMC; right bars: Galaxy \citep[from][]{mcs2005}.  Both are
similar and appear shaped by the initial mass function.  However, in
the SMC data the cut-off due to incompleteness sets in quite visibly
at earlier spectral types (brighter magnitudes).
}
\label{samplesSMCGalaxy}
\end{figure}

In both galaxies, the fraction of near-main sequence Be stars 
varies drastically from one
open cluster to the other but there is not even the beginning of an
explanation of this very conspicuous (and well-known) fact.  In order
not to be misled by such small-number instabilities, all WFI SMC 
and all Galactic emission-line stars were combined to one sample each.
The completeness with spectral type of these samples can be inferred 
from Fig.~\ref{samplesSMCGalaxy}, which confirms that the SMC 
sample is incomplete towards fainter stars, i.e.\ later B
sub-types, whereas in the Galaxy the increase toward later subtypes
basically follows the Initial Mass Function (IMF).  

\begin{figure}[h!]
\resizebox{\hsize}{!}{\includegraphics[angle=-90]{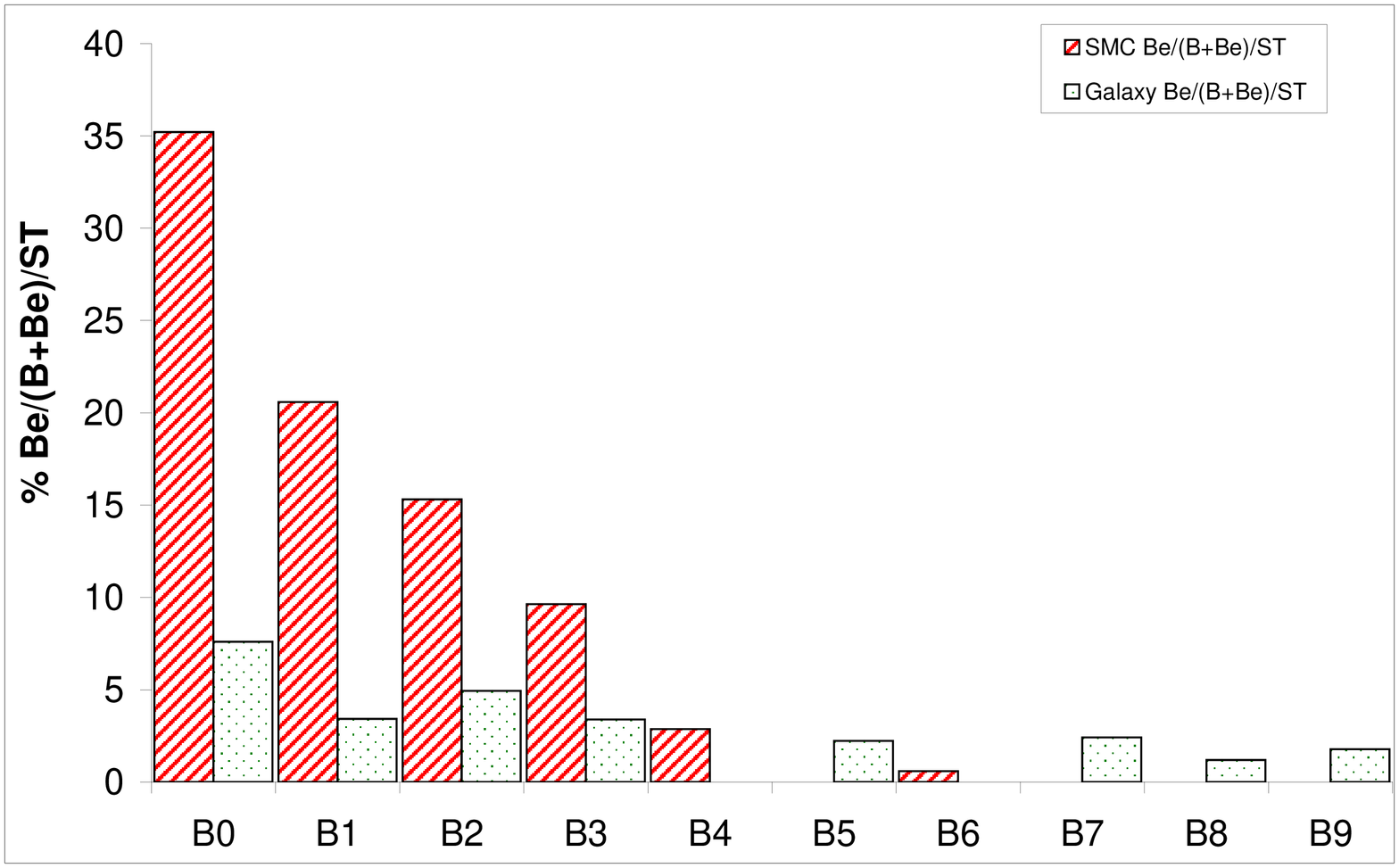}}
\caption{The percentage of definite Be stars among all B-type (definite+candidate Be + B) stars  
vs.\ spectral type. Shaded bars: SMC, light bars: Galaxy.}
\label{statBetoB}
\end{figure}

The global fractions of main-sequence Oe/Be/Ae stars per spectral
sub-class are provided in Table~\ref{propBe}.  As Fig.\
\ref{statBetoB} illustrates, the distributions have similar overall
shapes.  But with fractions reaching nearly 35\%, early-type Be stars
are more abundant in the SMC than in the Galaxy by a factor of 3-5.
This factor drops to 2-4 if the very populous cluster NGC\,330 is
removed from the sample so that the overfrequency of Be stars among
early spectral subtypes is robust.  The recent slitless study of
\citet{mathew08} reports similar Be-to-B star ratios in Galactic
clusters as \citet{mcs2005} do.

Note that these three studies may be compared because they are all 
single-epoch studies.  Since the Be phenomenon is transient, the true
frequency of Be stars must be higher than apparent from such surveys.
The study by \citet{fabregat03} suggests that up to one-third of all
Be stars may be missed at any one epoch.  Studying the Galactic cluster 
NGC\,3766, \citet{mcs2008} even suggest that 25 to 50 \%
of the Be stars could be missed in a single-epoch spectroscopic survey. 
This is, of course, very much dependent on the nature and quality 
of the data.  Note that is not known whether the volatily of emission lines 
is different in Galaxy and SMC.  In the Galactic field, the 
variability of classical Be stars is significantly higher among the 
early spectral subtypes, to which the present study is limited.

The inclusion of candidate Be stars (Table~\ref{propBe}) does not much
affect the distribution function in the SMC.  However, for the
Galactic late-type candidate-Be stars from \citet{mcs2005} there
is a large increase. It is conceivably due to the difficulty of
photometrically distinguishing pre-main sequence or Herbig Ae/Be stars
from classical Be stars. The more frequent occurrence of these stars
in young open clusters supports this interpretation.

\begin{table}[h]
\centering
\caption[]{Number ratios of Be to (B+Be) stars as a function of spectral 
type in SMC and Galaxy.  For each range in spectral type, the
fractions of definite emission-line stars and the combination of
definite and candidate emission-line stars are listed. }
\centering
\begin{tabular}{llll}
\hline
\hline	
Spectral type      & SMC & SMC \% &  Galaxy  \\
                   & \%  & without NGC\,330 &  \% \\
\hline	
O8-O9e	& 23.8	& 20.8 & 14.3 \\
B0e	& 35.2	& 26.3 & 7.6  \\  
all B0e & 36.1  &      & 12.7 \\
B1e     & 20.6  & 16.7 & 3.4  \\
all B1e & 27.0  &      & 7.7  \\
B2e     & 15.3  & 13.9 & 4.9  \\
all B2e & 19.9  &      & 7.8  \\
B3e     & 9.6   & 8.9  & 3.4  \\
all B3e & 14.0  &      & 6.8  \\
B4e     & 2.9   & 3.0  & 0.0  \\
all B4e & 7.6   &      & 0.0  \\
B5e     & 0.0   & 0.0  & 2.2  \\
all B5e & 1.8   &      & 9.2  \\
B6e     & 0.6   & 0.6  & 0.0  \\
all B6e & 1.2   &      & 0.0  \\
B7e     & 0.0   & 0.0  & 2.4  \\
all B7e & 0.0   &      & 7.2  \\
B8e     & 0.0   & 0.0  & 1.2  \\
all B8e & 0.0   &      & 9.3  \\
B9e     & 0.0   & 0.0  & 1.8  \\
all B9e & 0.0   &      & 11.1 \\
A0e     & 2.5   & 2.5  & 0.9  \\
\hline	
\hline
\end{tabular}
\label{propBe}
\end{table}

\subsection{The Be phenomenon as a function of spectral type} 

\begin{figure}[h!]
\centering
\resizebox{\hsize}{!}{\includegraphics[angle=-90]{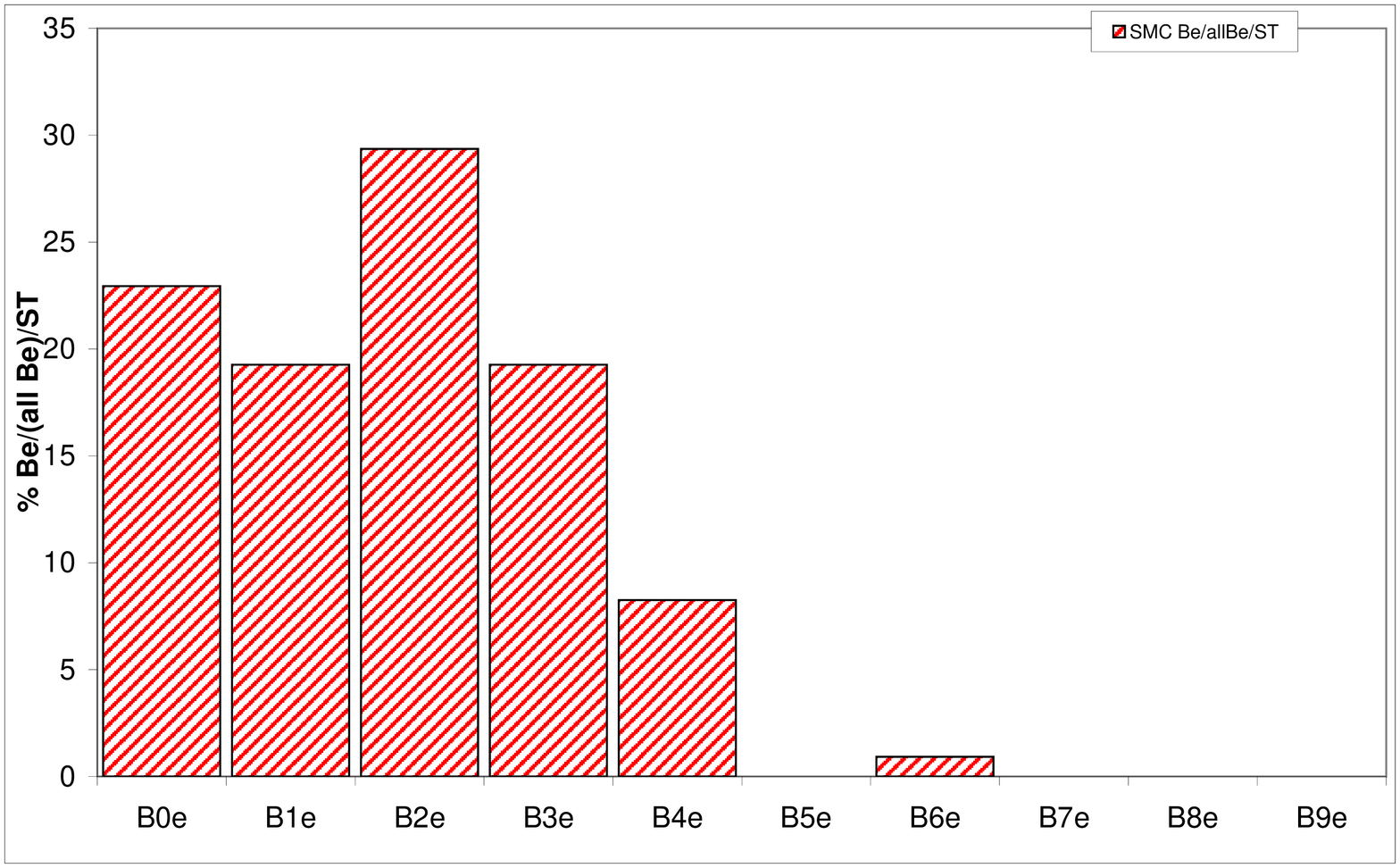}}
\resizebox{\hsize}{!}{\includegraphics[angle=-90]{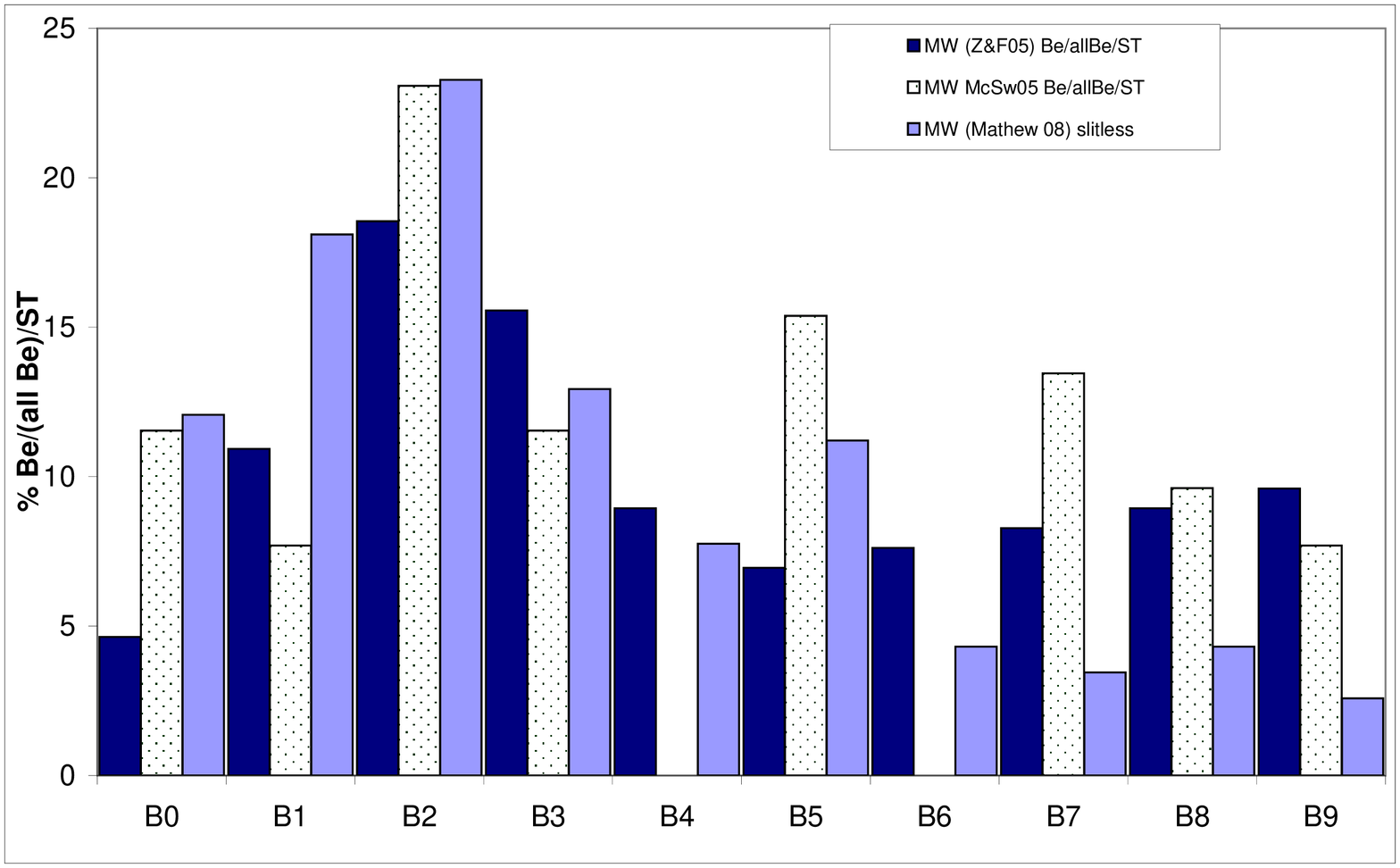}}
\caption{The distribution of definite Be stars, as a percentage 
of the total number of Be stars in the sample, with spectral type in 
SMC (top, this paper, red dashed bars) and Galaxy (bottom).  The data 
for the Galaxy arefrom \citet[][ left blue bars]{zf05}, \citet[][ middle white
bars]{mcs2005}, and \citet[][ right, light blue bars]{mathew08}.}
\label{BetoBe}
\end{figure}

Fig.\ \ref{BetoBe} presents the percentages per spectral sub-class,
referred to the total sample of Be stars, but separately for SMC and
Galaxy.  The Galactic data are from \citet{zf05}, \citet{mcs2005}, and
\citet{mathew08}.  The distributions for the two galaxies are similar but the
inclompleteness of the SMC data becomes rather apparent beyond B2 (cf.\
Sect.~\ref{deteclim}) and prevents a more detailed comparison.  The
highest number of Be stars is encountered at spectral type B2 in both
SMC and Galaxy.  Because this coincides with the maximum of the
H$\alpha$ emission-line strength \citep{zorec2007a} while for lower
line strengths the statistics are increasingly incomplete, it is
questionable whether Fig.\ \ref{BetoBe} reveals the real dependency of
the Be phenomenon on effective temperature.  At late spectral
sub-classes, the more complete Galactic data level off to a plateau as
already shown by \citet{kogure1982}. As lined out by
\citet{zf05} and \citet{zorec2007a}, this may be due to the combination of the
decreasing relative frequency of Be stars and the absolute increase in
the number of late B stars with the initial mass function. 

Unlike in Fig.\ \ref{statBetoB}, the distribution in Fig.\ \ref{BetoBe} 
does not drop very quickly with spectral type because the total number of 
Be stars per spectral bin is crudely constant to within a factor of 2-3.  
This is not true for B-type stars at large because the IMF lets their 
numbers increase rapidly towards lower temperatures.

\subsection{Evolution and age}
\label{Beevol}

\begin{figure}[h!]
\resizebox{\hsize}{!}{\includegraphics[angle=-90, width=7cm]{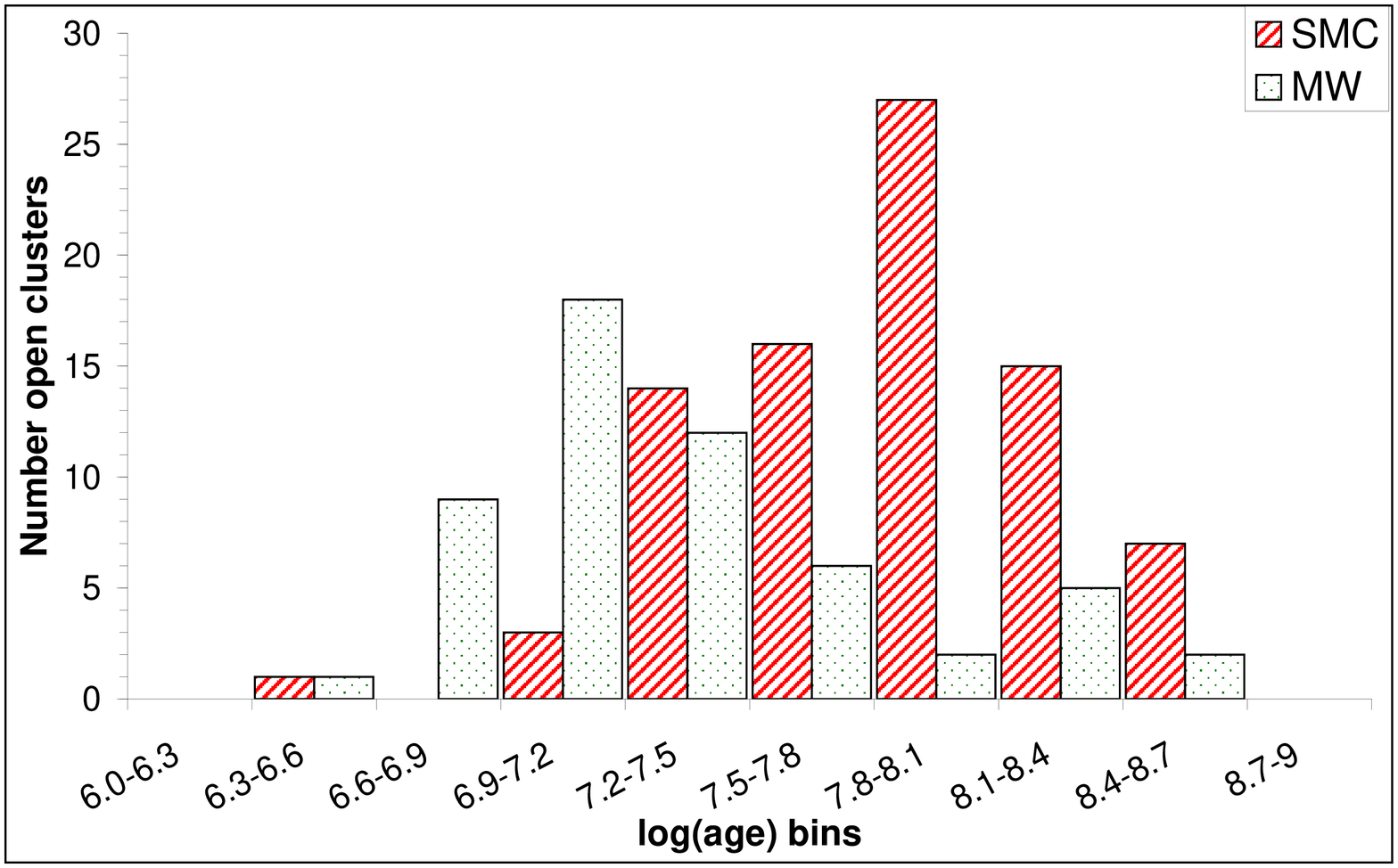}}
\caption{The log(age) distributions of open clusters in 
SMC (left bars) and Galaxy \citep[right bars, data from][]{mcs2005}.
}
\label{agesOcl}
\end{figure}

Fig.~\ref{agesOcl} shows the distributions in age of open clusters of
the SMC \citep{ageSMcocl} and Galactic samples.  In the SMC sample,
there are more old open clusters than in the Galactic one although 
the stellar population of the Galaxy at large is older than the one of the SMC.  
 
\begin{figure}[h!]
\centering
\resizebox{\hsize}{!}{\includegraphics[angle=-90]{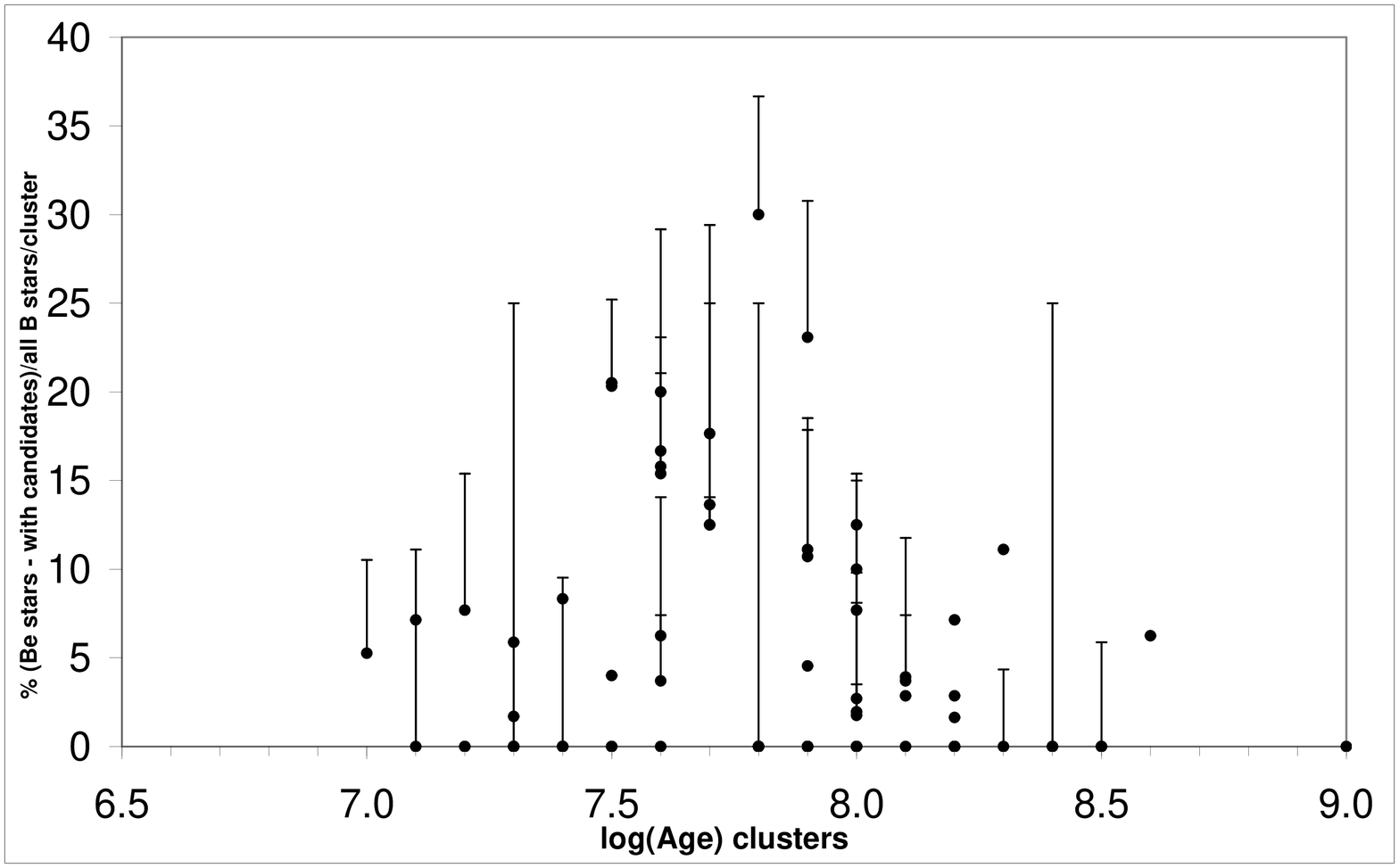}}
\caption{Number ratio of Be to B stars vs.\ cluster age in the SMC.
The circles correspond to definite Be stars only, and the 
vertical bars extend the values by the respective candidate Be stars.
In comparison, Fig.\ 4 of \citep{mcs2005}, which is the equivalent 
diagram for Galactic clusters, is basically flat with cluster age.
}
\label{pcBetoBOcl}
\end{figure}

The number ratios of Be to B stars as a function of the age of open
clusters in the SMC are shown in Fig.~\ref{pcBetoBOcl}.  For the
Galaxy, see Fig.\ 4 of \citet{mcs2005}.  There is a maximum at
log(age)$\sim$7.6 in the SMC, while in the Galaxy no clear trend is
seen.  If taken at face value, there may be a small evolutionary
enhancement of Be stars in the SMC.  But a similar distribution may
result already if the Be phenomenon peaks at a particular spectral
sub-type.

\begin{figure}[h!]
\centering
\resizebox{\hsize}{!}{\includegraphics[angle=-90]{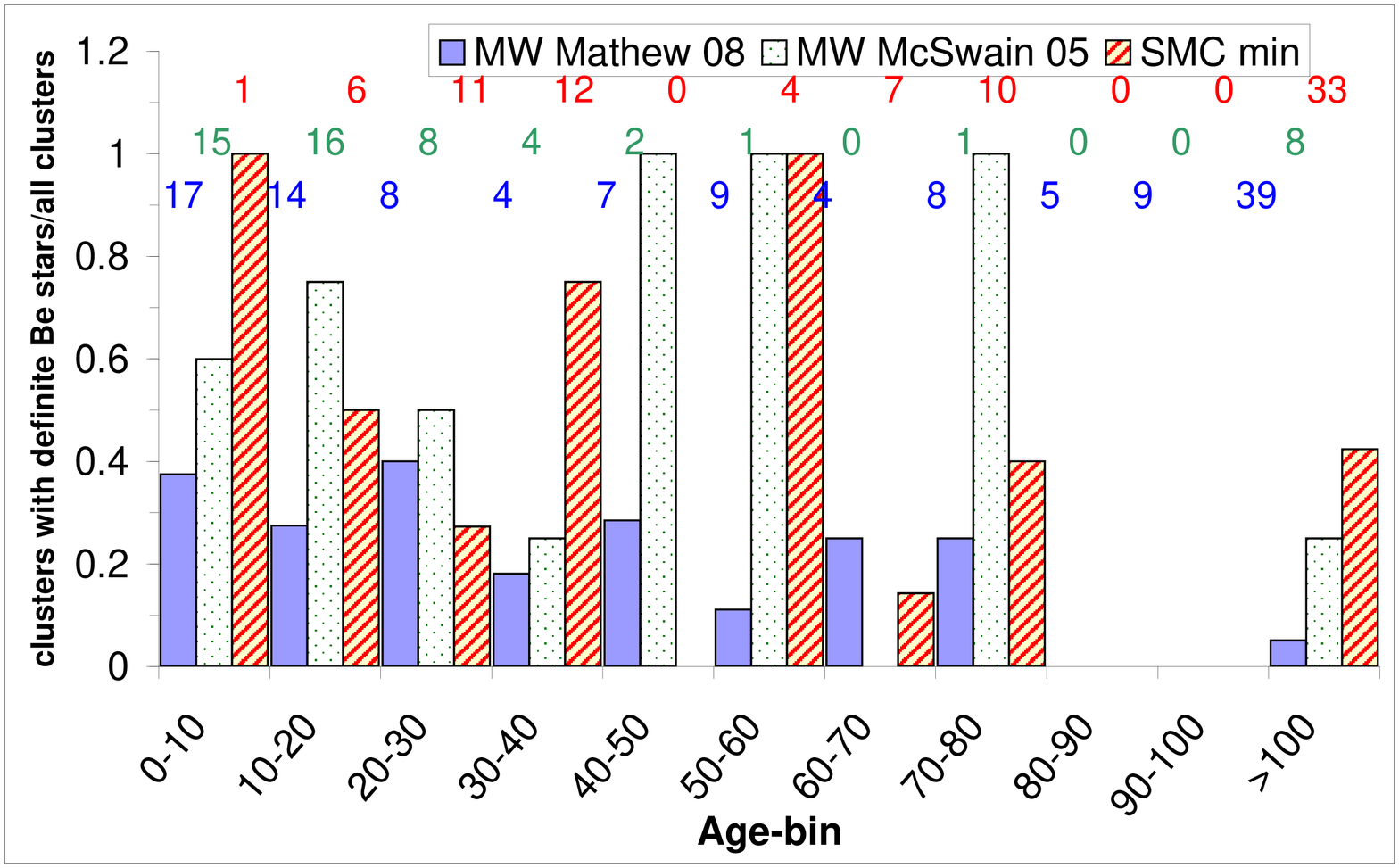}}
\resizebox{\hsize}{!}{\includegraphics[angle=-90]{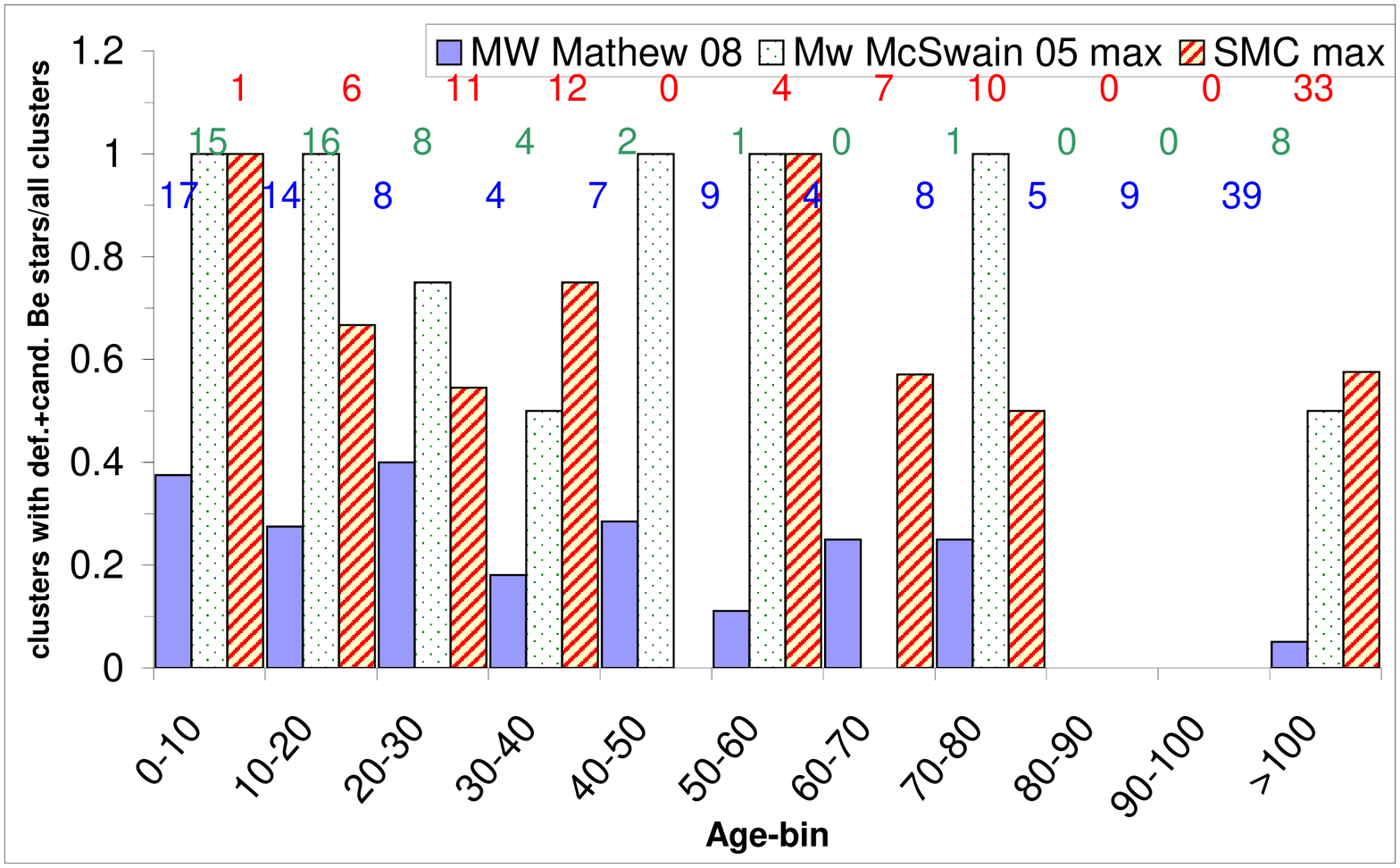}}
\caption{Fraction per age bin of open clusters with Be stars in the 
Galaxy and SMC \citep[following Fig.\ 8 of][]{mathew08}; 
top: definite Be stars, bottom: definite+candidate Be
stars. The data for the Galaxy are from \citet[][ left blue
bars]{mathew08}, \citet[][ middle white bars]{mcs2005}, and for the
SMC (this study, right red bars).  Numbers at the top show the
absolute numbers represented by each bar.
}
\label{OclwithBe}
\end{figure}

Similarly, Fig.~\ref{OclwithBe} presents the ratios of open clusters
with Be stars to all open clusters by age bins as defined by
\citet[][ their Fig.\ 8]{mathew08}; the top panel is for definite Be 
stars, and the bottom panel combines definite and candidate Be
stars.  The comparison is made with data in the Galaxy from
\citet{mathew08} and \citet{mcs2005}.  As noticed in the Galaxy by
\citet{mathew08}, there seems to be a first decrease of the number of open
clusters with Be stars towards 30-40 Myears (log(age)=7.5-7.6).  
Thereafter there is an increase during the evolution, 
and another decrease after 50-60 Myears
(log(age)=7.7-7.8).  These 2 Figures (\ref{agesOcl} and
\ref{pcBetoBOcl}) indicate that some Be stars could be born as Be
stars \citep{wis2007b}, while others only assume Be characteristics
during the evolution as mentioned by \citet{fabregat2000}.  The first
decrease, if real (the differences between the studies are large),
could be caused by Be stars reaching the terminal-age main sequence or
by an evolutionary change of the angular velocities so that not every initial 
Be star can sustain a high enough surface rotation rate to remain a Be star
throughout its entire main sequence life \citep{marta2007a}.

Moreover, for the Galaxy as well as low metallicity, respespectively, 
\citet{meynet2000} and \citet{maeder01} showed that, while the linear 
rotational velocity decreases with time, the fractional critical angular 
velocity (\omc) increases.  This holds for both medium and 
low-mass B-type stars over the range in metallicity studied and 
for massive B-O stars of low metallicity.  But in massive 
Galactic-metallicity stars \omc\ actually decreases with age due to 
the larger losses in mass and angular momentum.  

From observations of 
Galactic Be stars with \omc$\ge$70 \%, \citet[][ Fig.\ 11]{marta2007a} 
reckon that massive Be stars lose their emission-line characteristics  
after few million years, while late-type Be stars begin to appear 
at 40\% of their main sequence lifetime or nearly 
40 million years.  This is the age, at which (Fig.\ \ref{OclwithBe}) 
the fraction of open clusters with Be stars begins to rise again.
For SMC-like metallicity, \citet[][Fig.\ 11]{marta2007a} predict that 
massive and intermediate-mass Be stars appear between 3 and 5 million 
years.  At intermediate cluster ages, the fraction of
clusters hosting Be stars decreases because this massive population 
disappears when it reaches the TAMS.  Finally, at an age of about 35-45 
million years and SMC metallicity, late-type Be stars start to occur.  
This expectation, based on the 
evolution of \omc\ as a function of metallicity and mass, is 
qualitatively matched by the distribution in Fig.\ \ref{OclwithBe}.

\begin{figure}[h!]
\centering
\resizebox{\hsize}{!}{\includegraphics[angle=-90]{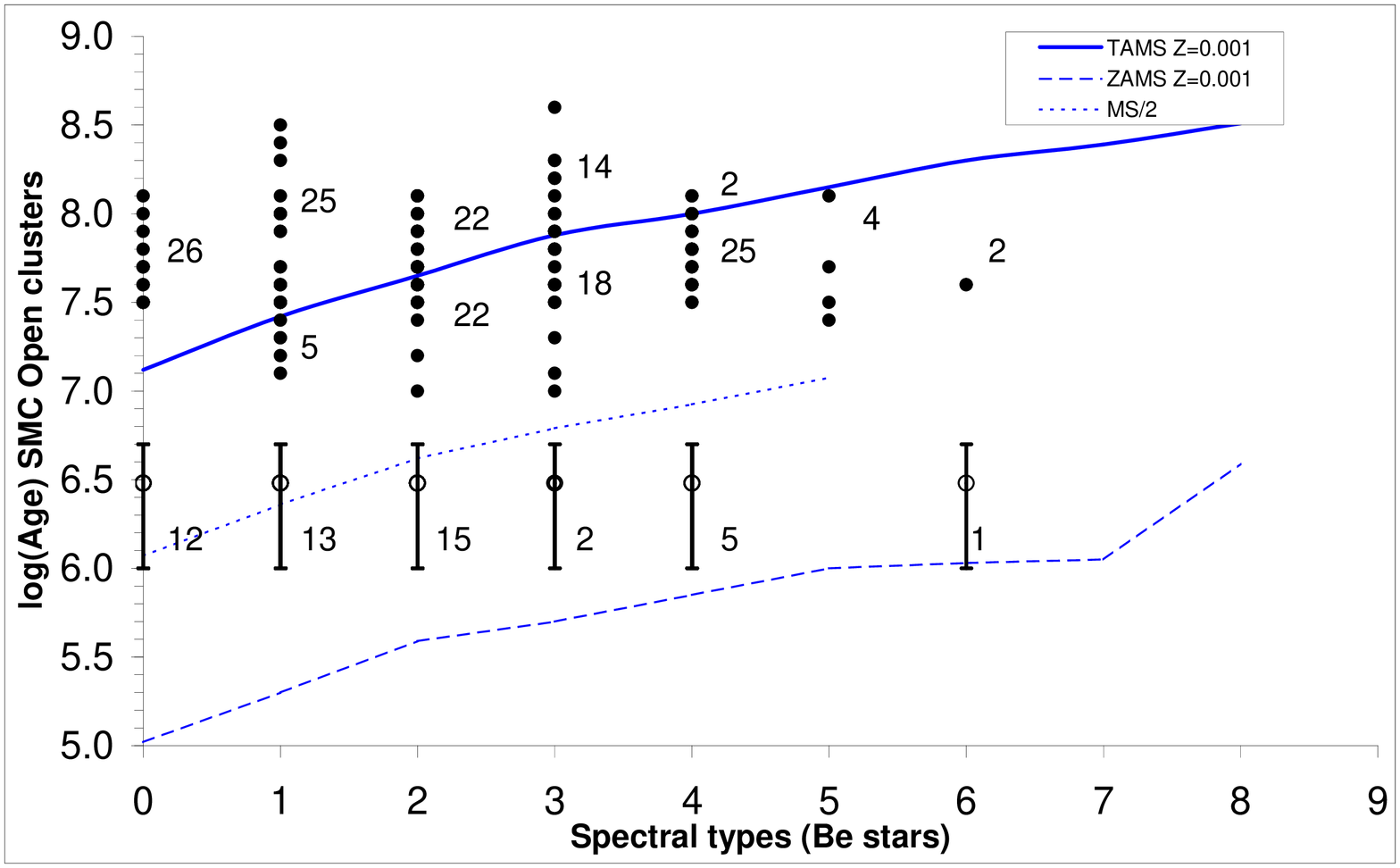}}
\resizebox{\hsize}{!}{\includegraphics[angle=-90]{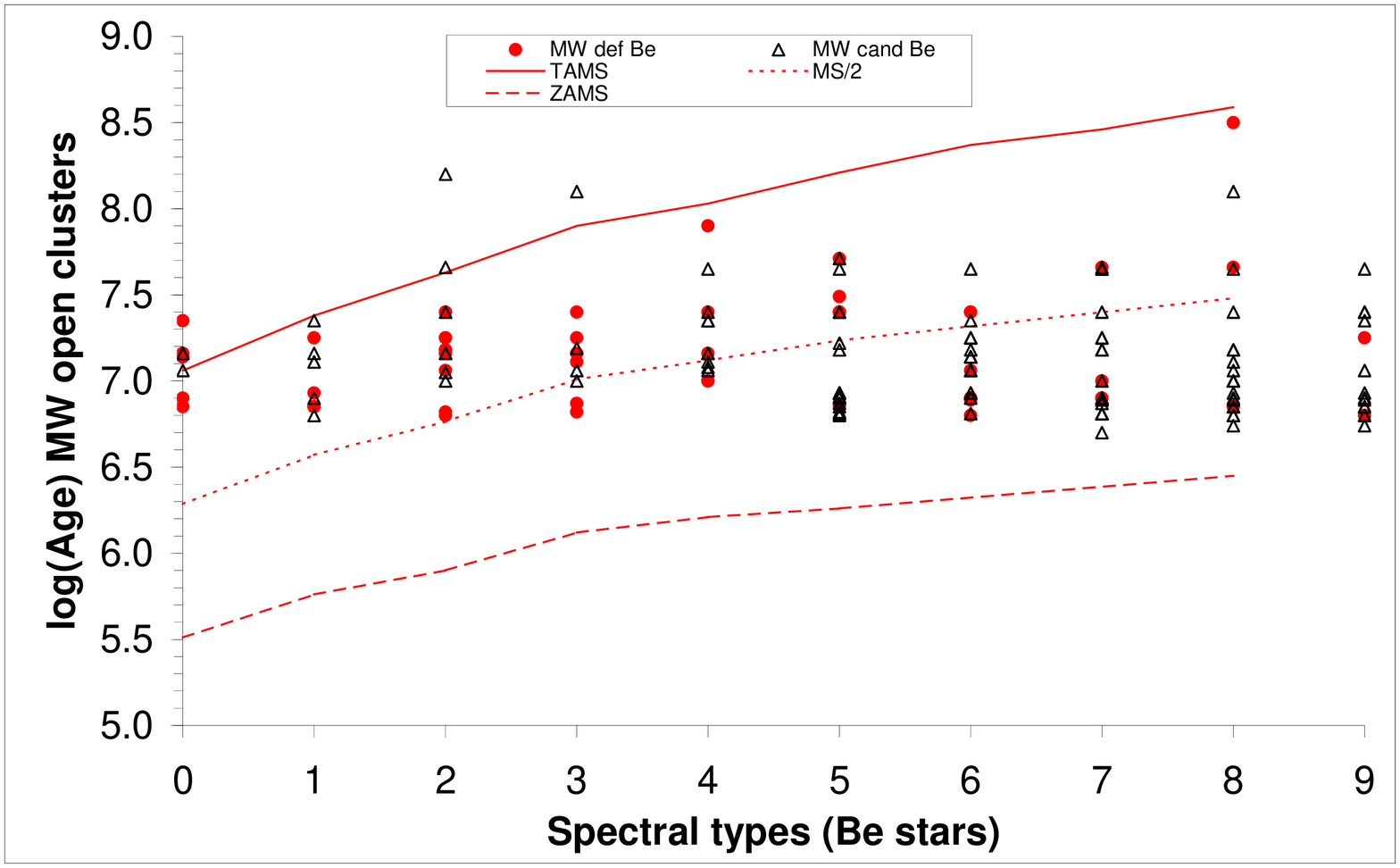}}
\caption{Spectral types of Be stars vs.\ log(age) of their host 
clusters in SMC (top) and Galaxy (bottom). The curves show terminal-
and zero-age main sequence and the main-sequence half-age following
\citet[][ with Z=0.001 for SMC and Z=0.020 for the 
Galaxy]{schaller1992}.  In the SMC, the numbers are the absolute
numbers of stars in each spectral sub-type, and the open circles with large age error-bars
correspond to the emission-line stars in NGC\,346.  In the Galaxy, the circles
denote the definite Be stars and the triangles identify the candidate
Be stars.}
\label{BeSTvsageocl}
\end{figure}

To investigate age dependencies further, Fig.~\ref{BeSTvsageocl}
presents the range in log(age) of the host clusters as a function of
spectral type (data from \citet{mcs2005}; recall that the Be stars in
the SMC were ``selected as main sequence stars'').  In the Galaxy,
early-type Be stars are found close to the terminal-age main-sequence
stars, intermediate-mass Be stars are mainly evolved, and definite
late-Be stars are also evolved. The late-type candidate Be stars could be
unevolved but there is a potential risk of confusion with pre-main
sequence objects.  Be stars seem to follow the evolutionary scheme
described in \citet{fabregat2000}, \citet{zorec2005}, and
\citet{marta2007a}, depending on their mass.

In the SMC, the OGLE ages \citep{ageSMcocl} of some clusters are
not really in agreement with their having very early-type stars (O-B0) as
members.  One example is Ogle-SMC72 with log(age)=7.6 $\pm$0.2 and
containing 3 B0e stars although B0 stars reach the terminal-age 
main-sequence already at log(t)=7. 

On the other hand, \citet{chiosi2006} derive log(age)=6.6$\pm$0.5 for this cluster,
which is fully consistent with B0e member stars.  There are differences between 
the ages from \citet{ageSMcocl} and from \citet{chiosi2006} also 
for other open clusters in the WFI sample, which could explain some of the 
discrepancies between individual spectral types and parent-cluster ages. 
Even though in some specific clusters the ages published by 
\citet{chiosi2006} are in better agreement with the presence of Be stars, this 
is not generally the case.
 
If all OGLE ages from \citet{ageSMcocl} are taken at face value, 55\% of the Be stars, which were 
selected as main sequence stars, could be younger than their host clusters. 
With the ages from \citet{chiosi2006}, this value reaches 62\%.
As the published error estimates are lower for \citet{ageSMcocl}, 
their estimates were adopted.
If the problem is not due to the assigned ages or to the TAMS calibration, 
all possible explanations for blue stragglers (e.g., multi-epoch or
continuous star formation, mass-transfer binaries, etc.) are potential
candidates.  This hypothesis could be reinforced by the finding that 
Be/X-ray binaries are more abundant in the SMC than in the Galaxy
according to \citet{haberl2000}, who explain this result by different
star-formation histories of the two galaxies. 

On the other hand, the color-magnitude diagrams in Sect.\ \ref{HRind} and the location 
of blue stragglers as delineated by \citet{ahumada2007} suggest that at most few 
Be stars lie in the blue-straggler zone while most Be stars are actually 
on the red side of the cluster main sequences.  As discussed in Sect.\ \ref{colourexc}, 
this is probably unrelated to their age.

The other main sequence 
Be stars, located below the TAMS, are found evolved to the second part of the main
sequence in agreement with the evolutionary picture sketched by
\citet{fabregat03} and \citet{marta2007a} for intermediate-mass Be stars in the SMC.  
The emission-line stars in NGC\,346 could be
classical Be stars but also pre-main sequence stars like Herbig Ae/Be
stars.  This would mostly affect the late spectral subtypes but at B0
some pre-main sequence stars are found as well \citep{nota2006}.

Both in SMC and Galaxy, diagrams like Fig.~\ref{BeSTvsageocl} but for
non-emission line B stars show a uniform distribution with age of the
open clusters.

In the SMC, Be stars are mostly located in the region corresponding 
to the second half of the main sequence
(except for some late-type stars in NGC\,346 but they could be pre-main sequence 
stars), in agreement with the results of \citet{fabregat2000} and 
\citet{marta2007a}.  But the lack of open clusters with ages
corresponding to the first half of the main sequence evolution of 
B-type stars makes it impossible to conclude that the Be phenomenon is 
{\it restricted} to the second half of the main-sequence evolution.  

In the Galaxy, definite Be stars (red circles in Fig.\ \ref{BeSTvsageocl}, 
bottom panel) have mostly evolved to the second half of the main-sequence band 
as reported before by \citet{fabregat2000}. The earliest (B0e) have 
alreached the TAMS.  The locations of candidate and definite Be stars 
with early spectral types largely overlap.  Towards later spectral sub-types, 
candidate Be stars could be less or even unevolved.  But there is a risk 
of confusion with pre-main sequence objects.  This uncertainty is smaller for 
definite Be stars.  

To confirm the nature and better determine the evolutionary status of these Be stars, 
spectra with higher resolution and spectral coverage are required of both young  
and medium-aged open clusters in the SMC as well as the Galaxy. 

A new result suggested by the present study is that in the SMC the Be
phenomenon appears particularly enhanced towards hotter and more
massive stars (O stars) than in the Galaxy.  Because the losses of
mass and angular momentum are lower in the SMC than in the Galaxy, this
result is plausible and qualitatively consistent with the theoretical
prediction from \citet{maeder01}.  However, these authors did not
quantify the expected fractions of Oe or Be stars as a function of
metallicity.

An important parameter to consider is \omc, the fractional critical 
angular rotation rate.  If low-metallicity stars form with about the same 
initial angular momentum as more metal-rich stars of equal mass but 
have smaller radii on the main sequence, their \omc~values must be higher. 

Since the, then, expected higher relative frequency of Be stars is,
in fact, higher in the SMC than in the Galaxy, \omc\ may be the  
parameter dominating the formation of rapidly rotating B stars. 
They may become Be stars just on account of their rapid rotation.  
Alternatively, the outbreak of the Be phenomenon may be helped by 
pulsation-assisted outbursts, which are triggered by the beating of 
two or more nonradial pulsations modes.  In the Galaxy, 
only one such case has been found so far
\citep[\object{$\mu$ Cen},~][]{rivi98}.  But several photometrically multiperiodic candidates 
were recently identified by \citet{marta2007b} 
and \citet{diago08} in the SMC.  Moreover, the latter authors report 
an order of magnitude larger incidence of pulsations among Be stars 
than in B stars without emission lines.  

In summary, the results of this work indicate that, at least for
single stars, the Be phenomenon is coupled to (initial) mass,
evolutionary stage, and metallicity.  But it is not evident that these
3 parameters are primary quantities determining the prevalence of the
Be phenomenon.  A more physical description, in accordance with the
above, is the one by \citet{marta2007a}, who submit that the Be
phenomenon depends primarily on the evolution of \omc.  \omc~is governed by
evolutionary stage and metallicity but the dependencies are different
in different mass domains, which leads to the confusing apparent lack
of consistency or uniqueness of empirical studies of the Be phenomenon
at large (the larger the area ``imaged'', the more single trees seem to
stand out).


\section{Summary and conclusions}
A slitless spectroscopic survey for emission-line objects in the SMC
was performed.  Fourteen fields covered most of the SMC.  From 3
million spectra, about 8,120 spectra of 4,437 stars in 84 clusters and
14 nearby comparison fields were automatically selected.  The final
database comprises 122 definite main-sequence Oe/Be/Ae stars and 54
candidate emission-line stars, 1,659 main-sequence O/B/A stars, 2,408
other normal stars, and 90 emission-line stars not near the main
sequence. Fifty-five emission-line stars in NGC\,346 were also found
and classified; but their nature - main sequence or pre-main sequence
stars - is not clear so that they were not included in the statistics
and analysis.

Cross-correlation with the OGLE database permitted these emission-line
objects to be associated with homogeneous photometric data.  For 49
additional emission-line stars photometric data could not be derived.
While the survey is spatially homogeneous, it is starting to become
incomplete around B3 (on the main sequence). For comparison, similar
Galactic (but photometric) data from the work of \citet{mcs2005} were
converted to the same scales in absolute luminosity and effective
temperature.

Careful analysis led to the following conclusions:
\begin{list}{--}{\itemsep=0mm\topsep=0mm}
\item
An intercomparison of clusters did not furnish any dependency of the
relative frequency of emission-line stars on spatial density. 
\item
In the SMC, the Be phenomenon is more strongly enhanced towards
early-type stars (O stars) than in the Galaxy.  Among early spectral
sub-types, the fraction of Be stars in the SMC exceeds the one in the
Galaxy by a factor of $\sim$3-5.
\item
The largest number of Be stars is found at spectral type B2 both in
SMC and Galaxy.  Since also the emission-line strength is largest near
B2, this result is difficult to interpret in the presence of low
sensitivity to weak emission lines.
\item
In color-magnitude diagrams, most Be stars are found off the zero-age 
main sequence, with many of them defining a separate red ``sequence''.  
\item
The age distribution of clusters hosting Be stars shows that the Be 
phenomenon does cover the second half of the main-sequence evolution.  
Some Be stars may have formed as Be stars while others may have acquired 
their Be nature only during the course of their evolution.  There are not 
enough young clusters in the SMC sample to say anything about the 
Be phenomenon during the first half of the main-sequence evolution of SMC 
stars. 
\item
The observations are consistent with \omc~being one of the main quantities
governing the statistics of emission-line stars in all sub-samples of
single stars.  \omc\ rises slowly with time for intermediate 
and late B stars of all metallicities for massive B and O stars.  The same 
holds for early-type B stars with SMC metallicity.  Only massive Galactic 
OB stars are different in that their \omc\ decreases with time.  
When also evolution, initial mass, and metallicity are considered, the 
relative abundance of Be stars takes on a multi-parametric appearance.  But 
\omc\ still dominates.

\end{list}
The above trends only stand out significantly in sufficiently large
samples.  Seemingly very similar, if not identical, small samples
(e.g., single open star clusters) can differ drastically in their
number of Be stars.  To date, there is not even a speculative
explanation for this.  Maybe, a large variation in the initial
distribution of rotation rates combined with a threshold in
\omc~plays a role.   

In forthcoming articles, we shall present results of WFI slitless
H$\alpha$ spectroscopy in the SMC field (outside clusters) and in open clusters
and the field of the LMC.  The nature of emission-line stars far from
the main sequence will also be expanded on in a future paper.


\begin{acknowledgements}
The authors acknowledge the referee for valuable comments 
that helped to present the essence of the paper more clearly.
C.M.\ thanks Drs A.-M.\ Hubert, M.\ Floquet, and Y.\ Fr\'emat for
sharing useful information during the preliminary analysis of our
data.  Dr E.\ Bertin's adaptation of his {\it SExtractor} package
proved most helpful for the mass reduction of the observations.  This
research has made use of the Simbad and Vizier databases maintained at
CDS, Strasbourg, France, of NASA's Astrophysics Data System
Bibliographic Services, and of the NASA/IPAC Infrared Science Archive,
which is operated by the Jet Propulsion Laboratory, California
Institute of Technology, under contract with the U.S.\ National
Aeronautics and Space Administration.  
This publication makes use of data products from the Two Micron All Sky Survey, 
which is a joint project of the University of Massachusetts and the Infrared 
Processing and Analysis Center/California Institute of Technology, 
funded by the U.S.\ National Aeronautics and Space Administration and the 
U.S.\ National Science Foundation.
C.M.\ is grateful for support from ESO's DGDF in 2006.
\end{acknowledgements}

\bibliographystyle{aa}
\bibliography{articlesbib}



\appendix
\onecolumn

\section{Comments on individual open clusters}
\label{comocl}

\subsection{NGC\,330 (Ogle-SMC107)}

NGC\,330 is a well-studied open cluster known for its high content of
Be stars \citep[see for example][]{keller99b}.  The area covered by 
NGC\,330 is the largest one of all clusters in this paper.  A total
of 400 spectra was examinded and 55 emission-line stars were
identified.  Cross-matching with OGLE photometry provided indications
that the majority of the latter are actually Be stars.

\subsection{Bruck\,60 (Ogle-SMC72)}

Bruck\,60 is one of the six SMC open clusters observed by
\citet{wis2006}. They found 26 Be stars, among them 6 candidates,
within a radius of 1.5\arcmin~while the present study extended over a
radius of 35\arcsec~centered on the cluster.  The two areas have 17
stars in common, of which {\it Album} rejected 8 highly blended
sources. The detection rate of 9/17
stars is consistent with the estimated general extraction efficiency
of 60\% in the region of Bruch\,60.  Of the 9 detected stars
(WBBe5, WBBe6, WBBe7, WBBe10, WBBe17, WBBe20, WBBe21, WBBe23, WBBe25),
2 (WBBe23 and WBBe25) are not properly separated and were also eliminated.
Of the 7 remaining emission-line stars, 4 (WBBe5,
WBBe10, WBBe17, WBBe20) are in common to both studies.  One star
(WBBe7), for which \citet{wis2007b} did not publish polarimetry, is
found without emission.  The two others (WBBe6 and WBBe21) are faint and
have a too low S/N to provide a reliable conclusion about the presence
of emission in their spectra.  The stars poorly separated and/or with
low S/N have V magnitudes of 17.4, 17.6, 18, and 18.  Finally, the
present study finds one candidate emission-line star not identified by 
\citet{wis2006}.

\subsection{NGC\,346}
\label{infosngc346}
This open cluster is highly complex with various sub-aggregates.  
\citet{bouret2003} published an age of 3 Myears and
a metallicity of 0.004. However, \citet{nota2006} found several
populations with different ages: 4.5 Gyears for stars in the field, a
young population with ages ranging from 3 to 5 Myears, in which stars
with a mass less than 3 M$_{\odot}$ are still pre-main sequence stars,
and a population with an intermediate age of 150 Myears.

In this open cluster and in its vicinity, we found 55 emission line
stars.  Dereddening with a global value of 0.008 mag for massive stars
from \citet{henne08} suggests that most of them are on the main sequence.  
The spectral classification obtained differs on average by 
1 spectral sub-type for the 12 stars shared with
spectral classifications by other authors.

Table~\ref{tab1ngc346} provides basic parameters for the 55
emission-line stars in NGC\,346 along with magnitudes from the EIS
pre-flames survey \citep{eis01}, from DENIS \citep{denis}, 
and from 2MASS \citep{2mass}. In Table~\ref{tab2ngc346}, other
parameters are given as well as spectral classifications from WFI and
other sources.  Owing to the large spread in age of the stars in this
cluster, some of them could be pre-main sequence stars.  Where
possible, cross-references to the studies by \citet{henne08},
\citet{wis2006}, \citet{wis2007b}, and \citet{hunter2008} and to the Simbad 
database are, therefore, also included.

Because the distinction between pre-main sequence (Herbig Ae/Be or 
T~Tauri stars), main-sequence (mostly classical Be stars), and post-main
sequence emission-line stars (e.g., WR) requires more spectroscopic
data with higher spectral resolution and coverage, but also membership
in the different sub-clusters, the stars from NGC\,346 were not
included in the overall analyses of this paper.  Their large number
could have introduced biases.

\begin{figure}[h!]
\centering
\includegraphics[width=18cm, height=23cm, angle=-180]{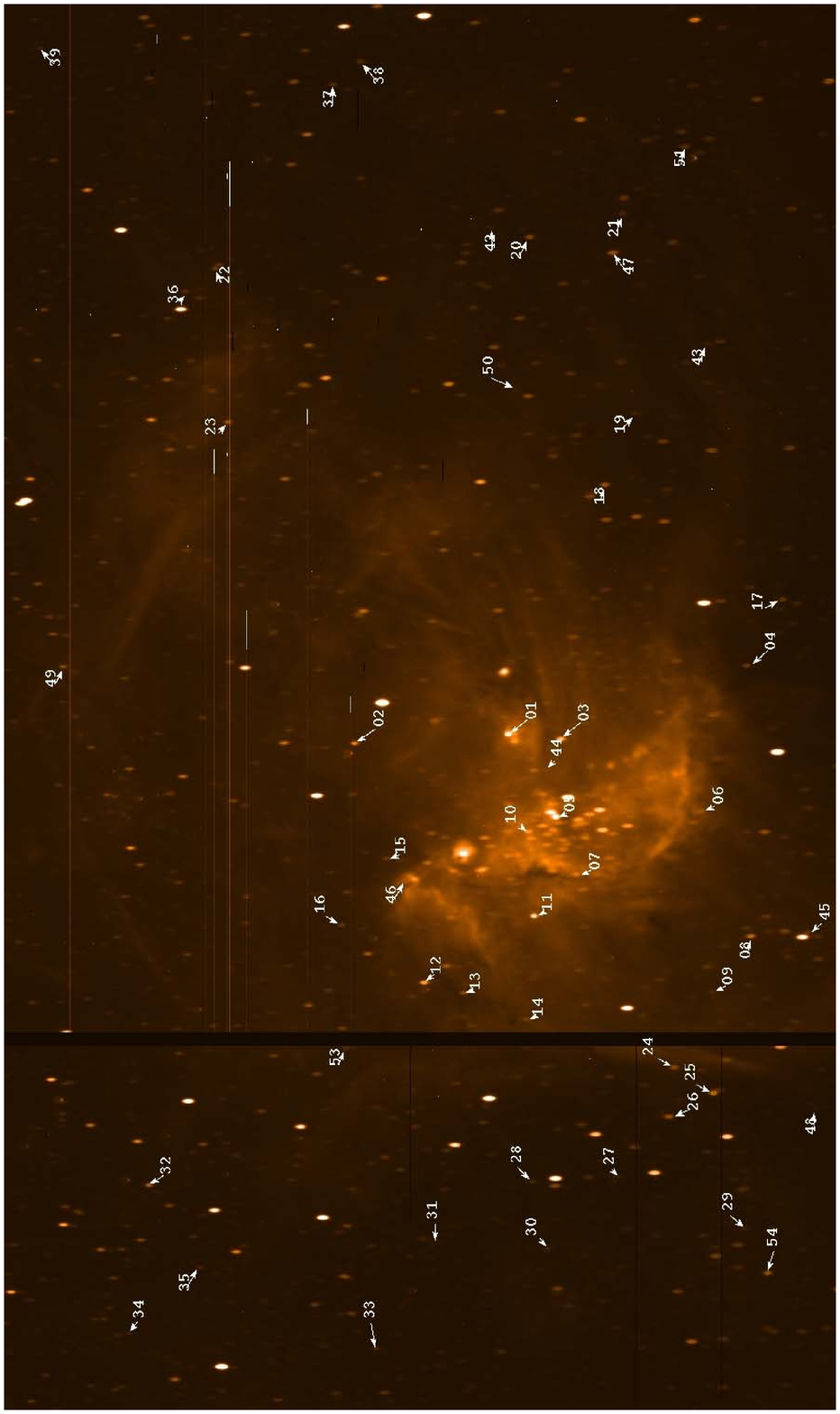}
\caption{Part of a WFI spectral image showing NGC\,346 and its vicinity. 
The sources appear elongated due their spectral nature.  North is at
the top, west to the right.  The broad black vertical line is due to a gap
between CCDs in the WFI mosaic.  Arrows with numbers identify
most of the emission-line stars, which can be found in
Table~\ref{tab1ngc346}. This Figure shows that the slitless 
spectroscopy allows to find circumstellar (CS) emission-line stars while slit-spectroscopy 
cannot disentangle CS and nebular emission lines in diffuse emission nebulae.
According to \citet{henne08}, there are many pre main sequence T~Tauri stars in 
the central regions of the cluster.  But they are too
faint to be detected here.}
\label{mapngc346}
\end{figure}

Fig.~\ref{mapngc346} shows part of a WFI image with NGC\,346. Most of
the emission-line stars are identified.  For the central parts of the cluster(s) 
\citet{henne08} report numerous pre-main sequence T~Tauri stars, which are unfortunately 
too faint (see Fig.~\ref{effdetect}) to be detected by the present study.
Fig.~\ref{mapngc346} also illustrates the ability of the WFI slitless spectroscopy 
to distinguish true circumstellar line emission from diffuse nebular emission.  
In low-resolution short-slit spectra, the difference can be marginal.  

\begin{flushleft}
\begin{landscape}

\tiny{
\begin{table*}[]
\caption{Table 1 of emission-line star in NGC\,346}
\centering
\begin{tabular}{@{\ }l@{\ \ }l@{\ \ }l@{\ \ }l@{\ \ }l@{\ \ }l@{\ \ }l@{\ \ }l@{\ \ }l@{\ \ }l@{\ \ }l@{\ \ }l@{\ \ }l@{\ \ }l@{\ \ }l@{\ \ }l@{\ \ }l@{\ \ }l@{\ \ }l@{\ }}
\hline
\hline
ID WFI & RA(2000)   & DEC(2000)   & RA(2000)   & DEC(2000)   & RA(2000)   & DEC(2000)   & B$_{EIS}$ & V$_{EIS}$ & I$_{Denis}$ & J$_{2M}$ & H$_{2M}$ & K$_{2M}$ \\
 & WFI   & WFI   & 2MASS   & 2MASS   & EIS   & EIS   &  &  &  &  &  &  \\
\hline
\hline
WFI[S11] NGC\,346-01 & 0 59 12.284 & -72 9 58.45 & 0 59 12.23 & -72 9 58.5 & 0 59 11.615 & -72 9 57.52 & 14.818 & 15.072 & \_ & 14.152 & 14.381 & 13.046 \\
WFI[S11] NGC\,346-02 & 0 59 28.866 & -72 10 16.68 & 0 59 28.76 & -72 10 16.7 & 0 59 28.750 & -72 10 16.55 & 15.041 & 15.122 & 15.035 & 14.958 & 14.879 & 14.651 \\
WFI[S11] NGC\,346-03 & 0 59 06.430 & -72 9 56.35 & 0 59 06.33 & -72 9 56.1 & 0 59 6.337 & -72 9 56.06 & 14.131 & 14.281 & 14.496 & 14.551 & 14.525 & 14.283 \\
WFI[S11] NGC\,346-04 & 0 58 47.641 & -72 9 03.11 & 0 58 47.47 & -72 9 03.0 & 0 58 47.525 & -72 9 2.70 & 15.580 & 15.729 & 15.722 & 15.773 & 15.946 & 15.815 \\
WFI[S11] NGC\,346-05 & 0 59 05.547 & -72 10 35.68 & 0 59 05.43 & -72 10 35.5 & \_ \_ \_ & \_ \_ \_ & 15.38$^{1}$ & 15.28$^{1}$ & \_ & 13.013 & 13.262 & 11.949 \\
WFI[S11] NGC\,346-06 & 0 58 49.709 & -72 10 19.86 & 0 58 49.55 & -72 10 19.6 & 0 58 49.609 & -72 10 19.52 & 15.780 & 15.889 & 16.098 & 15.498 & 15.368 & 15.327 \\
WFI[S11] NGC\,346-07 & 0 59 02.093 & -72 11 02.55 & 0 59 02.03 & -72 11 02.5 & 0 59 2.056 & -72 11 2.37 & 15.778 & 15.803 & 15.730 & 15.578 & 15.600 & 14.824 \\
WFI[S11] NGC\,346-08 & 0 58 41.925 & -72 11 18.16 & 0 58 41.80 & -72 11 17.9 & 0 58 41.861 & -72 11 17.55 & 14.849 & 14.919 & 14.982 & 14.891 & 14.593 & 14.707 \\
WFI[S11] NGC\,346-09 & 0 58 45.074 & -72 11 49.46 & \_ \_ \_ & \_ \_ \_ & 0 58 45.043 & -72 11 48.98 & 16.684 & 16.806 & \_ & \_ & \_ & \_ \\
WFI[S11] NGC\,346-10 & 0 59 08.214 & -72 10 45.57 & 0 59 08.13 & -72 10 45.4 & 0 59 8.141 & -72 10 45.25 & 15.619 & 15.652 & 15.257 & 15.226 & 14.713 & 14.407 \\
WFI[S11] NGC\,346-11 & 0 59 05.972 & -72 11 27.42 & 0 59 05.88 & -72 11 27.0 & 0 59 5.873 & -72 11 26.98 & 16.125 & 15.863 & 15.107 & 14.557 & 13.892 & 12.808 \\
WFI[S11] NGC\,346-12 & 0 59 16.683 & -72 12 10.36 & 0 59 16.65 & -72 12 10.2 & 0 59 16.613 & -72 12 10.04 & 14.687 & 16.638 & 14.480 & 14.273 & 14.280 & 14.059 \\
WFI[S11] NGC\,346-13 & 0 59 12.198 & -72 12 12.03 & 0 59 12.14 & -72 12 11.9 & 0 59 12.119 & -72 12 11.69 & 15.926 & 15.874 & 15.659 & 15.631 & 15.249 & 15.300 \\
WFI[S11] NGC\,346-14 & 0 59 04.794 & -72 12 19.78 & 0 59 04.65 & -72 12 19.7 & 0 59 4.613 & -72 12 19.46 & 16.352 & 16.471 & 16.141 & 16.339 & 16.175 & 15.551 \\
WFI[S11] NGC\,346-15 & 0 59 23.150 & -72 11 11.41 & 0 59 23.08 & -72 11 11.3 & 0 59 23.091 & -72 11 11.32 & 16.313 & 16.554 & \_ & 16.416 & 16.339 & 14.471 \\
WFI[S11] NGC\,346-16 & 0 59 26.818 & -72 11 48.86 & 0 59 26.77 & -72 11 48.8 & 0 59 26.718 & -72 11 48.56 & 16.415 & 16.567 & 16.680 & 16.628 & 16.695 & 15.369 \\
WFI[S11] NGC\,346-17 & 0 58 45.003 & -72 8 26.92 & 0 58 44.86 & -72 8 26.6 & 0 58 44.912 & -72 8 26.41 & 16.136 & 16.158 & 15.994 & 15.967 & 15.644 & 15.411 \\
WFI[S11] NGC\,346-18 & 0 59 06.540 & -72 7 45.18 & 0 59 06.39 & -72 7 45.2 & 0 59 6.382 & -72 7 44.97 & 14.748 & 14.919 & 14.962 & 15.097 & 15.073 & 14.899 \\
WFI[S11] NGC\,346-19 & 0 59 04.612 & -72 7 06.70 & 0 59 04.46 & -72 7 07.0 & 0 59 4.419 & -72 7 6.52 & 16.423 & 16.584 & 16.645 & 16.654 & 17.039 & 16.875 \\
WFI[S11] NGC\,346-20 & 0 59 19.462 & -72 5 47.83 & 0 59 19.33 & -72 5 48.0 & 0 59 19.283 & -72 5 47.80 & 15.559 & 15.684 & 15.617 & 15.571 & 15.544 & 15.128 \\
WFI[S11] NGC\,346-21 & 0 59 09.813 & -72 5 27.92 & 0 59 09.78 & -72 5 28.2 & 0 59 9.673 & -72 5 27.78 & 16.483 & 16.589 & 16.480 & 16.386 & 16.142 & 15.439 \\
WFI[S11] NGC\,346-22 & 0 59 53.038 & -72 6 31.07 & 0 59 52.90 & -72 6 31.1 & 0 59 52.925 & -72 6 31.06 & 16.404 & 16.458 & \_ & 15.518 & 15.638 & 15.158 \\
WFI[S11] NGC\,346-23 & 0 59 48.814 & -72 7 47.02 & 0 59 48.69 & -72 7 47.3 & 0 59 48.680 & -72 7 46.97 & 15.864 & 15.866 & 15.575 & 15.730 & 15.527 & 15.328 \\
WFI[S11] NGC\,346-24 & 0 58 47.609 & -72 12 36.66 & 0 58 47.43 & -72 12 36.8 & 0 58 47.499 & -72 12 36.57 & 15.263 & 15.471 & 15.735 & 15.393 & 15.193 & 14.938 \\
WFI[S11] NGC\,346-25 & 0 58 42.821 & -72 12 46.02 & \_ \_ \_ & \_ \_ \_ & 0 58 42.719 & -72 12 45.83 & 14.840 & 14.961 & 14.617 & \_ & \_ & \_ \\
WFI[S11] NGC\,346-26 & 0 58 47.215 & -72 13 01.68 & \_ \_ \_ & \_ \_ \_ & 0 58 47.104 & -72 13 1.57 & 14.662 & 14.781 & 14.562 & \_ & \_ & \_ \\
WFI[S11] NGC\,346-27 & 0 58 51.670 & -72 13 36.21 & \_ \_ \_ & \_ \_ \_ & 0 58 51.518 & -72 13 36.01 & 16.993 & 17.069 & 17.143 & \_ & \_ & \_ \\
WFI[S11] NGC\,346-28 & 0 59 00.971 & -72 13 46.43 & 0 59 00.87 & -72 13 46.5 & 0 59 0.873 & -72 13 46.39 & 16.960 & 16.873 & 16.578 & 16.241 & 15.895 & 15.066 \\
WFI[S11] NGC\,346-29 & 0 58 36.753 & -72 13 51.03 & \_ \_ \_ & \_ \_ \_ & 0 58 36.599 & -72 13 50.79 & 16.673 & 16.903 & 16.952 & \_ & \_ & \_ \\
WFI[S11] NGC\,346-30 & 0 58 58.031 & -72 14 18.79 & \_ \_ \_ & \_ \_ \_ & 0 58 57.922 & -72 14 18.69 & 17.116 & 17.201 & \_ & \_ & \_ & \_ \\
\hline
\end{tabular}
\label{tab1ngc346}
\end{table*}
}
\tiny{
\addtocounter{table}{-1}
\begin{table*}[tbph]
\caption{continued}
\centering
\begin{tabular}{@{\ }l@{\ \ }l@{\ \ }l@{\ \ }l@{\ \ }l@{\ \ }l@{\ \ }l@{\ \ }l@{\ \ }l@{\ \ }l@{\ \ }l@{\ \ }l@{\ \ }l@{\ \ }l@{\ \ }l@{\ \ }l@{\ \ }l@{\ \ }l@{\ }}
\hline
\hline
ID WFI & RA(2000)   & DEC(2000)   & RA(2000)   & DEC(2000)   & RA(2000)   & DEC(2000)   & B$_{EIS}$ & V$_{EIS}$ & I$_{Denis}$ & J$_{2M}$ & H$_{2M}$ & K$_{2M}$ \\
 & WFI   & WFI   & 2MASS   & 2MASS   & EIS   & EIS   &  &  &  &  &  &  \\
\hline
WFI[S11] NGC\,346-31 & 0 59 10.243 & -72 14 24.83 & \_ \_ \_ & \_ \_ \_ & 0 59 10.147 & -72 14 24.68 & 17.571 & 17.700 & 17.526 & \_ & \_ & \_ \\
WFI[S11] NGC\,346-32 & 0 59 42.773 & -72 14 22.03 & 0 59 42.73 & -72 14 22.2 & 0 59 42.713 & -72 14 21.88 & 15.370 & 15.526 & 15.205 & 15.193 & 15.041 & 14.753 \\
WFI[S11] NGC\,346-33 & 0 59 14.828 & -72 15 23.78 & 0 59 14.72 & -72 15 23.8 & 0 59 14.723 & -72 15 23.59 & 16.542 & 16.623 & 16.442 & 16.201 & 16.435 & 16.007 \\
WFI[S11] NGC\,346-34 & 0 59 42.097 & -72 15 37.94 & \_ \_ \_ & \_ \_ \_ & 0 59 42.048 & -72 15 37.68 & 17.300 & 17.412 & 17.315 & \_ & \_ & \_ \\
WFI[S11] NGC\,346-35 & 0 59 35.677 & -72 14 58.81 & \_ \_ \_ & \_ \_ \_ & 0 59 35.597 & -72 14 58.63 & 16.869 & 17.030 & 17.042 & \_ & \_ & \_ \\
WFI[S11] NGC\,346-36 & 0 59 55.900 & -72 6 45.42 & 0 59 55.79 & -72 6 45.5 & \_ \_ \_ & \_ \_ \_ & 14.61$^{1}$ & 14.39$^{1}$ & 15.934 & 15.969 & 15.938 & 15.183 \\
WFI[S11] NGC\,346-37 & 0 59 43.714 & -72 4 48.81 & 0 59 43.56 & -72 4 49.2 & 0 59 43.564 & -72 4 48.75 & 16.143 & 16.299 & 16.185 & 16.402 & 15.668 & 15.082 \\
WFI[S11] NGC\,346-38 & 0 59 41.155 & -72 4 34.93 & 0 59 41.12 & -72 4 34.9 & 0 59 41.004 & -72 4 34.88 & 16.290 & 16.402 & 16.138 & 16.324 & 15.714 & 15.806 \\
WFI[S11] NGC\,346-39 & 1 00 16.196 & -72 04 56.22 & \_ \_ \_ & \_ \_ \_ & 1 0 16.125 & -72 4 55.81 & 16.966 & 17.151 & \_ & \_ & \_ & \_ \\
WFI[S11] NGC\,346-40 & 0 59 58.161 & -72 04 04.95 & \_ \_ \_ & \_ \_ \_ & 0 59 58.015 & -72 4 4.83 & 16.925 & 17.047 & 17.040 & \_ & \_ & \_ \\
WFI[S11] NGC\,346-41 & 0 59 33.396 & -72 02 21.80 & \_ \_ \_ & \_ \_ \_ & 0 59 33.203 & -72 2 21.50 & 16.616 & 16.786 & 16.888 & \_ & \_ & \_ \\
WFI[S11] NGC\,346-42 & 0 59 23.963 & -72 05 46.42 & \_ \_ \_ & \_ \_ \_ & 0 59 23.787 & -72 5 46.37 & \_ & 17.625 & 17.689 & \_ & \_ & \_ \\
WFI[S11] NGC\,346-43 & 0 58 58.183 & -72 06 25.49 & \_ \_ \_ & \_ \_ \_ & 0 58 58.002 & -72 6 25.13 & 16.571 & 16.723 & 16.556 & \_ & \_ & \_ \\
WFI[S11] NGC\,346-44 & 0 59 07.577 & -72 10 12.96 & \_ \_ \_ & \_ \_ \_ & 0 59 7.413 & -72 10 12.82 & 17.623 & 17.749 & \_ & \_ & \_ & \_ \\
WFI[S11] NGC\,346-45 & 0 58 35.686 & -72 11 10.86 & \_ \_ \_ & \_ \_ \_ & \_ \_ \_ & \_ \_ \_ & 16.20$^{1}$ & 16.29$^{1}$ & \_ & \_ & \_ & \_ \\
WFI[S11] NGC\,346-46 & 0 59 20.713 & -72 11 21.51 & \_ \_ \_ & \_ \_ \_ & \_ \_ \_ & \_ \_ \_ & 18.37$^{1}$ & 18.63$^{1}$ & \_ & \_ & \_ & \_ \\
WFI[S11] NGC\,346-47 & 0 59 10.076 & -72 5 48.40 & 0 59 09.98 & -72 5 48.5 & 0 59 9.933 & -72 5 48.21 & 14.184 & 14.479 & 14.593 & 14.933 & 14.978 & 15.031 \\
WFI[S11] NGC\,346-48 & 0 58 31.435 & -72 12 44.87 & \_ \_ \_ & \_ \_ \_ & 0 58 31.241 & -72 12 44.81 & 17.640 & 17.746 & \_ & \_ & \_ & \_ \\
WFI[S11] NGC\,346-49 & 1 0 02.121 & -72 10 03.75 & 1 0 02.09 & -72 10 03.8 & 1 0 2.062 & -72 10 3.59 & 15.784 & 15.947 & 15.821 & 16.003 & 15.697 & 15.115 \\
WFI[S11] NGC\,346-50 & 0 59 18.221 & -72 07 04.26 & \_ \_ \_ & \_ \_ \_ & 0 59 18.103 & -72 7 4.32 & 17.843 & 17.973 & \_ & \_ & \_ & \_ \\
WFI[S11] NGC\,346-51 & 0 59 4.141 & -72 4 48.72 & 0 59 4.2 & -72 4 48.8 & 0 59 4.141 & -72 4 48.72 & 15.601 & 15.772 & 16.077 & 16.2 & 15.882 & 15.332 \\
WFI[S11] NGC\,346-52 & 0 59 6.487 & -72 2 37.27 & 0 59 6.51 & -72 2 37.5 & 0 59 6.487 & -72 2 37.27 & 14.987 & 14.415 & 13.674 & 13.234 & 12.92 & 12.824 \\
WFI[S11] NGC\,346-53 & 0 59 26.460 & -72 13 11.78 & 0 59 26.51 & -72 13 11.9 & 0 59 26.460 & -72 13 11.78 & 15.262 & 15.382 & 15.713 & 15.883 & 15.947 & 16.793 \\
WFI[S11] NGC\,346-54 & 0 58 33.256 & -72 14 10.95 & 0 58 33.21 & -72 14 11.2 & 0 58 33.253 & -72 14 10.92 & 15.627 & 15.28$^{1}$ & 15.266 & 15.233 & 15.26 & 14.834 \\
WFI[S11] NGC\,346-55 & 0 59 26.523 & -72 9 54.06 & 0 59 26.57 & -72 9 54.1 & 0 59 26.513 & -72 9 53.84 & 11.619 & 11.5$^{1}$ & 11.029 & 11.111 & 11.006 & 10.769 \\
\hline
\end{tabular}
\label{tab1ngc346}
\begin{flushleft}
$^{1}$: from Simbad
\end{flushleft}

\end{table*}
}

\tiny{
\begin{table*}[]
\caption{Table 2 of emission-line star in NGC\,346: the other classification of stars come from \citet[VFS]{hunter2008}, \citet[WB07]{wis2007b},
\citet[spitzer]{nota2006}, and \citet[H08]{henne08}}
\centering
\begin{tabular}{@{\ }l@{\ \ }l@{\ \ }l@{\ \ }l@{\ \ }l@{\ \ }l@{\ \ }l@{\ \ }l@{\ \ }l@{\ \ }l@{\ \ }l@{\ \ }l@{\ \ }l@{\ \ }l@{\ \ }l@{\ \ }l@{\ \ }l@{\ \ }l@{\ \ }l@{\ }}
\hline
\hline
Id WFI & Id EIS & Id Simbad & (B-V) & E[B-V] & (B-V)0 & V0 & Mv & ST & class Simbad & class VFS & class WB07 & class  & class H08 \\
 &  &  &  & from H08 &  &  &  & WFI &  &  &  & Spitzer &  \\
\hline
\hline
NGC\,346-01 & SMC5\_089286 & KWBBe448 & -0.254 & 0.08 & -0.334 & 14.82 & -3.93 & B0: & \_ &  &  & N28, class I &  \\
NGC\,346-02 & SMC5\_056563 & KWBBe248 & -0.081 & 0.22 & -0.301 & 14.44 & -4.31 & O9: & B0V &  & Be(2) &  & B0V (19) \\
NGC\,346-03 & SMC5\_030226 & MPG482 & -0.150 & 0.18 & -0.330 & 13.72 & -5.03 & O7: & B0.5V: &  &  &  & B0.5V (14) \\
NGC\,346-04 & SMC5\_083005 & KWBBe379 & -0.149 & 0.08 & -0.229 & 15.48 & -3.27 & B0: & \_ & B3e & not Be(3) & N69, class III &  \\
NGC\,346-05 & \_ & MPG454 & 0.100 & 0.08 & 0.020 & 15.03 & -3.72 & B0: & \_ &  &  &  &  \\
NGC\,346-06 & SMC5\_029906 & NMC47 & -0.109 & 0.08 & -0.189 & 15.64 & -3.11 & B1: & B1V & B1Ve &  & N6, class III &  \\
NGC\,346-07 & SMC5\_029282 & KWBBe191 & -0.025 & 0.08 & -0.105 & 15.56 & -3.20 & B1: & \_ &  & Be(1.5) &  &  \\
NGC\,346-08 & SMC5\_029130 & NMC45 & -0.070 & 0.08 & -0.150 & 14.67 & -4.08 & B0: & B0.2e & B0.2e & Be(2) &  &  \\
NGC\,346-09 & SMC5\_028674 & MPG208 & -0.122 & 0.08 & -0.202 & 16.56 & -2.19 & B2: & \_ &  &  &  &  \\
NGC\,346-10 & SMC5\_029509 & KWBBe212 & -0.033 & 0.08 & -0.113 & 15.40 & -3.35 & B0: & \_ &  &  & N90?, class II &  \\
NGC\,346-11 & SMC5\_081533 & KWBBe200 & 0.262 & 0.08 & 0.182 & 15.62 & -3.14 & B1: & Be &  & notBe, B[e] &  &  \\
NGC\,346-12 & SMC5\_055483 & KWBBe93 & -1.951 & 0.08 & -2.031 & 16.39 & -2.36 & B2$^{1}$: & \_ &  & Be(1) &  &  \\
NGC\,346-13 & SMC5\_055468 & KWBBe445 & 0.052 & 0.08 & -0.028 & 15.63 & -3.12 & B1: & B2e & B2e & Be(1.5) &  &  \\
NGC\,346-14 & SMC5\_078460 & KWBBe807 & -0.119 & 0.08 & -0.199 & 16.22 & -2.53 & B1: & \_ &  & Be(2) &  &  \\
NGC\,346-15 & SMC5\_188572 & KWBBe908 & -0.241 & 0.08 & -0.321 & 16.31 & -2.44 & B2: & \_ &  & Be(2) &  &  \\
NGC\,346-16 & SMC5\_028654 & KWBBe921 & -0.152 & 0.08 & -0.232 & 16.32 & -2.43 & B2: & \_ &  &  &  &  \\
NGC\,346-17 & SMC5\_006366 & KWBBe374 & -0.022 & 0.08 & -0.102 & 15.91 & -2.84 & B1: & \_ &  & Be(1.5) &  &  \\
NGC\,346-18 & SMC5\_078074 & KWBBe205 & -0.171 & 0.08 & -0.251 & 14.67 & -4.08 & B0: & \_ & B2esh & Be(2) &  &  \\
NGC\,346-19 & SMC5\_006627 & KWBBe814 & -0.161 & 0.08 & -0.241 & 16.34 & -2.41 & B2: & \_ &  & Be(2) &  &  \\
NGC\,346-20 & SMC5\_033513 & KWBBe236 & -0.125 & 0.08 & -0.205 & 15.44 & -3.31 & B0: & B0.5esh & B0.5esh & Be(1) &  &  \\
NGC\,346-21 & SMC5\_006934 & KWBBe856 & -0.106 & 0.08 & -0.186 & 16.34 & -2.41 & B2: & \_ &  & Be(2) &  &  \\
NGC\,346-22 & SMC5\_032946 & \_ & -0.054 & 0.08 & -0.134 & 16.21 & -2.54 & B1: & \_ &  &  &  &  \\
NGC\,346-23 & SMC5\_031915 & KWBBe266 & -0.002 & 0.08 & -0.082 & 15.62 & -3.13 & B1: & \_ &  & Be(2) &  &  \\
NGC\,346-24 & SMC5\_028019 & KWBBe377 & -0.208 & 0.08 & -0.288 & 15.22 & -3.53 & B0: & Be & B3esh & Be(2) &  &  \\
NGC\,346-25 & SMC5\_079464 & KWBBe171 & -0.121 & 0.08 & -0.201 & 14.71 & -4.04 & B0: & Em &  & Be(2) &  &  \\
NGC\,346-26 & SMC5\_038701 & MPG217 & -0.119 & 0.08 & -0.199 & 14.53 & -4.22 & O9: & O9.5IIIe & 09.5IIIe &  &  &  \\
NGC\,346-27 & SMC5\_054519 & \_ & -0.076 & 0.08 & -0.156 & 16.82 & -1.93 & B2: & \_ &  &  &  &  \\
NGC\,346-28 & SMC5\_054474 & KWBBe778 & 0.087 & 0.08 & 0.007 & 16.63 & -2.13 & B2: & \_ &  &  &  &  \\
NGC\,346-29 & SMC5\_092842 & \_ & -0.230 & 0.08 & -0.310 & 16.66 & -2.10 & B2: & \_ &  &  &  &  \\
NGC\,346-30 & SMC5\_026531 & \_ & -0.085 & 0.08 & -0.165 & 16.95 & -1.80 & B3: & \_ &  &  &  &  \\
\hline
\end{tabular}
\label{tab2ngc346}
\end{table*}
}
\tiny{
\addtocounter{table}{-1}
\begin{table*}[tbph]
\caption{continued}
\centering
\begin{tabular}{@{\ }l@{\ \ }l@{\ \ }l@{\ \ }l@{\ \ }l@{\ \ }l@{\ \ }l@{\ \ }l@{\ \ }l@{\ \ }l@{\ \ }l@{\ \ }l@{\ \ }l@{\ \ }l@{\ \ }l@{\ \ }l@{\ \ }l@{\ \ }l@{\ }}
\hline
\hline
Id WFI & Id EIS & Id Simbad & (B-V) & E[B-V] & (B-V)0 & V0 & Mv & ST & class Simbad & class VFS & class WB07 & class  & class H08 \\
 &  &  &  & from H08 &  &  &  & WFI &  &  &  & Spitzer &  \\
\hline
NGC\,346-31 & SMC5\_061961 & \_ & -0.129 & 0.08 & -0.209 & 17.45 & -1.30 & B4: & \_ &  &  &  &  \\
NGC\,346-32 & SMC5\_005173 & KWBBe259 & -0.156 & 0.08 & -0.236 & 15.28 & -3.47 & B0: & Be &  & Be(2) &  &  \\
NGC\,346-33 & SMC5\_068413 & \_ & -0.081 & 0.08 & -0.161 & 16.38 & -2.38 & B2: & \_ &  &  &  &  \\
NGC\,346-34 & SMC5\_068336 & \_ & -0.112 & 0.08 & -0.192 & 17.16 & -1.59 & B3: & \_ &  &  &  &  \\
NGC\,346-35 & SMC5\_053538 & \_ & -0.161 & 0.08 & -0.241 & 16.78 & -1.97 & B2: & \_ &  &  &  &  \\
NGC\,346-36 & \_ & SMC47551 & 0.220 & 0.08 & 0.140 & 14.14 & -4.61 & O8: & \_ &  &  &  &  \\
NGC\,346-37 & SMC5\_007036 & [MA93]1161 & -0.156 & 0.08 & -0.236 & 16.05 & -2.70 & B1: & \_ &  &  &  &  \\
NGC\,346-38 & SMC5\_081044 & [MA93]1160 & -0.112 & 0.08 & -0.192 & 16.15 & -2.60 & B1: & \_ &  &  &  &  \\
NGC\,346-39 & SMC5\_034165 & \_ & -0.185 & 0.08 & -0.265 & 16.90 & -1.85 & B2: & \_ &  &  &  &  \\
NGC\,346-40 & SMC5\_034799 & \_ & -0.122 & 0.08 & -0.202 & 16.80 & -1.95 & B2: & \_ &  &  &  &  \\
NGC\,346-41 & SMC5\_007493 & [MA93]1154 & -0.170 & 0.08 & -0.250 & 16.54 & -2.21 & B2: & \_ &  &  &  &  \\
NGC\,346-42 & SMC5\_228705 & \_ & \_ & 0.08 & \_ & 17.38 & -1.37 & B4:? & \_ &  &  &  &  \\
NGC\,346-43 & SMC5\_033051 & \_ & -0.152 & 0.08 & -0.232 & 16.48 & -2.28 & B2: & \_ &  &  &  &  \\
NGC\,346-44 & SMC5\_069688 & MPG498 & -0.126 & 0.08 & -0.206 & 17.50 & -1.25 & B4: & \_ &  &  &  &  \\
NGC\,346-45 & \_ & KWBBe350 & -0.090 & 0.08 & -0.170 & 16.04 & -2.71 & B1: & \_ &  & Be(2) &  &  \\
NGC\,346-46 & \_ & MPG698 & -0.260 & 0.08 & -0.340 & 18.38 & -0.37 & B6: & \_ &  &  &  & HBe?(12) \\
NGC\,346-47 & SMC5\_033514 & KWBBe89 & -0.295 & 0.08 & -0.375 & 14.23 & -4.52 & O8: & O7Iab &  & Be(2) &  &  \\
NGC\,346-48 & SMC5\_055109 & MPG108 & -0.106 & 0.08 & -0.186 & 17.50 & -1.25 & B4: & \_ &  &  &  &  \\
NGC\,346-49 & SMC5\_030040 & KWBBe543 & -0.163 & 0.08 & -0.243 & 15.70 & -3.05 & B1: & Em &  &  &  &  \\
NGC\,346-50 & SMC5\_058373 & \_ & -0.130 & 0.08 & -0.210 & 17.73 & -1.03 & B4: & \_ &  &  &  &  \\
NGC\,346-51 & SMC5\_076404 & \_ & -0.171 & 0.08 & -0.251 & 15.52 & -3.23 & B1: & \_ & B0Ve-bin &  &  &  \\
NGC\,346-52 & SMC5\_035903 & \_ & 0.572 & 0.08 & 0.492 & 14.17 & -4.58 & O8$^{1}$: & \_ &  &  &  &  \\
NGC\,346-53 & SMC5\_081656 & MPG753 & -0.120 & 0.08 & -0.200 & 15.13 & -3.62 & B0: & B1II & B1II &  &  &  \\
NGC\,346-54 & SMC5\_179751 & KWBBe152 & 0.347 & 0.08 & 0.267 & 15.03 & -3.72 & B0: & Em &  & Be(2) &  &  \\
NGC\,346-55 & SMC5\_207758 & HD5980 & 0.119 & 0.08 & 0.039 & 11.25 & -7.50 & WR & WR, EB, WNp &  & notBe &  &  \\
\hline
\end{tabular}
\label{tab2ngc346}
\begin{flushleft}
$^{1}$: not main sequence
\end{flushleft}

\end{table*}
}

\end{landscape}
\end{flushleft}
\twocolumn

\onecolumn

\section{Open clusters: tables}
Table~\ref{indcl} lists all open SMC clusters with their basic
properties as well as the WFI field(s) covering it.  Also shown are
the numbers and types of emission-line stars found in each WFI image.
The results after merger of multiple observations are contained in
Table~\ref{mergcl}.

\begin{flushleft}
\begin{landscape}

\tiny{
\begin{table*}[]
\caption{Data for individual observations of open clusters in the SMC. Several open clusters were observed twice or three times in different
images at different locations. For each cluster, we give the central
coordinates, the log(Age) from \citet{ageSMcocl}. The radii used in
this study is generally equal to 1.5$\times$ the radii from
\citet{ageSMcocl}. The area of metallicity from \citet{cioni06} is
given. For each observations, we provide in cols. 8, 9, 10, the number
of all kind of stars, of emission-line star, and candidate emission-line star. In col. 11, there is
the sum of emission-line star and candidate emission-line star. In col. 12, there is the ratio of
all emission-line star to all stars. The other next columns give the number of stars
for each category found in Ogle catalogues
\citep{oglemapsphoto}. After a comment about the open clusters, we
provide the ratio of number of WFI stars found in Ogle and emission-line star from
WFI found in Ogle}
\centering
\begin{tabular}{@{\ }l@{\ \ }l@{\ \ }l@{\ \ }l@{\ \ }l@{\ \ }l@{\ \ }l@{\ \ }l@{\ \ }l@{\ \ }l@{\ \ }l@{\ \ }l@{\ \ }l@{\ \ }l@{\ \ }l@{\ \ }l@{\ \ }l@{\ \ }l@{\ }}
\hline
\hline
Cluster & Image & RA(2000) & DEC(2000) & r & log(t) & Z & $^{a}$N* & $^{c}$ & $^{d}$ & $^{b}$ & $^{b/a}$  & Ntot & NELS & NELS? & comments & N* & ELS \\
        & WFI   &          &           & \arcsec &  &   &  & NELS & NELS? & c+d & \% & ogle & ogle & ogle  &  & ogle/wfi & ogle/wfi \\
\hline
\hline
SMC002 & 2 & 0 37 33.1 & -73 36 42.6 & 58 & 8.4 & 0.003-0.006 & 26 & 0 & 1 & 1 & 3.8 & 26 & 0 & 1 &  & 100.0 & 100.0 \\
SMC008 & 2 & 0 40 30.54 & -73 24 10.4 & 53 & 8.0 & 0.003-0.006 & 103 & 3 & 3 & 6 & 5.8 & 102 & 3 & 3 &  & 99.0 & 100.0 \\
SMC008 & 4 & 0 40 30.54 & -73 24 10.4 & 53 & 8.0 & 0.003-0.006 & 59 & 1 & 1 & 2 & 3.4 & 59 & 1 & 1 &  & 100.0 & 100.0 \\
SMC009 & 2 & 0 40 44.11 & -73 23 0.2 & 45 & 8.0 & 0.003-0.006 & 73 & 2 & 2 & 4 & 5.5 & 64 & 1 & 2 &  & 87.7 & 75.0 \\
SMC011 & 1 & 0 41 6.16 & -73 21 7.1 & 45 & 7.9 & 0.003-0.006 & 36 & 0 & 2 & 2 & 5.6 & 33 & 0 & 2 &  & 91.7 & 100.0 \\
SMC011 & 2 & 0 41 6.16 & -73 21 7.1 & 45 & 7.9 & 0.003-0.006 & 65 & 3 & 2 & 5 & 7.7 & 64 & 3 & 2 &  & 98.5 & 100.0 \\
SMC012 & 1 & 0 41 23.78 & -72 53 27.1 & 76 & $>$9 & 0.006-0.009 & 151 & 0 & 2 & 2 & 1.3 & 65 & 0 & 0 & very old & 43.0 & 0.0 \\
SMC015 & 1 & 0 42 54.13 & -73 17 37 & 37 & 8.1 & 0.003-0.006 & 88 & 1 & 3 & 4 & 4.5 & 78 & 1 & 3 &  & 88.6 & 100.0 \\
SMC015 & 2 & 0 42 54.13 & -73 17 37 & 38 & 8.1 & 0.003-0.006 & 79 & 1 & 1 & 2 & 2.5 & 52 & 1 & 1 &  & 65.8 & 100.0 \\
SMC016 & 1 & 0 42 58.46 & -73 10 7.2 & 53 & 8.3 & 0.003-0.006 & 119 & 0 & 1 & 1 & 0.8 & 115 & 0 & 1 &  & 96.6 & 100.0 \\
SMC017 & 2 & 0 43 32.74 & -73 26 25.4 & 32 & 7.9 & 0.003-0.006 & 74 & 1 & 0 & 1 & 1.4 & 64 & 1 & 0 & $^{1}$ & 86.5 & 100.0 \\
SMC018 & 2 & 0 43 37.57 & -73 26 37.9 & 32 & 7.9 & 0.003-0.006 & 64 & 1 & 0 & 1 & 1.6 & 57 & 1 & 0 & $^{1}$ & 89.1 & 100.0 \\
SMC019 & 1 & 0 43 37.59 & -72 57 30.9 & 15 & 8.6 & 0.003-0.006 & 10 & 0 & 0 & 0 & 0.0 & 9 & 0 & 0 &  & 90.0 & \_ \\
SMC020 & 1 & 0 43 37.89 & -72 58 48.3 & 12 & 8.6 & 0.003-0.006 & 7 & 0 & 0 & 0 & 0.0 & 7 & 0 & 0 &  & 100.0 & \_ \\
SMC025 & 4 & 0 45 13.88 & -73 13 9.2 & 19 & 8.0 & 0.003-0.006 & 30 & 0 & 0 & 0 & 0.0 & 29 & 0 & 0 &  & 96.7 & \_ \\
SMC032 & 4 & 0 45 54.33 & -73 30 24.2 & 37 & 8.0 & 0.003-0.006 & 41 & 0 & 0 & 0 & 0.0 & 41 & 0 & 0 &  & 100.0 & \_ \\
SMC033 & 4 & 0 46 12.26 & -73 23 34 & 23 & 7.2 & 0.003-0.006 & 10 & 0 & 1 & 1 & 10.0 & 9 & 0 & 1 &  & 90.0 & 100.0 \\
SMC038 & 4 & 0 47 6.15 & -73 15 24.9 & 26 & 8.1 & 0.003-0.006 & 64 & 0 & 0 & 0 & 0.0 & 64 & 0 & 0 &  & 100.0 & \_ \\
SMC039 & 4 & 0 47 11.61 & -73 28 38.1 & 61 & 8.0 & 0.003-0.006 & 197 & 3 & 1 & 4 & 2.0 & 176 & 3 & 1 &  & 89.3 & 100.0 \\
SMC039 & 6 & 0 47 11.61 & -73 28 38.1 & 61 & 8.0 & 0.003-0.006 & 93 & 4 & 6 & 10 & 10.8 & 93 & 4 & 6 &  & 100.0 & 100.0 \\
SMC043 & 4 & 0 47 52.38 & -73 13 20.3 & 27 & 8.5 & 0.003-0.006 & 63 & 1 & 1 & 2 & 3.2 & 55 & 1 & 1 &  & 87.3 & 100.0 \\
SMC047 & 5 & 0 48 28.14 & -72 59 0.3 & 45 & 7.8 & 0.003-0.006 & 24 & 1 & 2 & 3 & 12.5 & 19 & 0 & 2 &  & 79.2 & 66.7 \\
SMC049 & 6 & 0 48 37.47 & -73 24 53.2 & 47 & 7.0 & 0.003-0.006 & 88 & 2 & 5 & 7 & 8.0 & 62 & 2 & 4 &  & 70.5 & 85.7 \\
SMC054 & 6 & 0 49 17.6 & -73 22 19.8 & 34 & 8.0 & 0.003-0.006 & 46 & 3 & 2 & 5 & 10.9 & 43 & 3 & 2 &  & 93.5 & 100.0 \\
SMC059 & 5 & 0 50 16.06 & -73 1 59.6 & 31 & 7.8 & 0.003-0.006 & 25 & 0 & 0 & 0 & 0.0 & 25 & 0 & 0 &  & 100.0 & \_ \\
SMC061 & 6 & 0 50 0.26 & -73 15 17.7 & 26 & 7.4 & 0.003-0.006 & 44 & 1 & 4 & 5 & 11.4 & 42 & 1 & 4 &  & 95.5 & 100.0 \\
SMC064 & 5 & 0 50 39.55 & -72 57 54.8 & 45 & 8.1 & 0.003-0.006 & 117 & 4 & 4 & 8 & 6.8 & 112 & 3 & 4 &  & 95.7 & 87.5 \\
SMC066 & 6 & 0 50 55.39 & -73 12 11 & 21 & 7.8 & 0.003-0.006 & 3 & 0 & 1 & 1 & 33.3 & 3 & 0 & 1 &  & 100.0 & 100.0 \\
SMC067 & 5 & 0 50 55.54 & -72 43 39.7 & 53 & 8.2 & 0.003-0.006 & 107 & 1 & 2 & 3 & 2.8 & 97 & 1 & 2 & & 90.7 & 100.0 \\
SMC068 & 6 & 0 50 56.26 & -73 17 21.1 & 70 & 7.7 & 0.003-0.006 & 134 & 9 & 3 & 12 & 9.0 & 133 & 9 & 3 & & 99.3 & 100.0 \\
SMC068 & 8 & 0 50 56.26 & -73 17 21.1 & 70 & 7.7 & 0.003-0.006 & 121 & 4 & 0 & 4 & 3.3 & 106 & 4 & 0 &  & 87.6 & 100.0 \\
SMC069 & 5 & 0 51 14.13 & -73 9 41.5 & 45 & 7.6 & 0.003-0.006 & 139 & 1 & 5 & 6 & 4.3 & 114 & 1 & 4 &  & 82.0 & 83.3 \\
SMC069 & 6 & 0 51 14.13 & -73 9 41.5 & 45 & 7.6 & 0.003-0.006 & 108 & 5 & 4 & 9 & 8.3 & 102 & 5 & 4 & & 94.4 & 100.0 \\
SMC069 & 8 & 0 51 14.13 & -73 9 41.5 & 45 & 7.6 & 0.003-0.006 & 65 & 1 & 0 & 1 & 1.5 & 47 & 1 & 0 &  & 72.3 & 100.0 \\
SMC070 & 6 & 0 51 26.15 & -73 16 59.8 & 18 & 7.8 & $\le$0.003 & 23 & 1 & 0 & 1 & 4.3 & 22 & 1 & 0 &  & 95.7 & 100.0 \\
SMC070 & 8 & 0 51 26.15 & -73 16 59.8 & 18 & 7.8 & $\le$0.003 & 16 & 0 & 0 & 0 & 0.0 & 15 & 0 & 0 &  & 93.8 & \_ \\
\hline
\end{tabular}
\label{indcl}
\end{table*}
}
\tiny{
\addtocounter{table}{-1}
\begin{table*}[tbph]
\caption{continued}
\centering
\begin{tabular}{@{\ }l@{\ \ }l@{\ \ }l@{\ \ }l@{\ \ }l@{\ \ }l@{\ \ }l@{\ \ }l@{\ \ }l@{\ \ }l@{\ \ }l@{\ \ }l@{\ \ }l@{\ \ }l@{\ \ }l@{\ \ }l@{\ \ }l@{\ \ }l@{\ }}
\hline
\hline
Cluster & Image & RA(2000) & DEC(2000) & r & log(t) & Z & $^{a}$N* & $^{c}$ & $^{d}$ & $^{b}$ & $^{b/a}$ & Ntot & NELS & NELS? & comments & N* & ELS \\
        & WFI   &          &           & \arcsec &  &   &  & NELS & NELS? & c+d & \% & ogle & ogle & ogle  &  & ogle/wfi & ogle/wfi \\
\hline
SMC071 & 5 & 0 51 31.78 & -73 0 38.3 & 40 & 7.5 & 0.003-0.006 & 96 & 2 & 1 & 3 & 3.1 & 79 & 2 & 1 & & 82.3 & 100.0 \\
SMC071 & 8 & 0 51 31.78 & -73 0 38.3 & 40 & 7.5 & 0.003-0.006 & 24 & 1 & 0 & 1 & 4.2 & 5 & 0 & 0 &  & 20.8 & 0.0 \\
SMC072 & 6 & 0 51 41.69 & -73 13 46.8 & 35 & 7.6 & $\le$0.003 & 58 & 4 & 2 & 6 & 10.3 & 57 & 4 & 2 &  & 98.3 & 100.0 \\
SMC072 & 8 & 0 51 41.69 & -73 13 46.8 & 35 & 7.6 & $\le$0.003 & 44 & 1 & 2 & 3 & 6.8 & 41 & 1 & 2 &  & 93.2 & 100.0 \\
SMC073 & 5 & 0 51 44.03 & -72 50 25.1 & 53 & 8.2 & $\le$0.003 & 162 & 1 & 1 & 2 & 1.2 & 138 & 1 & 1 &  & 85.2 & 100.0 \\
SMC073 & 7 & 0 51 44.03 & -72 50 25.1 & 53 & 8.2 & $\le$0.003 & 97 & 0 & 0 & 0 & 0.0 & 39 & 0 & 0 & & 40.2 & \_ \\
SMC074 & 5 & 0 51 52.91 & -72 57 13.9 & 53 & 8.1 & $\le$0.003 & 155 & 1 & 1 & 2 & 1.3 & 130 & 1 & 1 &  & 83.9 & 100.0 \\
SMC074 & 7 & 0 51 52.91 & -72 57 13.9 & 53 & 8.1 & $\le$0.003 & 85 & 2 & 0 & 2 & 2.4 & 79 & 2 & 0 &  & 92.9 & 100.0 \\
SMC074 & 8 & 0 51 52.91 & -72 57 13.9 & 53 & 8.1 & $\le$0.003 & 90 & 1 & 0 & 1 & 1.1 & 37 & 0 & 0 &  & 41.1 & 0.0 \\
SMC075 & 8 & 0 51 54.32 & -73 5 52.9 & 19 & 8.4 & $\le$0.003 & 18 & 0 & 0 & 0 & 0.0 & 10 & 0 & 0 &  & 55.6 & \_ \\
SMC076 & 7 & 0 52 12.47 & -72 31 51.2 & 36 & 7.4 & $\le$0.003 & 19 & 1 & 0 & 1 & 5.3 & 5 & 1 & 0 &  & 26.3 & 100.0 \\
SMC077 & 5 & 0 52 13.34 & -73 0 12.2 & 23 & 7.9 & $\le$0.003 & 21 & 0 & 0 & 0 & 0.0 & 19 & 0 & 0 &  & 90.5 & \_ \\
SMC077 & 8 & 0 52 13.34 & -73 0 12.2 & 23 & 7.9 & $\le$0.003 & 19 & 0 & 0 & 0 & 0.0 & 6 & 0 & 0 &  & 31.6 & \_ \\
SMC078 & 5 & 0 52 16.56 & -73 1 4 & 45 & 7.9 & $\le$0.003 & 33 & 0 & 0 & 0 & 0.0 & 29 & 0 & 0 &  & 87.9 & \_ \\
SMC078 & 8 & 0 52 16.56 & -73 1 4 & 45 & 7.9 & $\le$0.003 & 52 & 1 & 0 & 1 & 1.9 & 23 & 0 & 0 &  & 44.2 & 0.0 \\
SMC081 & 5 & 0 52 33.65 & -72 40 53.6 & 39 & 7.4 & $\le$0.003 & 55 & 0 & 0 & 0 & 0.0 & 16 & 0 & 0 &  & 29.1 & \_ \\
SMC081 & 7 & 0 52 33.65 & -72 40 53.6 & 39 & 7.4 & $\le$0.003 & 55 & 0 & 1 & 1 & 1.8 & 23 & 0 & 0 &  & 41.8 & 0.0 \\
SMC082 & 7 & 0 52 42.12 & -72 55 31.6 & 45 & 7.8 & $\le$0.003 & 35 & 0 & 1 & 1 & 2.9 & 30 & 0 & 1 &  & 85.7 & 100.0 \\
SMC082 & 8 & 0 52 42.12 & -72 55 31.6 & 45 & 7.8 & $\le$0.003 & 54 & 1 & 1 & 2 & 3.7 & 20 & 0 & 0 &  & 37.0 & 0.0 \\
SMC083 & 8 & 0 52 44.27 & -72 58 47.8 & 36 & 7.8 & $\le$0.003 & 40 & 2 & 0 & 2 & 5.0 & 15 & 0 & 0 &  & 37.5 & 0.0 \\
SMC089 & 7 & 0 53 5.28 & -72 37 27.8 & 143 & 7.3 & $\le$0.003 & 478 & 11 & 9 & 20 & 4.2 & 106 & 2 & 2 & & 22.2 & 20.0 \\
SMC090 & 8 & 0 53 5.59 & -73 22 49.4 & 47 & 8.5 & $\le$0.003 & 100 & 1 & 1 & 2 & 2.0 & 98 & 1 & 1 &  & 98.0 & 100.0 \\
SMC092 & 7 & 0 53 17.9 & -72 45 59.5 & 42 & 7.4 & $\le$0.003 & 48 & 0 & 1 & 1 & 2.1 & 18 & 0 & 0 &  & 37.5 & 0.0 \\
SMC098 & 8 & 0 54 46.73 & -73 13 24.5 & 45 & 8.0 & $\le$0.003 & 36 & 1 & 3 & 4 & 11.1 & 36 & 1 & 3 &  & 100.0 & 100.0 \\
SMC099 & 7 & 0 54 48.24 & -72 27 57.8 & 45 & 7.6 & $\le$0.003 & 64 & 2 & 3 & 5 & 7.8 & 54 & 2 & 3 &  & 84.4 & 100.0 \\
SMC104 & 7 & 0 55 32.98 & -72 49 58.1 & 46 & 8.6 & $\le$0.003 & 129 & 2 & 0 & 2 & 1.6 & 120 & 2 & 0 & & 93.0 & 100.0 \\
SMC105 & 7 & 0 55 42.99 & -72 52 48.4 & 55 & 8.0 & $\le$0.003 & 66 & 3 & 2 & 5 & 7.6 & 61 & 3 & 2 &  & 92.4 & 100.0 \\
SMC107 & 9 & 0 56 18.68 & -72 27 50.4 & 115 & 7.5 & $\le$0.003 & 404 & 36 & 7 & 43 & 10.6 & 242 & 33 & 7 &  & 59.9 & 93.0 \\
SMC109 & 9 & 0 57 29.8 & -72 15 51.9 & 24 & 7.7 & $\le$0.003 & 32 & 2 & 4 & 6 & 18.8 & 28 & 2 & 3 &  & 87.5 & 83.3 \\
SMC112 & 9 & 0 57 57.14 & -72 26 42 & 29 & 7.5 & $\le$0.003 & 16 & 0 & 0 & 0 & 0.0 & 8 & 0 & 0 &  & 50.0 & \_ \\
SMC115 & 9 & 0 58 33.64 & -72 16 51.6 & 15 & 7.3 & $\le$0.003 & 10 & 0 & 1 & 1 & 10.0 & 5 & 0 & 1 &  & 50.0 & 100.0 \\
SMC115 & 11 & 0 58 33.64 & -72 16 51.6 & 15 & 7.3 & $\le$0.003 & 13 & 0 & 1 & 1 & 7.7 & 9 & 0 & 1 &  & 69.2 & 100.0 \\
SMC117 & 9 & 0 59 13.86 & -72 36 29.3 & 45 & 8.3 & $\le$0.003 & 46 & 1 & 4 & 5 & 10.9 & 40 & 1 & 4 &  & 87.0 & 100.0 \\
SMC117 & 12 & 0 59 13.86 & -72 36 29.3 & 45 & 8.3 & $\le$0.003 & 56 & 1 & 0 & 1 & 1.8 & 45 & 1 & 0 &  & 80.4 & 100.0 \\
SMC120 & 9 & 1 0 1.33 & -72 22 8.7 & 27 & 7.7 & $\le$0.003 & 34 & 3 & 5 & 8 & 23.5 & 31 & 2 & 3 &  & 91.2 & 62.5 \\
SMC120 & 11 & 1 0 1.33 & -72 22 8.7 & 27 & 7.7 & $\le$0.003 & 33 & 4 & 1 & 5 & 15.2 & 14 & 4 & 0 &  & 42.4 & 80.0 \\
SMC121 & 9 & 1 0 13.03 & -72 27 43.8 & 30 & 7.9 & $\le$0.003 & 22 & 2 & 1 & 3 & 13.6 & 15 & 2 & 1 &  & 68.2 & 100.0 \\
SMC121 & 11 & 1 0 13.03 & -72 27 43.8 & 30 & 7.9 & $\le$0.003 & 23 & 4 & 1 & 5 & 21.7 & 19 & 4 & 1 &  & 82.6 & 100.0 \\
SMC121 & 12 & 1 0 13.03 & -72 27 43.8 & 30 & 7.9 & $\le$0.003 & 24 & 4 & 0 & 4 & 16.7 & 19 & 4 & 0 &  & 79.2 & 100.0 \\
SMC124 & 9 & 1 0 34.41 & -72 21 55.8 & 34 & 7.6 & $\le$0.003 & 44 & 4 & 4 & 8 & 18.2 & 41 & 4 & 4 &  & 93.2 & 100.0 \\
SMC124 & 11 & 1 0 34.41 & -72 21 55.8 & 34 & 7.6 & $\le$0.003 & 35 & 1 & 3 & 4 & 11.4 & 12 & 1 & 1 &  & 34.3 & 50.0 \\
SMC126 & 12 & 1 1 2.01 & -72 45 5.2 & 50 & 8.0 & $\le$0.003 & 37 & 2 & 1 & 3 & 8.1 & 34 & 1 & 1 &  & 91.9 & 66.7 \\
\hline
\end{tabular}
\label{indcl}
\end{table*}
}
\tiny{
\addtocounter{table}{-1}
\begin{table*}[tbph]
\caption{continued}
\centering
\begin{tabular}{@{\ }l@{\ \ }l@{\ \ }l@{\ \ }l@{\ \ }l@{\ \ }l@{\ \ }l@{\ \ }l@{\ \ }l@{\ \ }l@{\ \ }l@{\ \ }l@{\ \ }l@{\ \ }l@{\ \ }l@{\ \ }l@{\ \ }l@{\ \ }l@{\ }}
\hline
\hline
Cluster & Image & RA(2000) & DEC(2000) & r & log(t) & Z & $^{a}$N* & $^{c}$ & $^{d}$ & $^{b}$ & $^{b/a}$ & Ntot & NELS & NELS? & comments & N* & ELS \\
        & WFI   &          &           & \arcsec &  &   &  & NELS & NELS? & c+d & \% & ogle & ogle & ogle  &  & ogle/wfi & ogle/wfi \\
\hline
SMC128 & 11 & 1 1 37.15 & -72 24 24.7 & 45 & 7.1 & 0.003-0.006 & 33 & 2 & 1 & 3 & 9.1 & 9 & 1 & 0 &  & 27.3 & 33.3 \\
SMC128 & 12 & 1 1 37.15 & -72 24 24.7 & 45 & 7.1 & 0.003-0.006 & 23 & 0 & 0 & 0 & 0.0 & 21 & 0 & 0 &  & 91.3 & \_ \\
SMC129 & 12 & 1 1 45.08 & -72 33 51.8 & 36 & 7.3 & 0.003-0.006 & 38 & 0 & 2 & 2 & 5.3 & 35 & 0 & 2 &  & 92.1 & 100.0 \\
SMC134 & 11 & 1 3 11.52 & -72 16 21 & 35 & 7.8 & 0.003-0.006 & 41 & 9 & 3 & 12 & 29.3 & 39 & 9 & 2 & & 95.1 & 91.7 \\
SMC134 & 13 & 1 3 11.52 & -72 16 21 & 35 & 7.8 & 0.003-0.006 & 26 & 1 & 2 & 3 & 11.5 & 21 & 1 & 2 &   & 80.8 & 100.0 \\
SMC137 & 12 & 1 3 22.67 & -72 39 5.6 & 45 & 7.6 & 0.003-0.006 & 48 & 6 & 3 & 9 & 18.8 & 37 & 4 & 2 & & 77.1 & 66.7 \\
SMC138 & 13 & 1 3 53.02 & -72 6 10.5 & 23 & 7.4 & 0.003-0.006 & 21 & 1 & 0 & 1 & 4.8 & 17 & 1 & 0 &  & 81.0 & 100.0 \\
SMC139 & 12 & 1 3 53.44 & -72 49 34.2 & 25 & 7.5 & 0.003-0.006 & 78 & 10 & 3 & 13 & 16.7 & 67 & 8 & 2 & & 85.9 & 76.9 \\
SMC140 & 12 & 1 4 14.1 & -72 38 49.1 & 31 & 7.2 & 0.003-0.006 & 26 & 2 & 1 & 3 & 11.5 & 22 & 1 & 1 &  & 84.6 & 66.7 \\
SMC140 & 14 & 1 4 14.1 & -72 38 49.1 & 31 & 7.2 & 0.003-0.006 & 18 & 0 & 0 & 0 & 0.0 & 4 & 0 & 0 &  & 22.2 & \_ \\
SMC141 & 12 & 1 4 30.18 & -72 37 9.4 & 43 & 8.2 & 0.003-0.006 & 46 & 2 & 0 & 2 & 4.3 & 42 & 2 & 0 &  & 91.3 & 100.0 \\
SMC141 & 14 & 1 4 30.18 & -72 37 9.4 & 43 & 8.2 & 0.003-0.006 & 7 & 0 & 0 & 0 & 0.0 & 2 & 0 & 0 &  & 28.6 & \_ \\
SMC142 & 11 & 1 4 36.21 & -72 9 38.5 & 51 & 7.3 & 0.003-0.006 & 108 & 4 & 1 & 5 & 4.6 & 91 & 1 & 1 & $^{2}$ & 84.3 & 40.0 \\
SMC142 & 13 & 1 4 36.21 & -72 9 38.5 & 51 & 7.3 & 0.003-0.006 & 26 & 0 & 1 & 1 & 3.8 & 7 & 0 & 0 & & 26.9 & 0.0 \\
SMC143 & 11 & 1 4 39.61 & -72 32 59.7 & 34 & 8.2 & 0.003-0.006 & 15 & 0 & 0 & 0 & 0.0 & 3 & 0 & 0 &  & 20.0 & \_ \\
SMC143 & 12 & 1 4 39.61 & -72 32 59.7 & 34 & 8.2 & 0.003-0.006 & 28 & 1 & 0 & 1 & 3.6 & 26 & 1 & 0 &  & 92.9 & 100.0 \\
SMC144 & 11 & 1 4 5.23 & -72 7 14.6 & 23 & 7.6 & 0.003-0.006 & 15 & 2 & 0 & 2 & 13.3 & 12 & 2 & 0 &  & 80.0 & 100.0 \\
SMC144 & 13 & 1 4 5.23 & -72 7 14.6 & 23 & 7.6 & 0.003-0.006 & 12 & 0 & 0 & 0 & 0.0 & 2 & 0 & 0 &  & 16.7 & \_ \\
SMC145 & 13 & 1 5 4.3 & -71 59 24.8 & 23 & 7.9 & 0.003-0.006 & 9 & 1 & 0 & 1 & 11.1 & 4 & 1 & 0 &  & 44.4 & 100.0 \\
SMC146 & 13 & 1 5 13.4 & -71 59 41.8 & 18 & 7.3 & 0.003-0.006 & 7 & 0 & 0 & 0 & 0.0 & 5 & 0 & 0 &  & 71.4 & \_ \\
SMC147 & 13 & 1 5 7.95 & -71 59 45.1 & 27 & 7.1 & 0.003-0.006 & 13 & 0 & 1 & 1 & 7.7 & 11 & 0 & 1 &  & 84.6 & 100.0 \\
SMC149 & 11 & 1 5 21.51 & -72 2 34.7 & 45 & 8.2 & 0.003-0.006 & 162 & 3 & 1 & 4 & 2.5 & 20 & 1 & 0 &  & 12.3 & 25.0 \\
SMC149 & 13 & 1 5 21.51 & -72 2 34.7 & 45 & 8.2 & 0.003-0.006 & 144 & 1 & 1 & 2 & 1.4 & 112 & 1 & 1 &  & 77.8 & 100.0 \\
SMC153 & 13 & 1 6 47.74 & -72 16 24.5 & 34 & 7.2 & 0.003-0.006 & 37 & 2 & 1 & 3 & 8.1 & 12 & 0 & 0 &  & 32.4 & 0.0 \\
SMC154 & 14 & 1 7 2.27 & -72 37 18.2 & 41 & 8.2 & 0.003-0.006 & 14 & 0 & 0 & 0 & 0.0 & 2 & 0 & 0 &  & 14.3 & \_ \\
SMC155 & 14 & 1 7 27.83 & -72 29 35.5 & 51 & 7.7 & 0.003-0.006 & 41 & 3 & 0 & 3 & 7.3 & 38 & 3 & 0 &  & 92.7 & 100.0 \\
SMC156 & 14 & 1 7 28.47 & -72 46 9.5 & 51 & 8.2 & 0.003-0.006 & 51 & 0 & 2 & 2 & 3.9 & 48 & 0 & 2 &  & 94.1 & 100.0 \\
SMC160 & 13 & 1 8 37.48 & -72 26 20.9 & 25 & 7.6 & 0.003-0.006 & 16 & 0 & 1 & 1 & 6.3 & 7 & 0 & 0 &  & 43.8 & 0.0 \\
SMC160 & 14 & 1 8 37.48 & -72 26 20.9 & 25 & 7.6 & 0.003-0.006 & 7 & 0 & 1 & 1 & 14.3 & 4 & 0 & 0 &  & 57.1 & 0.0 \\
SMC177 & 4 & 0 44 55.05 & -73 10 27.4 & 11 & 7.9 & 0.003-0.006 & 14 & 0 & 0 & 0 & 0.0 & 13 & 0 & 0 &  & 92.9 & \_ \\
SMC187 & 4 & 0 47 5.87 & -73 22 16.6 & 17 & 8.2 & 0.003-0.006 & 12 & 1 & 0 & 1 & 8.3 & 12 & 1 & 0 &  & 100.0 & 100.0 \\
SMC187 & 6 & 0 47 5.87 & -73 22 16.6 & 18 & 8.2 & 0.003-0.006 & 9 & 1 & 0 & 1 & 11.1 & 9 & 1 & 0 &  & 100.0 & 100.0 \\
\hline
\end{tabular}
\label{indcl}
\end{table*}
}
\tiny{
\addtocounter{table}{-1}
\begin{table*}[tbph]
\caption{continued}
\centering
\begin{tabular}{@{\ }l@{\ \ }l@{\ \ }l@{\ \ }l@{\ \ }l@{\ \ }l@{\ \ }l@{\ \ }l@{\ \ }l@{\ \ }l@{\ \ }l@{\ \ }l@{\ \ }l@{\ \ }l@{\ \ }l@{\ \ }l@{\ \ }l@{\ \ }l@{\ }}
\hline
\hline
Cluster & Image & RA(2000) & DEC(2000) & r & log(t) & Z & $^{a}$N* & $^{c}$ & $^{d}$ & $^{b}$ & $^{b/a}$ & Ntot & NELS & NELS? & comments & N* & ELS \\
        & WFI   &          &           & \arcsec &  &   &  & NELS & NELS? & c+d & \% & ogle & ogle & ogle  &  & ogle/wfi & ogle/wfi \\
\hline
SMC194 & 6 & 0 49 5.58 & -73 21 9.8 & 19 & 7.9 & 0.003-0.006 & 14 & 0 & 1 & 1 & 7.1 & 10 & 0 & 1 &  & 71.4 & 100.0 \\
SMC195 & 6 & 0 49 16.45 & -73 14 56.8 & 15 & 7.3 & 0.003-0.006 & 17 & 0 & 0 & 0 & 0.0 & 12 & 0 & 0 & & 70.6 & \_ \\
SMC198 & 5 & 0 50 7.51 & -73 11 25.9 & 30 & 7.9 & 0.003-0.006 & 51 & 0 & 2 & 2 & 3.9 & 49 & 0 & 2 &  & 96.1 & 100.0 \\
SMC198 & 6 & 0 50 7.51 & -73 11 25.9 & 30 & 7.9 & 0.003-0.006 & 45 & 3 & 1 & 4 & 8.9 & 45 & 3 & 1 &  & 100.0 & 100.0 \\
SMC200 & 5 & 0 50 38.98 & -72 58 43.6 & 9 & 8.0 & 0.003-0.006 & 10 & 0 & 0 & 0 & 0.0 & 10 & 0 & 0 &  & 100.0 & \_ \\
SMC210 & 8 & 0 52 30.3 & -73 2 59 & 15 & 8.2 & $\le$0.003 & 10 & 0 & 0 & 0 & 0.0 & 2 & 0 & 0 &  & 20.0 & \_ \\
SMC230 & 11 & 1 0 33.15 & -72 15 30.5 & 9 & 7.5 & $\le$0.003 & 5 & 0 & 0 & 0 & 0.0 & 1 & 0 & 0 &  & 20.0 & \_ \\
\hline
\end{tabular}
\label{indcl}
\begin{flushleft}
$^{1}$: SMC17 \& SMC18 have common stars since it appears to be a binary-cluster. \\
$^{2}$: The extracted spectrum of the candidate emission-line star is a blend of 2 emission-line stars spectra.\\
a:N*: This column correspond to the total number of stars/spectra extracted. \\
c:NELS: This column gives the number of definite emission-line stars found.\\
d:NELS?: this column gives the number of candidate emission-line stars found.\\
b:c+d: This column gives the total number of (definite+candidate) emission line stars.\\
b/a \%: This column gives the proportion in \% of emission-line stars to the whole sample by open cluster. 
\end{flushleft}

\end{table*}
}


%
\tiny{
\begin{table*}[]
\caption{Details about the individual open clusters with different observations merged. The redundant stars are removed. Note that, here
only the WFI stars found in Ogle catalogues \citep{oglemapsphoto} are taken into account.
In cols. 2, 3, 4, 5, the number of normal stars, emission-line star, candidate emission-line star, sum of emission-line star are given. In col. 6, the ratio of emission-line star to all stars
(normal+all emission-line star) is given. The next columns give the same indications than before but for classified stars (Be, B, Oe, O, etc). 
The last column gives the number of emission-line stars, which are in the Main Sequence but that are not O or B or A-type stars.}
\centering
\begin{tabular}{@{\ }l@{\ \ }l@{\ \ }l@{\ \ }l@{\ \ }l@{\ \ }l@{\ \ }l@{\ \ }l@{\ \ }l@{\ \ }l@{\ \ }l@{\ \ }l@{\ \ }l@{\ \ }l@{\ \ }l@{\ \ }l@{\ \ }l@{\ \ }l@{\ \ }l@{\ \ }l@{\ \ }l@{\ \ }l@{\ }}
\hline
\hline
Cluster & Nabs & NELS & NELS? & NELS & $\frac{ELS}{all *}$ & Be & Be? & B & $\frac{Be}{all B}$ & $\frac{Be+Be?}{allB}$ & Oe & O & $\frac{Oe}{all O}$ & Ae & A & $\frac{Ae}{all A}$ & OeBeAe & OBA & $\frac{OeBeAe}{all OBA}$ & ELS main sequence \\
 & ogle & ogle & ogle & no ogle & \% &  &  &  & \% & \% &  &  &  &  &  &  &  &  &  & not OBA \\
\hline
\hline
SMC002 & 25 & 0 & 1 & 0 & 3.8 & 0 & 1 & 3 & 0.00 & 25.00 &  &  &  &  & 3 & 0.00 & 1 & 6 & 14.29 &  \\
SMC008 & 110 & 4 & 2 & 0 & 5.2 & 1 & 1 & 55 & 1.75 & 3.51 &  & 1 & 0.00 &  & 9 & 0.00 & 2 & 65 & 2.99 & 4 \\
SMC009 & 60 & 1 & 2 & 0 & 4.8 & 1 & 2 & 34 & 2.70 & 8.11 &  &  &  &  & 2 & 0.00 & 3 & 36 & 7.69 &  \\
SMC011 & 64 & 3 & 3 & 0 & 8.6 & 3 & 2 & 22 & 11.11 & 18.52 &  & 1 & 0.00 & 1 & 4 & 20.00 & 6 & 27 & 18.18 &  \\
SMC012 & 65 & 0 & 0 & 2 & 3.0 & 0 &  & 3 & 0.00 & 0.00 &  &  &  &  & 2 & 0.00 & 0 & 5 & 0.00 &  \\
SMC015 & 94 & 3 & 2 & 0 & 5.1 & 1 & 1 & 25 & 3.70 & 7.41 &  &  &  &  & 15 & 0.00 & 2 & 40 & 4.76 & 3 \\
SMC016 & 111 & 0 & 1 & 0 & 0.9 & 0 & 1 & 22 & 0.00 & 4.35 &  & 1 & 0.00 &  & 11 & 0.00 & 1 & 34 & 2.86 &  \\
SMC017 & 62 & 1 & 0 & 0 & 1.6 & 0 &  & 23 & 0.00 & 0.00 & 1 & 1 & 50.00 &  & 6 & 0.00 & 1 & 30 & 3.23 &  \\
SMC018 & 55 & 1 & 0 & 0 & 1.8 & 1 &  & 21 & 4.55 & 4.55 &  & 2 & 0.00 &  &  &  & 1 & 23 & 4.17 & 1 \\
SMC019 & 7 & 0 & 0 & 0 & 0.0 &  &  &  &  & \_ &  &  &  &  &  &  & 0 & 0 &  & 2 \\
SMC020 & 7 & 0 & 0 & 0 & 0.0 &  &  &  &  & \_ &  &  &  &  & 1 & 0.00 & 0 & 1 & 0.00 &  \\
SMC025 & 29 & 0 & 0 & 0 & 0.0 &  &  & 15 & 0.00 & 0.00 &  &  &  &  &  &  & 0 & 15 & 0.00 &  \\
SMC032 & 39 & 0 & 0 & 0 & 0.0 &  &  & 11 & 0.00 & 0.00 &  &  &  &  & 1 & 0.00 & 0 & 12 & 0.00 &  \\
SMC033 & 8 & 0 & 1 & 0 & 11.1 &  &  & 6 & 0.00 & 0.00 & 1 &  & 100.00 &  &  &  & 1 & 6 & 14.29 &  \\
SMC038 & 64 & 0 & 0 & 0 & 0.0 &  &  & 17 & 0.00 & 0.00 &  &  &  &  & 4 & 0.00 & 0 & 21 & 0.00 &  \\
SMC039 & 182 & 4 & 7 & 0 & 5.7 & 1 & 4 & 46 & 1.96 & 9.80 &  &  &  & 2 & 15 & 11.76 & 7 & 61 & 10.29 & 4 \\
SMC043 & 53 & 1 & 1 & 0 & 3.6 &  & 1 & 16 & 0.00 & 5.88 &  &  &  &  & 6 & 0.00 & 1 & 22 & 4.35 & 1 \\
SMC047 & 17 & 0 & 2 & 1 & 15.0 &  & 1 & 3 & 0.00 & 25.00 &  &  &  &  &  &  & 1 & 3 & 25.00 & 1 \\
SMC049 & 55 & 2 & 4 & 1 & 11.3 & 1 & 1 & 17 & 5.26 & 10.53 &  &  &  & 1 & 5 & 16.67 & 3 & 22 & 12.00 & 3 \\
SMC054 & 38 & 3 & 2 & 0 & 11.6 & 2 & 1 & 17 & 10.00 & 15.00 &  & 1 & 0.00 &  & 4 & 0.00 & 3 & 22 & 12.00 & 2 \\
SMC059 & 25 & 0 & 0 & 0 & 0.0 &  &  & 9 & 0.00 & 0.00 &  &  &  &  &  &  & 0 & 9 & 0.00 &  \\
SMC061 & 37 & 1 & 4 & 0 & 11.9 &  & 2 & 19 & 0.00 & 9.52 &  &  &  &  & 2 & 0.00 & 2 & 21 & 8.70 & 3 \\
SMC064 & 105 & 3 & 4 & 1 & 7.1 & 2 & 4 & 45 & 3.92 & 11.76 &  & 1 & 0.00 &  &  &  & 6 & 46 & 11.54 & 1 \\
SMC066 & 2 & 0 & 1 & 0 & 33.3 &  & 1 &  & 0.00 & 100.00 &  & 1 & 0.00 &  &  &  & 1 & 1 &  &  \\
SMC067 & 94 & 1 & 2 & 0 & 3.1 & 1 &  & 34 & 2.86 & 2.86 &  &  &  &  & 1 & 0.00 & 1 & 35 & 2.78 & 2 \\
SMC068 & 142 & 11 & 3 & 0 & 9.0 & 8 & 1 & 55 & 12.50 & 14.06 &  & 1 & 0.00 &  & 4 & 0.00 & 9 & 60 & 13.04 & 5 \\
SMC069 & 164 & 5 & 7 & 0 & 6.8 & 4 & 5 & 55 & 6.25 & 14.06 &  & 2 & 0.00 &  & 13 & 0.00 & 9 & 70 & 11.39 & 3 \\
SMC070 & 24 & 1 & 0 & 0 & 4.0 &  &  & 9 & 0.00 & 0.00 &  &  &  &  & 1 & 0.00 & 0 & 10 & 0.00 & 1 \\
SMC071 & 81 & 2 & 1 & 0 & 3.6 & 1 &  & 24 & 4.00 & 4.00 & 1 & 1 & 50.00 &  &  &  & 2 & 25 & 7.41 & 1 \\
SMC072 & 60 & 4 & 2 & 0 & 9.1 & 3 & 1 & 15 & 15.79 & 21.05 &  &  &  &  & 4 & 0.00 & 4 & 19 & 17.39 & 2 \\
SMC073 & 161 & 1 & 1 & 0 & 1.2 & 1 &  & 60 & 1.64 & 1.64 &  &  &  &  & 10 & 0.00 & 1 & 70 & 1.41 & 1 \\
SMC074 & 183 & 1 & 2 & 0 & 1.6 & 1 &  & 34 & 2.86 & 2.86 &  & 1 & 0.00 &  & 17 & 0.00 & 1 & 52 & 1.89 & 2 \\
SMC075 & 10 & 0 & 0 & 0 & 0.0 &  &  & 5 & 0.00 & 0.00 &  &  &  &  & 2 & 0.00 & 0 & 7 & 0.00 &  \\
SMC076 & 4 & 1 & 0 & 0 & 20.0 &  &  & 1 & 0.00 & 0.00 &  &  &  &  &  &  & 0 & 1 & 0.00 & 1 \\
SMC077 & 24 & 0 & 0 & 0 & 0.0 &  &  & 8 & 0.00 & 0.00 &  &  &  &  & 1 & 0.00 & 0 & 9 & 0.00 &  \\
SMC078 & 52 & 0 & 0 & 1 & 1.9 &  &  & 17 & 0.00 & 0.00 &  &  &  &  & 7 & 0.00 & 0 & 24 & 0.00 &  \\
SMC081 & 38 & 0 & 0 & 1 & 2.6 &  &  & 4 & 0.00 & 0.00 &  &  &  &  & 7 & 0.00 & 0 & 11 & 0.00 &  \\
SMC082 & 45 & 0 & 2 & 1 & 6.3 &  &  & 9 & 0.00 & 0.00 &  &  &  &  & 4 & 0.00 & 0 & 13 & 0.00 & 2 \\
\hline
\end{tabular}
\label{mergcl}
\end{table*}
}
\tiny{
\addtocounter{table}{-1}
\begin{table*}[]
\caption{Individual clusters, continued}
\centering
\begin{tabular}{@{\ }l@{\ \ }l@{\ \ }l@{\ \ }l@{\ \ }l@{\ \ }l@{\ \ }l@{\ \ }l@{\ \ }l@{\ \ }l@{\ \ }l@{\ \ }l@{\ \ }l@{\ \ }l@{\ \ }l@{\ \ }l@{\ \ }l@{\ \ }l@{\ \ }l@{\ \ }l@{\ \ }l@{\ \ }l@{\ }}
\hline
\hline
Cluster & Nabs & NELS & NELS? & NELS & $\frac{ELS}{all *}$ & Be & Be? & B & $\frac{Be}{all B}$ & $\frac{Be+Be?}{allB}$ & Oe & O & $\frac{Oe}{all O}$ & Ae & A & $\frac{Ae}{all A}$ & OeBeAe & OBA & $\frac{OeBeAe}{all OBA}$ & ELS main sequence \\
 & ogle & ogle & ogle & no ogle & \% &  &  &  & \% & \% &  &  &  &  &  &  &  &  &  & not OBA \\
\hline
SMC083 & 15 & 0 & 0 & 2 & 11.8 &  &  & 3 & 0.00 & 0.00 &  &  &  &  & 4 & 0.00 & 0 & 7 & 0.00 &  \\
SMC089 & 102 & 2 & 2 & 16 & 16.4 & 1 &  & 16 & 5.88 & 5.88 &  &  &  & 1 & 6 & 14.29 & 2 & 22 & 8.33 & 2 \\
SMC090 & 96 & 1 & 1 & 0 & 2.0 &  &  & 9 & 0.00 & 0.00 &  &  &  &  & 1 & 0.00 & 0 & 10 & 0.00 & 2 \\
SMC092 & 18 & 0 & 0 & 1 & 5.3 &  &  & 7 & 0.00 & 0.00 &  &  &  &  & 1 & 0.00 & 0 & 8 & 0.00 &  \\
SMC098 & 32 & 1 & 3 & 0 & 11.1 &  &  & 6 & 0.00 & 0.00 &  &  &  &  & 2 & 0.00 & 0 & 8 & 0.00 & 4 \\
SMC099 & 49 & 2 & 3 & 0 & 9.3 & 1 & 1 & 25 & 3.70 & 7.41 &  &  &  &  & 1 & 0.00 & 2 & 26 & 7.14 & 3 \\
SMC104 & 118 & 2 & 0 & 0 & 1.7 & 1 &  & 15 & 6.25 & 6.25 &  &  &  &  & 6 & 0.00 & 1 & 21 & 4.55 & 1 \\
SMC105 & 56 & 3 & 2 & 0 & 8.2 & 1 &  & 7 & 12.50 & 12.50 & 1 & 1 & 50.00 &  & 2 & 0.00 & 2 & 10 & 16.67 & 3 \\
SMC107 & 202 & 33 & 7 & 3 & 17.6 & 25 & 6 & 92 & 20.33 & 25.20 & 1 &  & 100.00 &  & 16 & 0.00 & 32 & 108 & 22.86 & 8 \\
SMC109 & 23 & 2 & 3 & 0 & 17.9 & 1 & 1 & 6 & 12.50 & 25.00 &  &  &  &  &  &  & 2 & 6 & 25.00 & 3 \\
SMC112 & 8 & 0 & 0 & 0 & 0.0 &  &  & 4 & 0.00 & 0.00 &  &  &  &  & 1 & 0.00 & 0 & 5 & 0.00 &  \\
SMC115 & 11 & 0 & 1 & 0 & 8.3 &  & 1 & 3 & 0.00 & 25.00 &  &  &  &  & 1 & 0.00 & 1 & 4 & 20.00 &  \\
SMC117 & 60 & 2 & 4 & 0 & 9.1 & 2 &  & 16 & 11.11 & 11.11 &  &  &  &  & 4 & 0.00 & 2 & 20 & 9.09 & 4 \\
SMC120 & 30 & 3 & 3 & 0 & 16.7 & 3 & 2 & 12 & 17.65 & 29.41 &  &  &  &  & 2 & 0.00 & 5 & 14 & 26.32 & 1 \\
SMC121 & 23 & 5 & 2 & 0 & 23.3 & 3 & 1 & 9 & 23.08 & 30.77 &  &  &  &  & 4 & 0.00 & 4 & 13 & 23.53 & 3 \\
SMC124 & 36 & 4 & 5 & 0 & 20.0 & 4 & 3 & 17 & 16.67 & 29.17 &  & 2 & 0.00 &  & 1 & 0.00 & 7 & 20 & 25.93 & 2 \\
SMC126 & 32 & 1 & 1 & 1 & 8.6 & 1 & 1 & 11 & 7.69 & 15.38 &  & 1 & 0.00 &  & 1 & 0.00 & 2 & 13 & 13.33 &  \\
SMC128 & 27 & 1 & 0 & 2 & 10.0 & 1 &  & 13 & 7.14 & 7.14 &  &  &  &  &  &  & 1 & 13 & 7.14 &  \\
SMC129 & 33 & 0 & 2 & 0 & 5.7 &  &  & 17 & 0.00 & 0.00 &  & 2 & 0.00 &  &  &  & 0 & 19 & 0.00 & 2 \\
SMC134 & 30 & 9 & 2 & 1 & 28.6 & 9 & 2 & 19 & 30.00 & 36.67 &  &  &  &  & 2 & 0.00 & 11 & 21 & 34.38 &  \\
SMC137 & 31 & 4 & 2 & 3 & 22.5 & 4 & 2 & 20 & 15.38 & 23.08 &  &  &  &  &  &  & 6 & 20 & 23.08 &  \\
SMC138 & 16 & 1 & 0 & 0 & 5.9 & 1 &  & 11 & 8.33 & 8.33 &  &  &  &  &  &  & 1 & 11 & 8.33 &  \\
SMC139 & 57 & 8 & 2 & 3 & 18.6 & 8 &  & 31 & 20.51 & 20.51 &  & 1 & 0.00 &  &  &  & 8 & 32 & 20.00 & 2 \\
SMC140 & 22 & 1 & 1 & 1 & 12.0 & 1 & 1 & 11 & 7.69 & 15.38 &  & 1 & 0.00 &  &  &  & 2 & 12 & 14.29 &  \\
SMC141 & 42 & 2 & 0 & 0 & 4.5 &  &  & 18 & 0.00 & 0.00 &  &  &  & 1 &  & 100.00 & 1 & 18 & 5.26 & 1 \\
SMC142 & 90 & 1 & 1 & 3 & 5.3 & 1 &  & 58 & 1.69 & 1.69 &  & 1 & 0.00 &  & 3 & 0.00 & 1 & 62 & 1.59 & 1 \\
SMC143 & 28 & 1 & 0 & 0 & 3.4 & 1 &  & 13 & 7.14 & 7.14 &  &  &  &  & 2 & 0.00 & 1 & 15 & 6.25 &  \\
SMC144 & 10 & 2 & 0 & 0 & 16.7 & 2 &  & 8 & 20.00 & 20.00 &  &  &  &  &  &  & 2 & 8 & 20.00 &  \\
SMC145 & 3 & 1 & 0 & 0 & 25.0 &  &  & 3 & 0.00 & 0.00 &  &  &  &  &  &  & 0 & 3 & 0.00 & 1 \\
SMC146 & 5 & 0 & 0 & 0 & 0.0 &  &  & 4 & 0.00 & 0.00 &  &  &  &  &  &  & 0 & 4 & 0.00 &  \\
SMC147 & 10 & 0 & 1 & 0 & 9.1 &  & 1 & 8 & 0.00 & 11.11 &  &  &  &  &  &  & 1 & 8 & 11.11 &  \\
SMC149 & 127 & 2 & 1 & 0 & 2.3 &  &  & 48 & 0.00 & 0.00 &  &  &  & 1 & 15 & 6.25 & 1 & 63 & 1.56 & 2 \\
SMC153 & 12 & 0 & 0 & 3 & 20.0 &  &  & 3 & 0.00 & 0.00 &  &  &  &  & 2 & 0.00 & 0 & 5 & 0.00 &  \\
SMC154 & 2 & 0 & 0 & 0 & 0.0 &  &  &  &  & \_ &  &  &  &  &  &  & 0 & 0 &  &  \\
SMC155 & 35 & 3 & 0 & 0 & 7.9 & 3 &  & 19 & 13.64 & 13.64 &  &  &  &  &  &  & 3 & 19 & 13.64 &  \\
SMC156 & 46 & 0 & 2 & 0 & 4.2 &  &  & 13 & 0.00 & 0.00 &  & 1 & 0.00 &  &  &  & 0 & 14 & 0.00 & 2 \\
\hline
\end{tabular}
\label{mergcl}
\end{table*}
}
\tiny{
\addtocounter{table}{-1}
\begin{table*}[]
\caption{Individual clusters, continued}
\centering
\begin{tabular}{@{\ }l@{\ \ }l@{\ \ }l@{\ \ }l@{\ \ }l@{\ \ }l@{\ \ }l@{\ \ }l@{\ \ }l@{\ \ }l@{\ \ }l@{\ \ }l@{\ \ }l@{\ \ }l@{\ \ }l@{\ \ }l@{\ \ }l@{\ \ }l@{\ \ }l@{\ \ }l@{\ \ }l@{\ \ }l@{\ }}
\hline
\hline
Cluster & Nabs & NELS & NELS? & NELS & $\frac{ELS}{all *}$ & Be & Be? & B & $\frac{Be}{all B}$ & $\frac{Be+Be?}{allB}$ & Oe & O & $\frac{Oe}{all O}$ & Ae & A & $\frac{Ae}{all A}$ & OeBeAe & OBA & $\frac{OeBeAe}{all OBA}$ & ELS main sequence \\
 & ogle & ogle & ogle & no ogle & \% &  &  &  & \% & \% &  &  &  &  &  &  &  &  &  & not OBA \\
\hline
SMC160 & 11 & 0 & 0 & 2 & 15.4 &  &  & 6 & 0.00 & 0.00 &  &  &  &  &  &  & 0 & 6 & 0.00 &  \\
SMC177 & 13 & 0 & 0 & 0 & 0.0 &  &  & 5 & 0.00 & 0.00 &  &  &  &  & 1 & 0.00 & 0 & 6 & 0.00 &  \\
SMC187 & 13 & 1 & 0 & 0 & 7.1 &  &  & 3 & 0.00 & 0.00 & 1 &  & 100.00 &  &  &  & 1 & 3 & 25.00 & 2 \\
SMC194 & 9 & 0 & 1 & 0 & 10.0 &  &  & 5 & 0.00 & 0.00 &  &  &  &  & 1 & 0.00 & 0 & 6 & 0.00 & 1 \\
SMC195 & 12 & 0 & 0 & 0 & 0.0 &  &  & 4 & 0.00 & 0.00 &  &  &  &  & 2 & 0.00 & 0 & 6 & 0.00 &  \\
SMC198 & 64 & 3 & 2 & 0 & 7.2 & 3 & 2 & 23 & 10.71 & 17.86 &  & 1 & 0.00 &  & 3 & 0.00 & 5 & 27 & 15.63 &  \\
SMC200 & 10 & 0 & 0 & 0 & 0.0 &  &  & 2 & 0.00 & 0.00 &  &  &  &  &  &  & 0 & 2 & 0.00 &  \\
SMC210 & 2 & 0 & 0 & 0 & 0.0 &  &  & 1 & 0.00 & 0.00 &  &  &  &  &  &  & 0 & 1 & 0.00 &  \\
SMC230 & 1 & 0 & 0 & 0 & 0.0 &  &  & 1 & 0.00 & 0.00 &  &  &  &  &  &  & 0 & 1 & 0.00 &  \\
\hline
\end{tabular}
\label{mergcl}
\end{table*}
}

\end{landscape}
\end{flushleft}


\section{Tables of Oe/Be/Ae stars, and other emission-line star}
\label{tablesBeOeAeemission-line star}

The following tables include information (astrometry, apparent and
dereddened photometry, classification, etc.) for each emission-line
star found.

\begin{flushleft}
\begin{landscape}

\tiny{
\begin{table*}[]
\caption{Table 1 of Be stars: cross-correlation WFI with Ogle \citep[][ cols. 3, 5, 8, 9]{oglemapsphoto}: astrometry, apparent photometry. 
The last column 'Mult' indicates whether the star is found several times as emission-line star in case of different observations in
WFI (different images). `1' means that the star was observed once (or the spectrum extracted once).
 }
\centering
\begin{tabular}{@{\ }l@{\ \ }l@{\ \ }l@{\ \ }l@{\ \ }l@{\ \ }l@{\ \ }l@{\ \ }l@{\ \ }l@{\ \ }l@{\ \ }l@{\ \ }l@{\ \ }l@{\ \ }l@{\ \ }l@{\ \ }l@{\ \ }l@{\ \ }l@{\ \ }l@{\ }}
\hline
\hline
Cluster & Image & Image & ID & ID & r & RA(2000)   & DEC(2000)   & RA(2000)   & DEC(2000)   & V & B & I & B-V & V-I & E[B-V] & log(t) & Mult \\
        & WFI & Ogle & WFI & Ogle & \arcsec & WFI   & WFI   & Ogle   & Ogle   &  &  &  &  &  &  &  &  \\
\hline
\hline
SMC8 & 2 & sc2 & 34940 & 35350 & 0.7 & 0 40 36.853 & -73 23 42.137 & 0 40 36.950 & -73 23 41.900 & 16.978 & 16.857 & 16.911 & -0.120 & 0.067 & 0.070 & 8 & 2 \\
SMC9 & 2 & sc2 & 35267 & 35310 & 0.7 & 0 40 48.606 & -73 23 39.151 & 0 40 48.700 & -73 23 38.900 & 16.420 & 16.360 & 16.226 & -0.060 & 0.194 & 0.070 & 8 & 1 \\
SMC11 & 2 & sc2 & 39854 & 61751 & 0.7 & 0 41 6.708 & -73 21 10.595 & 0 41 6.770 & -73 21 10.300 & 16.566 & 16.529 & 16.464 & -0.037 & 0.102 & 0.080 & 7.9 & 1 \\
SMC11 & 2 & sc2 & 40737 & 64032 & 0.7 & 0 41 8.835 & -73 20 43.105 & 0 41 8.910 & -73 20 42.800 & 17.113 & 17.053 & 17.046 & -0.060 & 0.067 & 0.080 & 7.9 & 2 \\
SMC11 & 2 & sc2 & 40443 & 64100 & 0.8 & 0 40 57.870 & -73 20 45.337 & 0 40 57.960 & -73 20 45.000 & 17.436 & 17.385 & 17.387 & -0.051 & 0.049 & 0.080 & 7.9 & 1 \\
SMC15 & 1 & sc3 & 8873 & 19972 & 1.0 & 0 42 54.158 & -73 17 47.769 & 0 42 54.110 & -73 17 47.200 & 16.104 & 16.222 & 15.864 & 0.118 & 0.240 & 0.100 & 8.1 & 2 \\
\hline
\end{tabular}
\label{tableBe1}
\end{table*}
}
\tiny{
\addtocounter{table}{-1}
\begin{table*}[]
\caption{Be stars table1, continued}
\centering
\begin{tabular}{@{\ }l@{\ \ }l@{\ \ }l@{\ \ }l@{\ \ }l@{\ \ }l@{\ \ }l@{\ \ }l@{\ \ }l@{\ \ }l@{\ \ }l@{\ \ }l@{\ \ }l@{\ \ }l@{\ \ }l@{\ \ }l@{\ \ }l@{\ \ }l@{\ \ }l@{\ }}
\hline
\hline
Cluster & Image & Image & ID & ID & r & RA(2000)   & DEC(2000)   & RA(2000)   & DEC(2000)   & V & B & I & B-V & V-I & E[B-V] & log(t) & Mult \\
 & WFI & Ogle & WFI & Ogle & \arcsec & WFI   & WFI   & Ogle   & Ogle   &  &  &  &  &  &  &  &  \\
\hline
SMC107 & 9 & sc7 & 36804 & 205918 & 1.4 & 0 56 14.797 & -72 28 47.857 & 0 56 14.900 & -72 28 47.400 & 15.350 & 15.296 & 15.221 & -0.054 & 0.129 & 0.100 & 7.5 & 1 \\
SMC107 & 9 & sc7 & 37317 & 205923 & 1.6 & 0 56 17.606 & -72 28 32.697 & 0 56 17.720 & -72 28 32.200 & 15.682 & 15.635 & 15.551 & -0.047 & 0.131 & 0.100 & 7.5 & 1 \\
SMC107 & 9 & sc7 & 38203 & 205925 & 1.3 & 0 56 26.511 & -72 28 9.711 & 0 56 26.600 & -72 28 9.200 & 15.702 & 15.652 & 15.561 & -0.050 & 0.140 & 0.100 & 7.5 & 1 \\
SMC107 & 9 & sc7 & 38679 & 205930 & 1.6 & 0 56 31.034 & -72 27 58.173 & 0 56 31.150 & -72 27 57.700 & 15.527 & 15.484 & 15.407 & -0.043 & 0.120 & 0.100 & 7.5 & 1 \\
SMC107 & 9 & sc7 & 38683 & 205931 & 1.1 & 0 56 22.946 & -72 27 54.233 & 0 56 23.020 & -72 27 53.800 & 15.263 & 15.175 & 15.189 & -0.088 & 0.074 & 0.100 & 7.5 & 1 \\
SMC107 & 9 & sc7 & 38865 & 206011 & 0.9 & 0 56 13.394 & -72 27 44.887 & 0 56 13.370 & -72 27 44.500 & 15.808 & 15.707 & 15.731 & -0.101 & 0.076 & 0.100 & 7.5 & 1 \\
\hline
\end{tabular}
\label{tableBe1}
\end{table*}
}
\tiny{
\addtocounter{table}{-1}
\begin{table*}[]
\caption{Be stars table1, continued}
\centering
\begin{tabular}{@{\ }l@{\ \ }l@{\ \ }l@{\ \ }l@{\ \ }l@{\ \ }l@{\ \ }l@{\ \ }l@{\ \ }l@{\ \ }l@{\ \ }l@{\ \ }l@{\ \ }l@{\ \ }l@{\ \ }l@{\ \ }l@{\ \ }l@{\ \ }l@{\ \ }l@{\ }}
\hline
\hline
Cluster & Image & Image & ID & ID & r & RA(2000)   & DEC(2000)   & RA(2000)   & DEC(2000)   & V & B & I & B-V & V-I & E[B-V] & log(t) & Mult \\
 & WFI & Ogle & WFI & Ogle & \arcsec & WFI   & WFI   & Ogle   & Ogle   &  &  &  &  &  &  &  &  \\
\hline
SMC126 & 12 & sc9 & 22724 & 13435 & 0.8 & 1 0 55.826 & -72 44 50.848 & 1 0 55.890 & -72 44 50.600 & 16.887 & 16.806 & 16.904 & -0.082 & -0.017 & 0.070 & 8 & 1 \\
SMC128 & 11 & sc9 & 10309 & 78904 & 1.3 & 1 1 36.265 & -72 24 17.857 & 1 1 36.360 & -72 24 17.500 & 17.361 & 17.256 & 17.485 & -0.105 & -0.124 & 0.090 & 7.1 & 1 \\
SMC134 & 11 & sc9 & 23627 & 168332 & 0.6 & 1 3 9.526 & -72 16 22.771 & 1 3 9.570 & -72 16 22.600 & 14.997 & 14.976 & 14.981 & -0.022 & 0.016 & 0.050 & 7.8 & 2 \\
SMC134 & 11 & sc9 & 23610 & 168408 & 0.5 & 1 3 11.782 & -72 16 24.363 & 1 3 11.790 & -72 16 24.200 & 16.924 & 16.839 & 16.853 & -0.086 & 0.071 & 0.050 & 7.8 & 2 \\
SMC134 & 11 & sc9 & 23713 & 168413 & 0.8 & 1 3 11.483 & -72 16 20.708 & 1 3 11.460 & -72 16 20.500 & 17.162 & 17.149 & 17.091 & -0.013 & 0.071 & 0.050 & 7.8 & 1 \\
SMC134 & 11 & sc9 & 23901 & 168419 & 0.5 & 1 3 11.254 & -72 16 18.228 & 1 3 11.260 & -72 16 18.100 & 17.244 & 17.202 & 17.154 & -0.042 & 0.088 & 0.050 & 7.8 & 1 \\
\hline
\end{tabular}
\label{tableBe1}
\end{table*}
}


%
\small{
\begin{table*}[]
\caption{Table 2 of Be stars: corrected photometry and classification using calibration from \citet{lang1992}.
If a ``?'' appears in the col. ID Simbad it indicates a poor degree of confidence on the identification in Simbad}
\centering
\begin{tabular}{@{\ }l@{\ \ }l@{\ \ }l@{\ \ }l@{\ \ }l@{\ \ }l@{\ \ }l@{\ \ }l@{\ \ }l@{\ \ }l@{\ \ }l@{\ \ }l@{\ \ }l@{\ \ }l@{\ }}
\hline
\hline
Cluster & Image & ID & (B-V)$_{0}$ & V$_{0}$ & Mv & (V-I)$_{0}$ & ST & ID Simbad & ST Simbad  \\
 & WFI & WFI &  &  &  &  &  &  &  \\
\hline
\hline
SMC8 & 2 & 34940 & -0.190 & 16.761 & -1.989 &   -0.034 & B2 &  &  \\
SMC9 & 2 & 35267 & -0.130 & 16.203 & -2.547 &   0.093 & B2 & [MA93] 20  & Em* \\
SMC11 & 2 & 39854 & -0.117 & 16.318 & -2.432 &  -0.013 & B2 &  &  \\
SMC11 & 2 & 40737 & -0.140 & 16.865 & -1.885 &  -0.048 & B2 &  &  \\
SMC11 & 2 & 40443 & -0.131 & 17.188 & -1.562 &  -0.066 & B3 &  &  \\
SMC15 & 1 & 8873 & 0.018 & 15.794 & -2.956 &    0.096 & B1 &  &  \\
\hline
\end{tabular}
\label{tableBe2}
\end{table*}
}
\small{
\addtocounter{table}{-1}
\begin{table*}[]
\caption{Be stars table2, continued}
\centering
\begin{tabular}{@{\ }l@{\ \ }l@{\ \ }l@{\ \ }l@{\ \ }l@{\ \ }l@{\ \ }l@{\ \ }l@{\ \ }l@{\ \ }l@{\ \ }l@{\ \ }l@{\ \ }l@{\ \ }l@{\ }}
\hline
\hline
Cluster & Image & ID & (B-V)$_{0}$ & V$_{0}$ & Mv & (V-I)$_{0}$ & ST & ID Simbad & ST Simbad \\
 & WFI & WFI &  &  &  &  &  &  &  \\
\hline
SMC107 & 9 & 36804 & -0.154 & 15.040 & -3.710 &  -0.015 & B0 & [MFH2007] SMC5-64327 & B3II-IIIe \\
SMC107 & 9 & 37317 & -0.147 & 15.372 & -3.378 &  -0.013 & B0 & Cl* NGC\, 330 ROB B36  & Em* \\
SMC107 & 9 & 38203 & -0.150 & 15.392 & -3.358 &  -0.004 & B0 & [MFH2007] SMC5-37137 & B2IIIe \\
SMC107 & 9 & 38679 & -0.143 & 15.217 & -3.533 &  -0.024 & B0 & [MFH2007] SMC5-13978  & B3IIIe \\
SMC107 & 9 & 38683 & -0.188 & 14.953 & -3.797 &  -0.070 & B0 & Cl* NGC\, 330 ROB A15  & Em* \\
SMC107 & 9 & 38865 & -0.201 & 15.498 & -3.252 &  -0.068 & B0 &  &  \\
\hline
\end{tabular}
\label{tableBe2}
\end{table*}
}
\small{
\addtocounter{table}{-1}
\begin{table*}[]
\caption{Be stars table2, continued}
\centering
\begin{tabular}{@{\ }l@{\ \ }l@{\ \ }l@{\ \ }l@{\ \ }l@{\ \ }l@{\ \ }l@{\ \ }l@{\ \ }l@{\ \ }l@{\ \ }l@{\ \ }l@{\ \ }l@{\ \ }l@{\ }}
\hline
\hline
Cluster & Image & ID & (B-V)$_{0}$ & V$_{0}$ & Mv & (V-I)$_{0}$ & ST & ID Simbad & ST Simbad  \\
 & WFI & WFI &  &  &  &  &  &  &  \\
\hline
SMC128 & 11 & 10309 & -0.195 & 17.082 & -1.668 & -0.254 & B3 &  &  \\
SMC134 & 11 & 23627 & -0.072 & 14.842 & -3.908 & -0.056 & B0 & [MA93] 1364 & Em* \\
SMC134 & 11 & 23610 & -0.136 & 16.769 & -1.981 & -0.001 & B2 &  &  \\
SMC134 & 11 & 24123 & 0.014 & 16.705 & -2.045 &  0.019 & B2 & [MB2000] 224 & Em* \\
SMC134 & 11 & 23713 & -0.063 & 17.007 & -1.743 & -0.001 & B3 &  &  \\
SMC134 & 11 & 23886 & -0.258 & 17.155 & -1.595 & 0.046 & B3 &  &  \\
\hline
\end{tabular}
\label{tableBe2}
\end{table*}
}


%
\tiny{
\begin{table*}[]
\caption{Table 1 of candidate Be stars: same as for table~\ref{tableBe1}}
\centering
\begin{tabular}{@{\ }l@{\ \ }l@{\ \ }l@{\ \ }l@{\ \ }l@{\ \ }l@{\ \ }l@{\ \ }l@{\ \ }l@{\ \ }l@{\ \ }l@{\ \ }l@{\ \ }l@{\ \ }l@{\ \ }l@{\ \ }l@{\ \ }l@{\ \ }l@{\ \ }l@{\ }}
\hline
\hline
Cluster & Image & Image & ID & ID  & r & RA(2000)   & DEC(2000)   & RA(2000)   & DEC(2000)   & V & B & I & B-V & V-I & E[B-V] & log(t) & Mult \\
 & WFI & Ogle & WFI & Ogle & \arcsec & WFI   & WFI   & Ogle   & Ogle   &  &  &  &  &  &  &  &  \\
\hline
\hline
SMC002 & 2 & sc1 & 11621 & 35532 & 0.9 & 0 37 41.567 & -73 37 4.805 & 0 37 41.720 & -73 37 4.300 & 15.941 & 15.886 & 15.790 & -0.055 & 0.151 & 0.060 & 8.4 & 1 \\
SMC008 & 2 & sc2 & 34493 & 35344 & 1.0 & 0 40 35.389 & -73 23 56.521 & 0 40 35.540 & -73 23 56.300 & 17.218 & 17.103 & 17.231 & -0.115 & -0.013 & 0.070 & 8 & 1 \\
SMC009 & 2 & sc2 & 35862 & 35312 & 0.8 & 0 40 44.245 & -73 23 14.832 & 0 40 44.350 & -73 23 14.600 & 16.393 & 16.291 & 16.447 & -0.102 & -0.054 & 0.070 & 8 & 1 \\
SMC009 & 2 & sc2 & 35791 & 35497 & 0.8 & 0 40 37.924 & -73 23 12.782 & 0 40 38.030 & -73 23 12.500 & 17.774 & 17.671 & 17.753 & -0.104 & 0.021 & 0.070 & 8 & 1 \\
SMC011 & 2 & sc2 & 40080 & 61911 & 0.7 & 0 41 6.174 & -73 21 4.138 & 0 41 6.250 & -73 21 3.800 & 17.586 & 17.345 & 17.276 & -0.241 & 0.309 & 0.080 & 7.9 & 1 \\
SMC011 & 2 & sc2 & 40465 & 64097 & 0.7 & 0 41 6.581 & -73 20 50.625 & 0 41 6.650 & -73 20 50.300 & 17.915 & 17.847 & 17.877 & -0.068 & 0.038 & 0.080 & 7.9 & 1 \\
\hline
\end{tabular}
\label{tableBecand1}
\end{table*}
}
\tiny{
\addtocounter{table}{-1}
\begin{table*}[]
\caption{Candidate Be stars table1, continued}
\centering
\begin{tabular}{@{\ }l@{\ \ }l@{\ \ }l@{\ \ }l@{\ \ }l@{\ \ }l@{\ \ }l@{\ \ }l@{\ \ }l@{\ \ }l@{\ \ }l@{\ \ }l@{\ \ }l@{\ \ }l@{\ \ }l@{\ \ }l@{\ \ }l@{\ \ }l@{\ \ }l@{\ }}
\hline
\hline
Cluster & Image & Image & ID & ID  & r & RA(2000)   & DEC(2000)   & RA(2000)   & DEC(2000)   & V & B & I & B-V & V-I & E[B-V] & log(t) & Mult \\
 & WFI & Ogle & WFI & Ogle & \arcsec & WFI   & WFI   & Ogle   & Ogle   &  &  &  &  &  &  &  &  \\
\hline
SMC99 & 7 & sc7 & 58955 & 70940 & 1.4 & 0 54 41.207 & -72 28 2.335 & 0 54 41.320 & -72 28 2.200 & 17.371 & 17.282 & 17.352 & -0.089 & 0.019 & 0.080 & 7.6 & 1 \\
SMC107 & 9 & sc7 & 38635 & 206127 & 1.3 & 0 56 20.764 & -72 27 53.744 & 0 56 20.870 & -72 27 54.300 & 17.003 & 16.972 & 16.903 & -0.031 & 0.100 & 0.100 & 7.5 & 1 \\
SMC107 & 9 & sc7 & 38660 & 206138 & 0.6 & 0 56 23.650 & -72 27 52.520 & 0 56 23.660 & -72 27 52.200 & 17.210 & 17.178 & 17.124 & -0.032 & 0.086 & 0.100 & 7.5 & 1 \\
SMC107 & 9 & sc7 & 39170 & 206197 & 0.8 & 0 56 2.796 & -72 27 20.533 & 0 56 2.800 & -72 27 20.100 & 17.117 & 17.007 & 17.145 & -0.110 & -0.028 & 0.100 & 7.5 & 1 \\
SMC107 & 9 & sc7 & 35214 & 202114 & 1.4 & 0 56 13.101 & -72 29 41.006 & 0 56 13.200 & -72 29 40.500 & 17.593 & 17.554 & 17.541 & -0.039 & 0.052 & 0.100 & 7.5 & 1 \\
SMC107 & 9 & sc7 & 35856 & 138111 & 1.1 & 0 55 59.550 & -72 29 6.606 & 0 55 59.610 & -72 29 6.000 & 18.434 & 18.374 & 18.397 & -0.059 & 0.037 & 0.100 & 7.5 & 1 \\
\hline
\end{tabular}
\label{tableBecand1}
\end{table*}
}


%
\small{
\begin{table*}[]
\caption{Table 2 of candidate Be stars:  same as for table~\ref{tableBe2}}
\centering
\begin{tabular}{@{\ }l@{\ \ }l@{\ \ }l@{\ \ }l@{\ \ }l@{\ \ }l@{\ \ }l@{\ \ }l@{\ \ }l@{\ \ }l@{\ \ }l@{\ \ }l@{\ \ }l@{\ \ }l@{\ }}
\hline
\hline
Cluster & Image & ID  & (B-V)$_{0}$ & V$_{0}$ & Mv &  (V-I)$_{0}$ & ST & ID Simbad & ST Simbad \\
        & WFI   & WFI &             &         &    &              &    &           &           \\
\hline
\hline
SMC2 & 2 & 11621 & -0.115 & 15.755 & -2.995 &  0.065 & B1 &  &  \\
SMC8 & 2 & 34493 & -0.185 & 17.001 & -1.749 &  -0.114 & B3 &  &  \\
SMC9 & 2 & 35862 & -0.172 & 16.176 & -2.574 &  -0.155 & B1 &  &  \\
SMC9 & 2 & 35791 & -0.174 & 17.557 & -1.193 &  -0.080 & B4 &  &  \\
SMC11 & 2 & 40080 & -0.321 & 17.338 & -1.412 & 0.194 & B3 &  &  \\
SMC11 & 2 & 40465 & -0.148 & 17.667 & -1.083 & -0.077 & B4 &  &  \\
\hline
\end{tabular}
\label{tableBecand2}
\end{table*}
}
\small{
\addtocounter{table}{-1}
\begin{table*}[]
\caption{Candidate Be stars table2, continued}
\centering
\begin{tabular}{@{\ }l@{\ \ }l@{\ \ }l@{\ \ }l@{\ \ }l@{\ \ }l@{\ \ }l@{\ \ }l@{\ \ }l@{\ \ }l@{\ \ }l@{\ \ }l@{\ \ }l@{\ \ }l@{\ }}
\hline
\hline
Cluster & Image & ID  & (B-V)$_{0}$ & V$_{0}$ & Mv &  (V-I)$_{0}$ & ST & ID Simbad & ST Simbad \\
        & WFI   & WFI &             &         &    &              &    &           &           \\
\hline
SMC99 & 7 & 58955 & -0.169 & 17.123 & -1.627 &  -0.096 & B3 &  &  \\
SMC107 & 9 & 38635 & -0.131 & 16.693 & -2.057 & -0.044 & B2 &  &  \\
SMC107 & 9 & 38660 & -0.132 & 16.900 & -1.850 & -0.058 & B2 &  &  \\
SMC107 & 9 & 39170 & -0.210 & 16.807 & -1.943 & -0.172 & B2 & Cl* NGC\, 330 ROB B18  & *iC \\
SMC107 & 9 & 35214 & -0.139 & 17.283 & -1.467 & -0.092 & B3 &  &  \\
SMC107 & 9 & 37709 & -0.198 & 17.662 & -1.088 & 0.013 & B4 &  &  \\
\hline
\end{tabular}
\label{tableBecand2}
\end{table*}
}


%
\tiny{
\begin{table*}[]
\caption{Table 1 of Ae and Oe stars: cross-correlation WFI with Ogle:  same as for table~\ref{tableBe1} }
\centering
\begin{tabular}{@{\ }l@{\ \ }l@{\ \ }l@{\ \ }l@{\ \ }l@{\ \ }l@{\ \ }l@{\ \ }l@{\ \ }l@{\ \ }l@{\ \ }l@{\ \ }l@{\ \ }l@{\ \ }l@{\ \ }l@{\ \ }l@{\ \ }l@{\ \ }l@{\ \ }l@{\ }}
\hline
\hline
Cluster & Image & Image & ID & ID & r & RA(2000)   & DEC(2000)   & RA(2000)   & DEC(2000)   & V & B & I & B-V & V-I & E[B-V] & log(t) & Mult \\
 & WFI & Ogle & WFI & Ogle & \arcsec & WFI   & WFI   & Ogle   & Ogle   &  &  &  &  &  &  &  &  \\
\hline
\hline
SMC11 & 2 & sc2 & 39734 & 63847 & 0.8 & 0 41 2.835 & -73 21 11.479 & 0 41 2.860 & -73 21 12.600 & 20.762 & 20.832 & 20.417 & 0.069 & 0.344 & 0.080 & 7.9 & 1 \\
SMC39 & 4 & sc4 & 18277 & 104655 & 1.3 & 0 47 7.353 & -73 28 24.163 & 0 47 7.390 & -73 28 23.200 & 19.937 & 19.822 & 19.993 & -0.115 & -0.056 & 0.110 & 8 & 1 \\
SMC39 & 6 & sc4 & 11532 & 106710 & 0.6 & 0 47 13.047 & -73 28 19.429 & 0 47 12.980 & -73 28 19.300 & 20.467 & 20.315 & 20.476 & -0.152 & -0.009 & 0.110 & 8 & 1 \\
SMC49 & 6 & sc5 & 20245 & 14801 & 0.9 & 0 48 43.036 & -73 24 39.905 & 0 48 42.950 & -73 24 40.800 & 20.820 & 21.054 & 20.668 & 0.233 & 0.151 & 0.060 & 7 & 1 \\
SMC89 & 7 & sc6 & 42527 & 169518 & 1.2 & 0 52 52.054 & -72 37 25.012 & 0 52 52.030 & -72 37 26.400 & 20.387 & 20.522 & 20.439 & 0.135 & -0.052 & 0.090 & 7.3 & 1 \\
SMC141 & 12 & sc10 & 41782 & 53127 & 1.4 & 1 4 35.572 & -72 36 58.556 & 1 4 35.690 & -72 36 59.300 & 19.840 & 19.945 & 19.684 & 0.104 & 0.155 & 0.090 & 8.2 & 1 \\
\hline
\end{tabular}
\label{tableAeOe1}
\end{table*}
}

\small{
\begin{table*}[]
\caption{Table 2 of Ae and Oe stars:  same as for table~\ref{tableBe2}}
\centering
\begin{tabular}{@{\ }l@{\ \ }l@{\ \ }l@{\ \ }l@{\ \ }l@{\ \ }l@{\ \ }l@{\ \ }l@{\ \ }l@{\ \ }l@{\ \ }l@{\ \ }l@{\ \ }l@{\ \ }l@{\ }}
\hline
\hline
Cluster & Image & ID & (B-V)$_{0}$ & V$_{0}$ & Mv & (V-I)$_{0}$ & ST & ID Simbad & ST Simbad \\
 & WFI & WFI &  &  &  &  &  &  &  \\
\hline
\hline
SMC11 & 2 & 39734 & -0.011 & 20.514 & 1.764   & 0.229 & A3-A4 &  &   \\
SMC39 & 4 & 18277 & -0.225 & 19.596 & 0.846   & -0.214 & A0 &  &  \\
SMC39 & 6 & 11532 & -0.262 & 20.126 & 1.376   & -0.167 & A2 &  &  \\
SMC49 & 6 & 20245 & 0.173 & 20.634 & 1.884    & 0.065 & A3-A4 &  &  \\
SMC89 & 7 & 42527 & 0.045 & 20.108 & 1.358    & -0.182 & A2 &  &  \\
SMC141 & 12 & 41782 & 0.014 & 19.561 & 0.811  & 0.025 & A0 &  &  \\
SMC149 & 11 & 44134 & -0.253 & 19.554 & 0.804 & 0.080 & A0 &  & \\
\hline                           
SMC17 & 2 & 33282 & 0.081 & 13.868 & -4.882    & 0.187 & O8 & LHA 115-S 3 & A7/HAeBe?  \\
SMC33 & 4 & 25961 & -0.211 & 14.081 & -4.669   & -0.215 & O8 &  &  \\
SMC71 & 5 & 26201 & -0.128 & 13.852 & -4.898   & -0.021 & O8 & [MA93] 470  & Em* \\
SMC105 & 7 & 18826 & -0.170 & 14.436 & -4.314  & -0.016 & O9 & 2MASS J00554860-7252073  & Em* \\
SMC107 & 9 & 37633 & -0.072 & 13.085 & -5.665  & 0.021 & O5 & [MFH2007] SMC5-82819 & B2IIIe \\
SMC187 & 4 & 29240 & -0.062 & 14.324 & -4.426  & 0.135 & O9 & LIN 88  & Em* \\
\hline
\end{tabular}
\label{tableAeOe2}
\end{table*}
}


%
\tiny{
\begin{table*}[]
\caption{Table 1 of other kind of emission line stars (not Main Sequence stars):  same as for table~\ref{tableBe1}.}
\centering
\begin{tabular}{@{\ }l@{\ \ }l@{\ \ }l@{\ \ }l@{\ \ }l@{\ \ }l@{\ \ }l@{\ \ }l@{\ \ }l@{\ \ }l@{\ \ }l@{\ \ }l@{\ \ }l@{\ \ }l@{\ \ }l@{\ \ }l@{\ \ }l@{\ \ }l@{\ \ }l@{\ }}
\hline
\hline
Cluster & Image & Image & ID & ID & r & RA(2000)   & DEC(2000)   & RA(2000)   & DEC(2000)   & V & B & I & B-V & V-I & E[B-V] & log(t) & Mult \\
 & WFI & Ogle & WFI & Ogle & \arcsec & WFI   & WFI   & Ogle   & Ogle   &  &  &  &  &  &  &  &  \\
\hline
\hline
SMC008 & 2 & sc2 & 32681 & 33368 & 1.2 & 0 40 22.224 & -73 24 49.108 & 0 40 22.420 & -73 24 48.900 & 17.847 & 18.481 & 17.071 & 0.631 & 0.775 & 0.070 & 8 & 1 \\
SMC008 & 2 & sc2 & 33371 & 35303 & 1.5 & 0 40 24.878 & -73 24 26.225 & 0 40 25.120 & -73 24 26.000 & 18.069 & 19.699 & 15.803 & 1.622 & 2.264 & 0.070 & 8 & 1 \\
SMC008 & 2 & sc2 & 32765 & 33369 & 1.3 & 0 40 23.181 & -73 24 46.718 & 0 40 23.390 & -73 24 46.500 & 18.371 & 19.231 & 17.328 & 0.856 & 1.042 & 0.070 & 8 & 1 \\
SMC008 & 2 & sc2 & 33200 & 34401 & 0.5 & 0 40 27.520 & -73 24 36.202 & 0 40 27.470 & -73 24 36.800 & 20.375 & 19.699 & 20.097 & -0.675 & 0.277 & 0.070 & 8 & 2 \\
SMC015 & 1 & sc3 & 9969 & 19977 & 1.0 & 0 42 54.809 & -73 17 15.972 & 0 42 54.770 & -73 17 15.400 & 16.644 & 18.242 & 15.152 & 1.592 & 1.491 & 0.100 & 8.1 & 1 \\
SMC015 & 1 & sc3 & 9516 & 19975 & 1.1 & 0 42 53.612 & -73 17 29.583 & 0 42 53.540 & -73 17 29.000 & 17.246 & 18.649 & 15.412 & 1.397 & 1.834 & 0.100 & 8.1 & 1 \\
\hline
\end{tabular}
\label{table1otherels}
\end{table*}
}
\tiny{
\addtocounter{table}{-1}
\begin{table*}[]
\caption{Other emission-line star table1, continued}
\centering
\begin{tabular}{@{\ }l@{\ \ }l@{\ \ }l@{\ \ }l@{\ \ }l@{\ \ }l@{\ \ }l@{\ \ }l@{\ \ }l@{\ \ }l@{\ \ }l@{\ \ }l@{\ \ }l@{\ \ }l@{\ \ }l@{\ \ }l@{\ \ }l@{\ \ }l@{\ \ }l@{\ }}
\hline
\hline
Cluster & Image & Image & ID & ID & r & RA(2000)   & DEC(2000)   & RA(2000)   & DEC(2000)   & V & B & I & B-V & V-I & E[B-V] & log(t) & Mult \\
 & WFI & Ogle & WFI & Ogle & \arcsec & WFI   & WFI   & Ogle   & Ogle   &  &  &  &  &  &  &  &  \\
\hline
SMC70 & 6 & sc5 & 39108 & 261948 & 1.1 & 0 51 25.509 & -73 17 9.571 & 0 51 25.520 & -73 17 8.800 & 19.636 & 20.363 & 18.654 & 0.724 & 0.981 & 0.090 & 7.8 & 1 \\
SMC71 & 5 & sc5 & 26433 & 288728 & 0.5 & 0 51 24.596 & -73 0 51.497 & 0 51 24.660 & -73 0 52.000 & 15.915 & 15.888 & 15.247 & -0.028 & 0.667 & 0.070 & 7.5 & 1 \\
SMC72 & 6 & sc6 & 47906 & 17311 & 0.9 & 0 51 42.821 & -73 13 27.500 & 0 51 42.780 & -73 13 27.100 & 16.113 & 16.113 & 15.721 & -0.001 & 0.392 & 0.080 & 7.6 & 2 \\
SMC72 & 6 & sc6 & 47440 & 18048 & 0.6 & 0 51 38.503 & -73 13 23.609 & 0 51 38.500 & -73 13 23.300 & 18.846 & 19.410 & 18.046 & 0.561 & 0.798 & 0.080 & 7.6 & 1 \\
SMC73 & 5 & sc6 & 47237 & 61592 & 0.7 & 0 51 44.171 & -72 49 51.499 & 0 51 44.140 & -72 49 51.500 & 18.021 & 18.971 & 16.946 & 0.947 & 1.074 & 0.110 & 8.2 & 1 \\
SMC74 & 5 & sc6 & 33275 & 49197 & 0.2 & 0 51 55.562 & -72 57 52.561 & 0 51 55.590 & -72 57 53.000 & 16.560 & 17.270 & 15.946 & 0.708 & 0.613 & 0.090 & 8.1 & 1 \\
\hline
\end{tabular}
\label{table1otherels}
\end{table*}
}
\tiny{
\addtocounter{table}{-1}
\begin{table*}[]
\caption{Other emission-line star table1, continued}
\centering
\begin{tabular}{@{\ }l@{\ \ }l@{\ \ }l@{\ \ }l@{\ \ }l@{\ \ }l@{\ \ }l@{\ \ }l@{\ \ }l@{\ \ }l@{\ \ }l@{\ \ }l@{\ \ }l@{\ \ }l@{\ \ }l@{\ \ }l@{\ \ }l@{\ \ }l@{\ \ }l@{\ }}
\hline
\hline
Cluster & Image & Image & ID & ID & r & RA(2000)   & DEC(2000)   & RA(2000)   & DEC(2000)   & V & B & I & B-V & V-I & E[B-V] & log(t) & Mult \\
 & WFI & Ogle & WFI & Ogle & \arcsec & WFI   & WFI   & Ogle   & Ogle   &  &  &  &  &  &  &  &  \\
\hline
SMC109 & 9 & sc7? & 61352 & 52936 & 0.9 & 0 57 31.114 & -72 16 0.786 & 0 57 31.170 & -72 16 0.400 & 17.537 & 17.554 & 17.094 & 0.016 & 0.443 & 0.040 & 7.7 & 1 \\
SMC109 & 9 & sc7? & 61387 & 54360 & 1.3 & 0 57 29.499 & -72 15 44.768 & 0 57 29.550 & -72 15 45.800 & 18.637 & 99.999 & 19.054 & 9.999 & -0.416 & 0.040 & 7.7 & 1 \\
SMC109 & 9 & sc7? & 61314 & 56075 & 1.3 & 0 57 30.225 & -72 15 58.665 & 0 57 30.190 & -72 15 59.800 & 19.285 & 18.798 & 18.822 & -0.487 & 0.463 & 0.040 & 7.7 & 1 \\
SMC117 & 9 & sc8 & 25911 & 139438 & 1.7 & 0 59 8.344 & -72 36 55.340 & 0 59 8.460 & -72 36 55.000 & 17.504 & 18.668 & 16.290 & 1.159 & 1.213 & 0.130 & 8.3 & 1 \\
SMC117 & 9 & sc8 & 27255 & 139451 & 1.7 & 0 59 10.979 & -72 36 17.866 & 0 59 11.100 & -72 36 17.600 & 17.114 & 18.590 & 15.712 & 1.471 & 1.400 & 0.130 & 8.3 & 1 \\
SMC117 & 9 & sc8 & 27276 & 139453 & 1.7 & 0 59 9.653 & -72 36 15.719 & 0 59 9.770 & -72 36 15.500 & 17.595 & 18.946 & 16.228 & 1.346 & 1.366 & 0.130 & 8.3 & 1 \\
\hline
\end{tabular}
\label{table1otherels}
\end{table*}
}

\small{
\begin{table*}[]
\caption{Table 2 of other kind of emission line stars (not Main Sequence stars):  same as for table~\ref{tableBe2}.
Note that for these stars, the classification (for main sequence stars) is less reliable than in other previous tables}
\centering
\begin{tabular}{@{\ }l@{\ \ }l@{\ \ }l@{\ \ }l@{\ \ }l@{\ \ }l@{\ \ }l@{\ \ }l@{\ \ }l@{\ \ }l@{\ \ }l@{\ }}
\hline
\hline
Cluster & Image & ID & (B-V)$_{0}$ & V$_{0}$ & Mv & (V-I)$_{0}$  & ST  & ID Simbad & ST Simbad \\
 & WFI & WFI &  &  &  &   &   &  &  \\
\hline
\hline
SMC008 & 2 & 32681 & 0.561 & 17.630 & -1.120 & 0.674  & G0-2III-II:  &  &  \\
SMC008 & 2 & 33371 & 1.552 & 17.852 & -0.898 & 2.163  & M3III:  &  &  \\
SMC008 & 2 & 32765 & 0.786 & 18.154 & -0.596 & 0.941  & G5III   &  &  \\
SMC008 & 2 & 33200 & -0.745 & 20.158 & 1.408 & 0.176  & A2  & [MA93] 11 & Em* \\
SMC015 & 1 & 9969 & 1.492 & 16.334 & -2.416 & 1.347  & M2III-II:  &  &  \\
SMC015 & 1 & 9516 & 1.297 & 16.936 & -1.814 & 1.690  & K9III  &  &  \\
\hline
\end{tabular}
\label{table2otherels}
\end{table*}
}
\small{
\addtocounter{table}{-1}
\begin{table*}[]
\caption{Other emission-line star table 2, continued}
\centering
\begin{tabular}{@{\ }l@{\ \ }l@{\ \ }l@{\ \ }l@{\ \ }l@{\ \ }l@{\ \ }l@{\ \ }l@{\ \ }l@{\ \ }l@{\ \ }l@{\ \ }l@{\ \ }l@{\ \ }l@{\ \ }l@{\ \ }l@{\ \ }l@{\ \ }l@{\ \ }l@{\ }}
\hline
\hline
Cluster & Image & ID & (B-V)$_{0}$ & V$_{0}$ & Mv & (V-I)$_{0}$  & ST  & ID Simbad & ST Simbad \\
 & WFI & WFI &  &  &  &   &   &  &  \\
\hline
SMC70 & 6 & 39108 & 0.634 & 19.357 & 0.607 & 0.851  & F1III:  &  &  \\
SMC71 & 5 & 26433 & -0.098 & 15.698 & -3.052 & 0.566  & B9III-II:  &  &  \\
SMC72 & 6 & 47906 & -0.081 & 15.865 & -2.885 & 0.277  & B8III:  & [WSH2003] d-129 ? & IR \\
SMC72 & 6 & 47440 & 0.481 & 18.598 & -0.152 & 0.683  & A8III:  &  &  \\
SMC73 & 5 & 47237 & 0.837 & 17.680 & -1.070 & 0.916  & G8III-II:  &  &  \\
SMC74 & 5 & 33275 & 0.618 & 16.281 & -2.469 & 0.483  & F9II:  & 2MASS J00515560-7257534  & IR, Cepheid \\
\hline
\end{tabular}
\label{table2otherels}
\end{table*}
}
\small{
\addtocounter{table}{-1}
\begin{table*}[]
\caption{Other emission-line star table 2, continued}
\centering
\begin{tabular}{@{\ }l@{\ \ }l@{\ \ }l@{\ \ }l@{\ \ }l@{\ \ }l@{\ \ }l@{\ \ }l@{\ \ }l@{\ \ }l@{\ \ }l@{\ \ }l@{\ \ }l@{\ \ }l@{\ \ }l@{\ \ }l@{\ \ }l@{\ \ }l@{\ \ }l@{\ }}
\hline
\hline
Cluster & Image & ID & (B-V)$_{0}$ & V$_{0}$ & Mv & (V-I)$_{0}$  & ST  & ID Simbad & ST Simbad \\
 & WFI & WFI &  &  &  &   &   &  &  \\
\hline
SMC117 & 9 & 25911 & 1.029 & 17.101 & -1.649 & 1.026  & K3III-II:  &  &  \\
SMC117 & 9 & 27255 & 1.341 & 16.711 & -2.039 & 1.213  & M1III-II:  &  &  \\
SMC117 & 9 & 27276 & 1.216 & 17.192 & -1.558 & 1.179  & K5III  &  &  \\
SMC117 & 9 & 27278 & \_ & 19.769 & 1.019 & 0.298  & A1  &  &  \\
SMC120 & 9 & 52998 & \_ & 18.102 & -0.648 & 1.165  & B5-B6  &  &  \\
SMC121 & 11 & 3908 & 1.214 & 17.392 & -1.358 & 1.214  & K5III  &  &  \\
\hline
\end{tabular}
\label{table2otherels}
\end{table*}
}

\end{landscape}
\end{flushleft}


%
\small{
\begin{longtable}{llllllllll}
\caption{\label{primels} emission-line star from WFI without identification in SC catalogues from Ogle \citep{oglemapsphoto}. The status indicates
whether the star is found as a definite emission-line star (ELS) or a candidate emission-line star (ELS?)}\\
\hline
\hline
Cluster & Image & ID  & RA(2000)   & DEC(2000)   & E[B-V] & log(t) & status  & ID Simbad & ST Simbad \\
        & WFI   & WFI & WFI        & WFI         &        &        &         &           &   \\
\hline
\endfirsthead
\caption{continued.}\\
\hline
\hline
Cluster & Image & ID  & RA(2000)   & DEC(2000)   & E[B-V] & log(t) & status  & ID Simbad & ST Simbad \\
        & WFI   & WFI & WFI        & WFI         &        &        &         &           &   \\
\hline
\hline		
\endhead
\hline
\endfoot
SMC12 & 1 & 45403 & 0 41 24.034 & -72 53 54.477 & 0.06 & 9 & ELS?  &  &  \\
SMC12 & 1 & 46376 & 0 41 18.777 & -72 53 8.976 & 0.06 & 9 & ELS?  &  &  \\
SMC47 & 5 & 27197 & 0 48 34.021 & -72 58 35.742 & 0.12 & 7.8 & ELS  &  &  \\
SMC49 & 6 & 21247 & 0 48 40.269 & -73 24 8.730 & 0.06 & 7 & ELS?  &  &  \\
SMC64 & 5 & 31661 & 0 50 39.163 & -72 57 51.239 & 0.1 & 8.1 & ELS  & OGLE J005039.05-725751.4 & Be? \\
SMC78 & 8 & 49392 & 0 52 16.683 & -73 1 24.842 & 0.08 & 7.9 & ELS  & LIN 202  & Em* \\
SMC81 & 7 & 36209 & 0 52 37.860 & -72 40 49.885 & 0.12 & 7.4 & ELS?  &  &  \\
SMC82 & 7 & 61803 & 0 52 39.276 & -72 55 29.603 & 0.12 & 7.4 & ELS  &  &  \\
\end{longtable}
}



\newpage

\twocolumn

\section{HR diagrammes for individual open clusters}
\label{HRind}
Colour/magnitude (M$_{v}$, (B-V)$_{0}$) diagrammes are presented
separately for each open cluster in the SMC with at least 10 members
and at least 1 WFI emission-line star with OGLE photometry
\citet{oglemapsphoto}

\begin{figure*}
\begin{tabular}{cc}
    \centering
    \includegraphics[angle=-90, width=8cm]{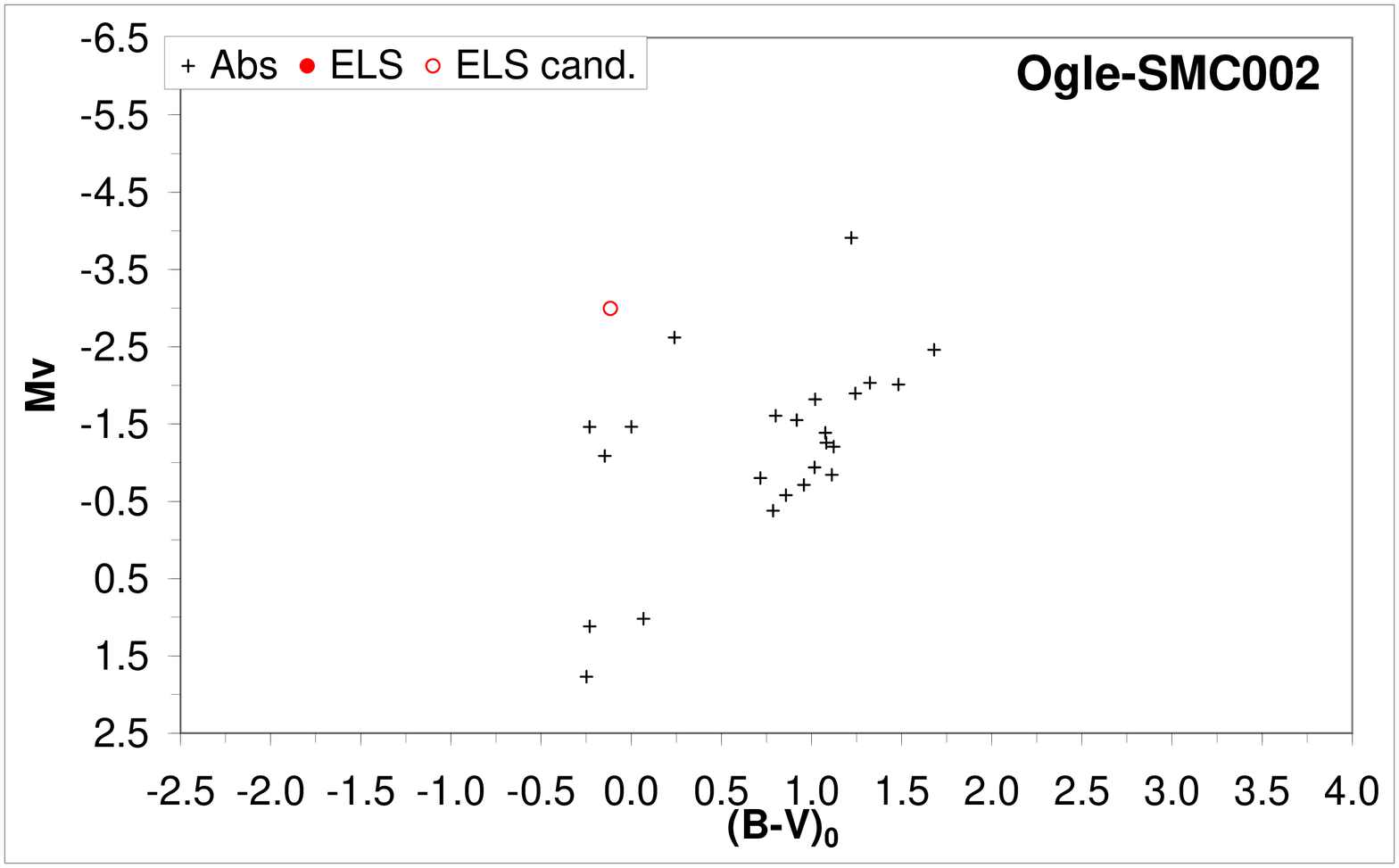} &  \includegraphics[angle=-90, width=8cm]{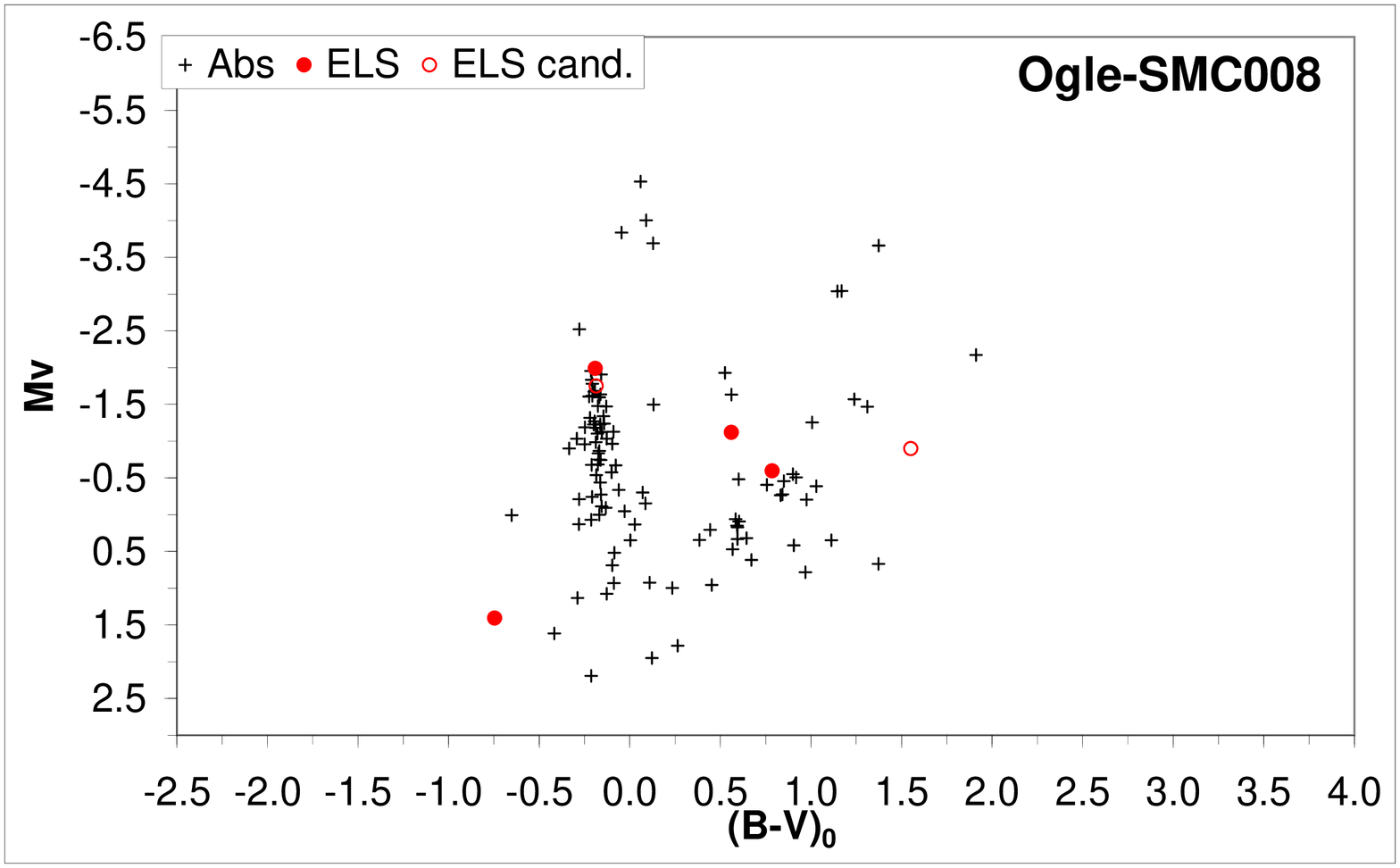} \\
    \includegraphics[angle=-90, width=8cm]{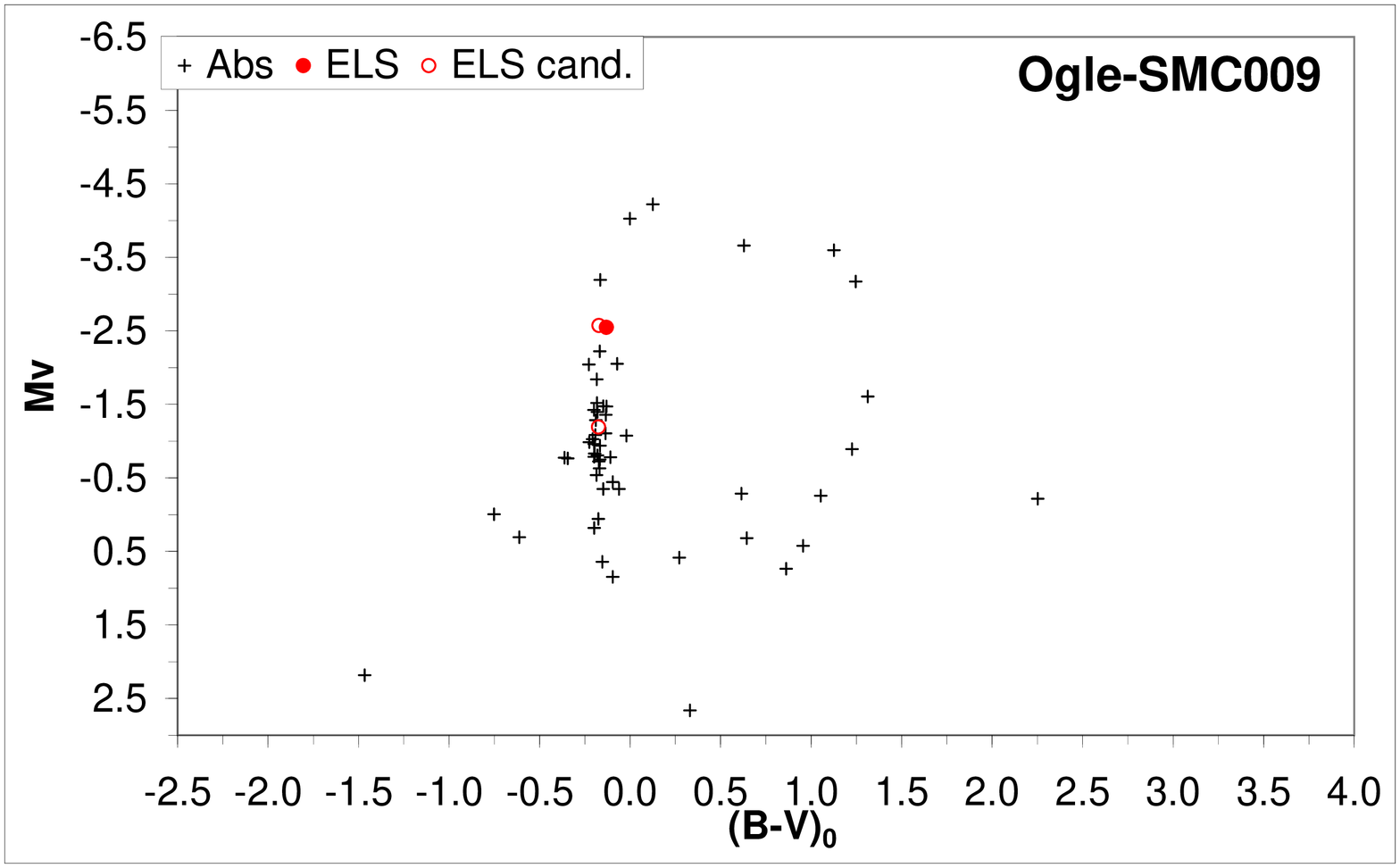} &  \includegraphics[angle=-90, width=8cm]{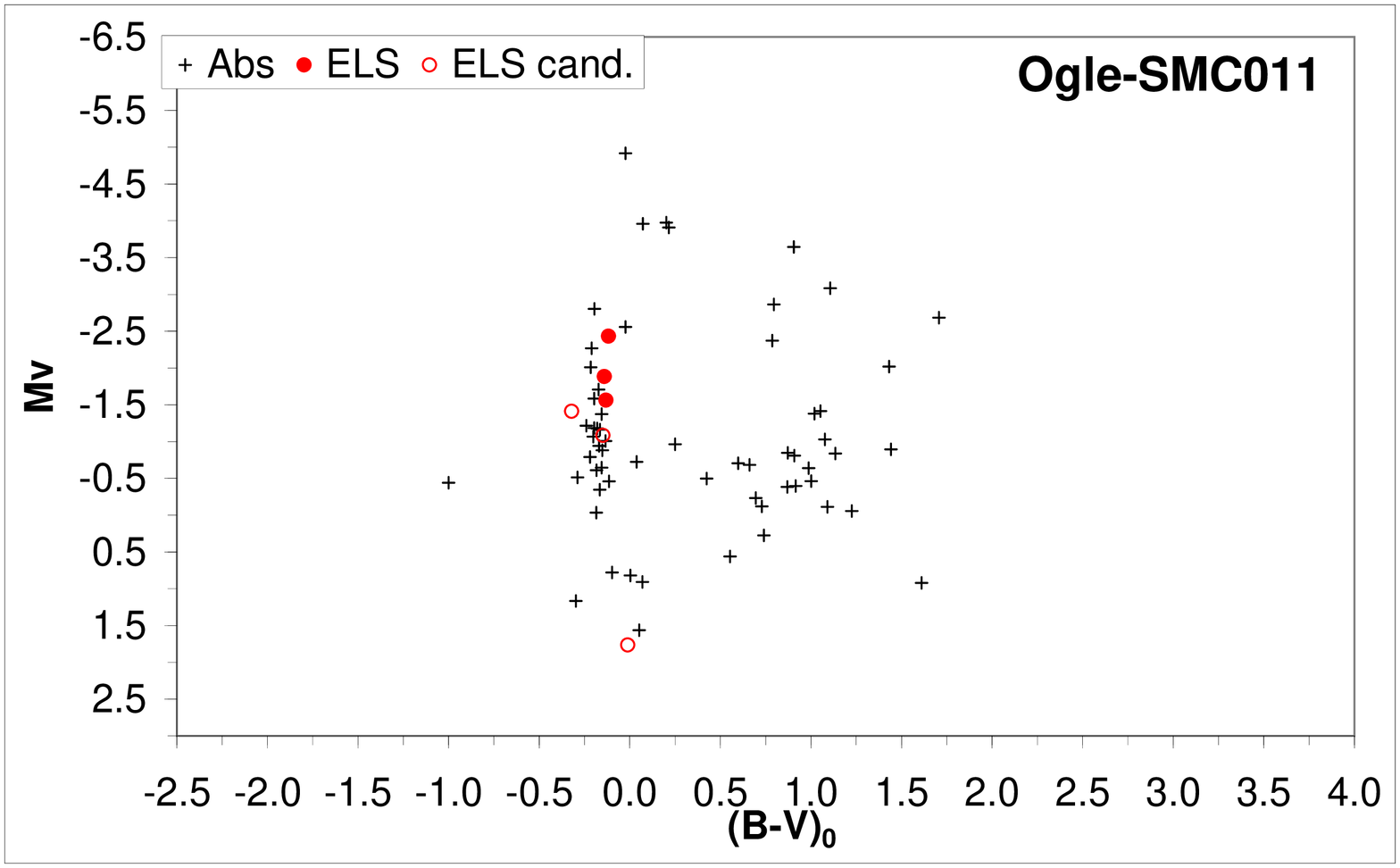} \\
    \includegraphics[angle=-90, width=8cm]{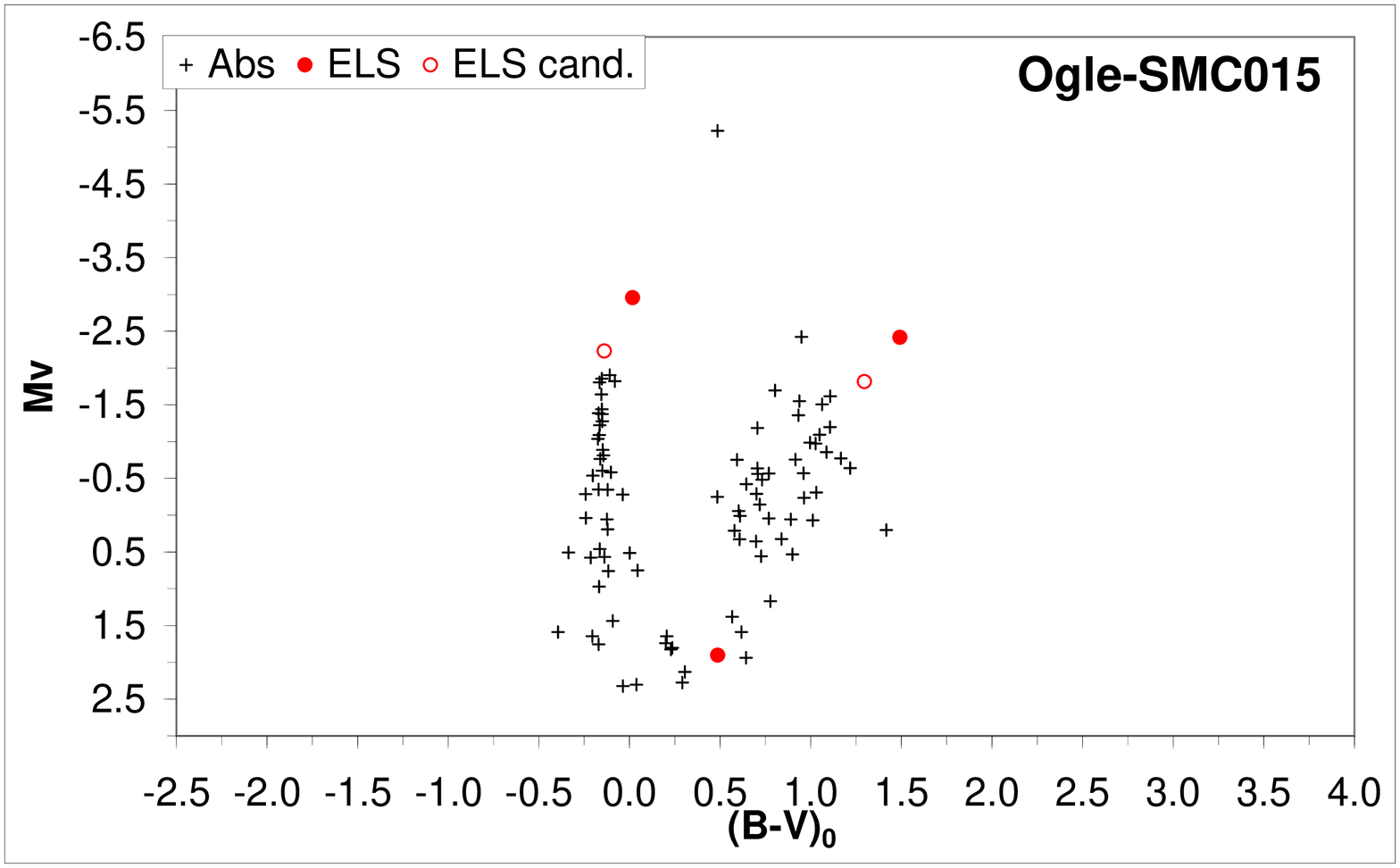} &  \includegraphics[angle=-90, width=8cm]{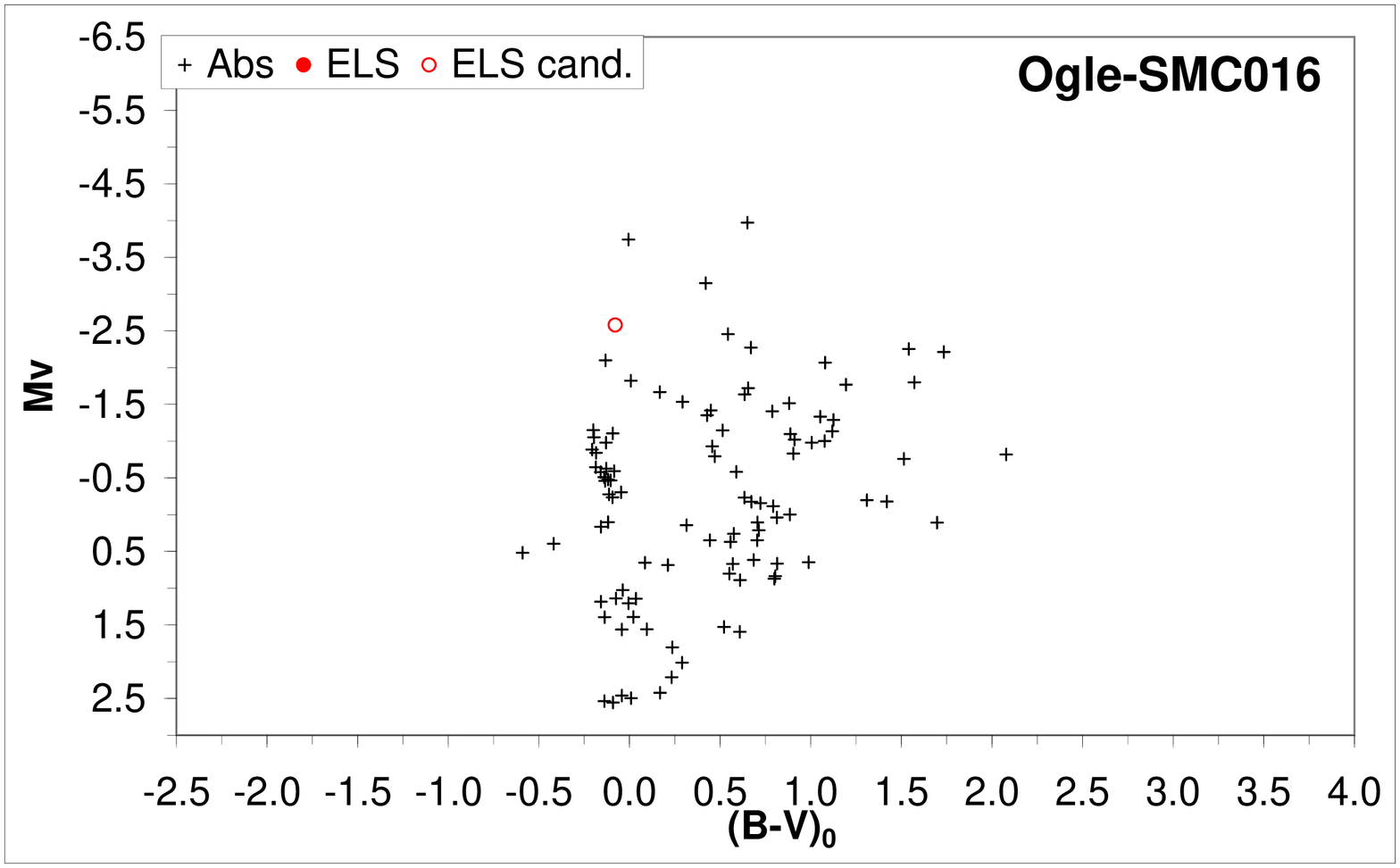} \\
    \includegraphics[angle=-90, width=8cm]{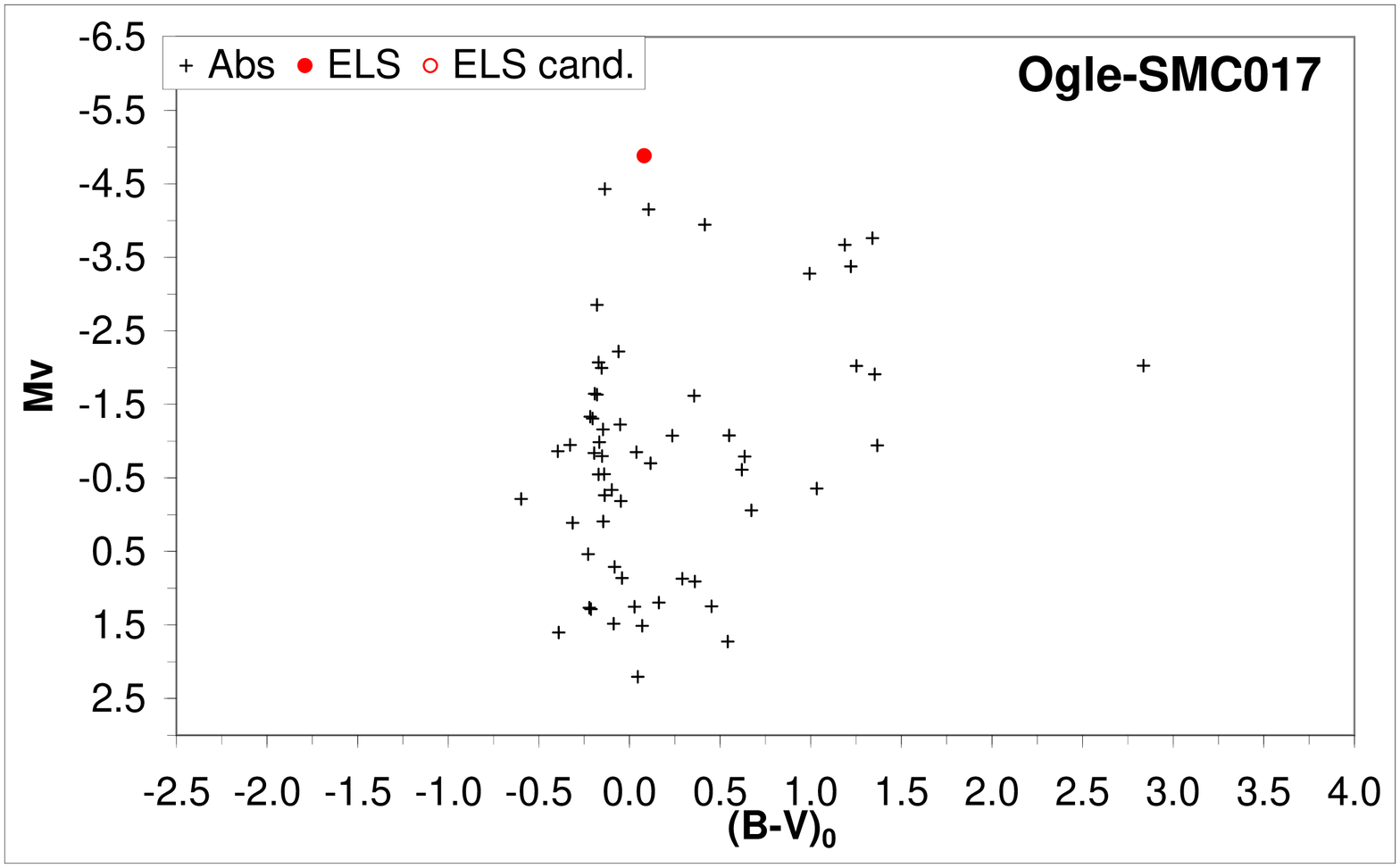} &  \includegraphics[angle=-90, width=8cm]{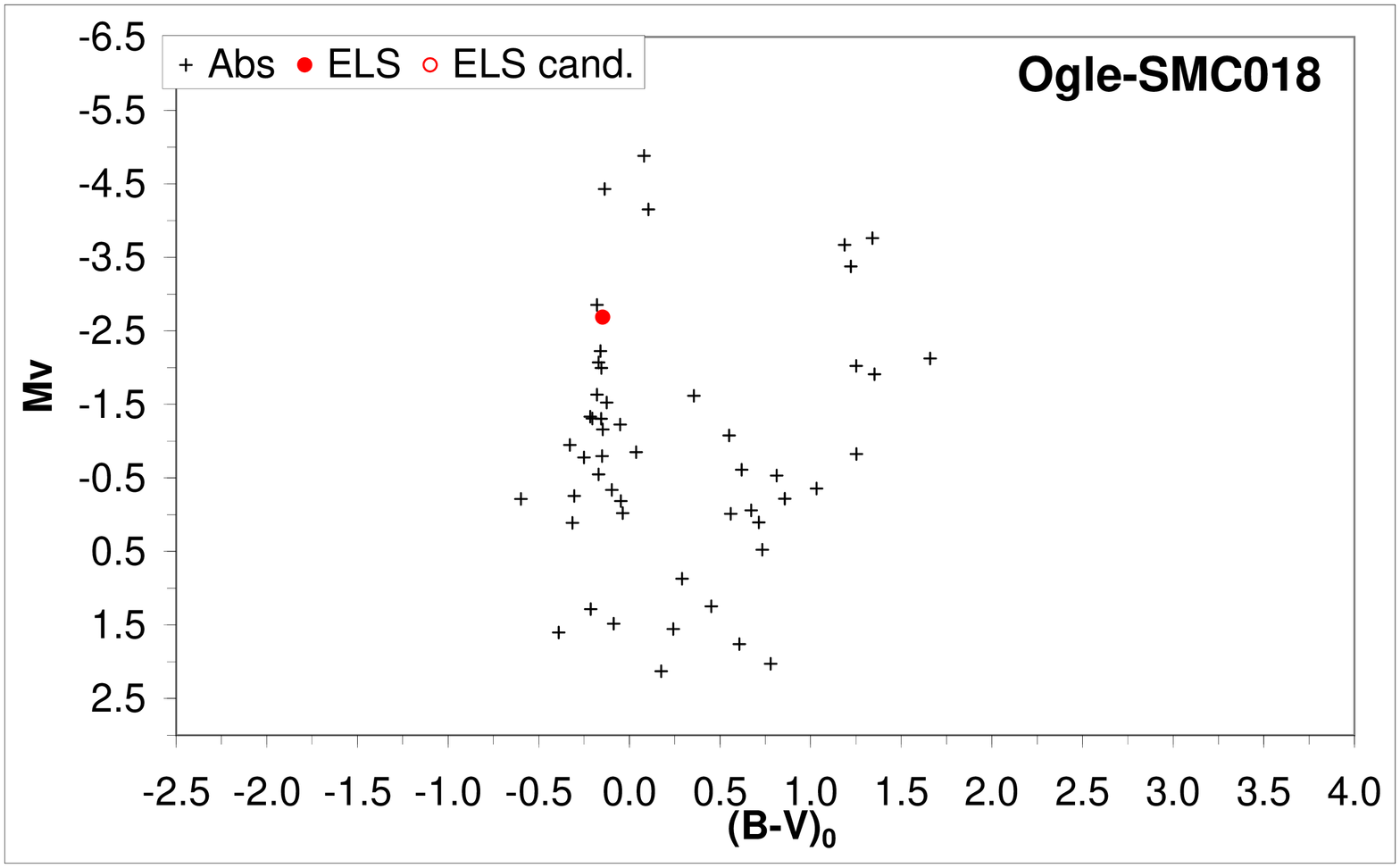} \\
\end{tabular}
\caption{Dereddened colour (B-V)$_{0}$, absolute magnitude Mv diagrammes for open clusters in the SMC with at least 
10 stars and at least 1 emission-line star (red filled circle) or candidate emission-line star (red opened circle). Normal stars appear as black '+'.
The identification of each open cluster is given in the right-top corner.}
\label{HR1}
\end{figure*}

\begin{figure*}
\begin{tabular}{cc}
    \centering
    \includegraphics[angle=-90, width=8cm]{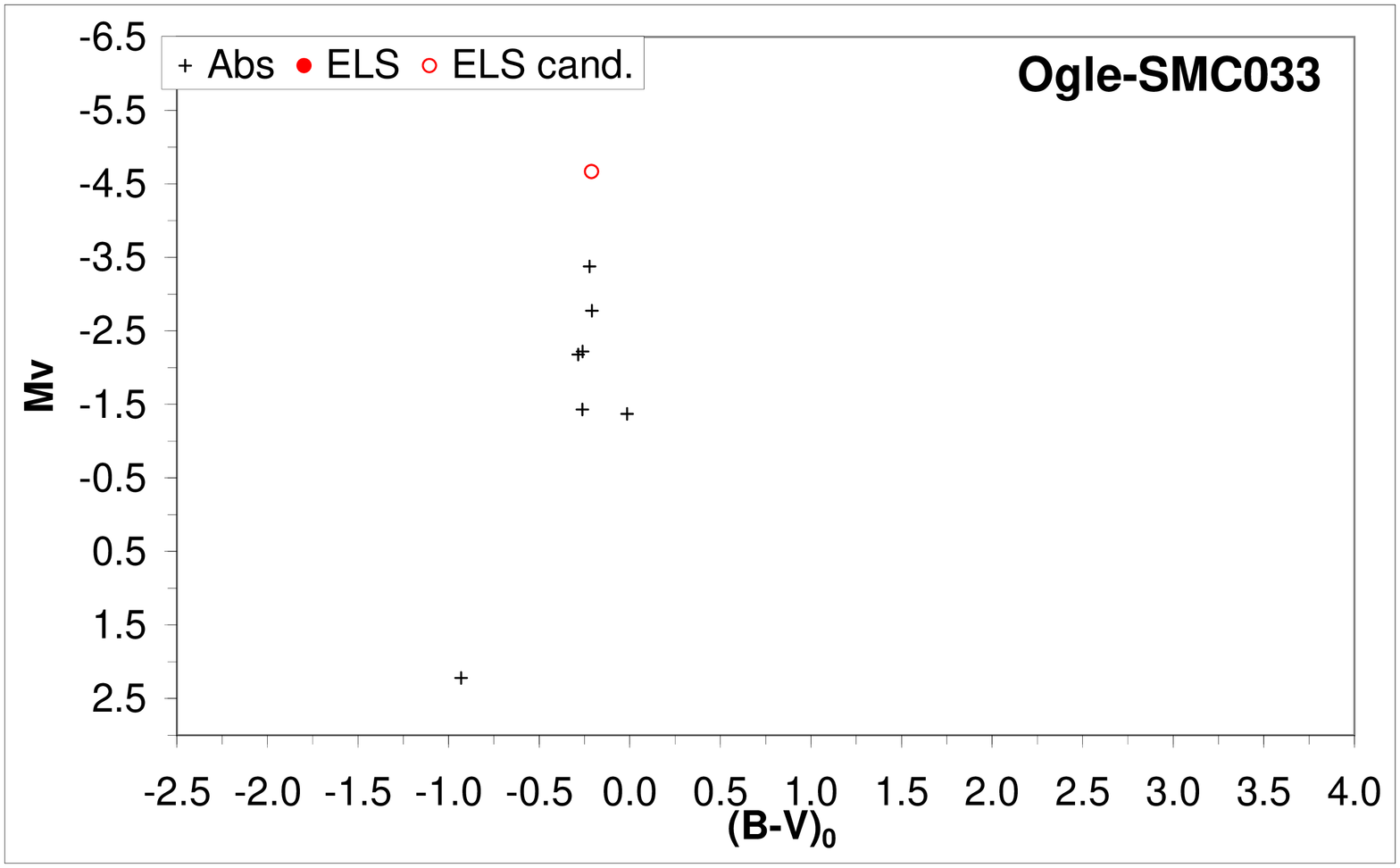} &  \includegraphics[angle=-90, width=8cm]{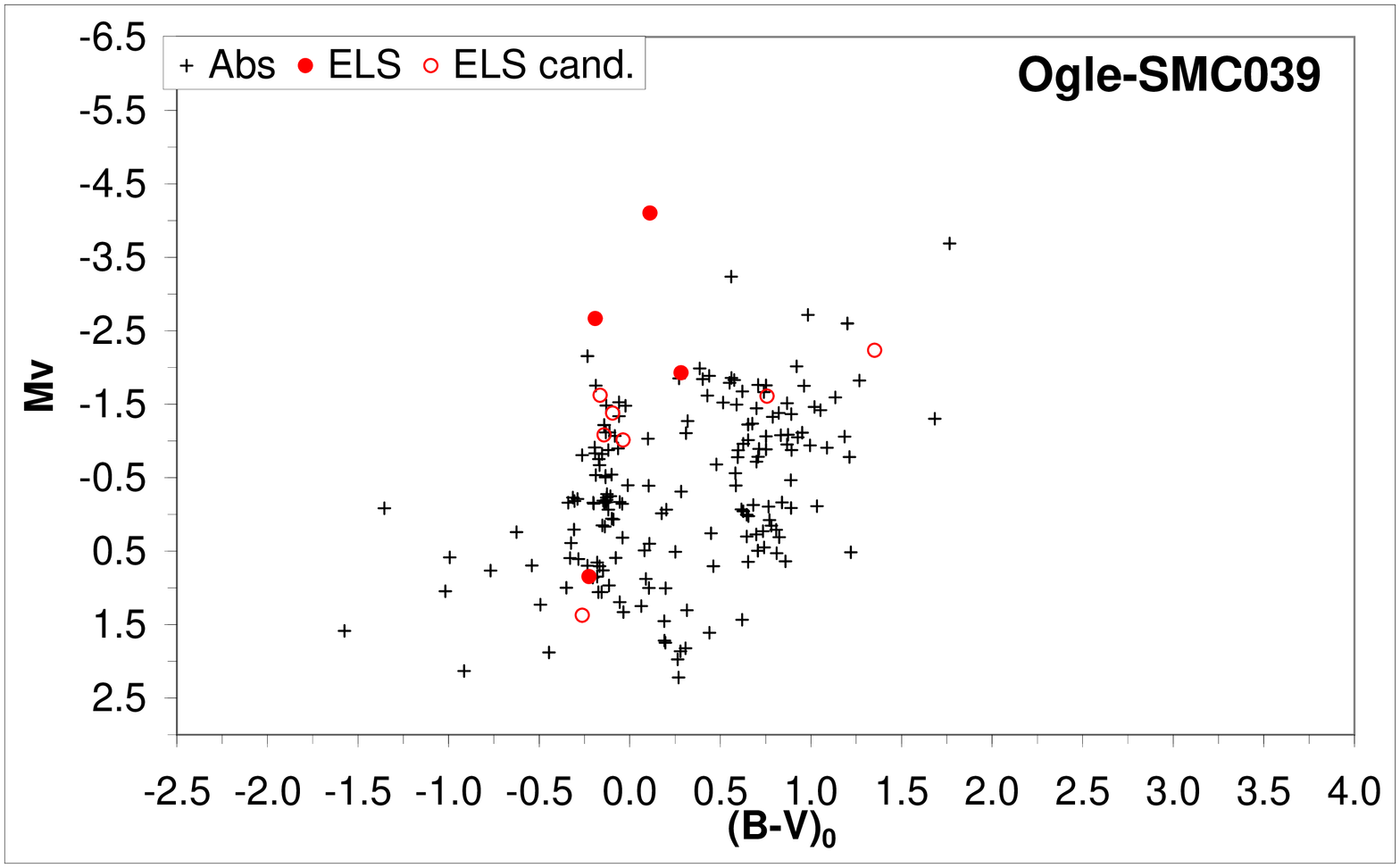} \\
    \includegraphics[angle=-90, width=8cm]{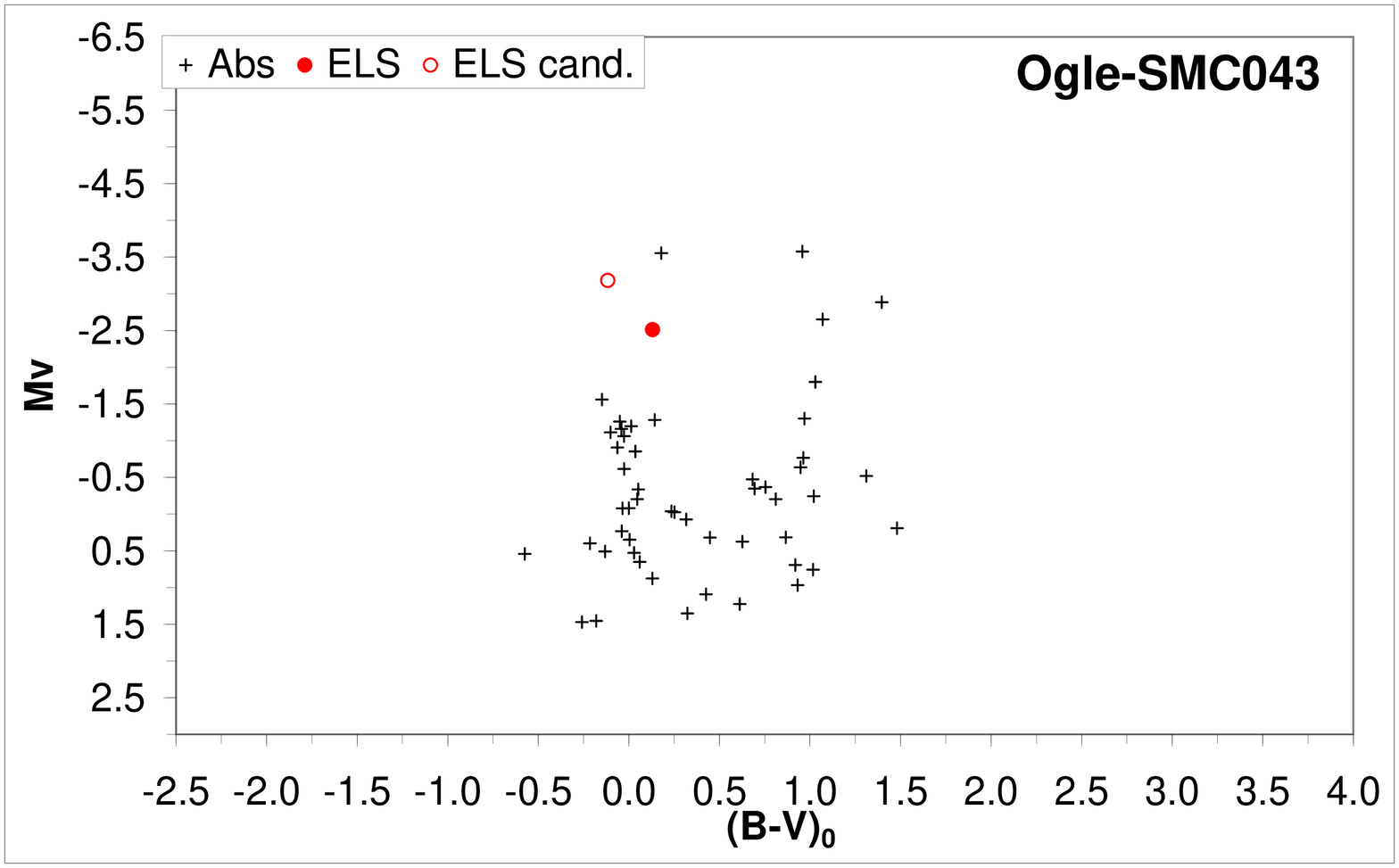} &  \includegraphics[angle=-90, width=8cm]{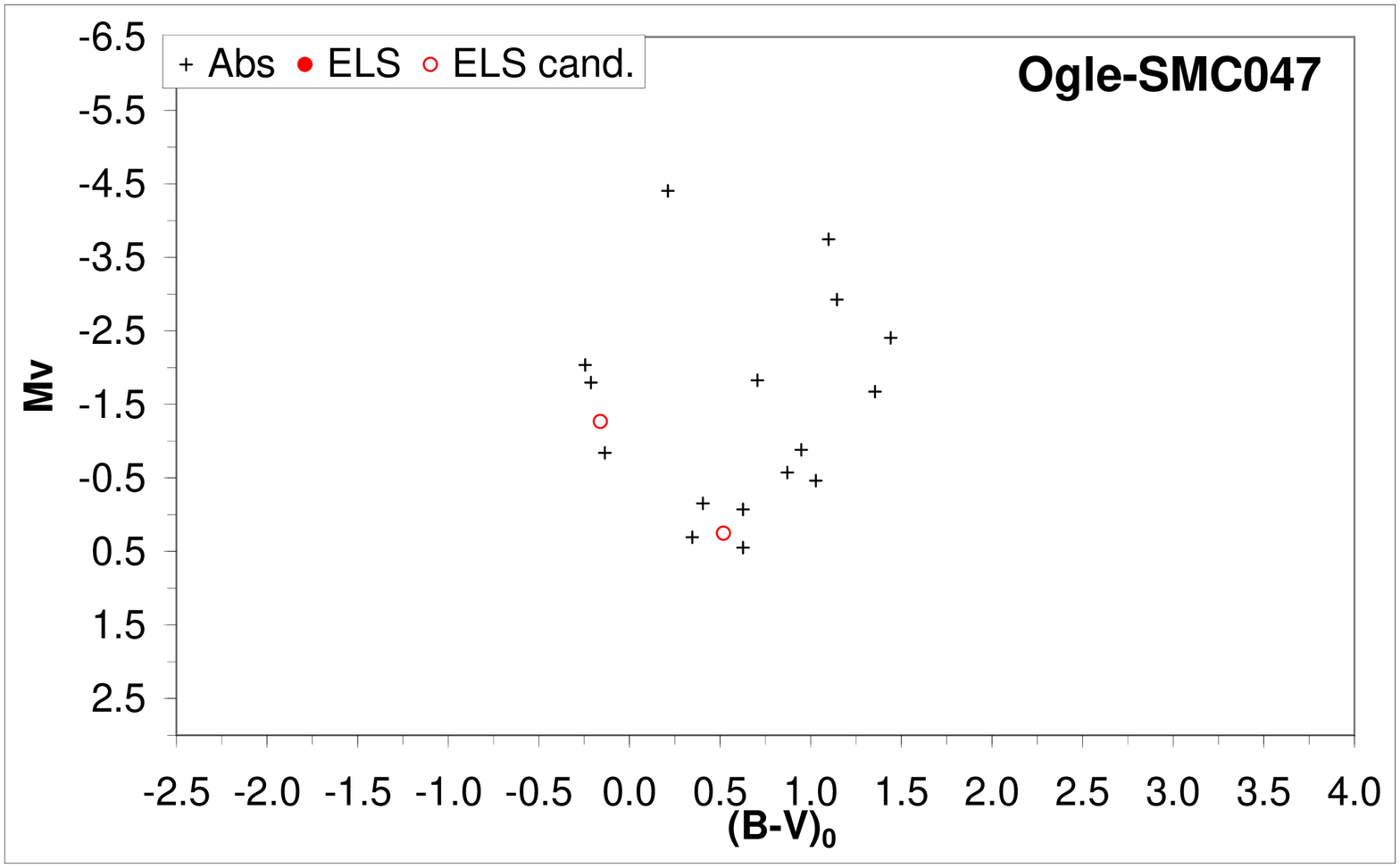} \\
    \includegraphics[angle=-90, width=8cm]{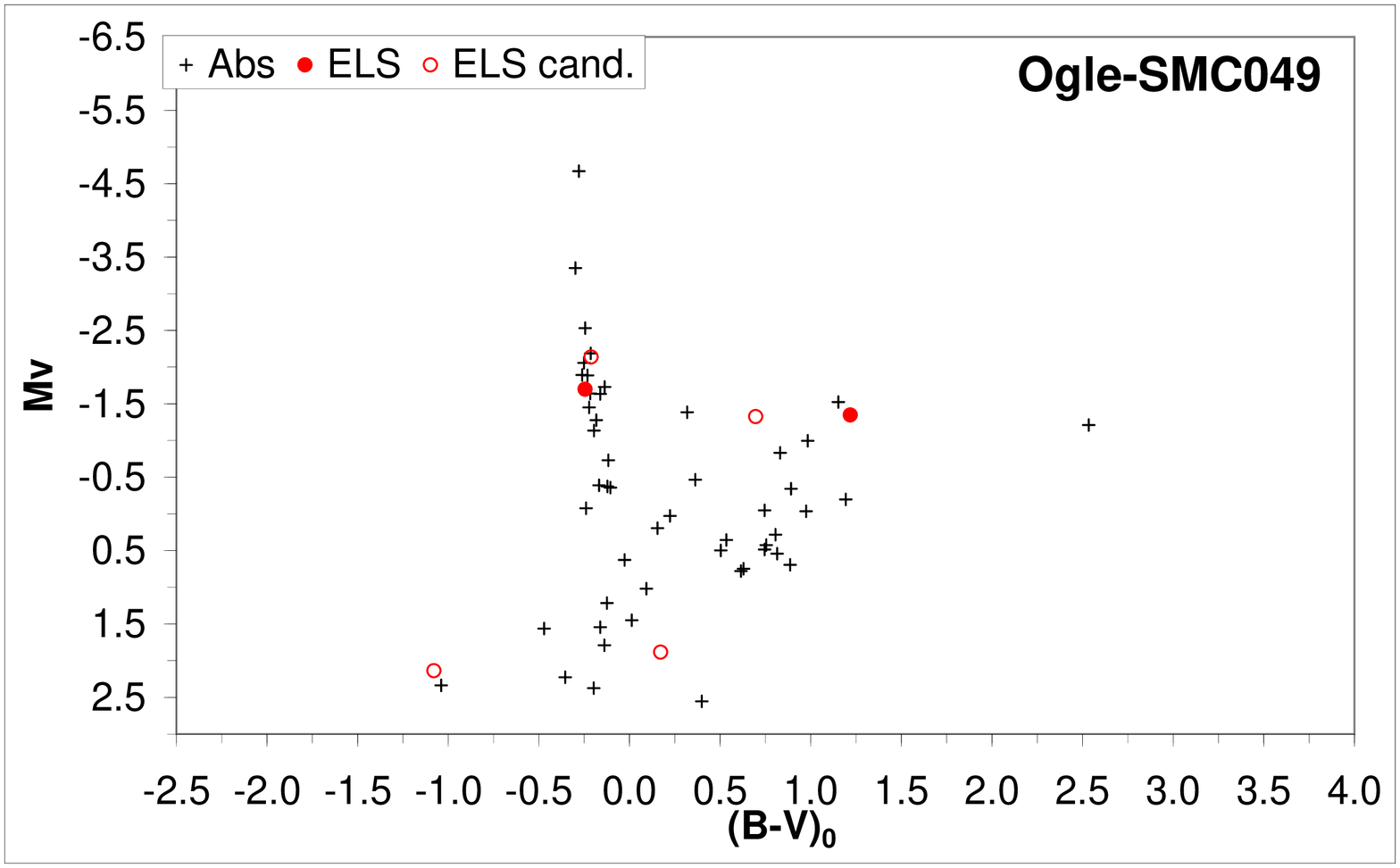} &  \includegraphics[angle=-90, width=8cm]{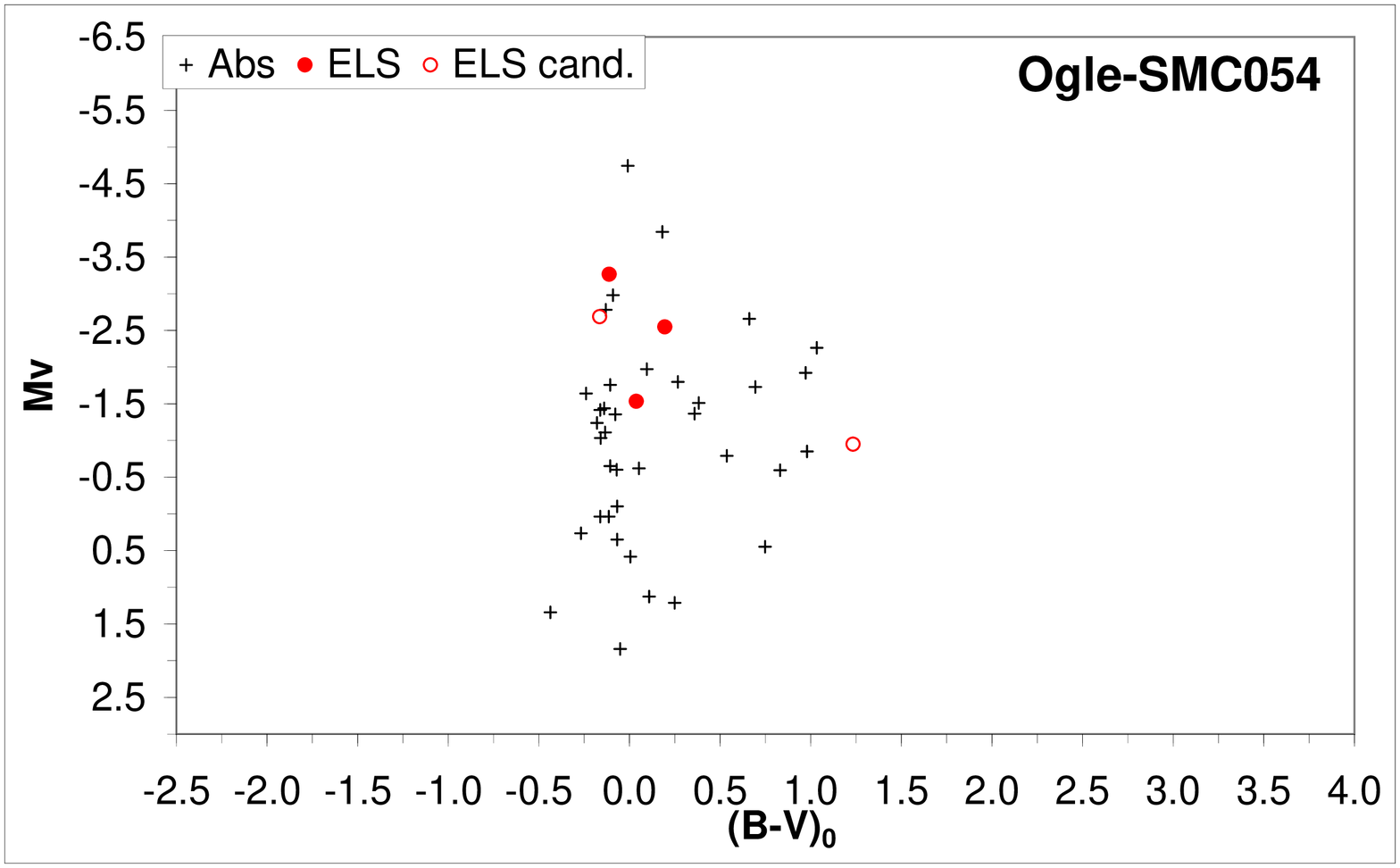} \\
    \includegraphics[angle=-90, width=8cm]{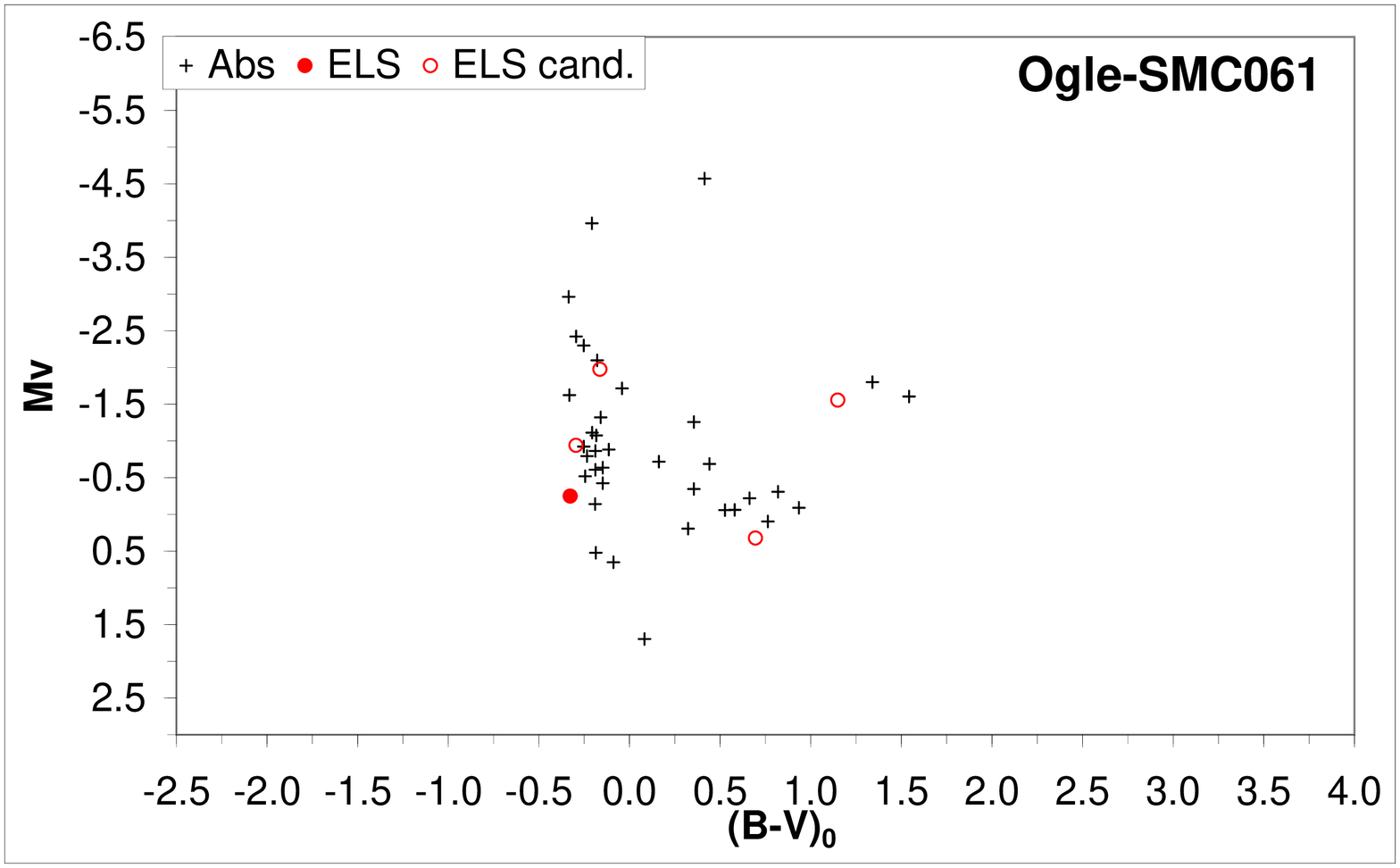} &  \includegraphics[angle=-90, width=8cm]{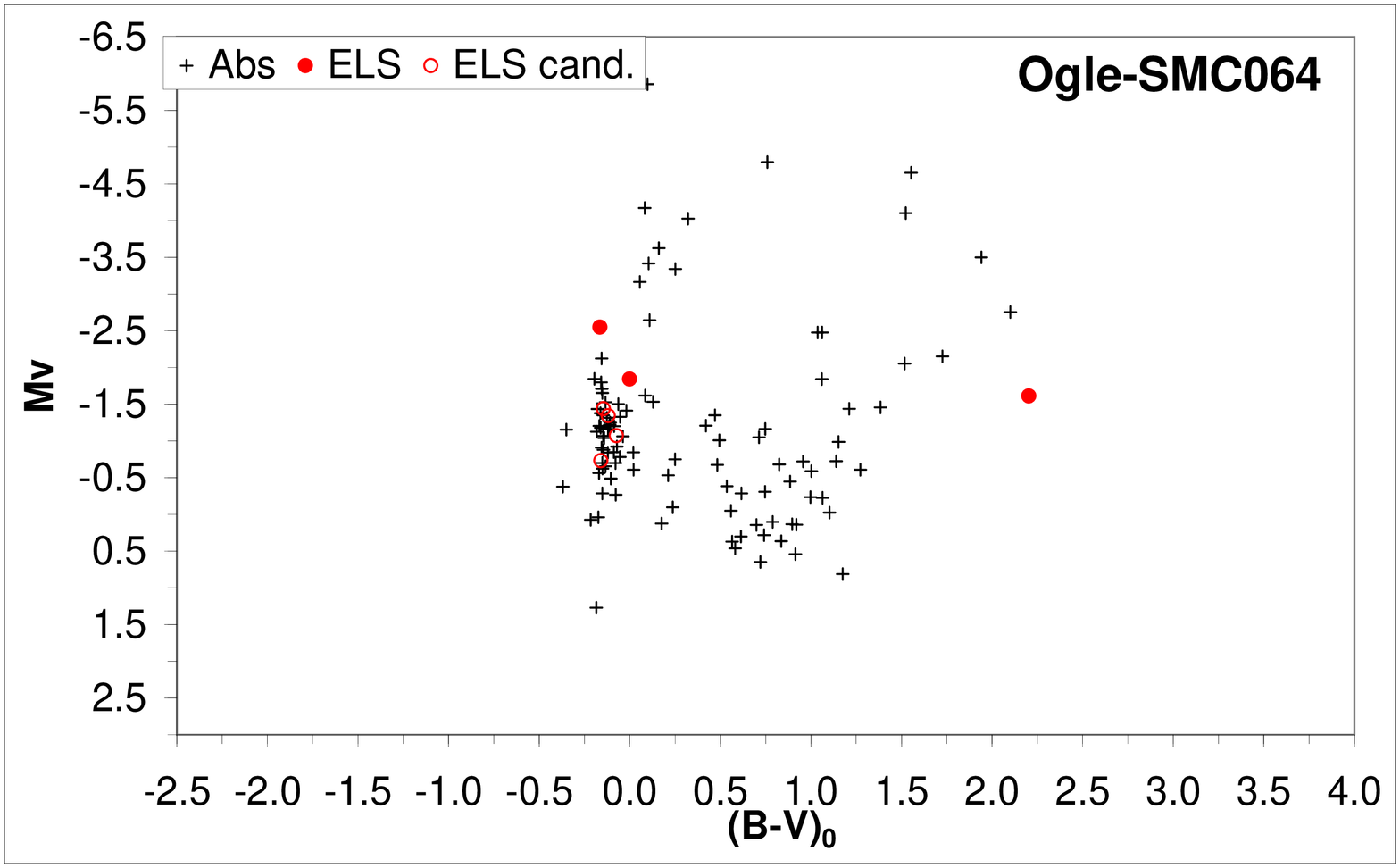} \\
\end{tabular}
\caption{Same caption as for fig.~\ref{HR1}}
\label{HR2}
\end{figure*} 

\begin{figure*}
\begin{tabular}{cc}
    \centering
    \includegraphics[angle=-90, width=8cm]{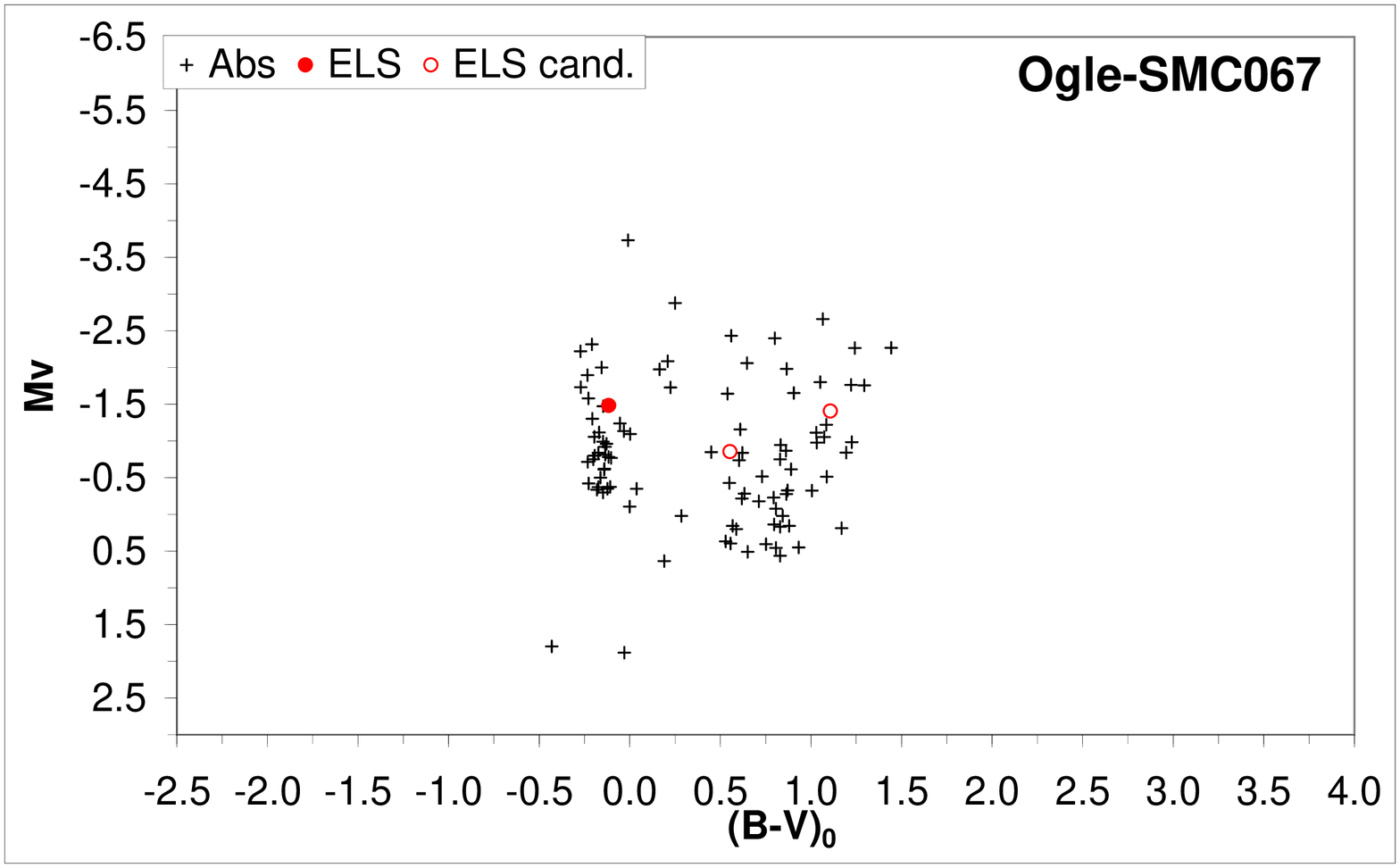} &  \includegraphics[angle=-90, width=8cm]{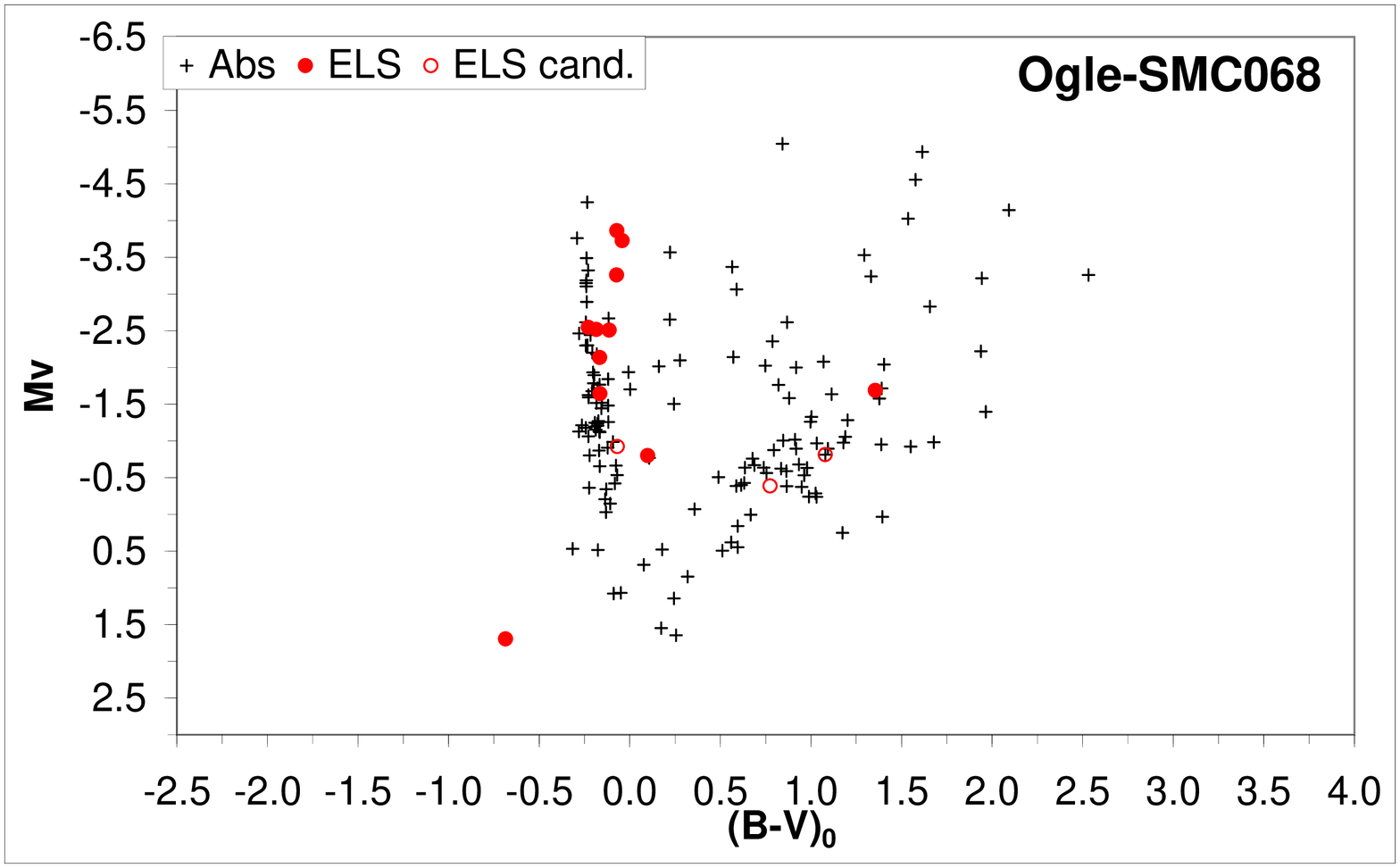} \\
    \includegraphics[angle=-90, width=8cm]{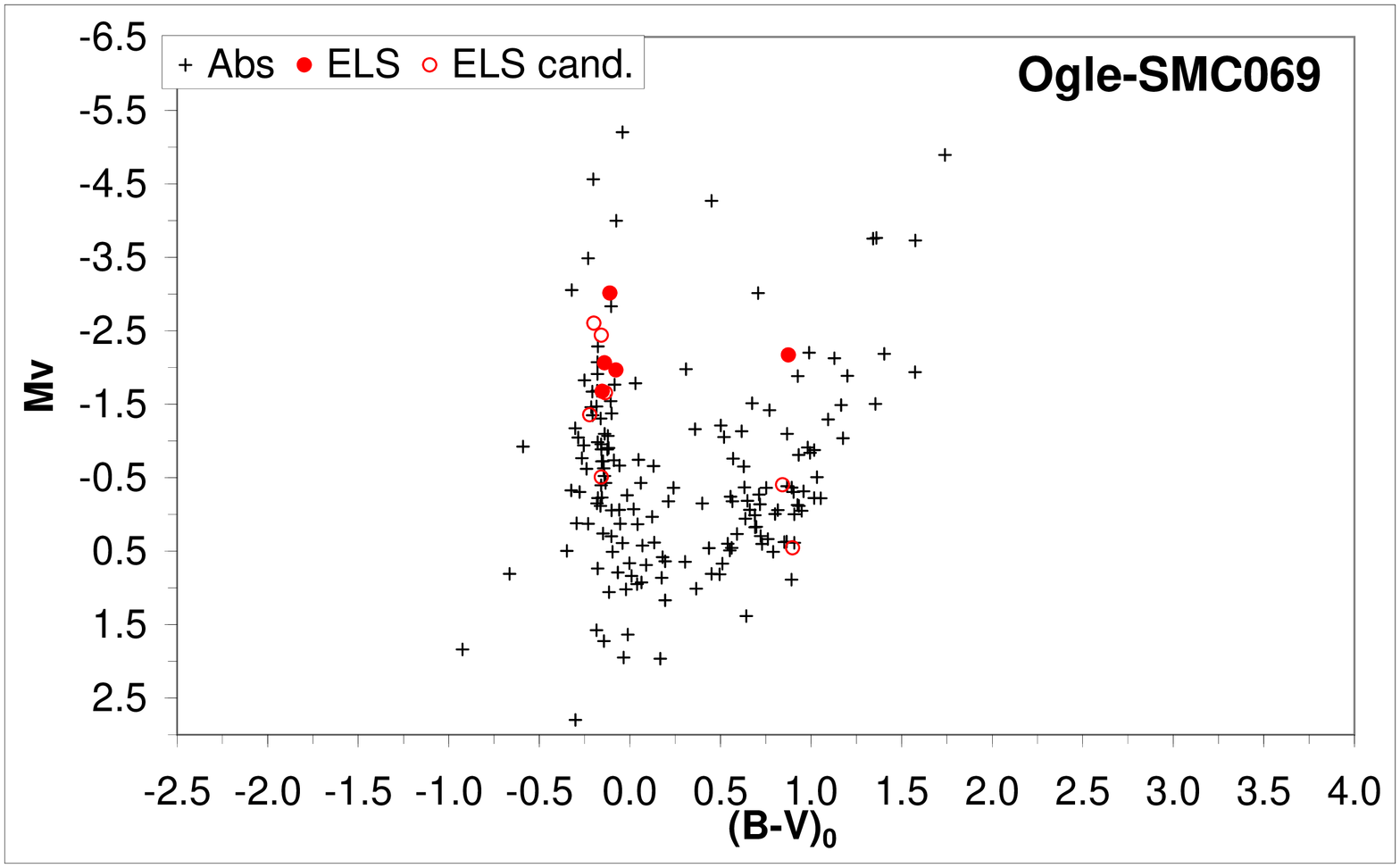} &  \includegraphics[angle=-90, width=8cm]{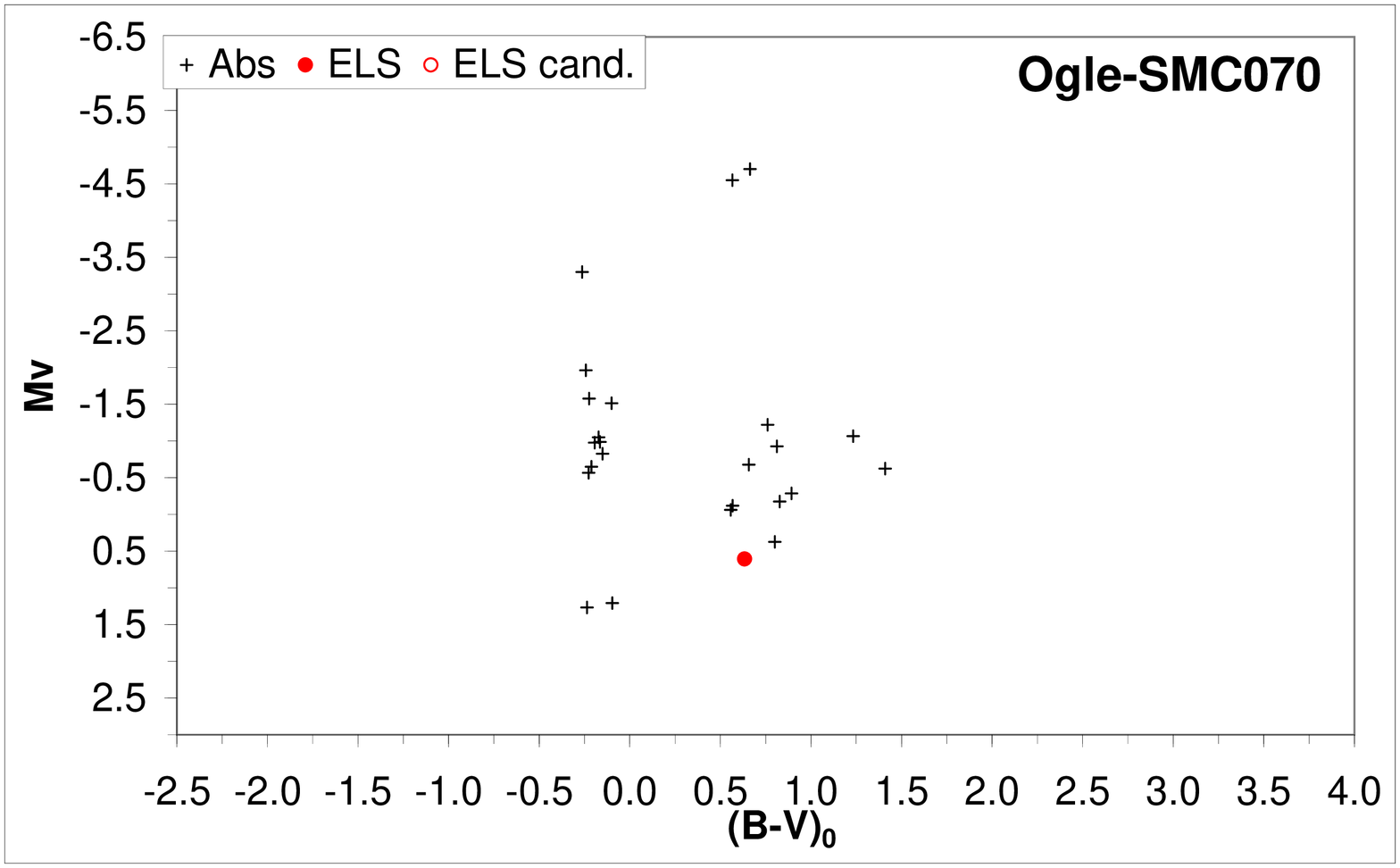} \\
    \includegraphics[angle=-90, width=8cm]{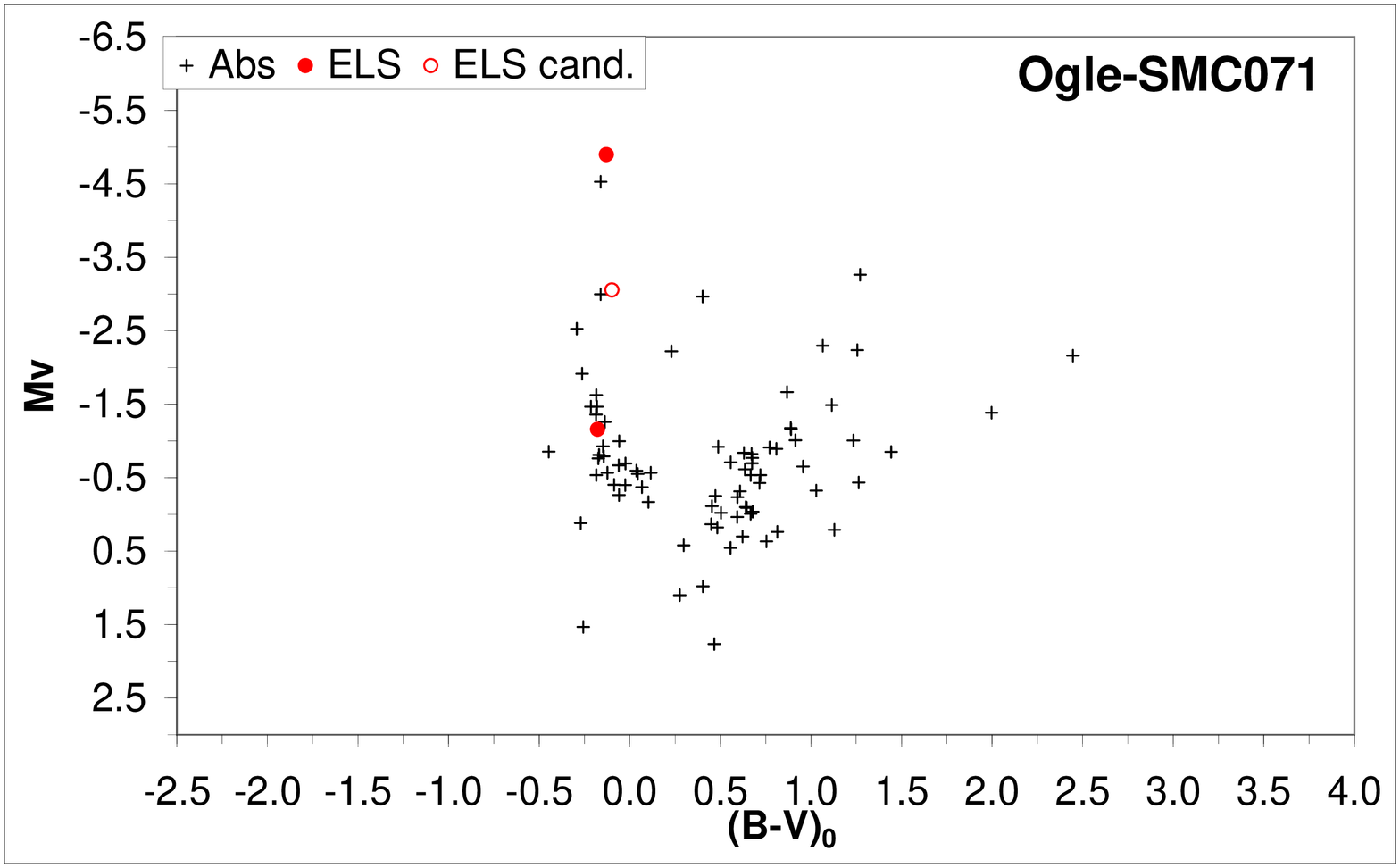} &  \includegraphics[angle=-90, width=8cm]{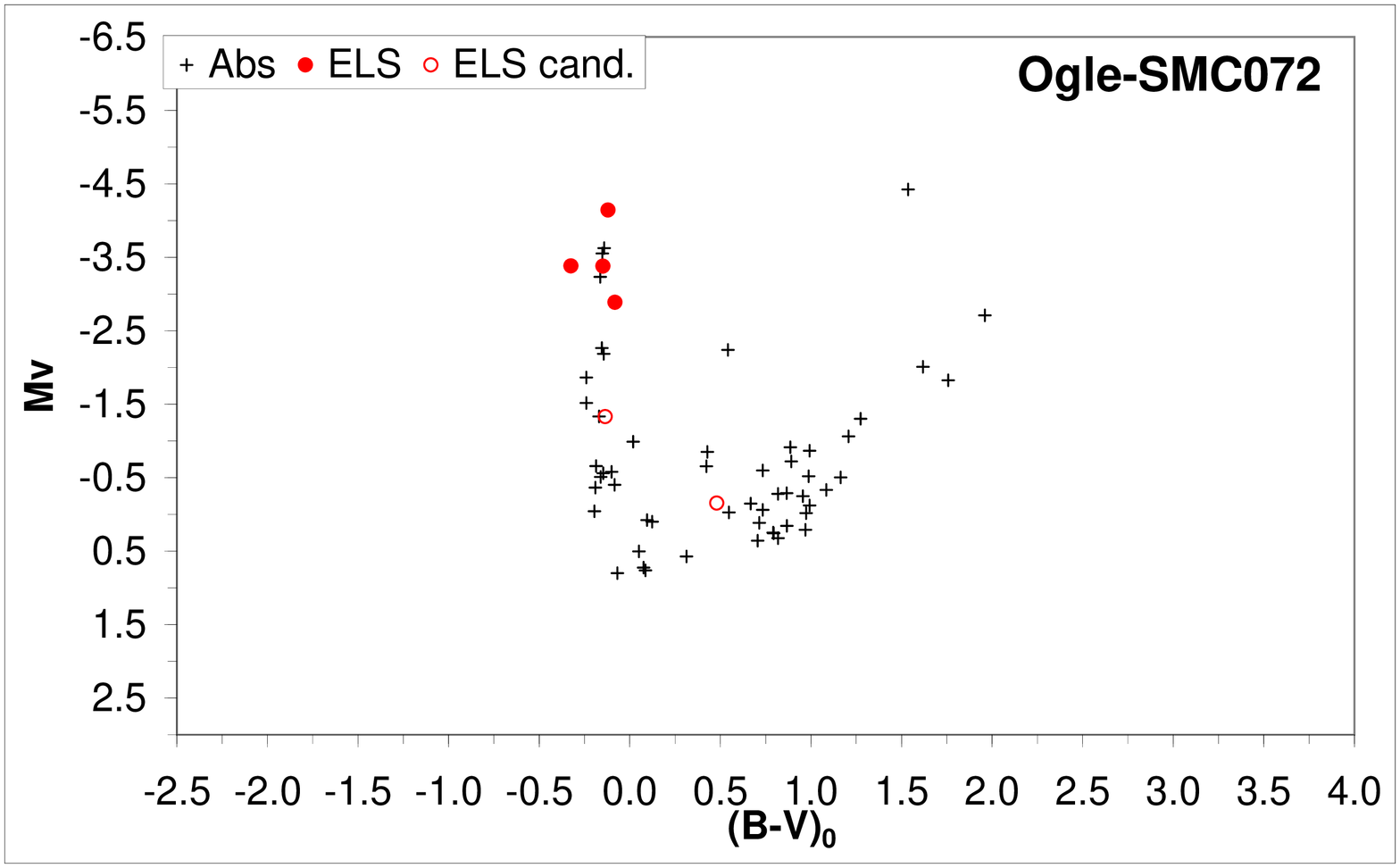} \\
    \includegraphics[angle=-90, width=8cm]{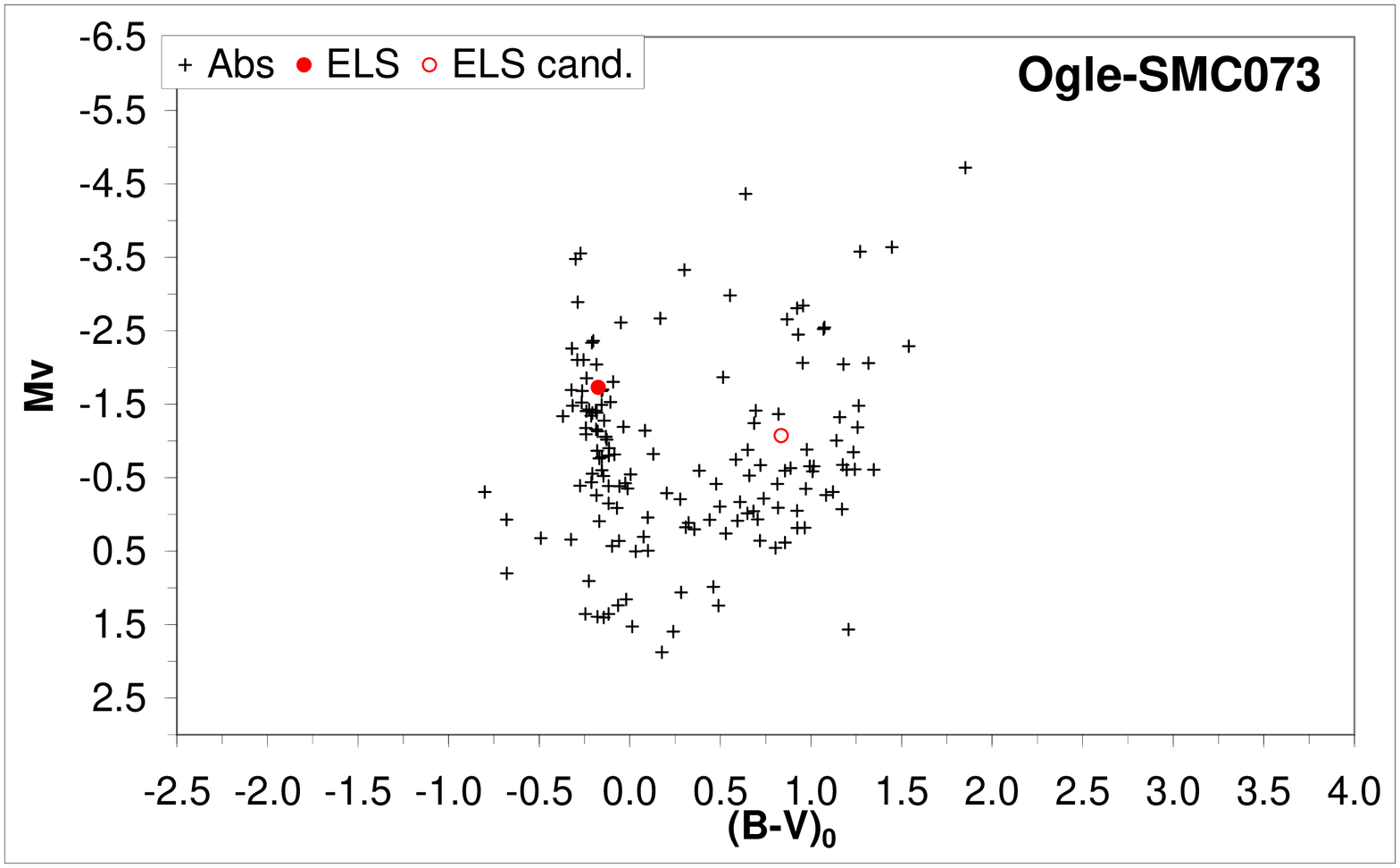} &  \includegraphics[angle=-90, width=8cm]{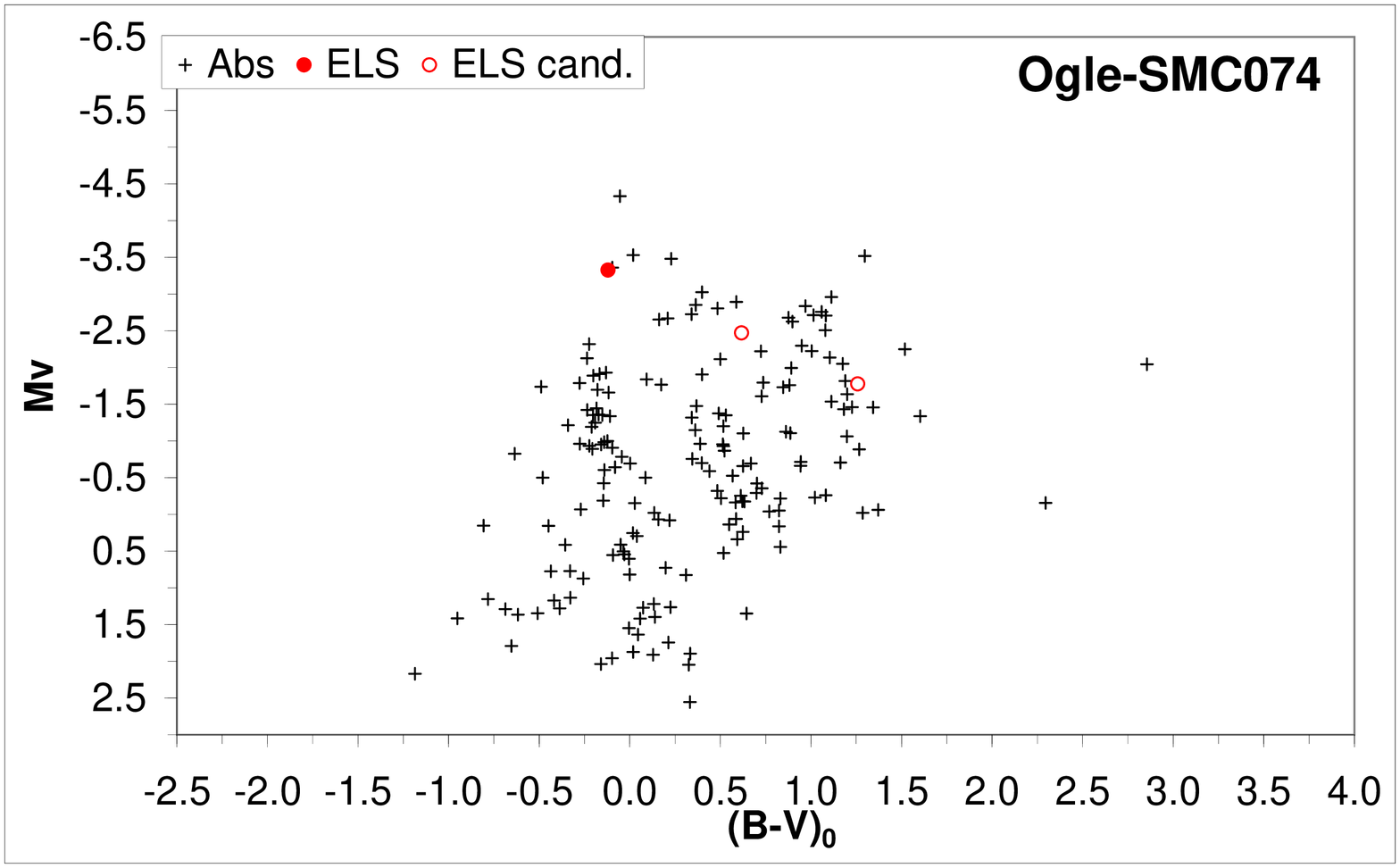} \\
\end{tabular}
\caption{Same caption as for fig.~\ref{HR1}}
\label{HR3}
\end{figure*}

\begin{figure*}
\begin{tabular}{cc}
    \centering
    \includegraphics[angle=-90, width=8cm]{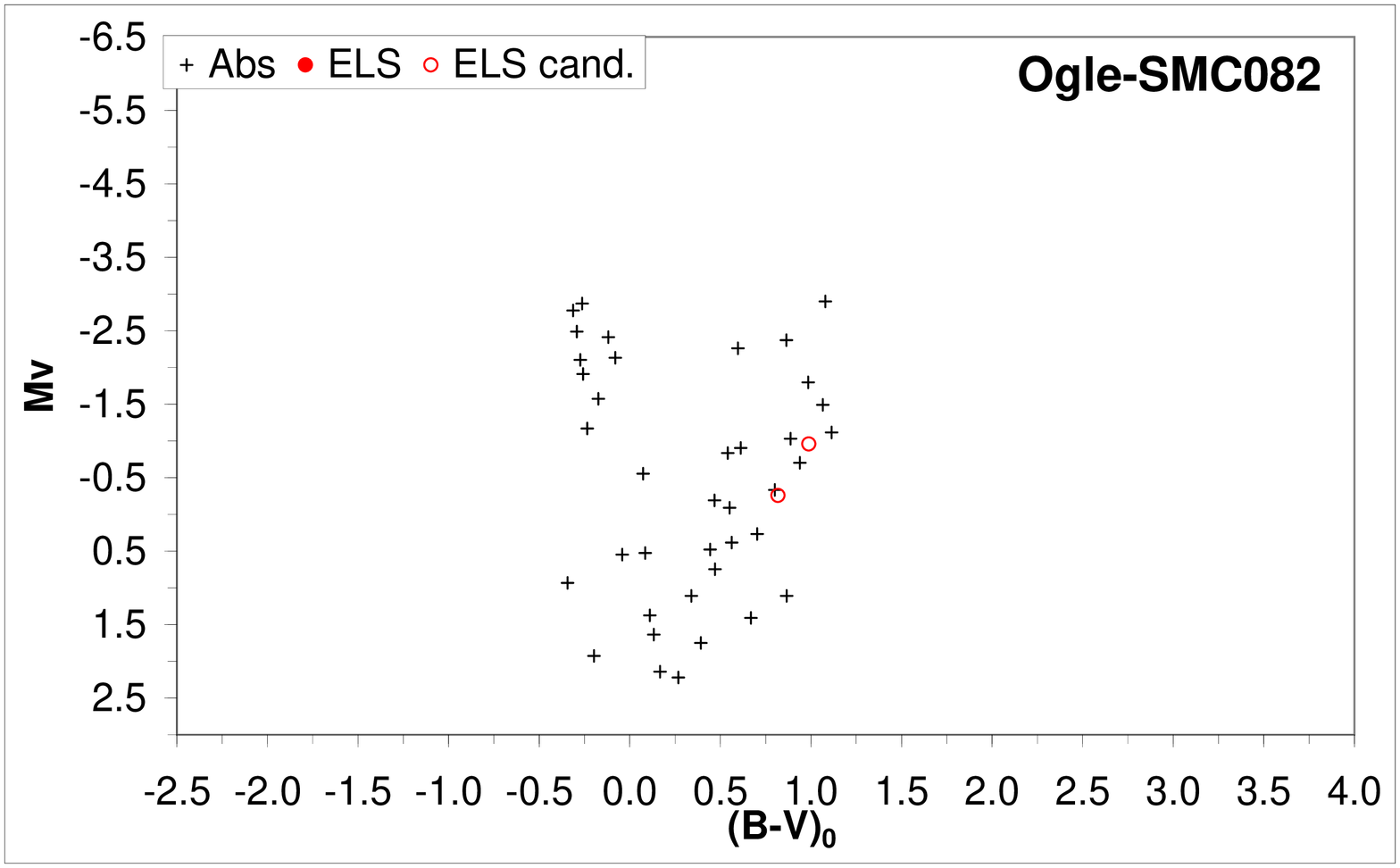} &  \includegraphics[angle=-90, width=8cm]{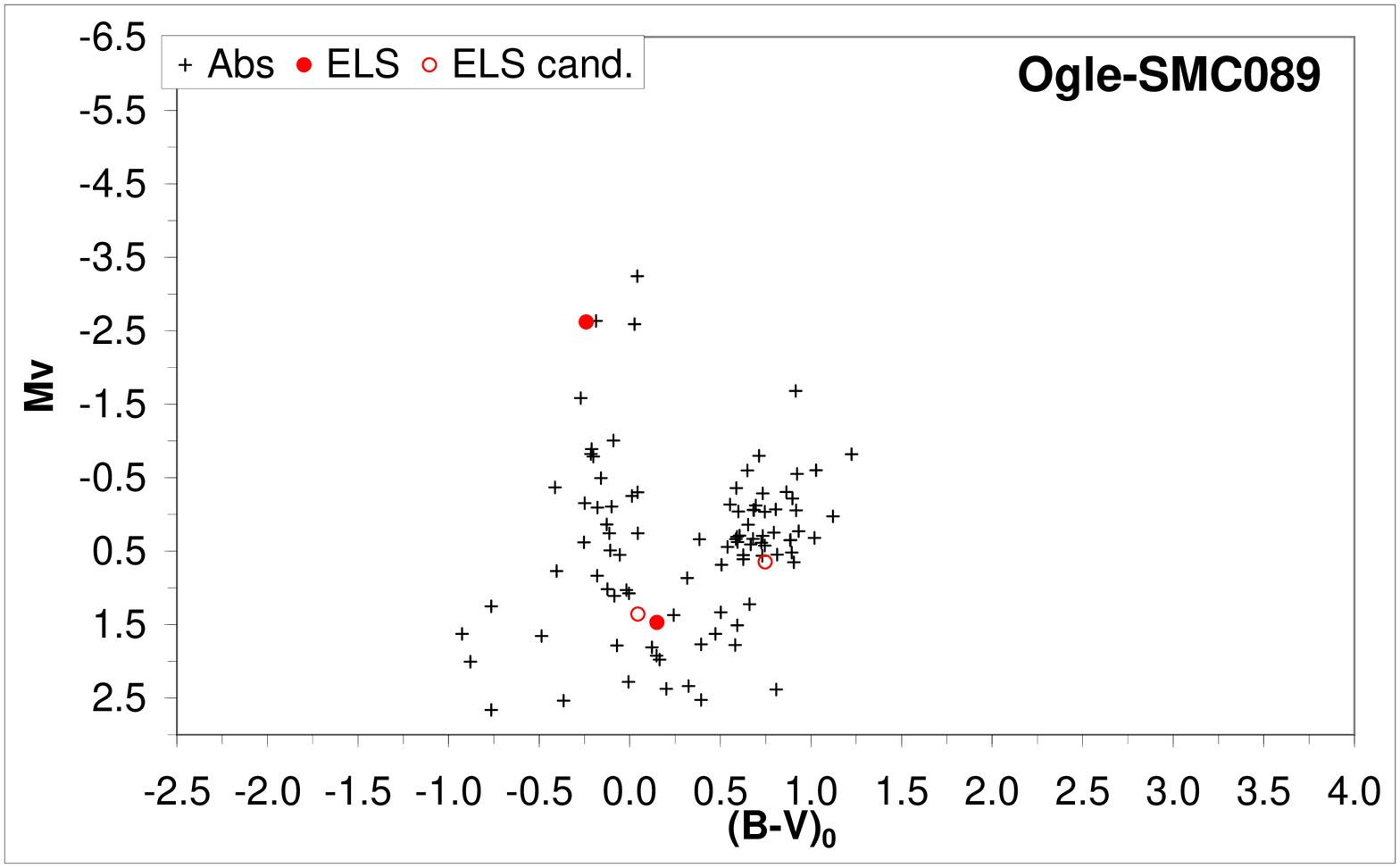} \\
    \includegraphics[angle=-90, width=8cm]{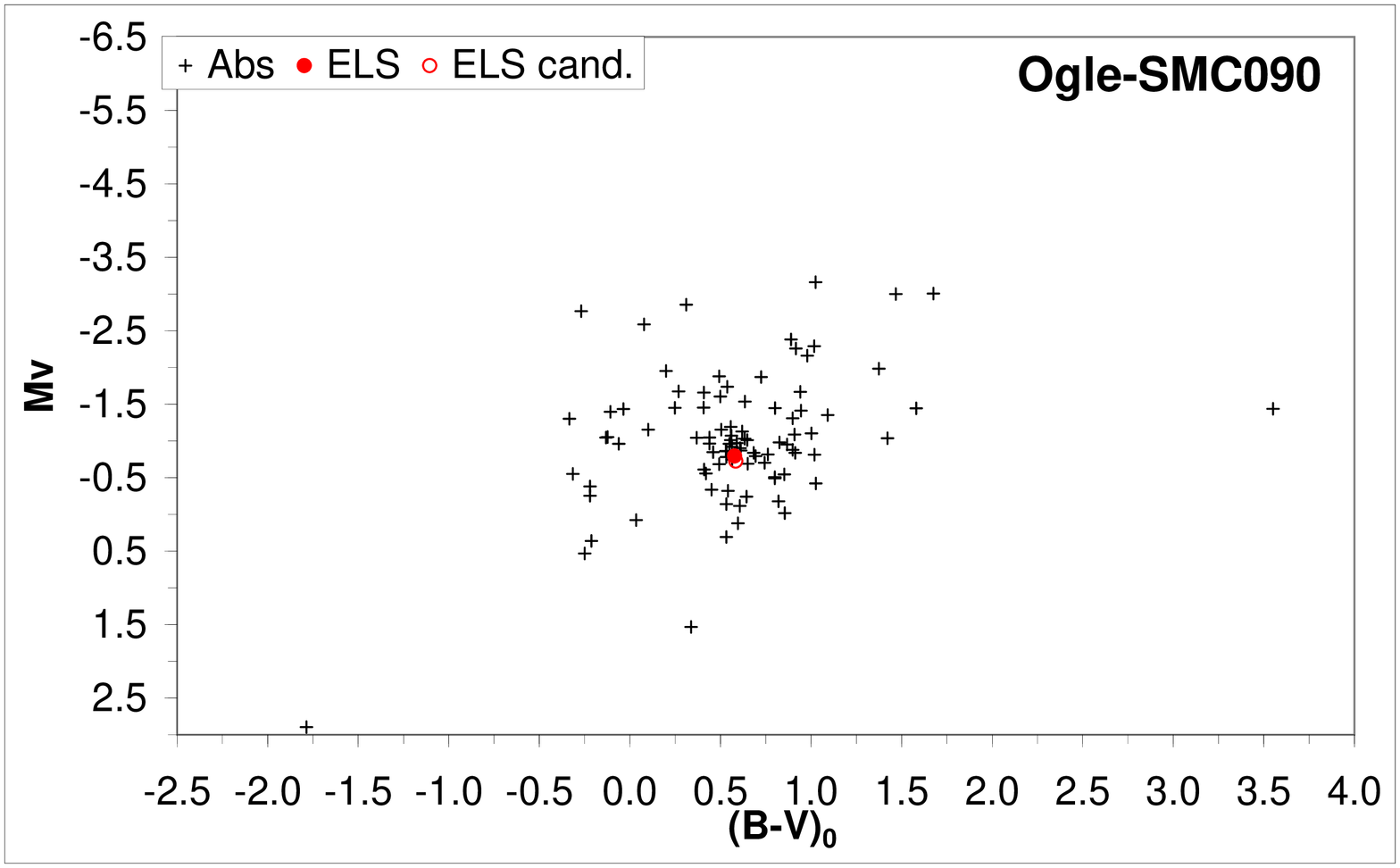} &  \includegraphics[angle=-90, width=8cm]{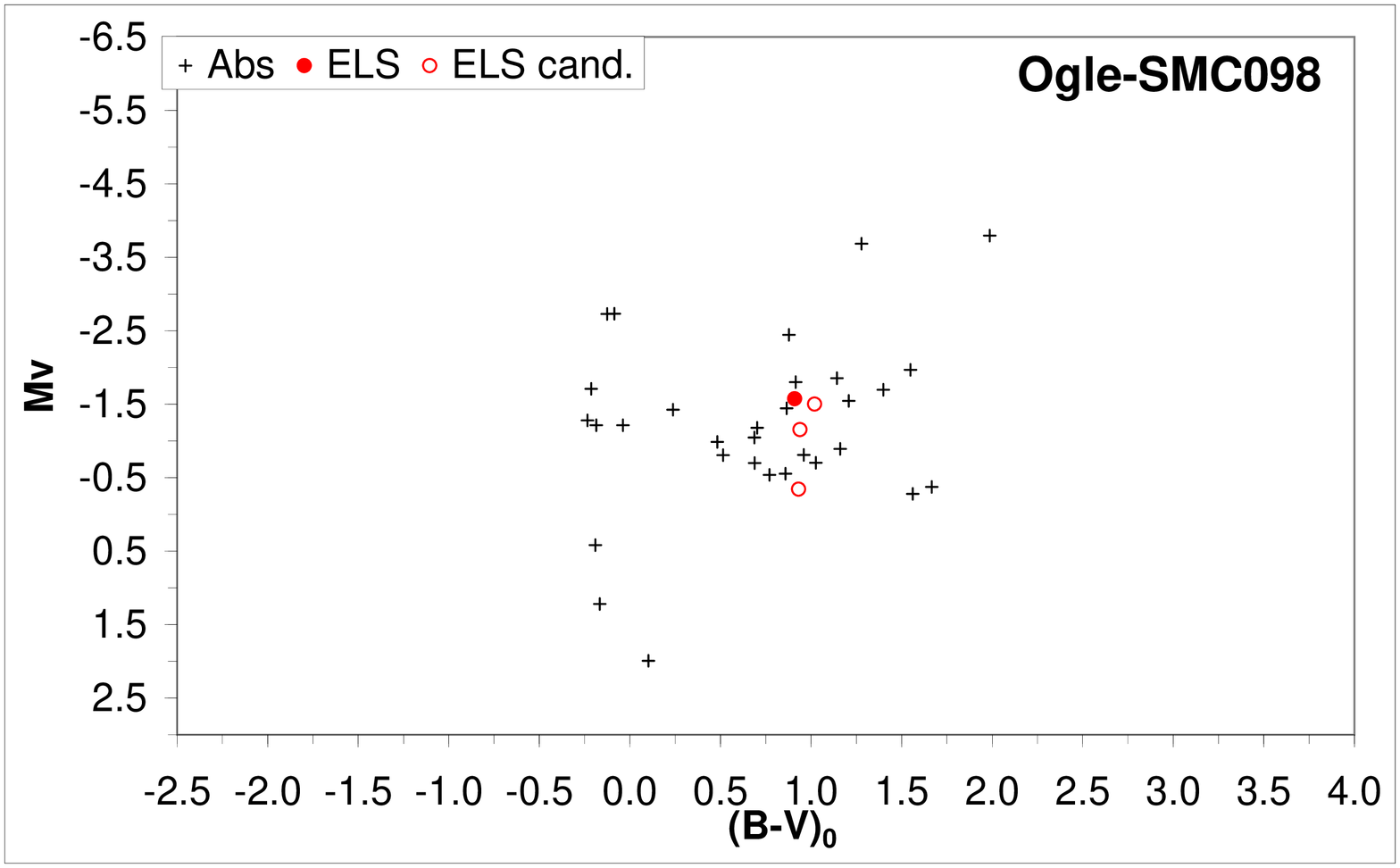} \\
    \includegraphics[angle=-90, width=8cm]{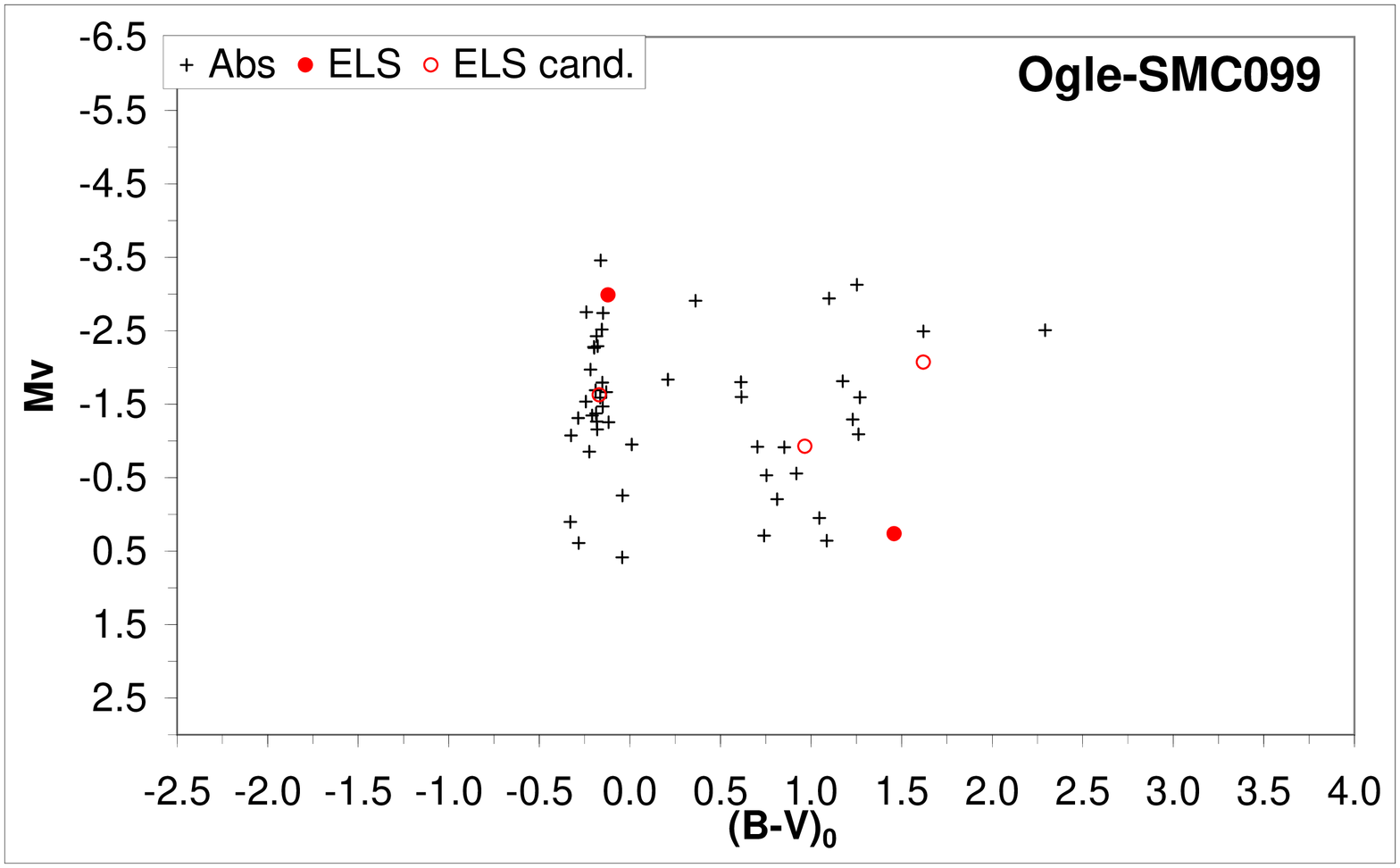} &  \includegraphics[angle=-90, width=8cm]{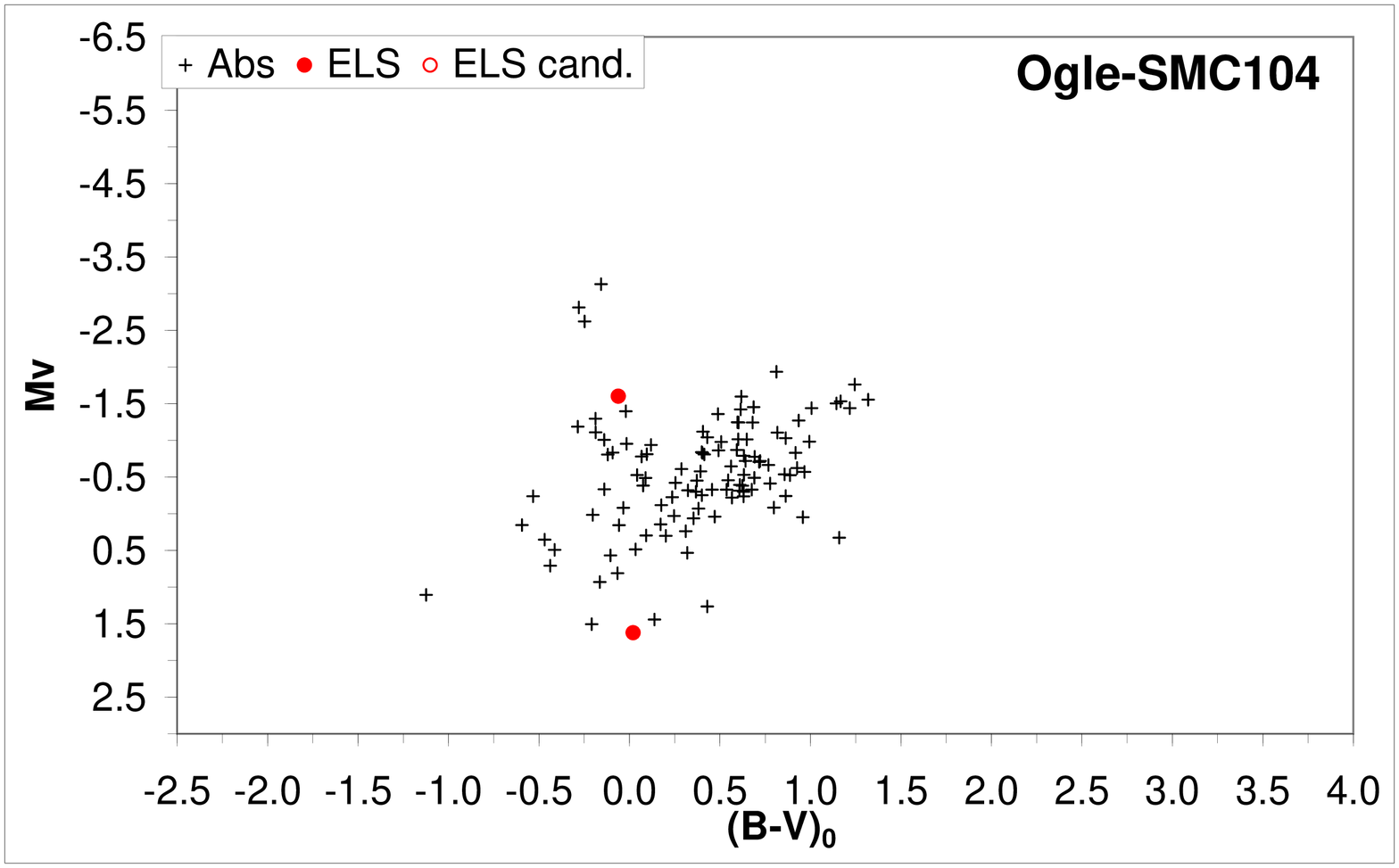} \\
    \includegraphics[angle=-90, width=8cm]{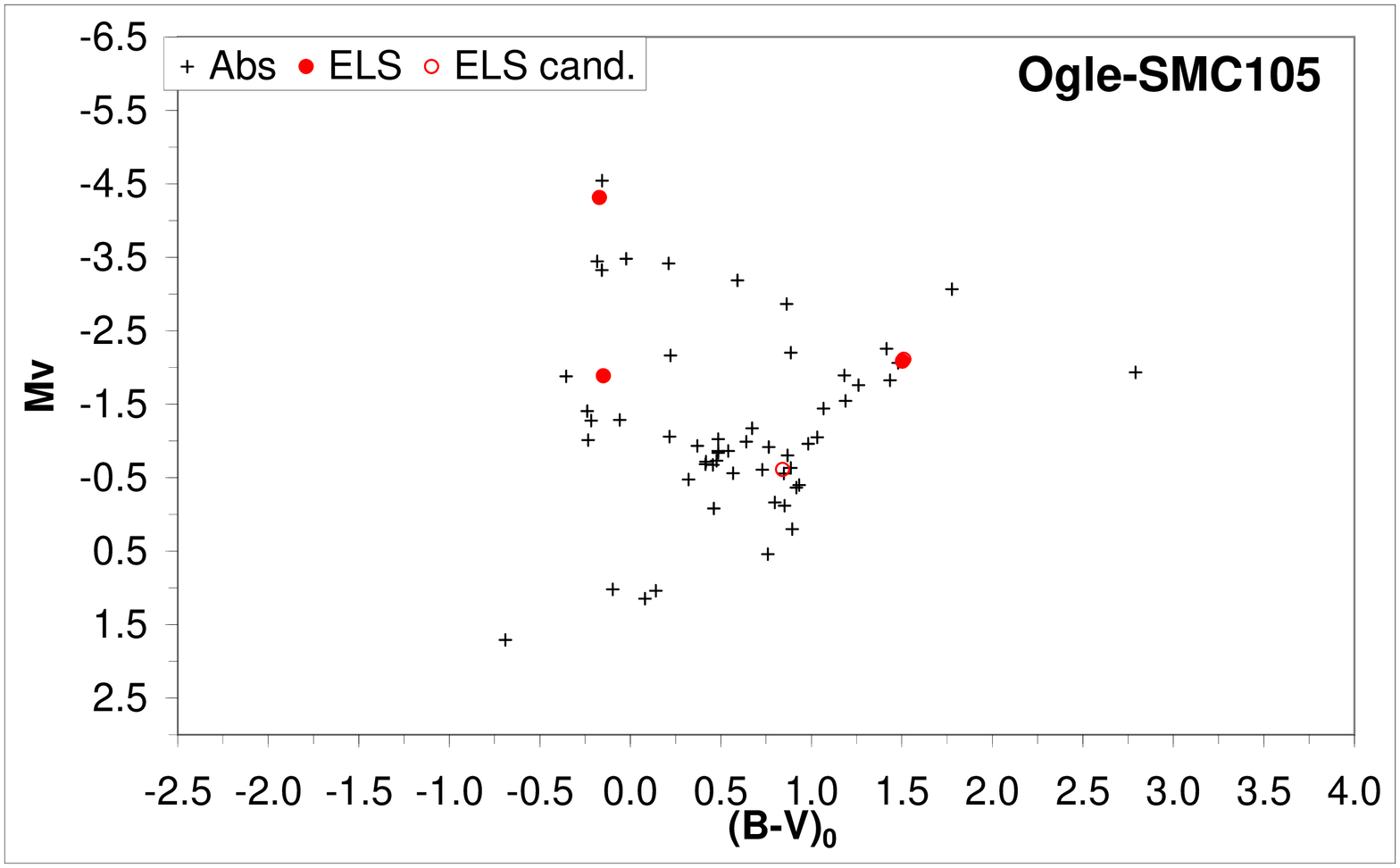} &  \includegraphics[angle=-90, width=8cm]{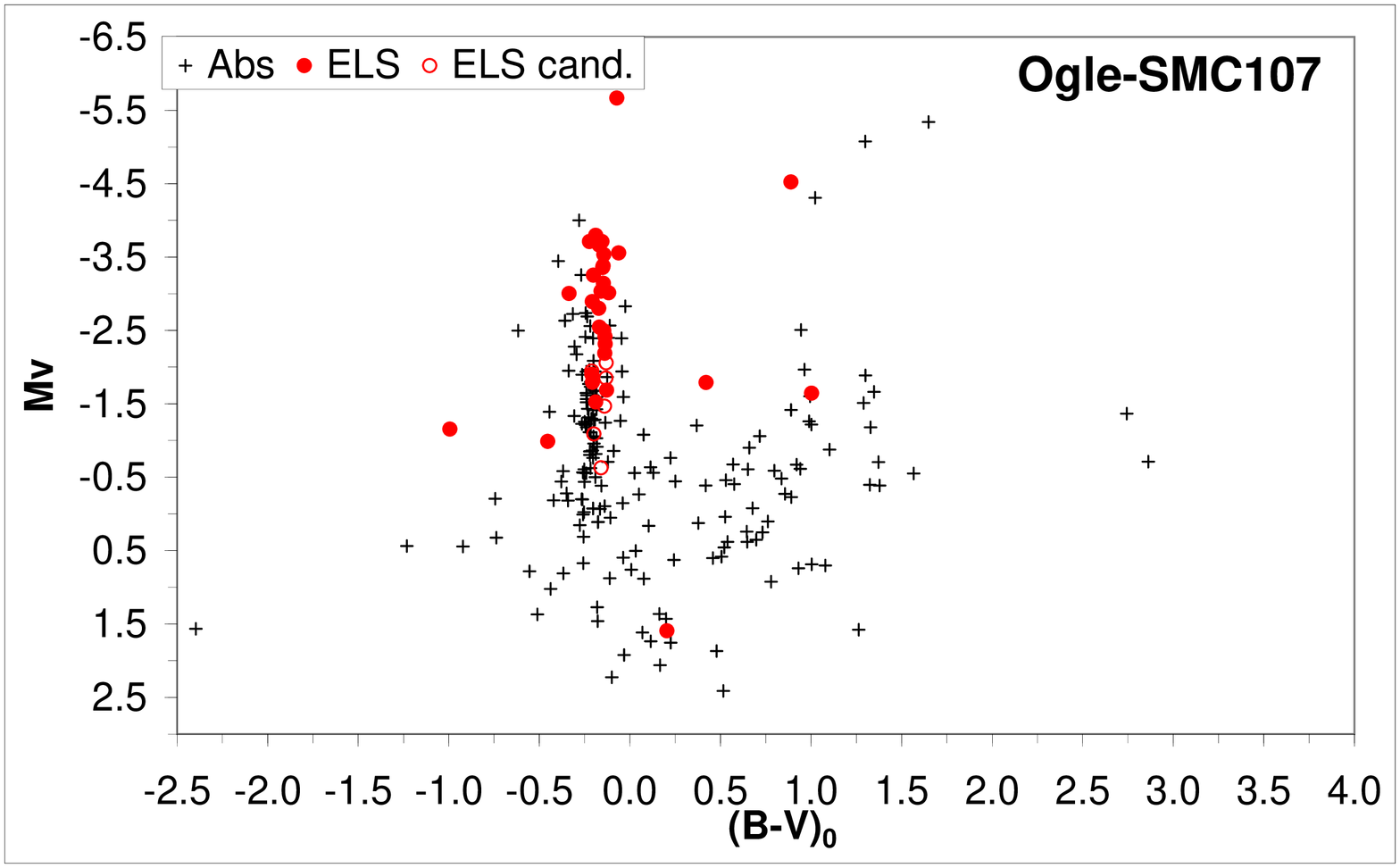} \\
\end{tabular}
\caption{Same caption as for fig.~\ref{HR1}}
\label{HR4}
\end{figure*} 

\begin{figure*}
\begin{tabular}{cc}
    \centering
    \includegraphics[angle=-90, width=8cm]{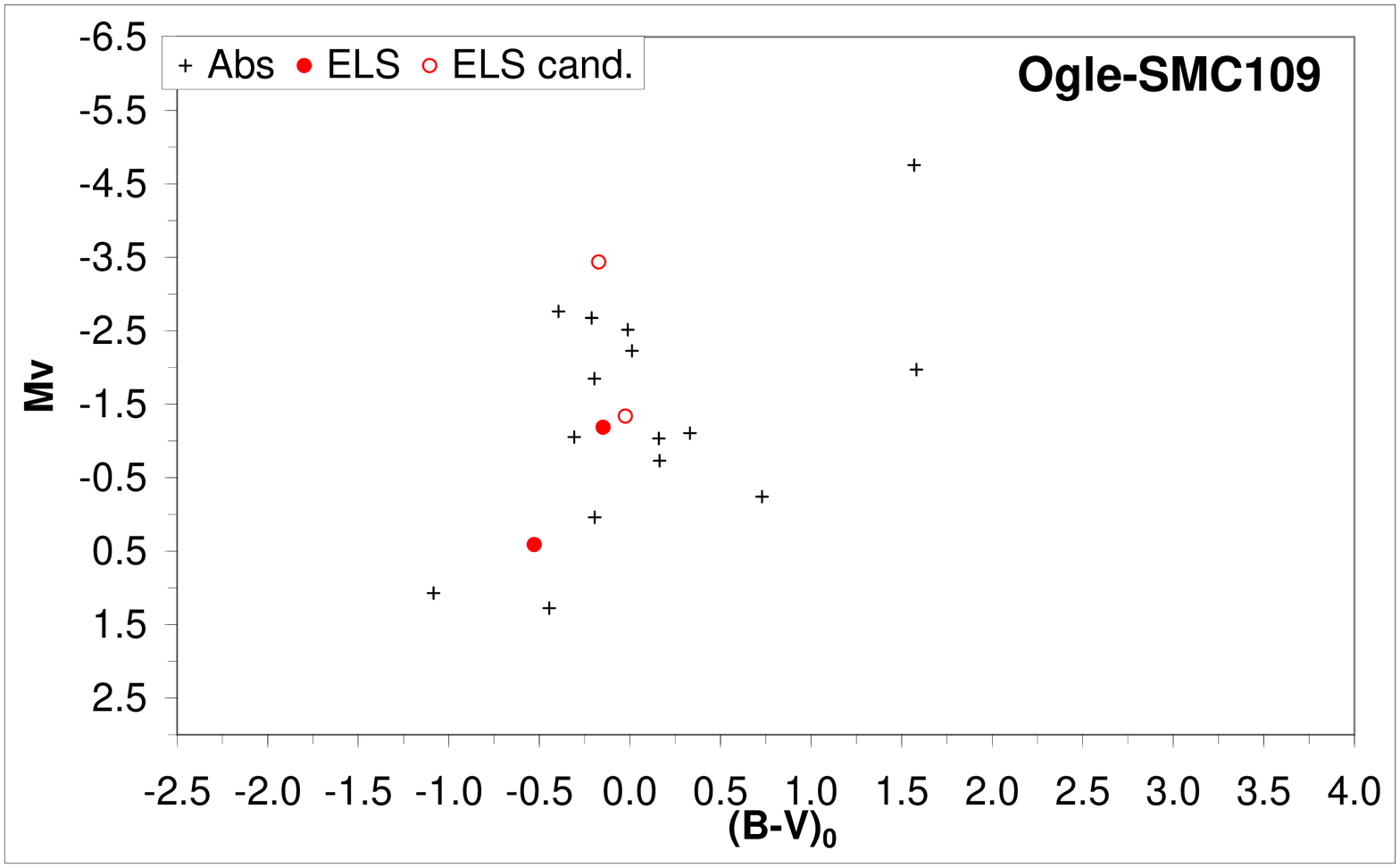} &  \includegraphics[angle=-90, width=8cm]{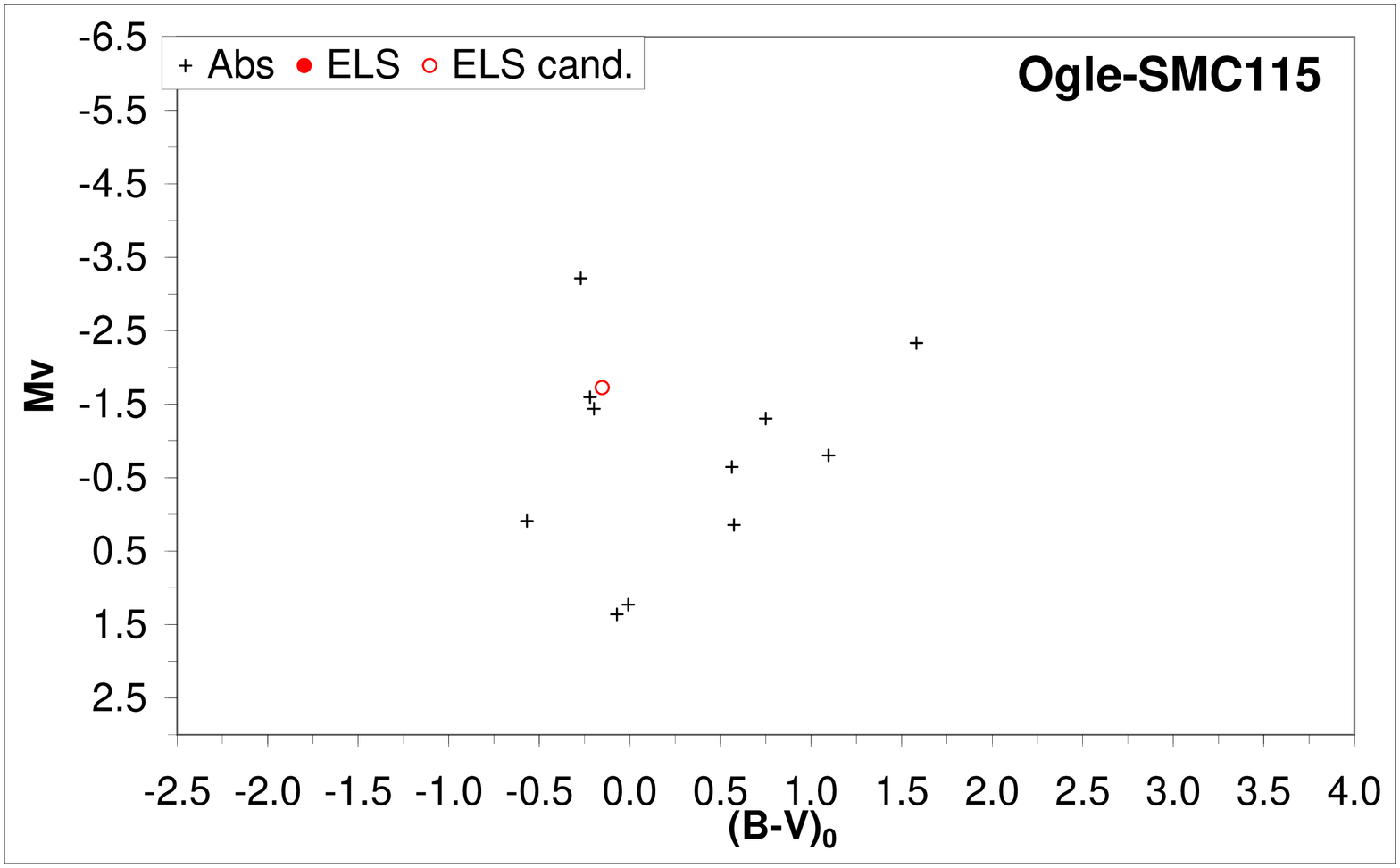} \\
    \includegraphics[angle=-90, width=8cm]{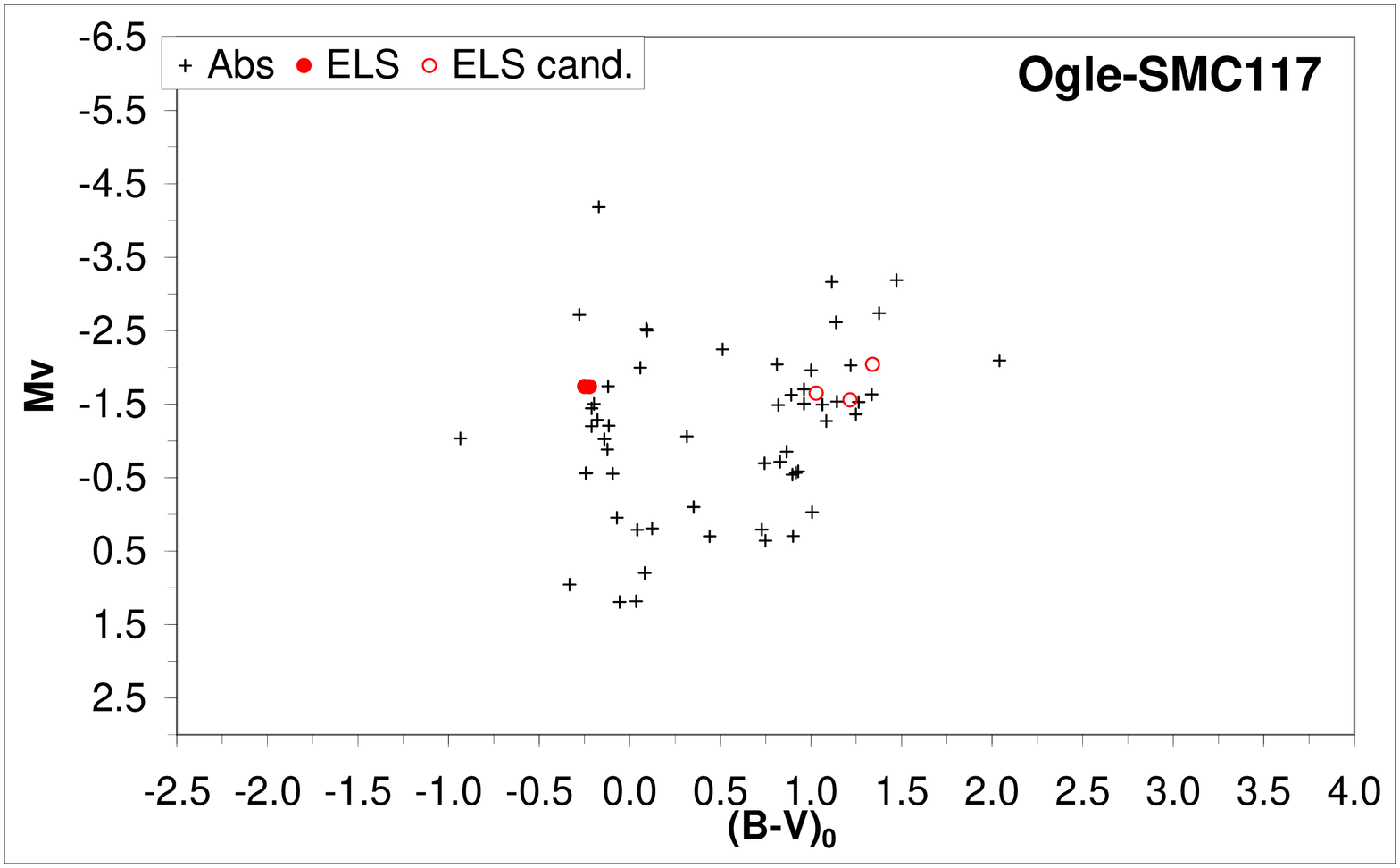} &  \includegraphics[angle=-90, width=8cm]{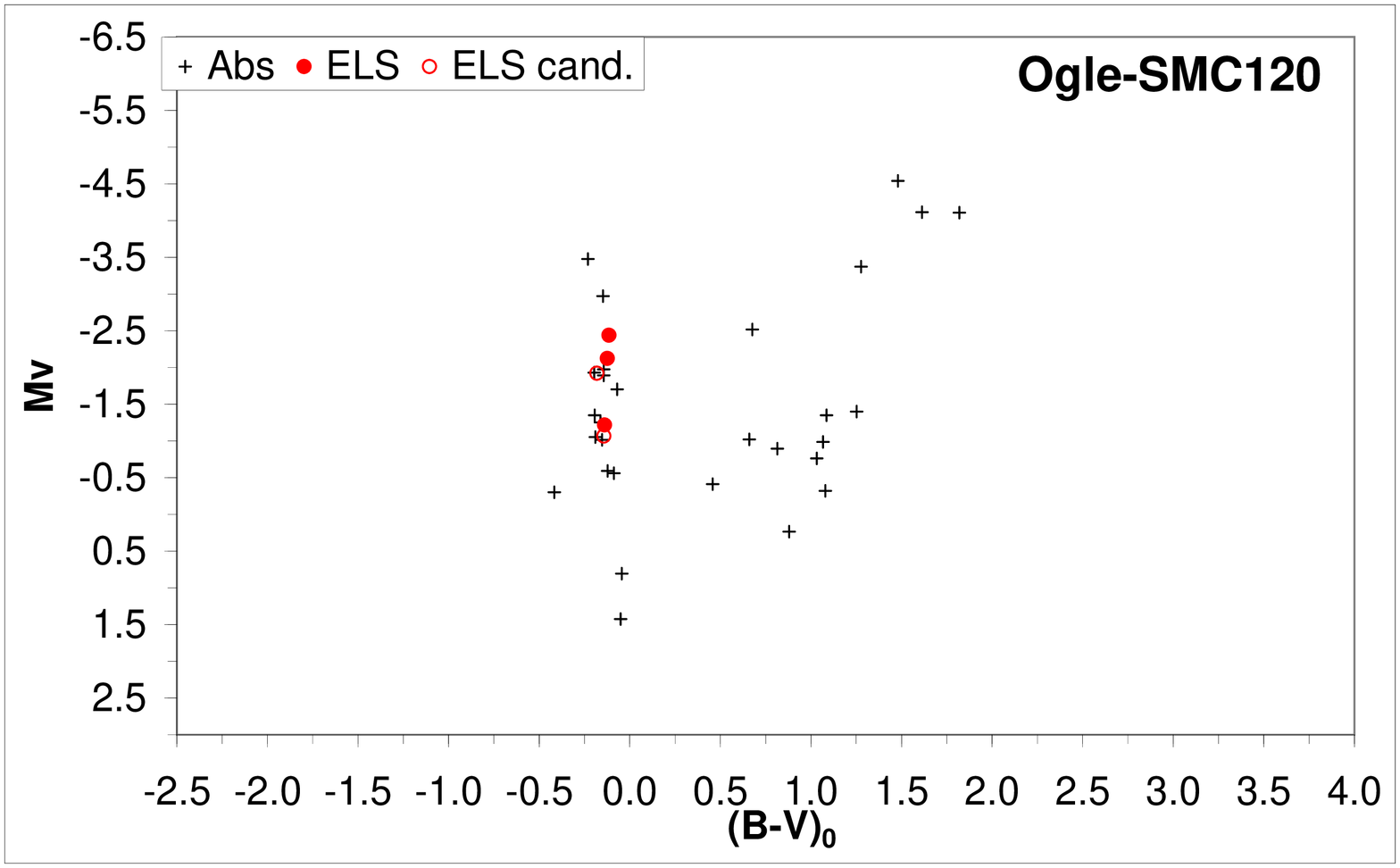} \\
    \includegraphics[angle=-90, width=8cm]{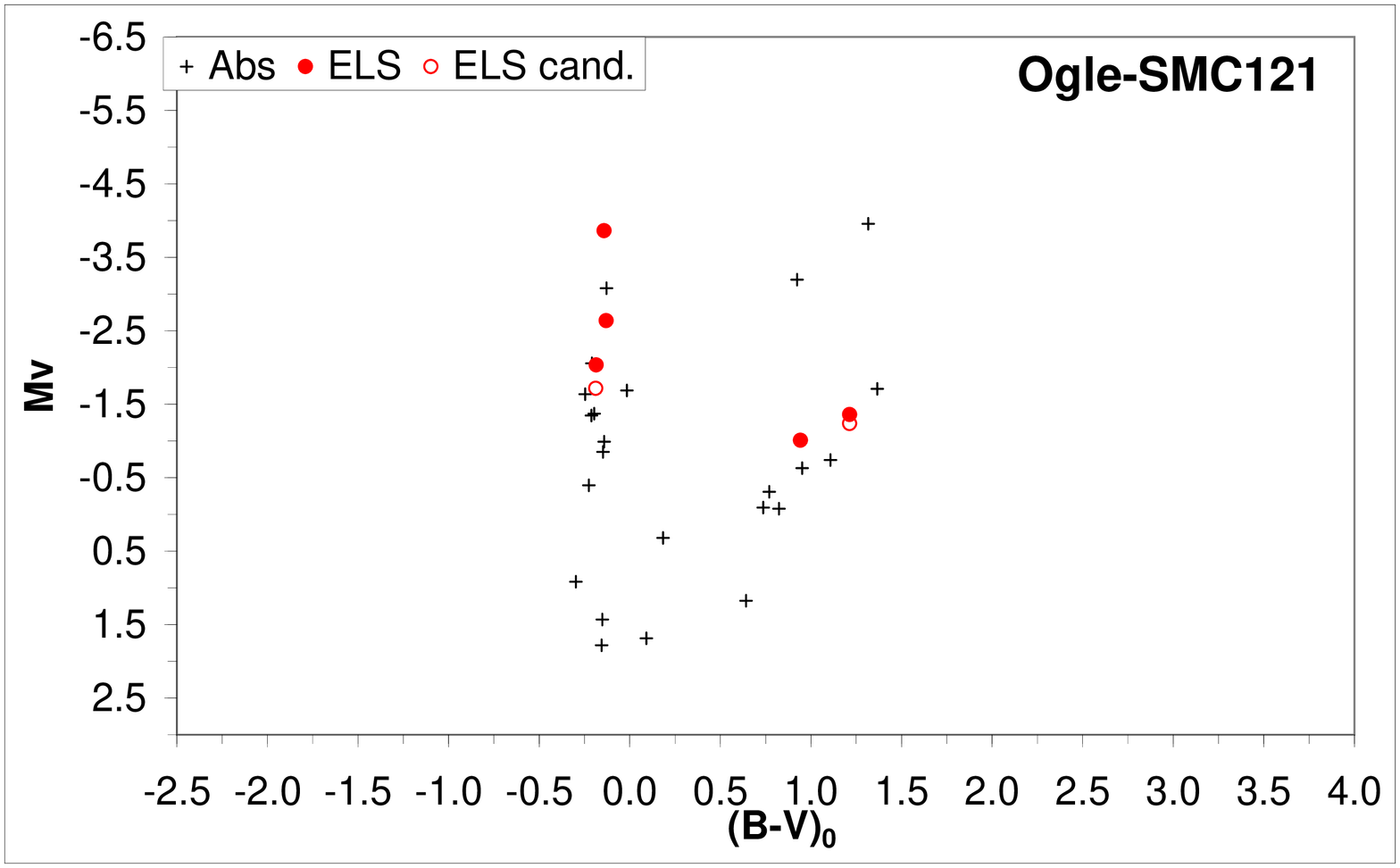} &  \includegraphics[angle=-90, width=8cm]{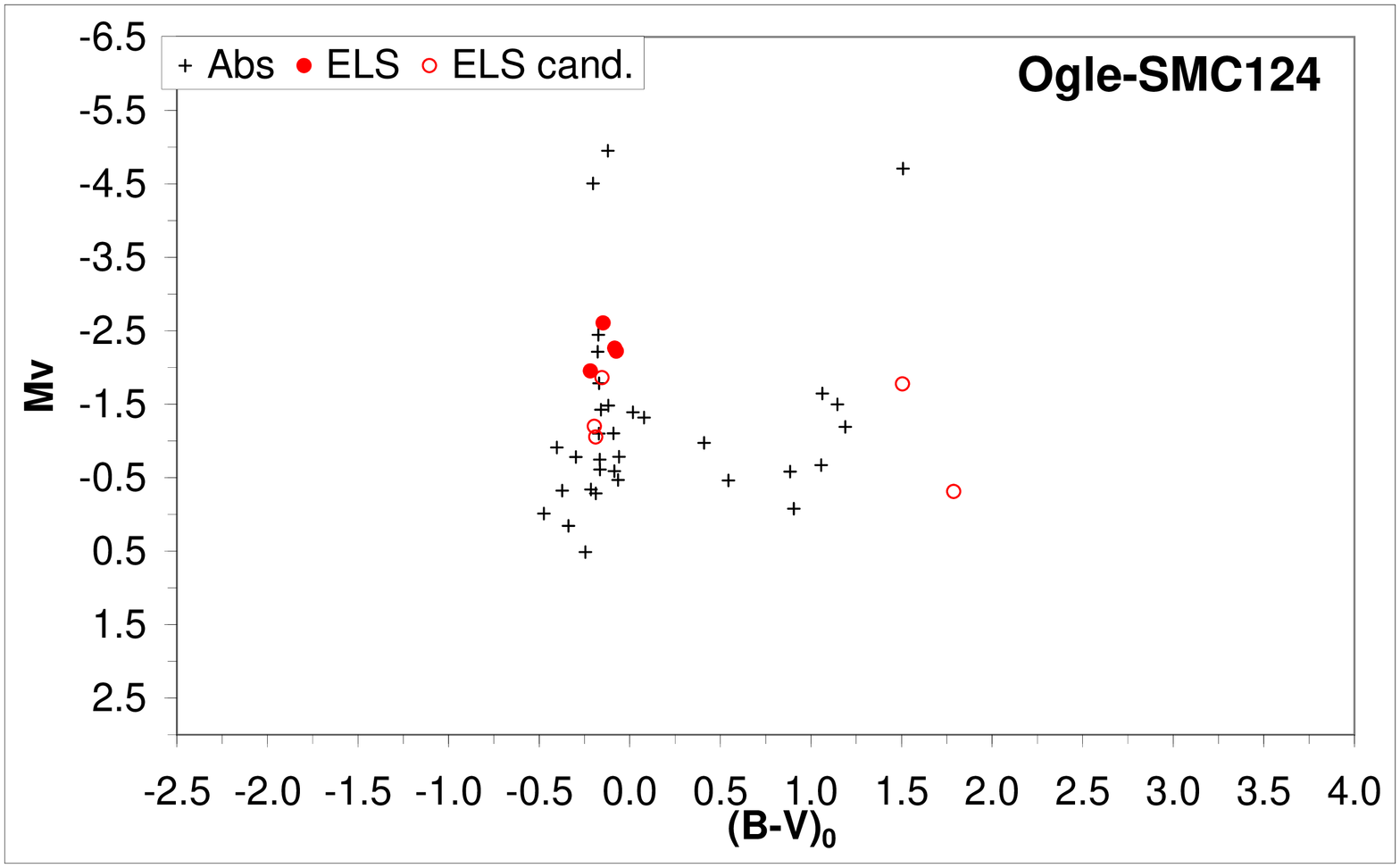} \\
    \includegraphics[angle=-90, width=8cm]{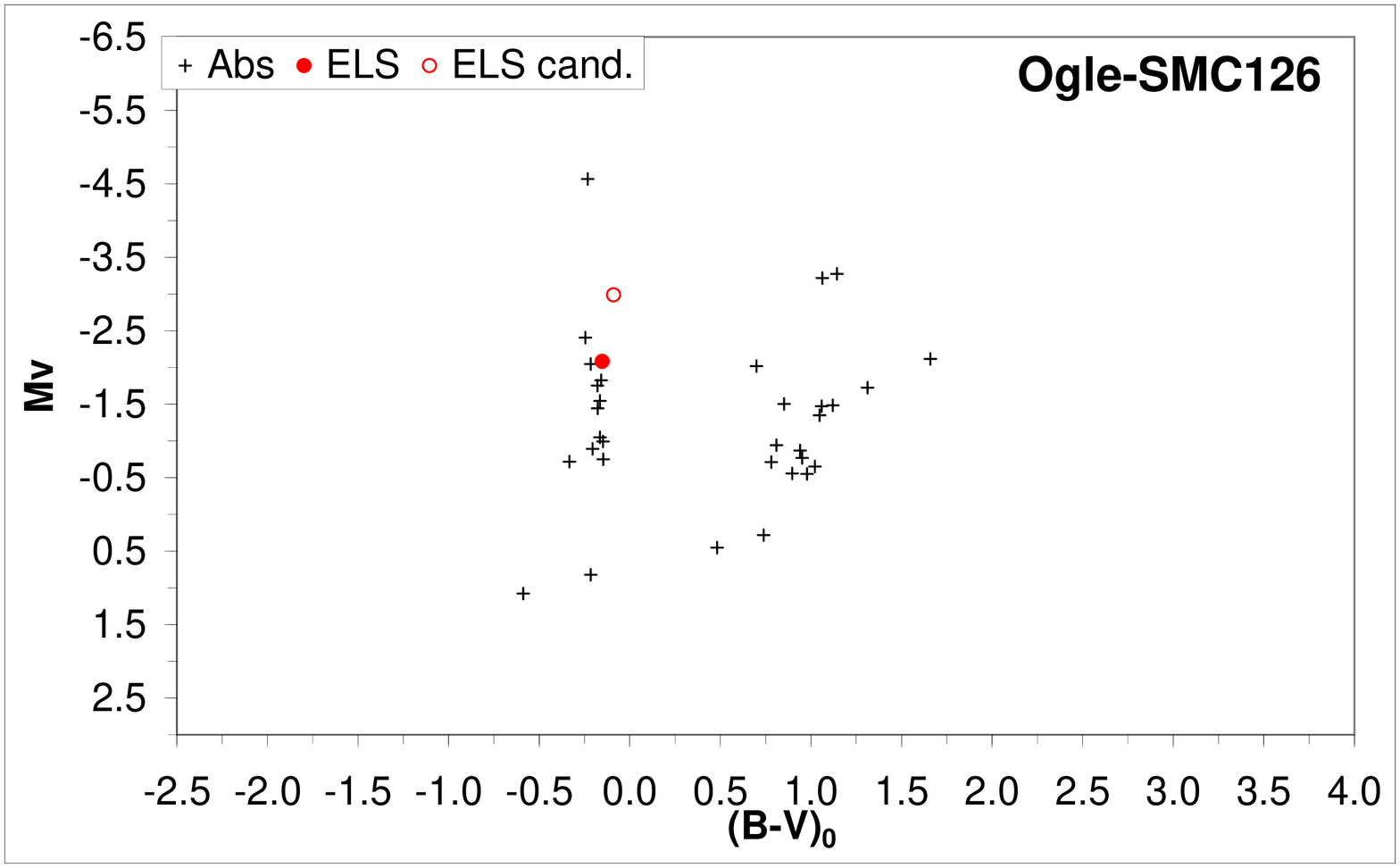} &  \includegraphics[angle=-90, width=8cm]{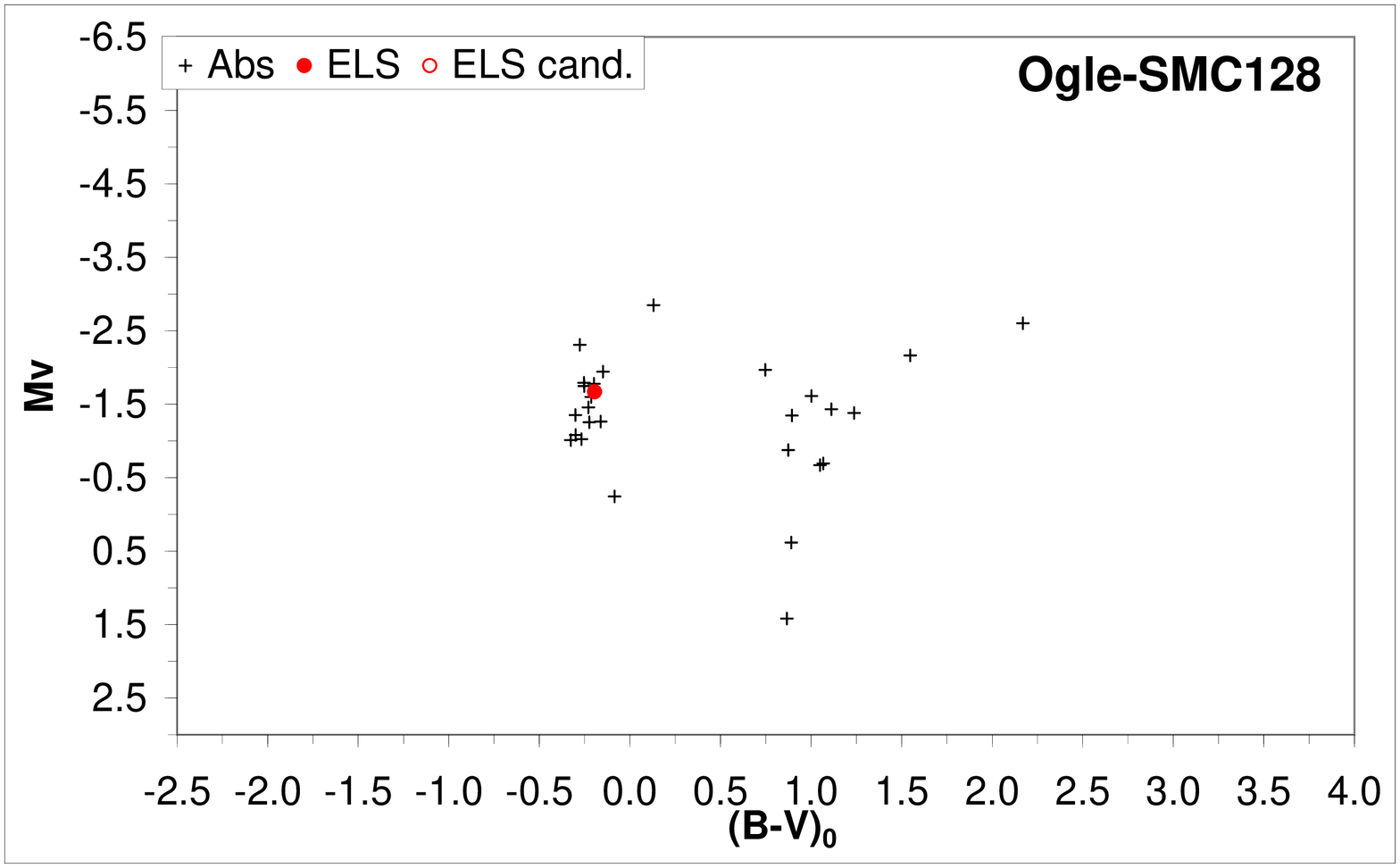} \\
\end{tabular}
\caption{Same caption as for fig.~\ref{HR1}}
\label{HR5}
\end{figure*} 

\begin{figure*}
\begin{tabular}{cc}
    \centering
    \includegraphics[angle=-90, width=8cm]{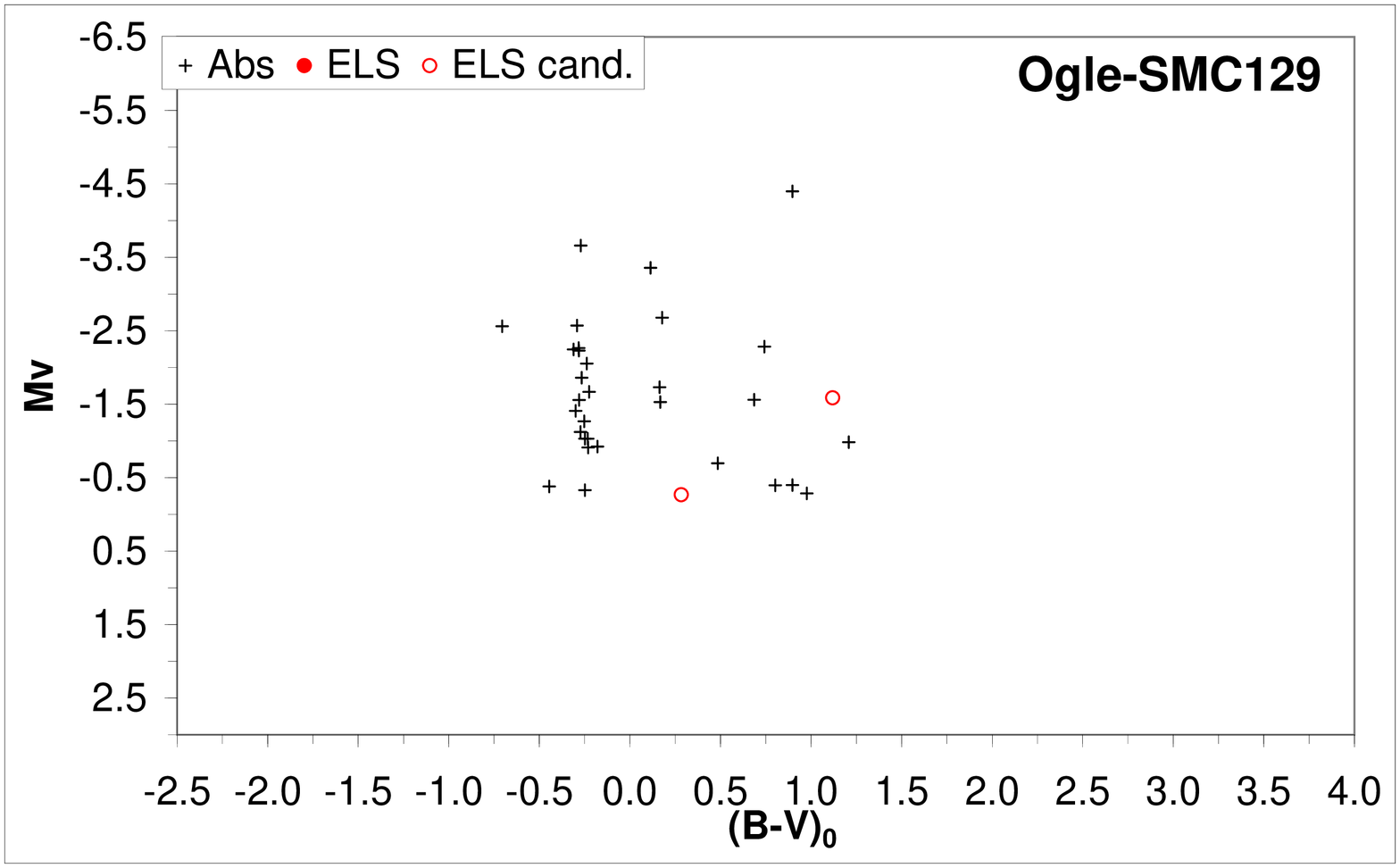} &  \includegraphics[angle=-90, width=8cm]{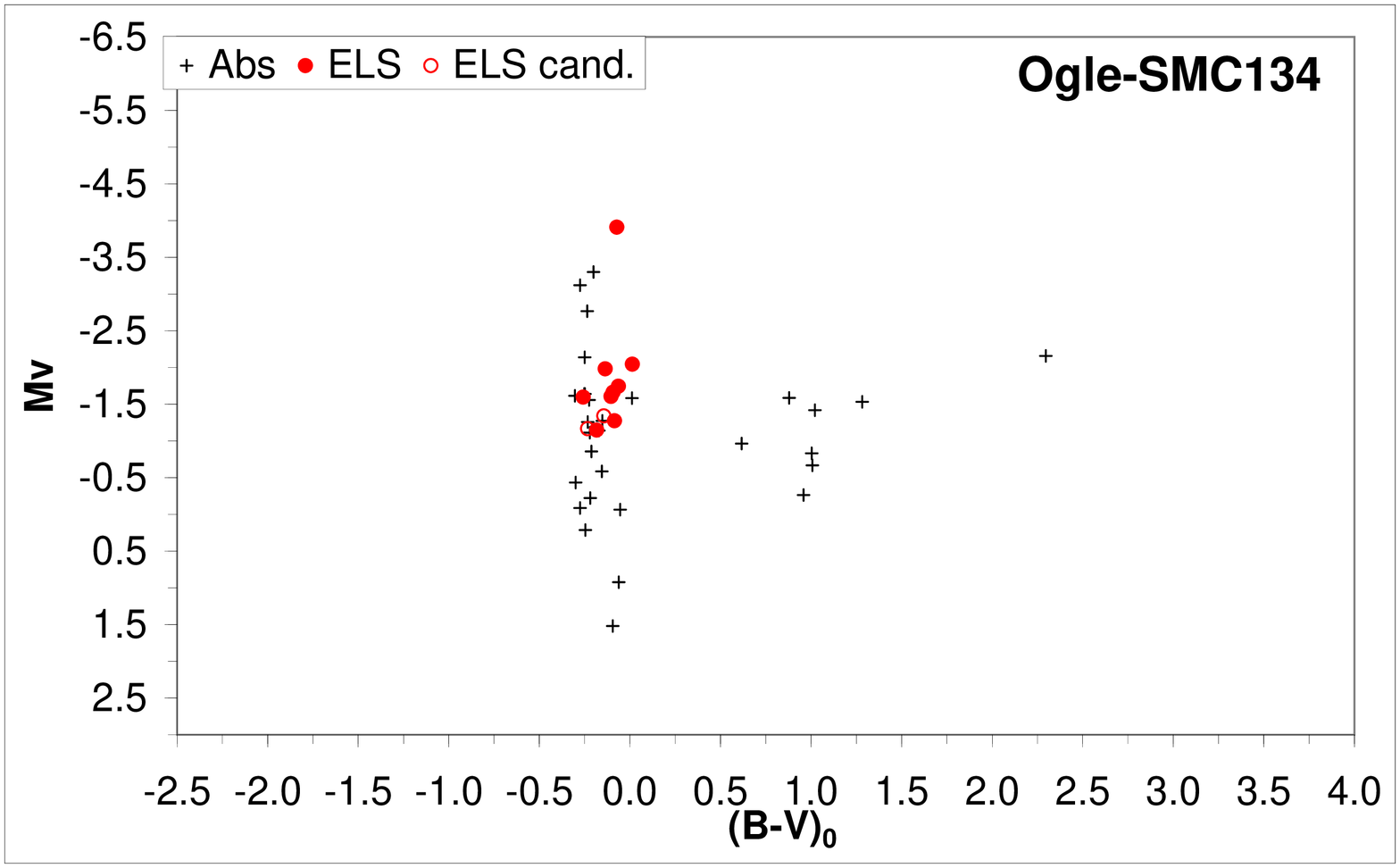} \\
    \includegraphics[angle=-90, width=8cm]{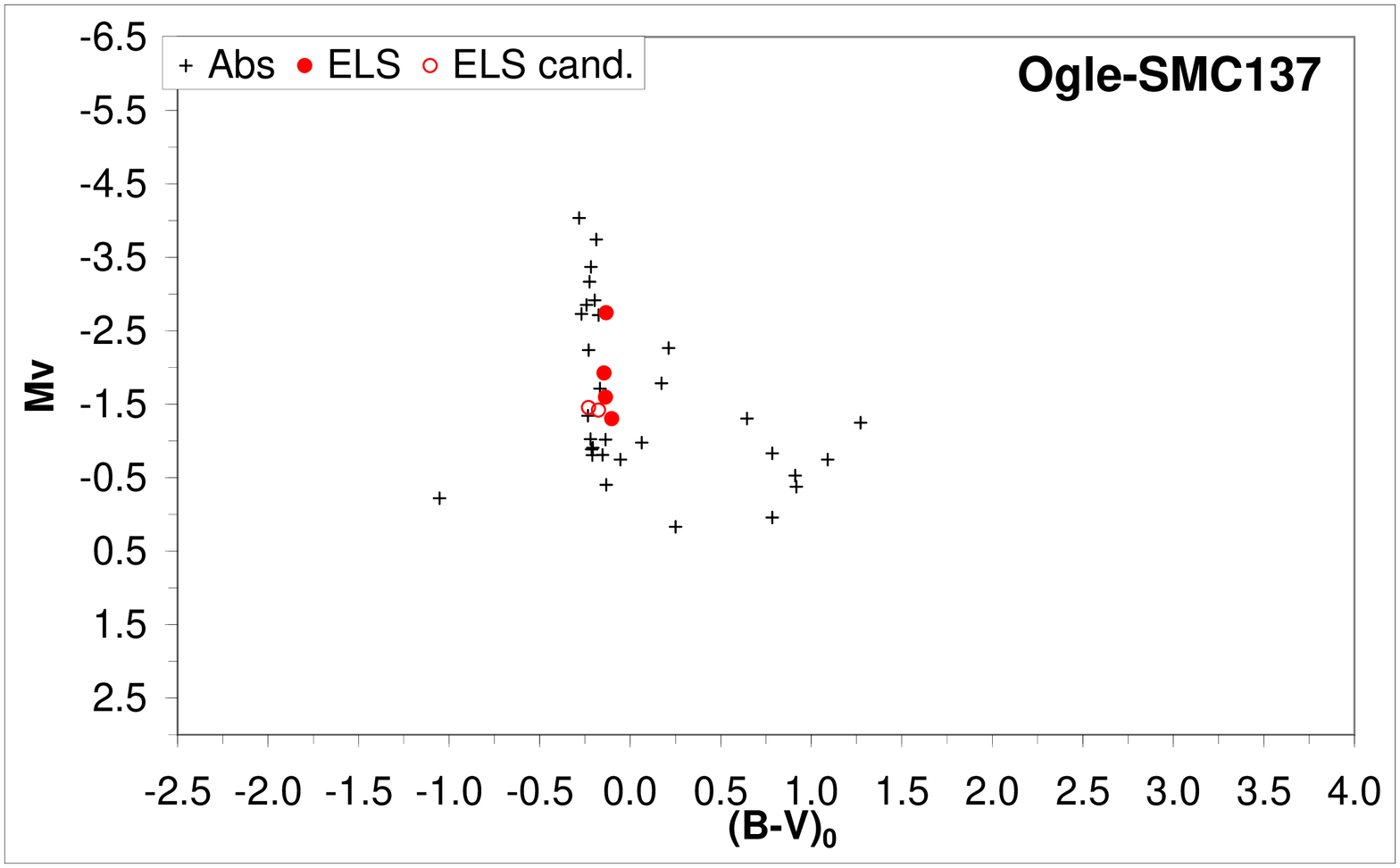} &  \includegraphics[angle=-90, width=8cm]{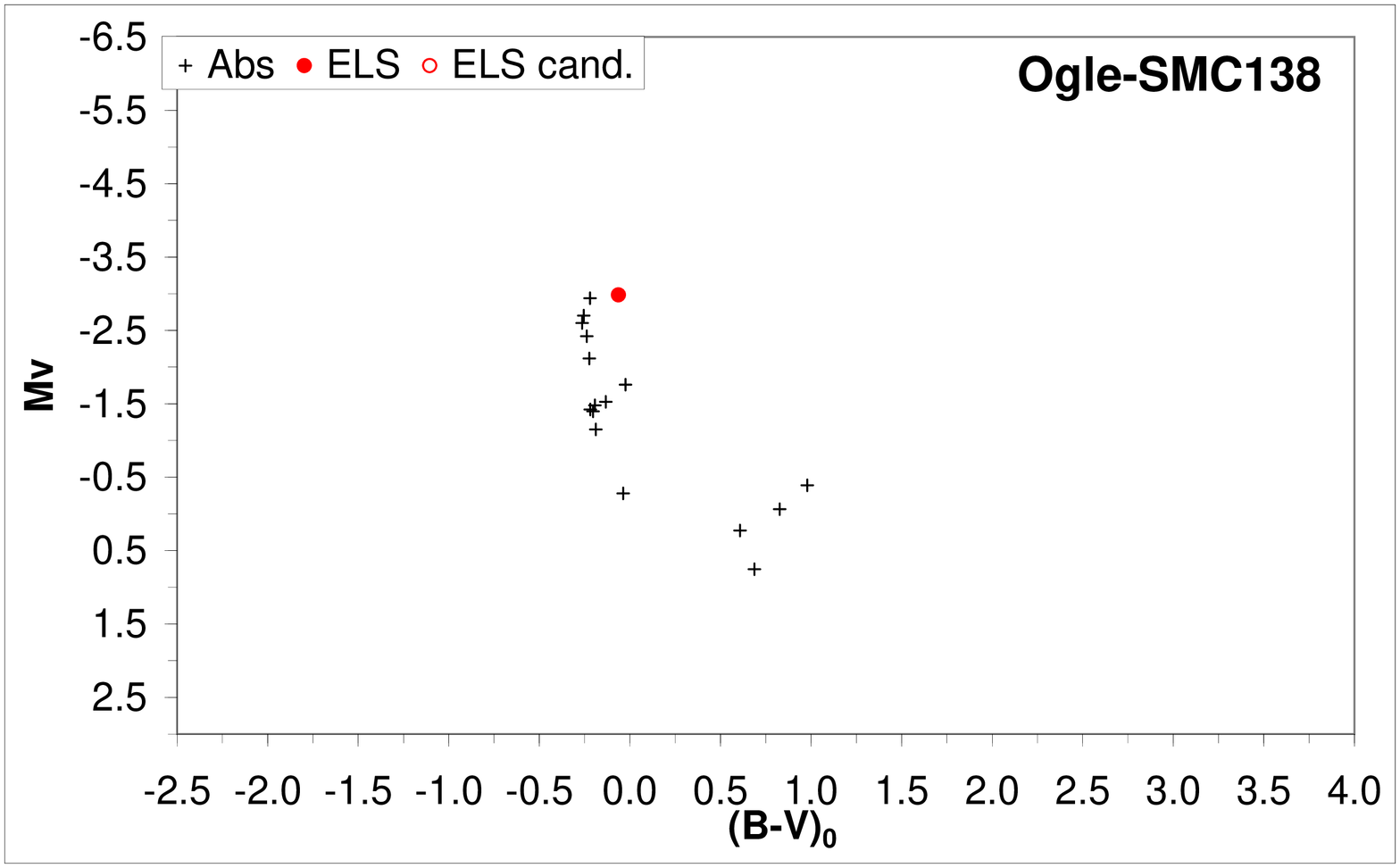} \\
    \includegraphics[angle=-90, width=8cm]{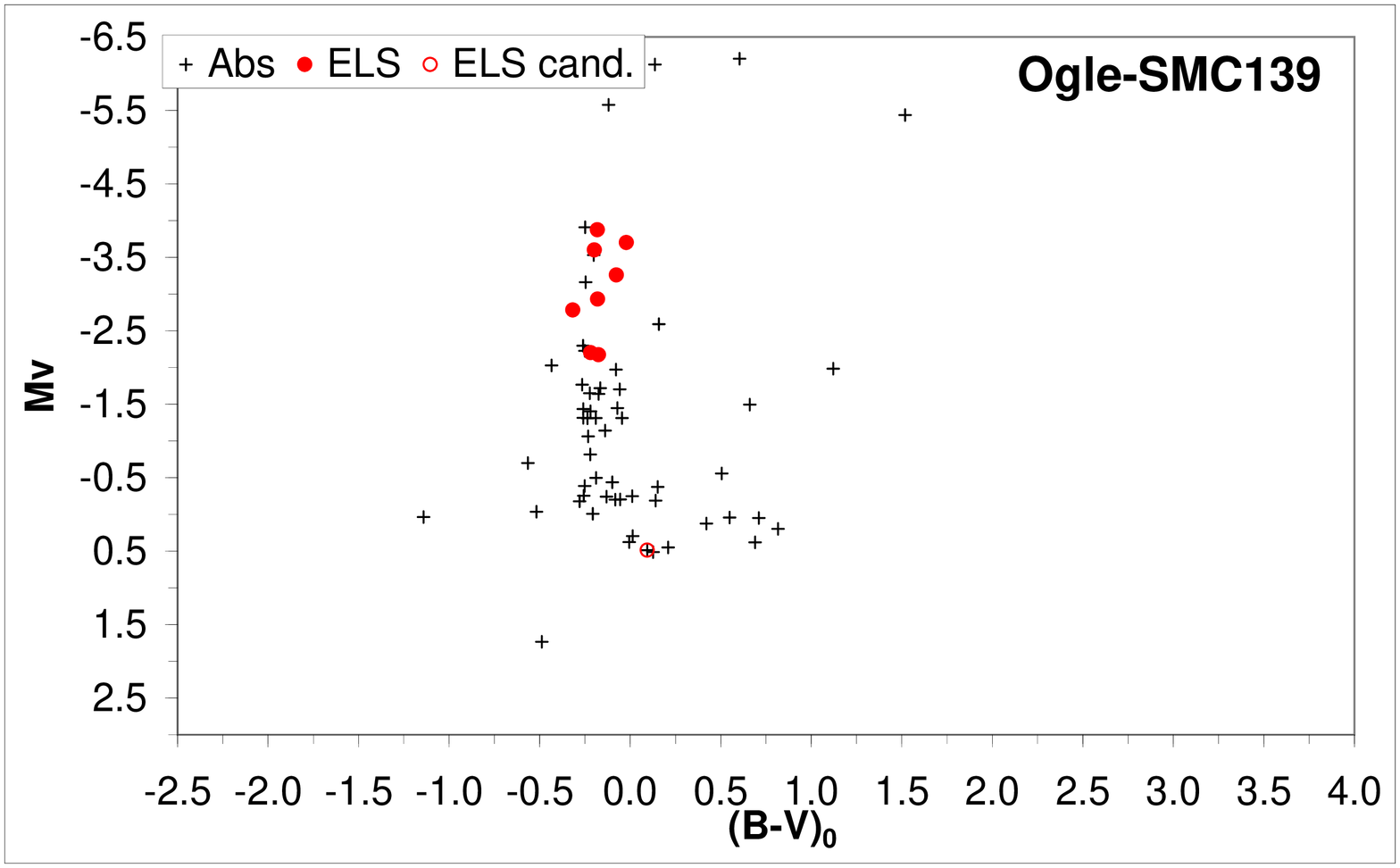} &  \includegraphics[angle=-90, width=8cm]{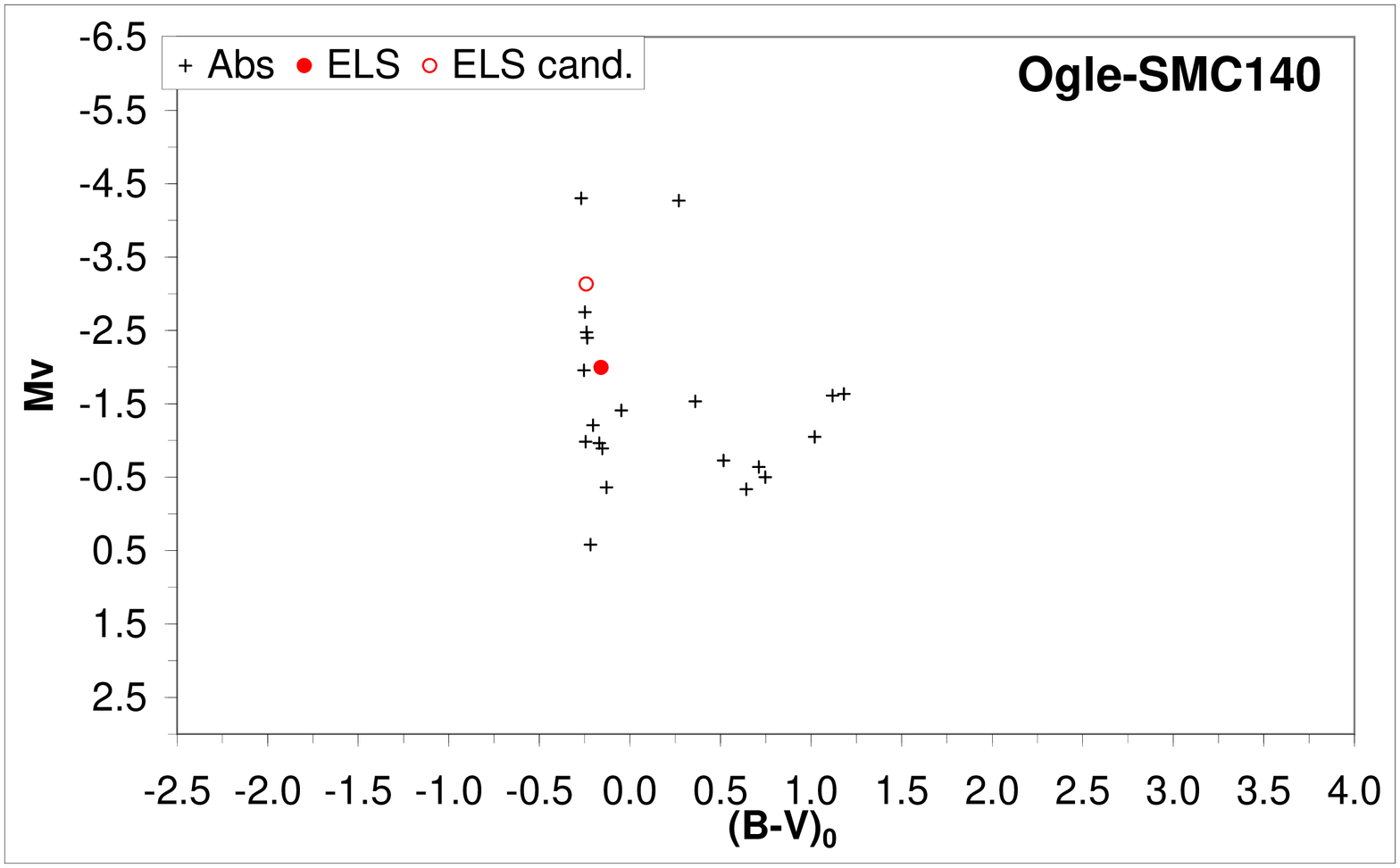} \\
    \includegraphics[angle=-90, width=8cm]{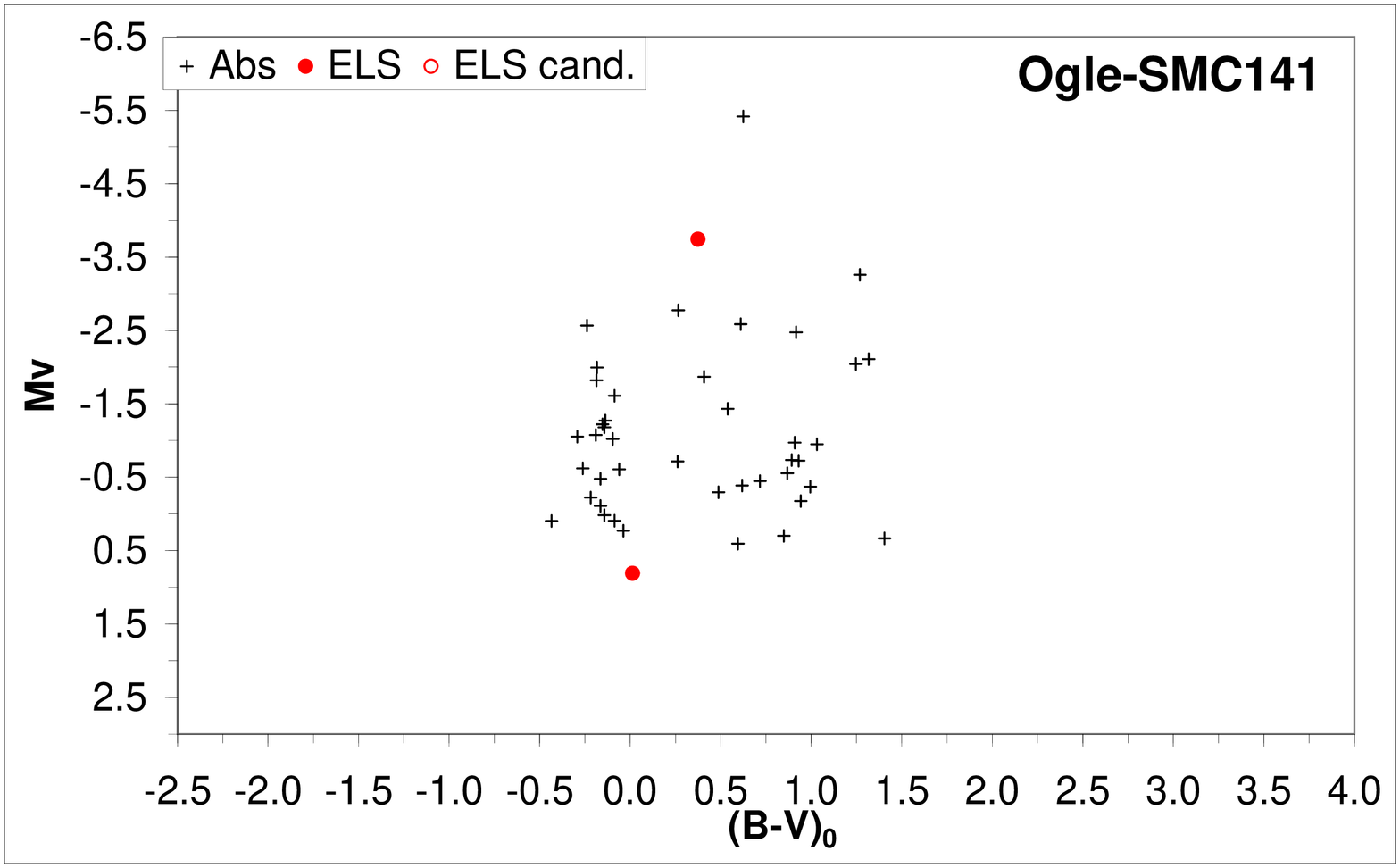} &  \includegraphics[angle=-90, width=8cm]{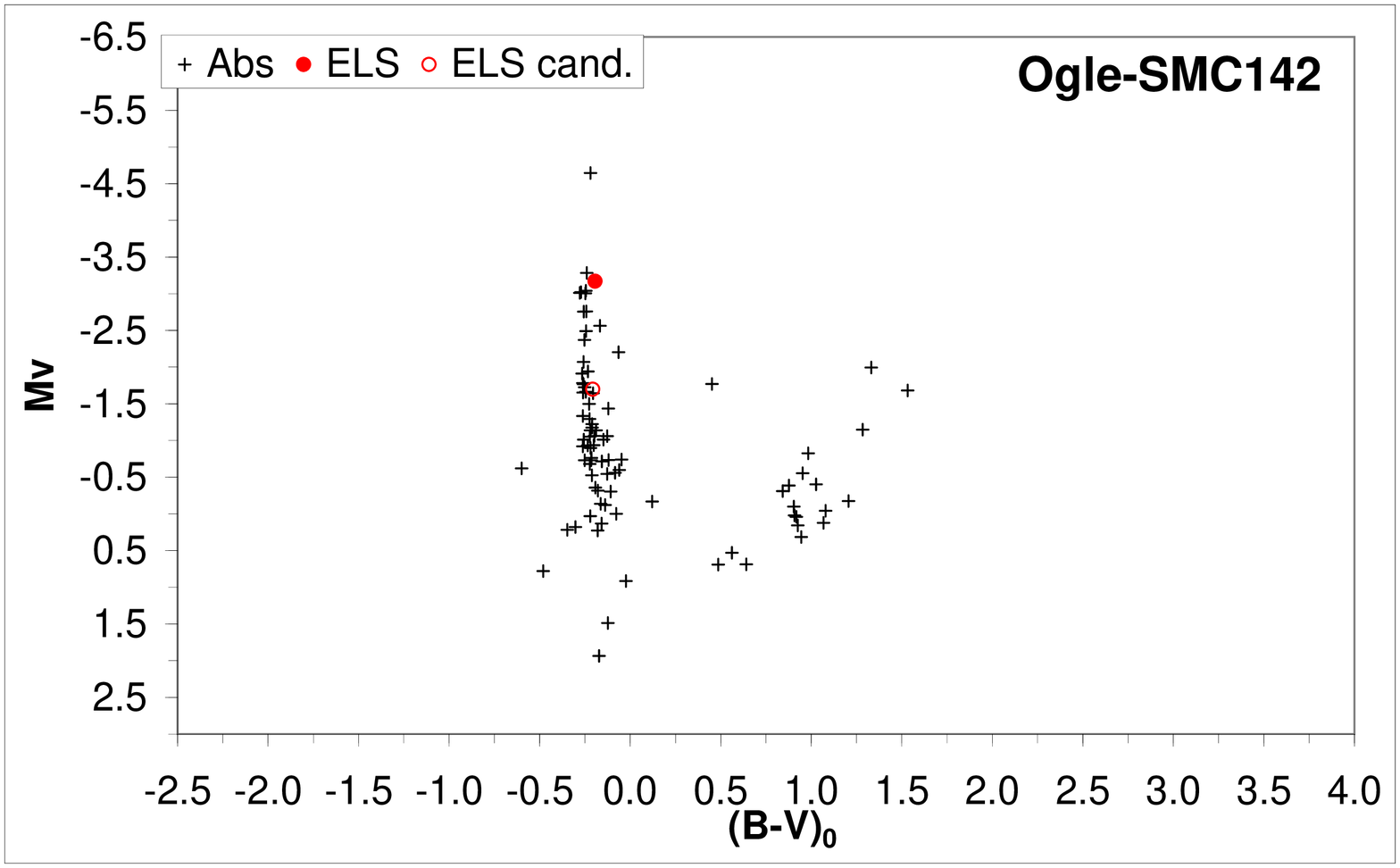} \\
\end{tabular}
\caption{Same caption as for fig.~\ref{HR1}}
\label{HR6}
\end{figure*}

\begin{figure*}
\begin{tabular}{cc}
    \centering
    \includegraphics[angle=-90, width=8cm]{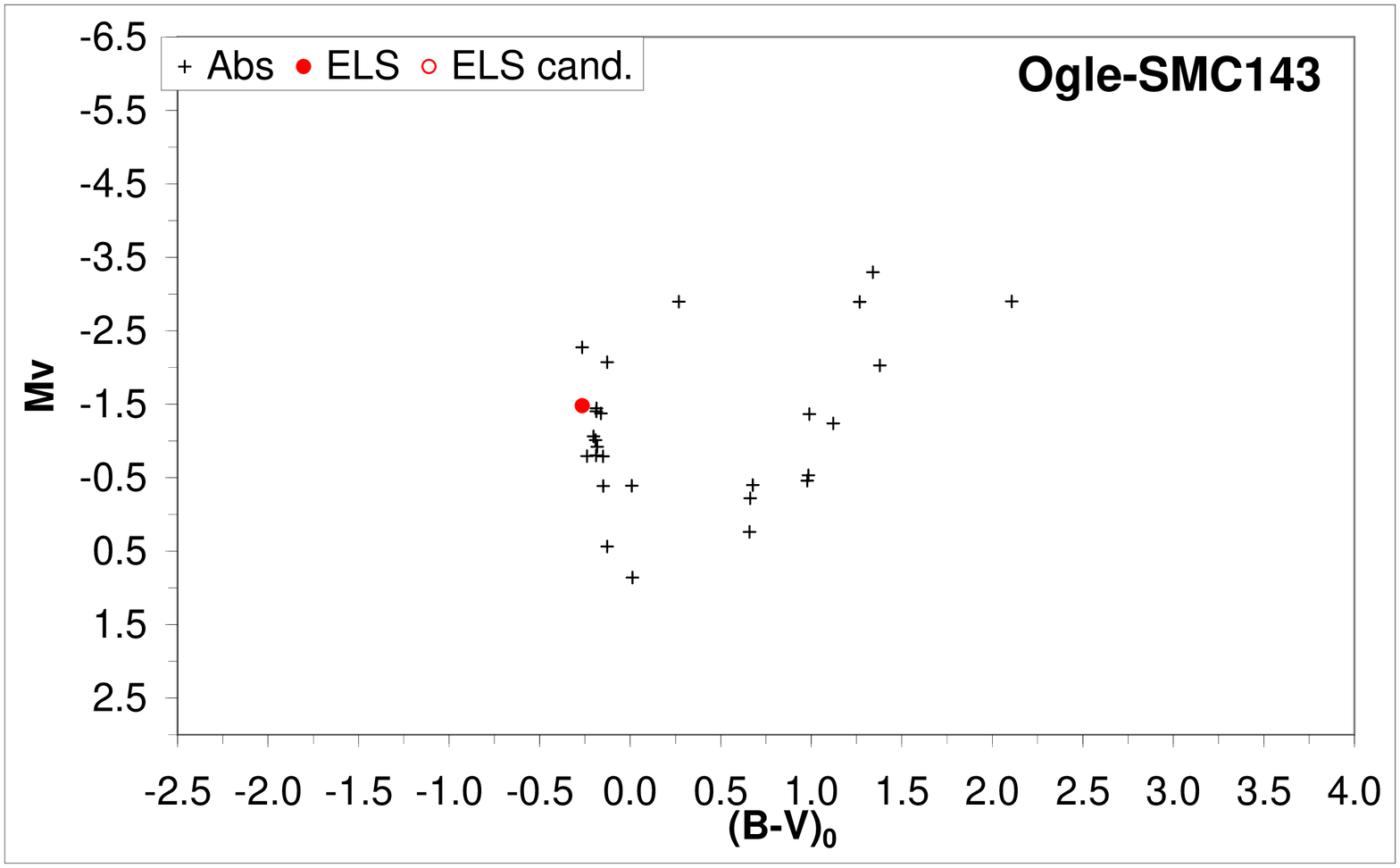} &  \includegraphics[angle=-90, width=8cm]{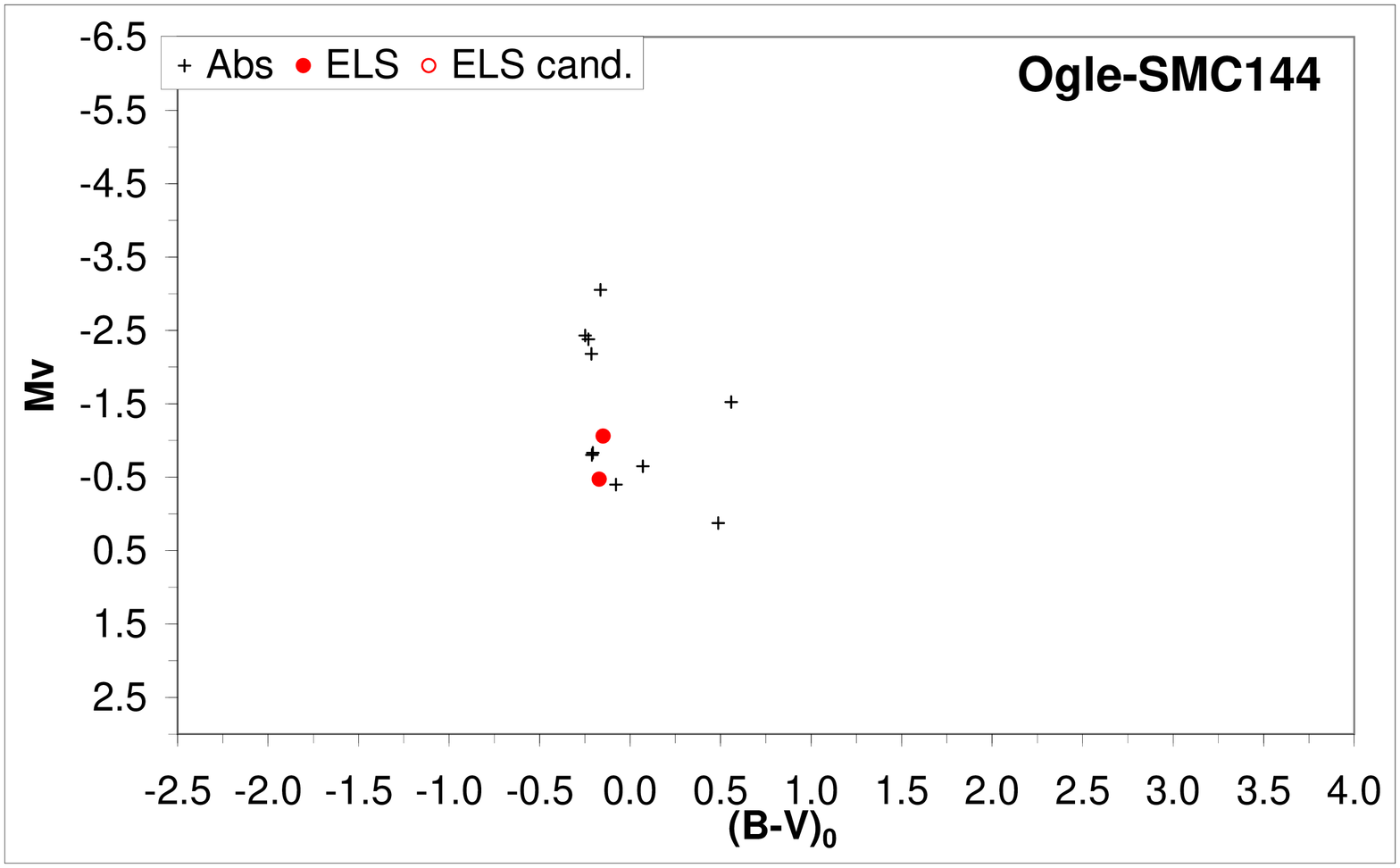} \\
    \includegraphics[angle=-90, width=8cm]{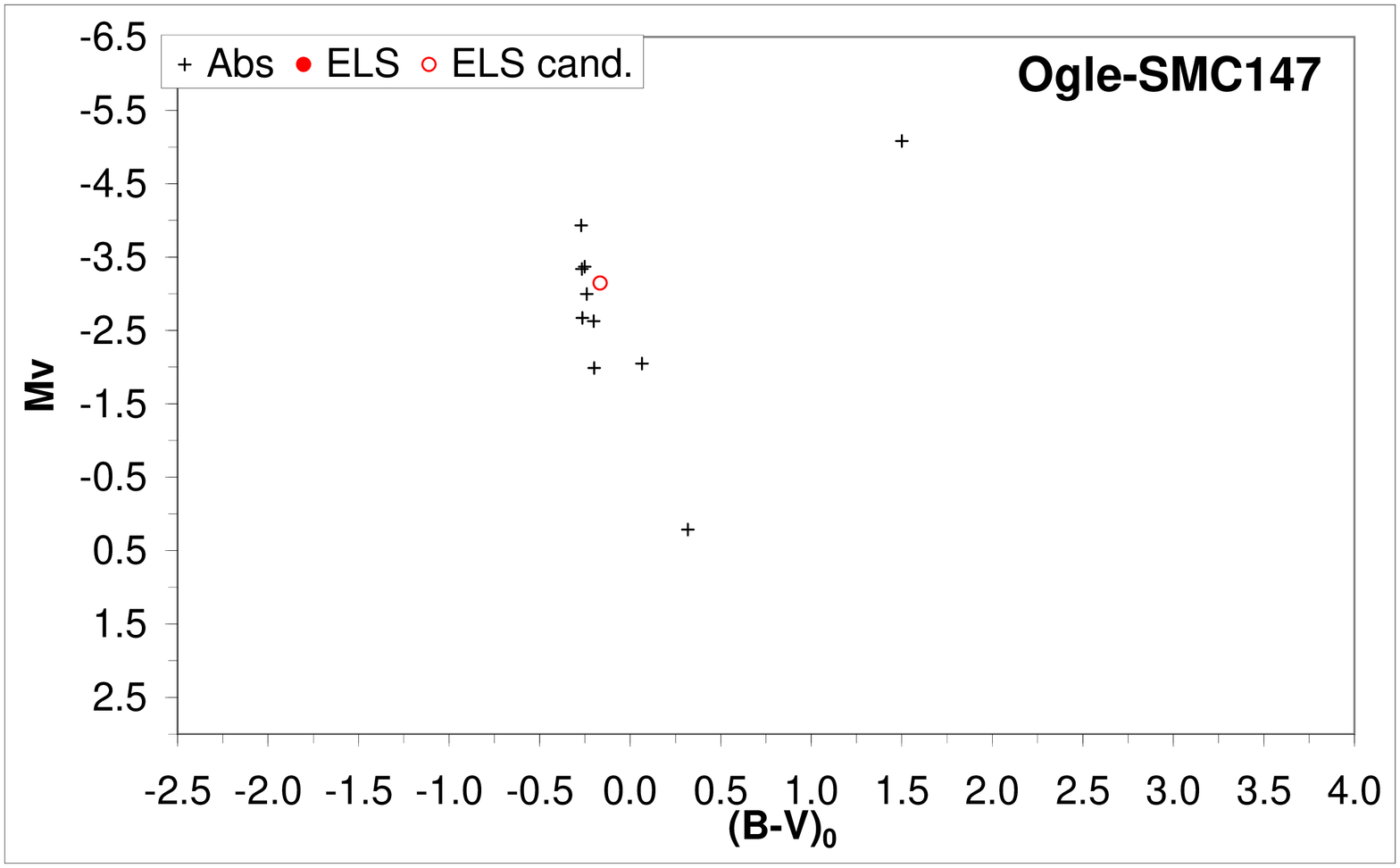} &  \includegraphics[angle=-90, width=8cm]{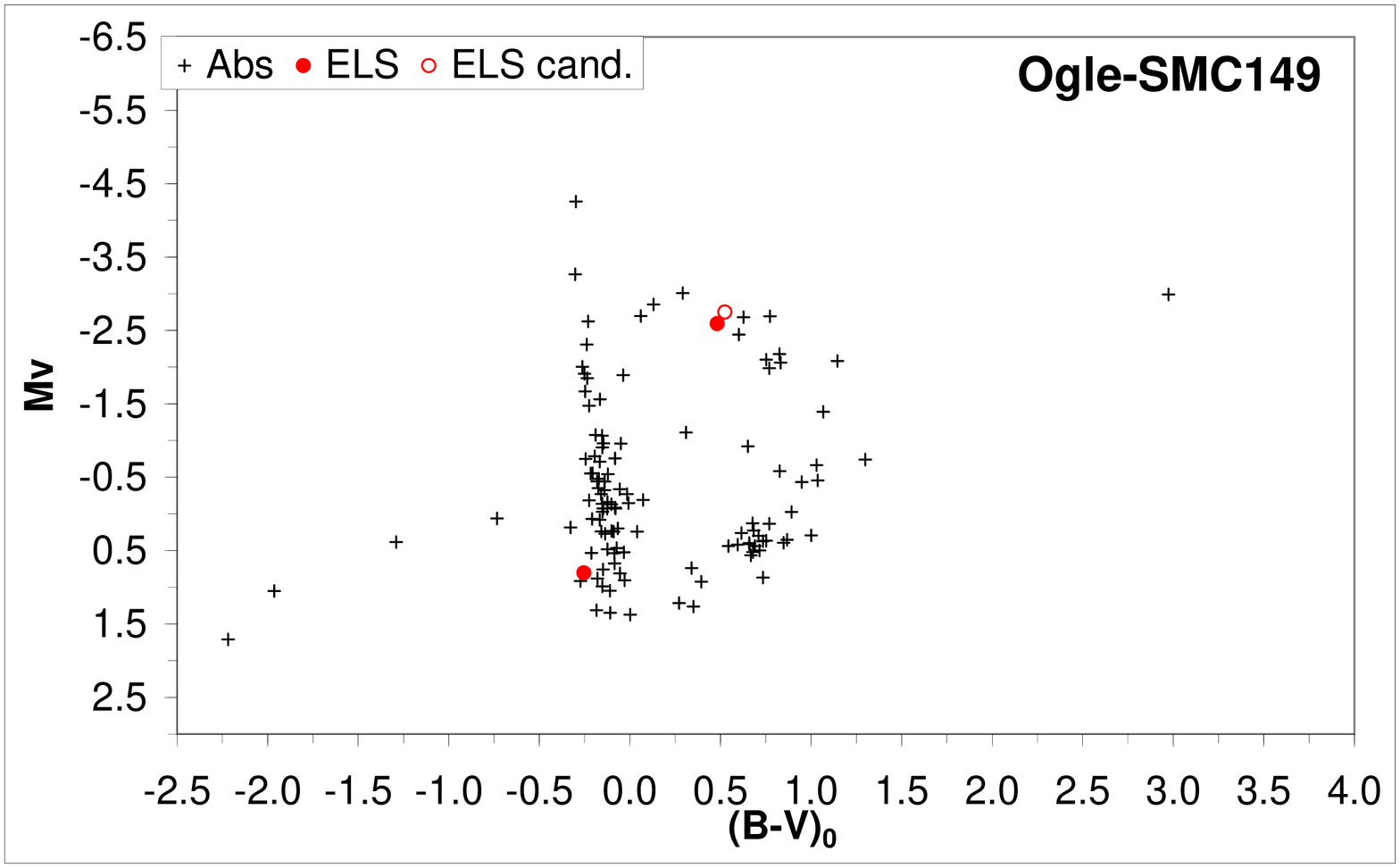} \\
    \includegraphics[angle=-90, width=8cm]{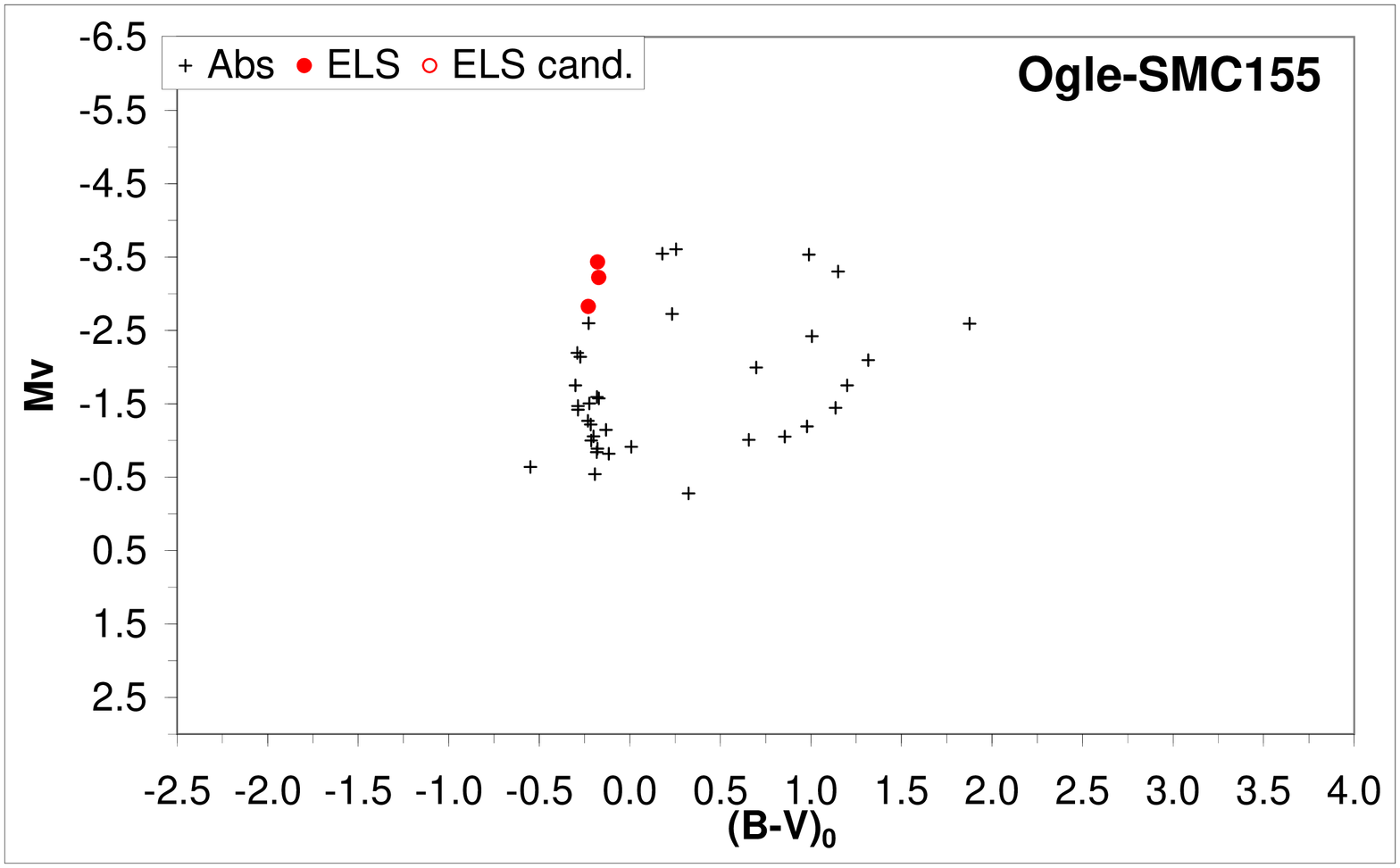} &  \includegraphics[angle=-90, width=8cm]{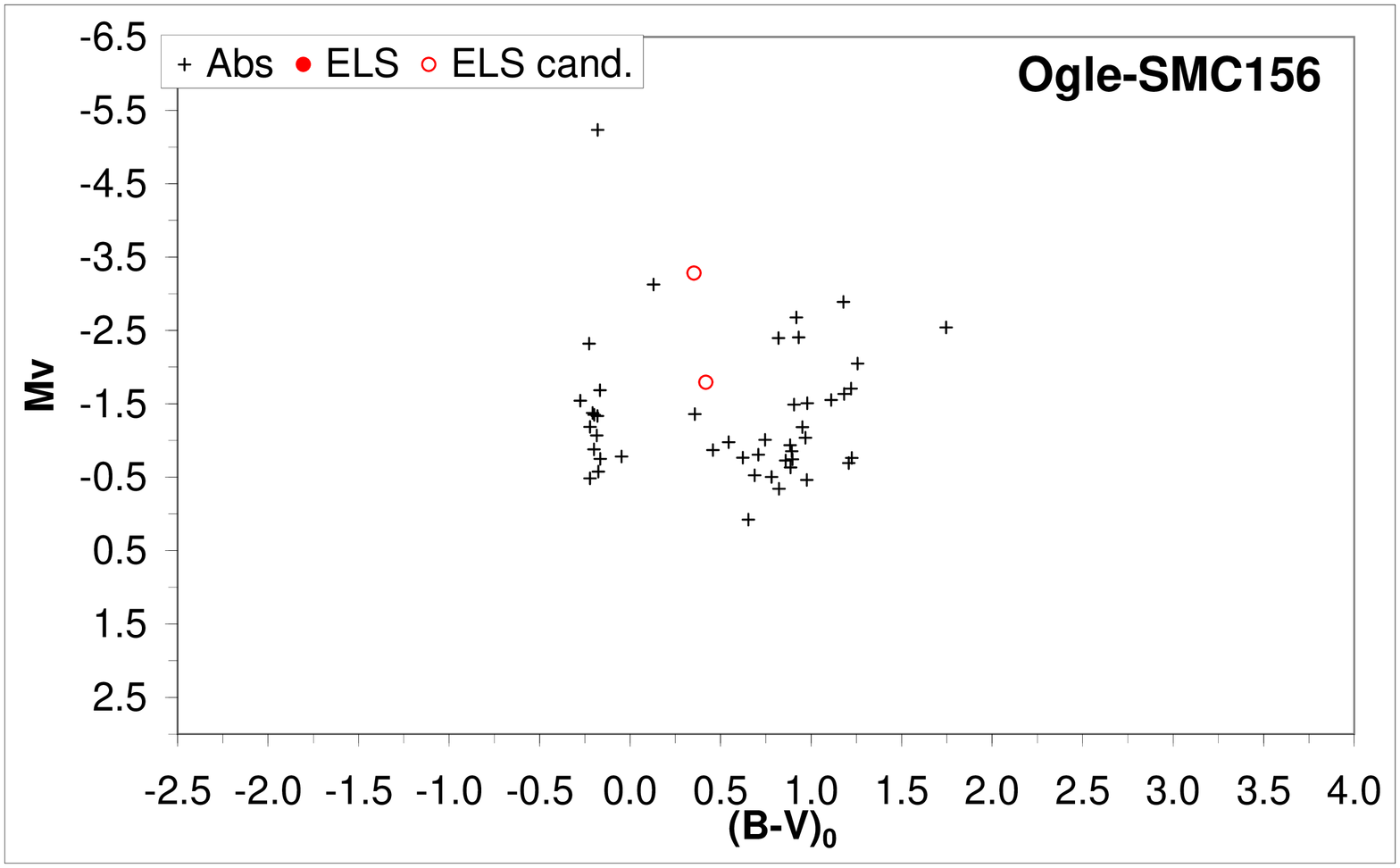} \\
\end{tabular}
\caption{Same caption as for fig.~\ref{HR1}}
\label{HR7}
\end{figure*} 

\begin{figure*}
\begin{tabular}{cc}
    \centering
    \includegraphics[angle=-90, width=8cm]{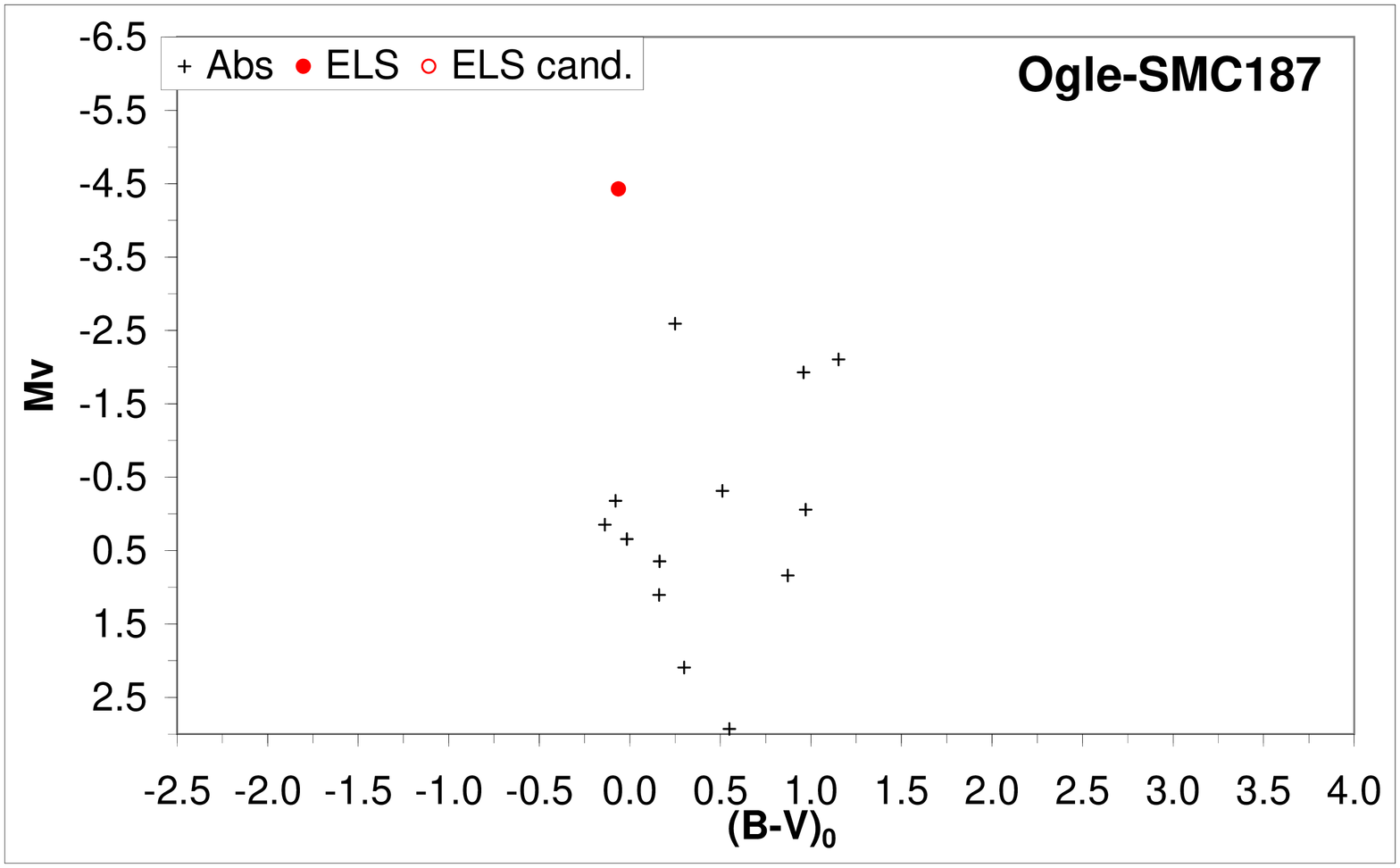} &  \includegraphics[angle=-90, width=8cm]{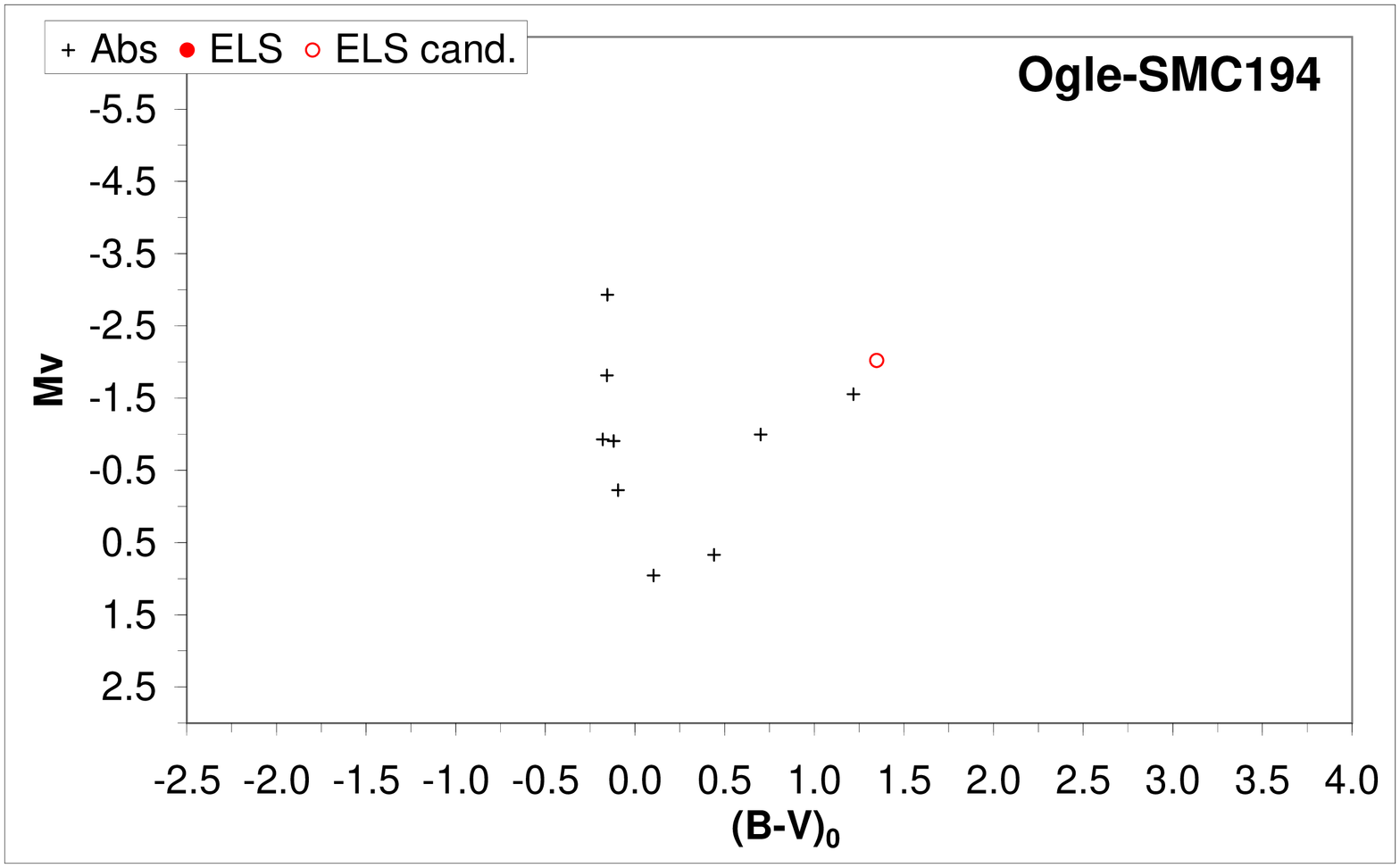} \\
    \includegraphics[angle=-90, width=8cm]{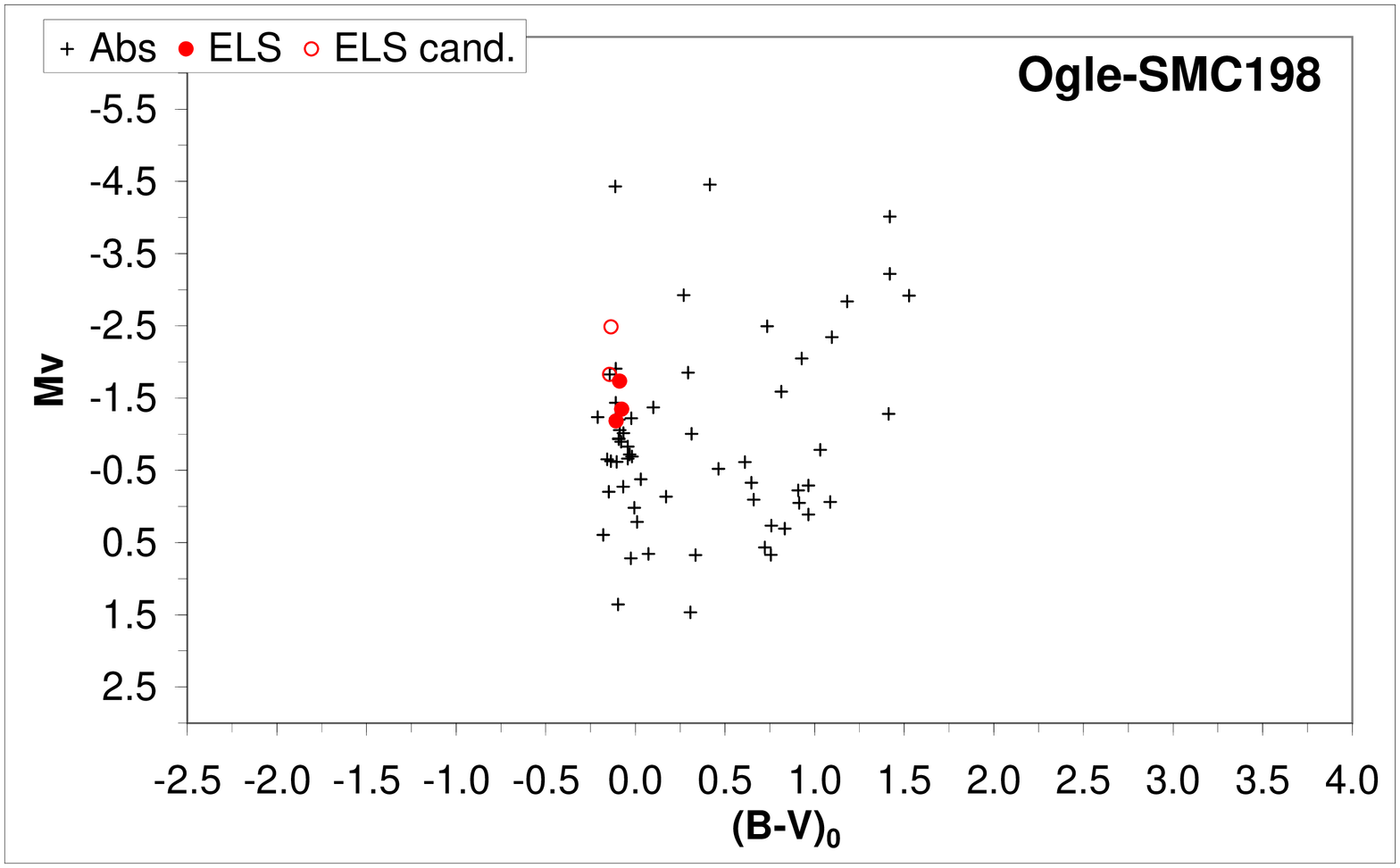} &   \\
\end{tabular}
\caption{Same caption as for fig.~\ref{HR1}}
\label{HR8}
\end{figure*}

\end{document}